# Technical Report

## Elements of style of BPMN language

Anacleto Correia[1,2]

[1] *CITI/Faculdade de Ciências e Tecnologia/Universidade Nova de Lisboa, Lisboa*

[2] *Escola Naval/CINAV*

anacleto.correia@gmail.com

## Abstract

Several BPMN graphical tools support, at least partly, the OMG's BPMN specification. The BPMN standard is an essential guide for tools' makers when implementing the rules regarding depiction of BPMN diagrammatic constructs. Process modelers should also know how to rigorously use BPMN constructs when depicting business processes either for business or IT purposes.

Several already published OMG's standards include the formal specification of well-formedness rules concerning the metamodels they address. However, the BPMN standard does not. Instead, the rules regarding BPMN elements are only informally specified in natural language throughout the overall BPMN documentation. Without strict rules concerning the correct usage of BPMN elements, no wonder that plenty of available BPMN tools fail to enforce BPMN process models' correctness.

To mitigate this problem, and therefore contribute for achieving BPMN models' correctness, we propose to supplement the BPMN metamodel with well-formedness rules expressed by OCL invariants. So, this document contributes to bring together a set of requirements that tools' makers must comply with, in order to claim a broader BPMN 2 compliance. For the regular process modeler, this report provides an extensive and pragmatic catalog of BPMN elements' usage, to be followed in order to attain correct BPMN process models.

## Resumo

Numerosas ferramentas de modelação BPMN suportam, pelo menos parcialmente, a especificação BPMN definida pela OMG. O estandarte BPMN é um guia essencial para fabricantes de ferramentas de modelação. Os modeladores de processos necessitam também de saber, como usar de forma rigorosa os elementos da notação, para representar processos relacionados com o negócio ou a infraestrutura tecnológica.

Diversos estandartes publicados pela OMG incluem a especificação formal de regras de boa-formação associadas ao respetivo metamodelo. Contudo, com o BPMN tal não acontece. As regras relativas aos elementos do BPMN estão apenas informalmente especificadas, em linguagem natural, ao longo do documento de especificação BPMN. Sem regras estritas, relativas à correta utilização do BPMN, não admira que muitas das ferramentas BPMN disponíveis falhem na verificação da correção dos modelos de processos.

Para obviar este problema, e assim contribuir para a obtenção de modelos BPMN corretos, propomos complementar o metamodelo BPMN, com regras de boa-formação expressas através de invariantes OCL. Desta forma, este trabalho contribui para sistematizar um conjunto de requisitos que os fabricantes devem assegurar que as suas ferramentas cumpram, de forma a poderem reivindicar a conformidade com o BPMN 2. Para o utilizador regular do BPMN, como linguagem de modelação, este relatório fornece um extenso catálogo prático, que mostra como utilizar os elementos da notação BPMN, de forma a obter modelos de processos corretos.

## Table of contents

**List of code snippets**

**List of figures**

**Acronyms**

| | |
|---|---|
| AD | Activity Diagram |
| ASM | Abstract State Machines |
| BASEL III | International regulatory framework for banks |
| BPEL | Business Process Execution Language |
| BPMN | Business Process Model and Notation |
| BPMS | Business Process Management System |
| BWW | Bunge-Wand-Weber ontology |
| CSP | Communicating Sequential Processes |
| CWM | The Common Warehouse Metamodel |
| MOF | Meta-Object Facility |
| OCL | Object Constraint Language |
| OMG | Object Management Group |
| OWL | Web Ontology Language |
| P/N | Petri Nets |
| QVT | Query/View/Transformation |
| SOX | Sarbanes–Oxley Act |
| UML | Unified Modeling Language |
| USE | UML based Specification Environment |
| XMI | XML Metadata Interchange |
| XML | Extensible Markup Language |
| XPDL | XML Process Definition Language |

# 1. Introduction

The BPMN 2 standard [1], published by the Object Management Group (OMG), has as one its main purposes to provide a notation understandable by different kinds of process modelers and users: (1) process analysts that sketch the initial documentation of business processes; (2) process implementers which are responsible for actually implementing business processes; (3) business users which are accountable for business processes' instantiation and monitoring.

The requirement for accurate specification of process models is relevant for several activities in the organization, namely when delivering processes' specifications to meet regulatory and legal conditions (e.g. SOX [2], BASEL III [3]), as well as for analysis, design and development of process-aware information systems [4], service-oriented architectures [5], and web services alike systems [6].

Although BPMN can be used for all these purposes, the business and technical models produced are quite different in nature. The main focus of BPMN business models, for documentation purposes, is on comprehension of basic process flow. Hence, the emphasis on the *happy path*, avoiding depicting excessive details [7]. Exception handling and abnormal situations are often bypassed. On other hand, execution capabilities of the BPMN language are relevant when technical models are produced. Process developers require translation of BPMN models into machine readable languages either for models' sharing across multiple domains, using different technologies, as well as enactment in distributed environment (e.g. integrating BPEL and web services standards). So, BPMN as a process modeling language joins different levels of abstractions. These perspectives encompass different sorts of constructs either for models' static graphical representation, as properties for providing information for processes' simulation and execution. To harness the potential of the BPMN standard, the different levels of abstraction of the process life cycle must be tighten up. Ensuring the compliance of models, made in primary stages of process modeling, enforces the correctness of those process models and therefore their reuse on enactment and monitoring.

Usually, the metamodels made available in OMG standards (e.g. UML2 [8], CWM [9], QVT [10]) are anchored on the Meta-Object Facility (MOF), a meta-metamodel framework (M3)[1] that uses a class diagram notation to depict the metaclass diagram (M2). However, sometimes associations and other structural constraints from UML class diagram - such as multiplicity and mandatory constraints of associations, as well as the type of relationships (e.g. aggregation, composition or inheritance) - are not enough to convey static semantics rules [11]. These are more complex rules that add restrictions to the structure of models, hard or impossible to express using graphical elements of class diagrams. These rules, also dubbed as "well-formedness rules", contribute significantly for enhancing correctness of models (M1) instantiated from a metamodel. That's why several OMG metamodels, are supplemented with additional OCL (Object Constraint Language) [12] clauses, formalizing static semantic rules. OCL is a precise textual language which provides constraint and object query expressions to the metamodel level (M2) that cannot otherwise be expressed by the graphical notation.

The BPMN standard describes the constructs of a process modeling language, and allowed relationships among those constructs, by means of a metamodel specification (M2) [13]. However, throughout the document, the modeling rules are described only in natural language. This raises problems for BPMN tools' implementers and regular process modelers. Tools' implementers, besides having to decrypt the rules, must translate those rules to a computable format to provide effective business process model validation. On the other hand, to apply modeling rules regarding constructs, process modelers must also interpret several hundred pages of the standard.

Furthermore, if process modelers are free to combine the large plethora of BPMN constructs available, without of a verification or recommendation facility embedded in the modeling tool, inconsistent or invalid models can be easily generated. So, the great amount of BPMN elements may lead non savvy process modelers producing faulty process models. Moreover, it is not straightforward that simple observation of process model could allow users to detect inconsistencies or well-formedness errors (e.g. a split element does not have a corresponding join element).

The main contribution of this work is on gathering an extensive catalog of BPMN well-formedness rules, visually illustrated and formally defined as OCL clauses. The catalog is intended either for tools' makers and process modelers: the former can take it as a requirements specification for a BPMN tool; the latter can refer to it as a best-practices guide for correct usage of BPMN modeling elements.

This report is structured as follows: the current section introduces the problem and motivation of this report. Section 2 analyze critically BPMN and its recent evolution to become the most widely used process modeling language. Section 3 presents a literature review highlighting previous contributions, on the central topic of this report: the compliance with BPMN modeling rules. The BPMN metamodel is overviewed in section 4 and the constructs of the language used in the process orchestration are described. Based on the abstract syntax of BPMN, defined as OCL clauses, a set of well-formedness rules are derived to supplement the BPMN metamodel (section 5). Also in section

[1] MOF is a four-layered architecture. It provides a meta-meta model at the top layer, the M3 layer. This M3-model is the language to build metamodels: the M2 models. The M2-models describe elements of the M1-layer, which are M1-models. The last layer is the M0-layer which describes real-world objects.

5, model snippets depicting the well-formedness rules are presented and collated with models misusing the rules. The verification process of proposed rules correctness is described in section 0. A survey is presented in section 7 to evaluate current BPMN tools support of modeling rules conform to the standard. Finally, section 0 concludes this report and provide directions for future work.

## 2. BPMN analysis

From BPMN 1.2 onwards (see Fig. 1), the number of changes to the BPMN standard increased substantially. Updates made include the addition of new constructs, and changes on elements' properties to fix inconsistencies and ambiguities [14]. As result, BPMN 2 specification was acknowledged as a step forward regarding the extension on scope and expressiveness of the business process modeling language. As previously referred, one of the relevant upgrades was the syntax of BPMN have been formalized through a metamodel representing the language's constructs and their relationships.

According Recker et al., BPMN as the most recent process modeling language is ontologically one of the most complete for process modeling [15, 16]. While, for instance, a UML Activity Diagram (AD) can be built from around 20 different modeling constructs [17], an orchestration BPMN process model has almost 100 different modeling constructs, including among others 5 types of sub-processes, 9 task types, 6 gateway types, 3 activity markers, 5 data types, 3 sequence flow types (see Fig. 3), and 51 event types (see Fig. 4). This gave to the language great expressiveness both for diagramming and enactment purposes [14].

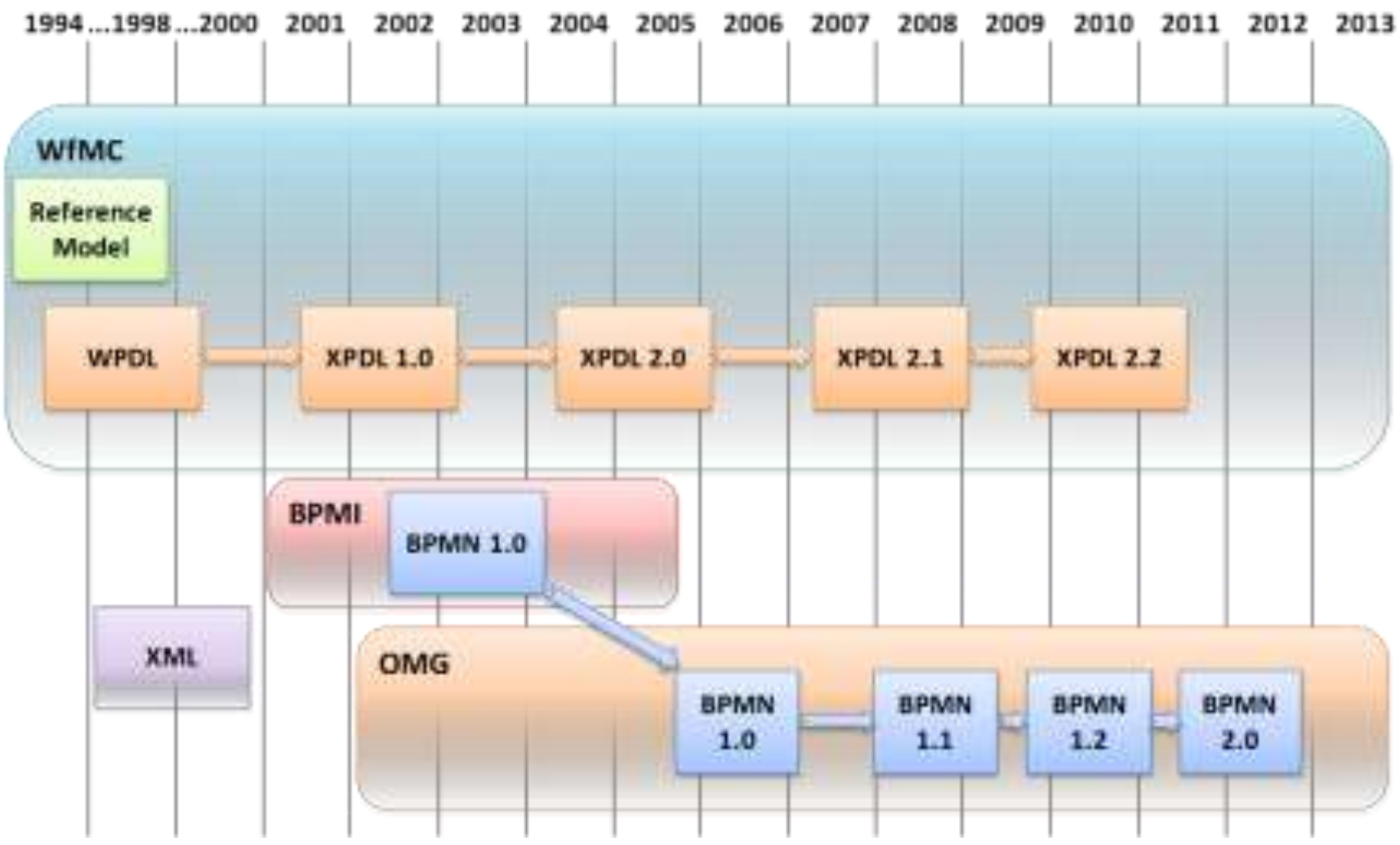

**Fig. 1.** - BPMN Standard Timeline - Releases (adapted from [18])

The semantics of many of the BPMN constructs came clearly from the field of executable process specifications. Among others, one can refer special constructs dealing with loops, exception handling, and transactions. Programmers and IT-specialists are more familiar with these sorts of elements. Process analysts typical only use a restricted part of the whole notation (see [19]) and normally omit items with more complex semantics. However, some BPMN experts [7] consider that these technical constructs should also be used by business analysts, in order that BPMN models are able to show business-relevant exceptions and their handling. In favor to this argument they recall the familiar 80-20 rule, which says that 80% of the costs, delays, and errors come from 20% of the cases - the ones that are the exceptions to the happy path. So, business analysts, as domain experts, should be the ones accountable for modeling exceptions.

According to assessments made [20], the standard is suitable for representing almost all control-flow elements distilled in Workflow Patterns [21]. Workflow Patterns are used as a template to compare the expressiveness of process modeling languages, since they are independent of any specific language [22]. Control-flow patterns are defined at process model level (with execution semantics applied to process instances). Some examples of control-flow patterns include *sequence*, *and split*, and *and join*, as well as *exclusive or split* and *exclusive or join*.

However, the large current number of diagrammatic symbols of BPMN have also increased its perceived complexity [23]. As acknowledged in [6], the complexity of process modeling languages affects the ability of process modelers to model the domain using a process-centric approach. In [24] it is claimed that the learning performance, depends on the person's characteristics, as well as by learning materials and the way concepts are presented. Language complexity is thus a significant issue, because it can affect the learnability, the ease of use and overall diffusion of a language.

Recent studies [19, 23, 25] on process modeling languages indicated also that the perceived complexity affects negatively the usage of those languages. However, paradoxically, also showed that languages such as BPMN, with a larger vocabulary, are used more frequently than others with a more restricted vocabulary (e.g. P/N, AD).

There are several approaches, from different fields, which have been used in studies (e.g. comparison and analysis) about modeling languages in general, and business process modeling languages in particular. Some of those fields are: semiotics [26-29], ontology [15, 16, 30-36], measures [37, 38], empirical studies [39-41], meta-modeling [42-46], and several other domains [47-50]. We highlight particularly here, the ontology perspective, given its broad usage. The ontology-based evaluations have as basis a model of real-world concepts (a representational ontology) and involve a mapping between the model and the constructs of the modeling language. The Bunge-Wand-Weber ontology (BWW) [51-53], has been widely used for assessing business process modeling languages. BWW allows the evaluation of the degree of coverage, by language constructs, of certain dimensions of process models (e.g. structural, dynamics, and components).

BWW was used in a study to quest the extent in which different subsets of BPMN vary in their complexity [25]. The findings show that BPMN language exhibits a large amount of complexity due to constructs and constraints not graphically depicted. This seems to indicate that the underlying rules and constraints of the language are a significant source of complexity. Those findings lead also to two relevant conclusions:

- Users tend to reduce the complexity of BPMN by ignoring the language's underlying rules, when making practical use of it. In the particular case of building process models, this means that process modelers violate BPMN underlying grammatical rules, using constructs outside their intended scope;
- Not all BPMN constructs are equally suitable for modeling purposes. The bulk of the hidden complexity comes from constructs that are seldom used in practice. Thus, most of the complexity of the language is disregarded by process modelers since they do not use the constructs whose semantic is more difficult to apprehend;

BPMN has found widespread acceptance by industry, since its inception with the BPMN 1.0, and a wide-ranging tool support. This suggests the importance given to the requirement of BPMN expressiveness even at the expense of its increased complexity. Although, researchers have shown that BPMN suffers from a number of shortcomings with impact in language's clarity, being complexity one of them as previously mentioned, this does not seems to affect the language's acceptance [6]. The findings also indicate that BPMN users, consciously or not, take active steps to reduce language's complexity. Anecdotal evidence suggests that a common way of such reduction is through development and enforcement of modeling conventions within organizations [54]. Modeling conventions typically mitigate the complexity of BPMN through a selection of constructs with an overall lower complexity and defining the patterns for better usage of those constructs. Users can also impose to themselves restrictions, in the usage of the BPMN, which deliberately lowers the level of language's complexity. The introduction of restrictive modeling conventions, i.e. modeling best-practices, makes the constructs easier to apply and the resulting models easier to communicate.

The catalog of BPMN well-formedness rules, proposed in this report, contributes for reducing the complexity of BPMN, by systematizing set of patterns for BPMN constructs' correct usage. These patterns can also be enforced as modeling best-practices easing the BPMN adoption by organizations. On the other hand, by embedding well-formedness rules in BPMN tools, the violation of BPMN modeling rules could be detected in real-time, reducing or even removing the discretionary use of constructs by process modelers.

## 3. Related work on BPMN rules verification

Model checking of BPMN rules in process diagrams has been a matter of some research. Work done includes approaches using languages with formal semantics. Due to the mathematical foundation of those languages, they allow several properties' formal verification, such as different classes of workflow control-flow errors [55]. In the following we summarize the well-known formal verification methods that have been used for checking of BPMN models, as well as their specific formalism.

- ***Communicating Sequential Processes* -** Wong and Gibbons [56, 57] use the CSP language and the behavioral semantics as a denotational model for augmenting the semantic model of BPMN. Relative timing information is introduced to allow the specification of timing constraints on concurrent activities. The processes' behavioral properties underlying BPMN diagrams are analyzed using the FDR model checker;
- ***Petri Nets* -** Dijkman et al. use a formal semantics for a subset of BPMN constructs in Petri nets [58]. The semantics they provide maps BPMN pools into Workflow nets. A Workflow net is a P/N such that there is a unique source place $i$ ($\bullet i = \emptyset$), a unique sink place $o$ ($o\bullet = \emptyset$), and every other place and transition is on a directed path from the unique source place to the unique sink place. The proposed semantics only includes a subset of BPMN, and does not model properly multiple instances, exception handling and message flows.
- ***Web Ontology Language* -** In [59] is defined an ontology that formally represents the BPMN specification (BPMN 2.0 Ontology), which can be used as a knowledge base for checking the syntax of concrete BPMN models. The description of an element is combined within the corresponding class and further explanations are provided in annotations. This approach claims allowing a much faster understanding of BPMN. In addition, the ontology is used as a syntax checker to validate concrete BPMN process models. A common property of ontologies and the Web Ontology Language (OWL) [60] semantics is the so-called open-world assumption [61], a form of partial description or under-specification as a means of abstraction, i.e., from the absence of statements, a deductive reasoner must not infer that the statement is false.
- ***Abstract State Machines* -** The approach regarding BPMN semantics in [62], uses Abstract State Machines (ASM) [63], a form of rigorous pseudo-code that keep up with the inheritance chain in the process modeling language class hierarchy. The abstract model of the dynamic semantics[2] of the BPMN language is attained, by inserting rules as behavioral elements at appropriate places in the class hierarchy. This enables the definition of the language's execution semantics. Eventually, the enhanced process modeling language can be used to check the conformance of process diagrams.

There are some limitations, regarding BPMN models' verification, on the mentioned approaches.

- For some methods, most of the verification rules are only able of being checked after process diagrams are at valid state. This hinders the conversion from BPMN language into the specific language and environment of the model checker.
- The approaches technically demanding w.r.t. the formalisms. Even advanced process modelers can hardly expected be acquainted with the above mentioned formalisms. Moreover, modelers must deal with a blended environment integrating the BPMN modeling tool and a model checker tool.
- The ontological approach proposed uses the open-world assumption [61], which advocates that what is not currently known to be true, is false. A different assumption is used by the Model-Driven Approach (MDA), which assumes that the model has complete information to restrict arbitrary extensions of the system that could lead to inconsistencies (closed-world assumption).

The underlying proposal of this report intends to supplement the BPMN standard by providing a catalog of rules that could enhance BPMN models' construction. Those rules being depicted and become an integral part of the BPMN metamodel, implemented as OCL invariants, could be enforced by the same tools that already support the BPMN modeling language. Therefore, during the BPMN process models' design, in addition to syntax error verification, which is available with limitations from current tools (see section 7), the modelers would have also, real-time checking about semantic violations in models (e.g. a throw event without a corresponding catch event; a mismatch between flows from a split parallel gateway and the incomings on the corresponding joint element, a possible cause of a deadlock situation). Similarly, BPMN modeling best-practices of an organization, could also be established and enforced. This kind of rules (standard or best-practices based) could greatly enhance the quality of BPMN diagrams resulting from the design phase of business processes' life cycle.

[2] The dynamic semantics (aka execution semantics) of a language defines how and when the various constructs of a language should produce a program behavior.

# 4. BPMN metamodel

Nowadays, BPMN is broadly acknowledged as filling the set of required criteria [64] for being a business process modeling standard, namely: (1) the notation is widely adopted either by business and IT people, due to be tool independent [65]; (2) has a meta-model, which a consistent vocabulary of concepts and relationships and thus an abstract syntax; (3) the notation and metamodel allow the focus and analysis on particular details of the process model (e.g. control-flow, data, organization); and (4) provide an exchange format for process models (e.g. XPDL [66], XMI [67]).

The BPMN metamodel combine three notations: a notation for modeling processes' orchestration and collaboration (see Fig. 2); a notation called conversation, a simplified version of collaboration diagrams; and a third notation for modeling participant interactions, called choreography. 151 meta-classes and 200 associations totalize the complete metamodel. Fig. 3 depicts the stereotypes of meta-classes defined in the specification, for orchestration and collaboration purposes. In Fig. 4 is depicted all possible graphical representation of events.

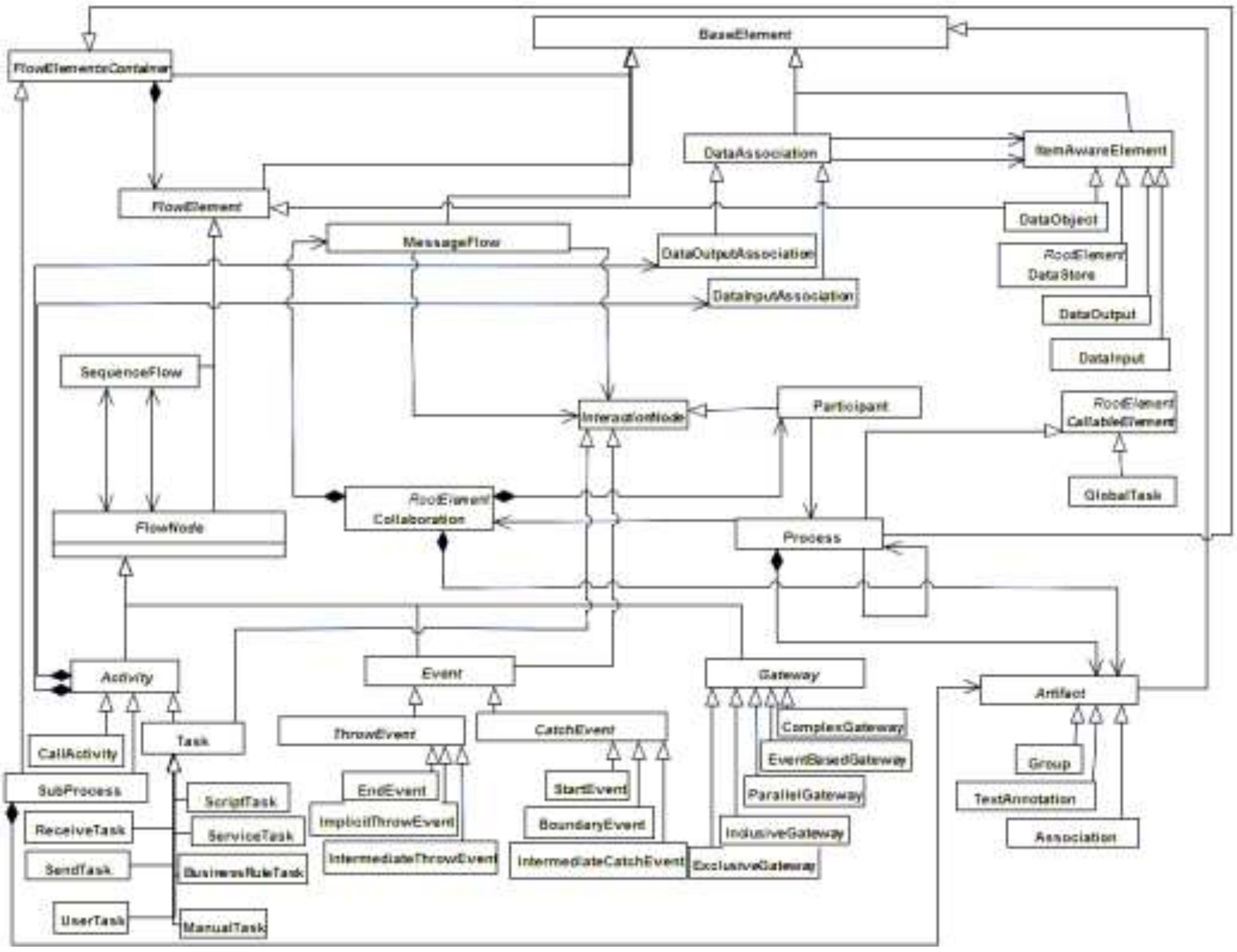

**Fig. 2.** - BPMN metamodel excerpt - processes' orchestration and collaboration

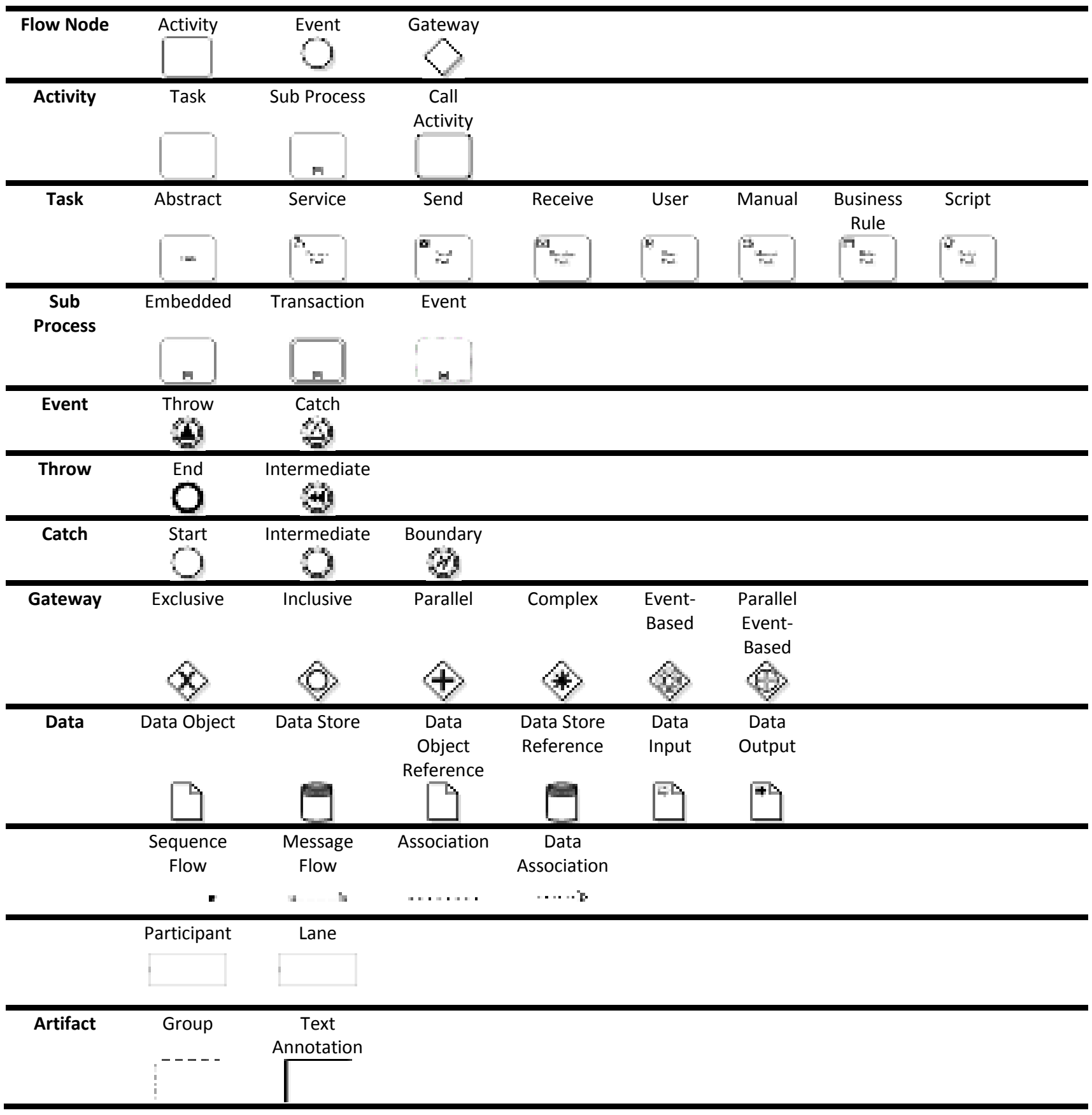

**Fig. 3.** - Stereotypes of BPMN elements

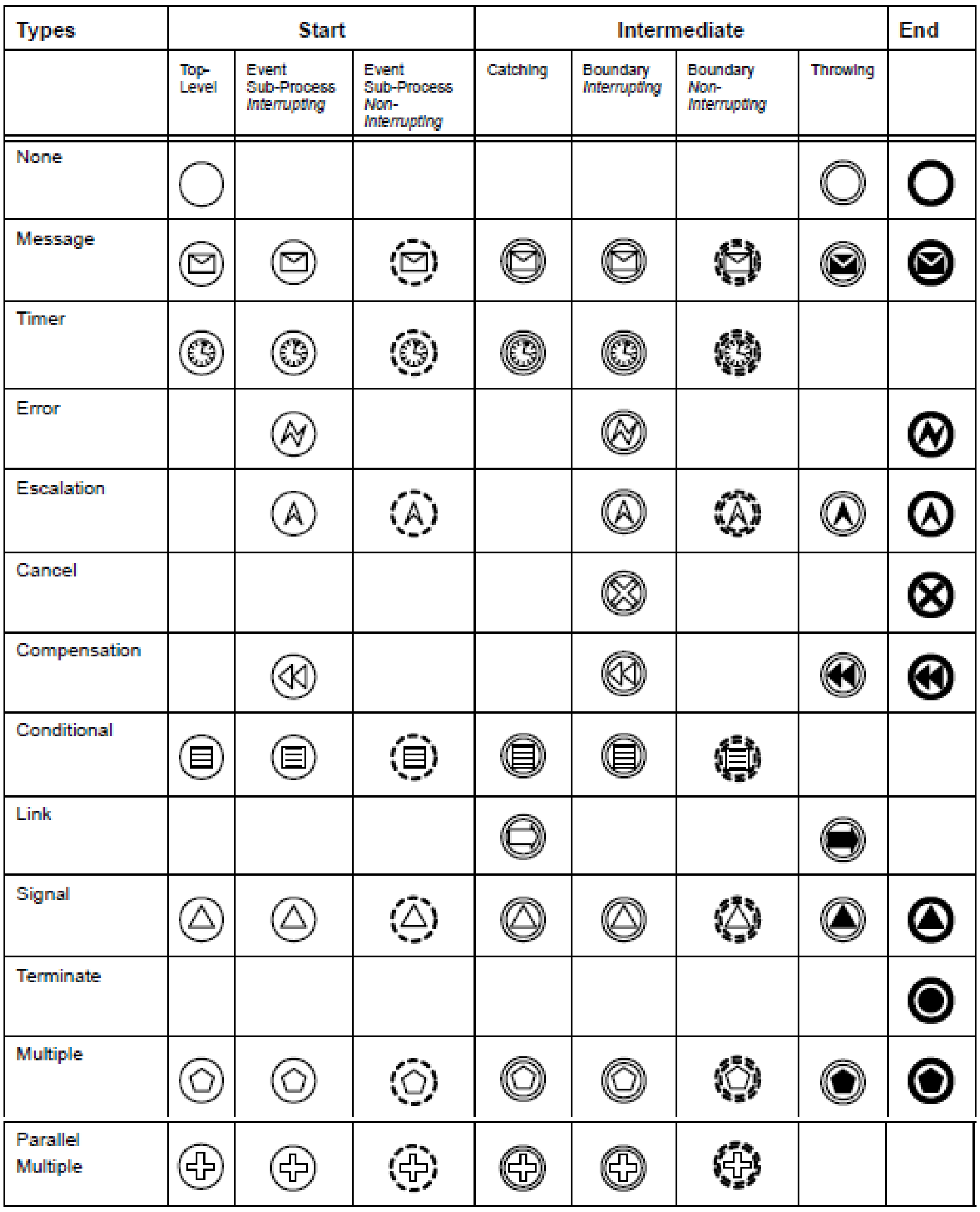

**Fig. 4.** - Event types in BPMN (source [1])

In this work we consider only, the part of the BPMN metamodel concerned with participants' collaboration and process orchestration. In the remaining of this section the metamodel is overviewed by introducing its main concepts and connections. For a formal and graphical presentation of BPMN modeling rules (later in section 5), the BPMN abstract syntax is defined. We illustrated here the abstract syntax definition by presenting some code snippets (Listing 1-Listing 4) of it[3]. The abstract syntax is the basis for the BPMN rules' coding and embedding into the metamodel. For the BPMN abstract syntax coding and verification, it was used the USE tool (UML based Specification Environment) [68] (described in section 0).

The *BaseElement* is the abstract super metaclass for most of elements on BPMN metamodel (see Listing 1). In Fig. 5 are depicted all the metaclasses that inherit from the top metaclass *BaseElement*:

- *Artifact* – allows to show additional information about a *process* and is linked to *flow objects* through *associations*;
- *FlowElement* – abstract super class for all elements that can appear in a Process flow, which are the *FlowNodes*;
- *MessageFlow* – used to show the flow of *Messages* between two *Participants* that are prepared to send and receive messages;
- *DataAssociation* – used to move data between *Data Objects*, *Properties*, and *Activities*, *Processes*, and *GlobalTasks*;
- *ItemAwareElement* – elements that are subject to store or convey items during process execution; and
- *FlowElementsContainer* – defines the superset of elements that group elements in the diagrams.

[3] The full script with BPMN abstract syntax and BPMN rules coded as OCL constraints is available at http://sdrv.ms/16EvDjG

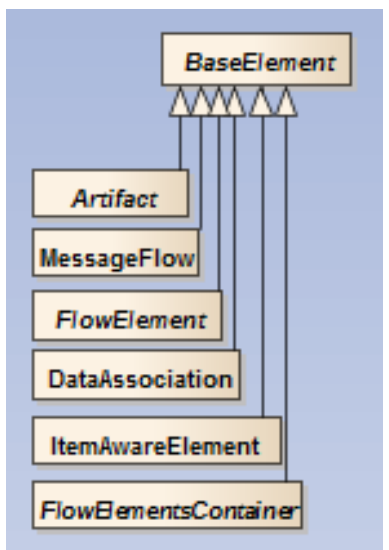

**Fig. 5.** - Metaclasses inheriting from *BaseElement*

```
abstract class BaseElement
   attributes
      id : String
   operations
-- ...
end --BaseElement
```

**Listing 1 – *BaseElement* class definition**

The metaclass *Process* (see Fig. 6 and Listing 2) describes a sequence of Activities carry out in an organization with some specific objective. If a process interacts with other processes, it must participate in a *Collaboration*. A collaboration groups several participants. Each *Participant* (aka Pool) must address only one process. Since a participant is an *InteractionNode*, it can send or receive *MessageFlow*.

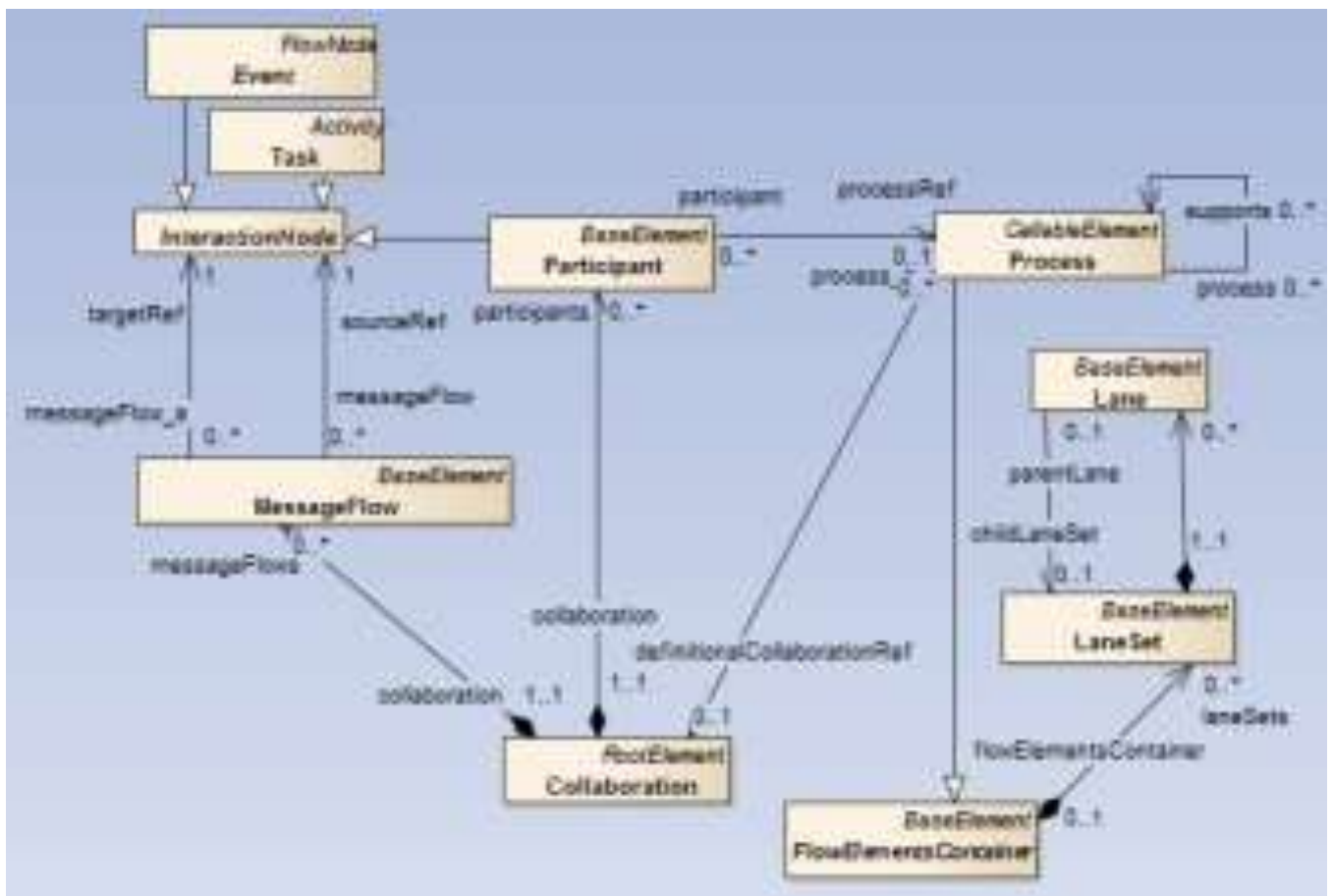

**Fig. 6.** - Process metaclass connections

```
class Process < FlowElementsContainer, CallableElement
   attributes
                processType : ProcessType
                isClosed : Boolean
                isExecutable : Boolean
   operations
   -- ...
 end -- Process
```

**Listing 2 – *Process* metaclass definition**

Fig. 7 depicts some of most instantiated metaclasses when a BPMN process model is drawn. A *FlowElementsContainer* (which can be a *Process* or a *SubProcess*), is a container of *FlowElement*. A flow element can be *FlowNode* (see Fig. 8), *SequenceFlow* or *DataObject*. A sequence flow link the various kind of flow node. The metaclass *ItemAwareElement* is the abstract class of the several kind of metaclasses, representing transient (*DataObject*), persistent (*DataStore*), input data or output data to/from *Activity* by mean of subclasses of *DataAssociation*.

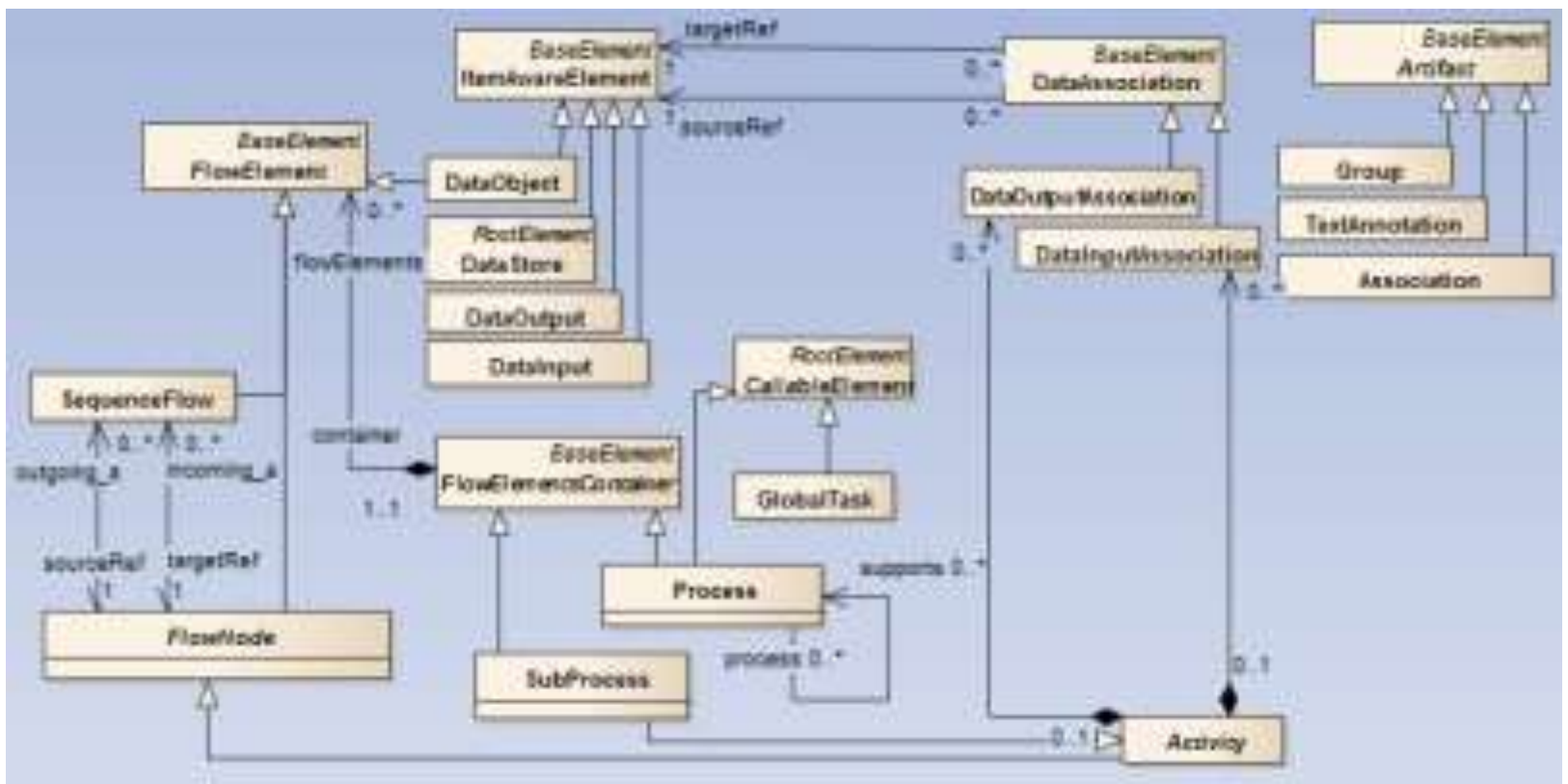

**Fig. 7.** - Metaclasses instantiate in a process orchestration

As one can see in Fig. 8 a *FlowNode* (Listing 3) can be one of several kinds: *Activity* (Listing 4), *Event* or *Gateway*.

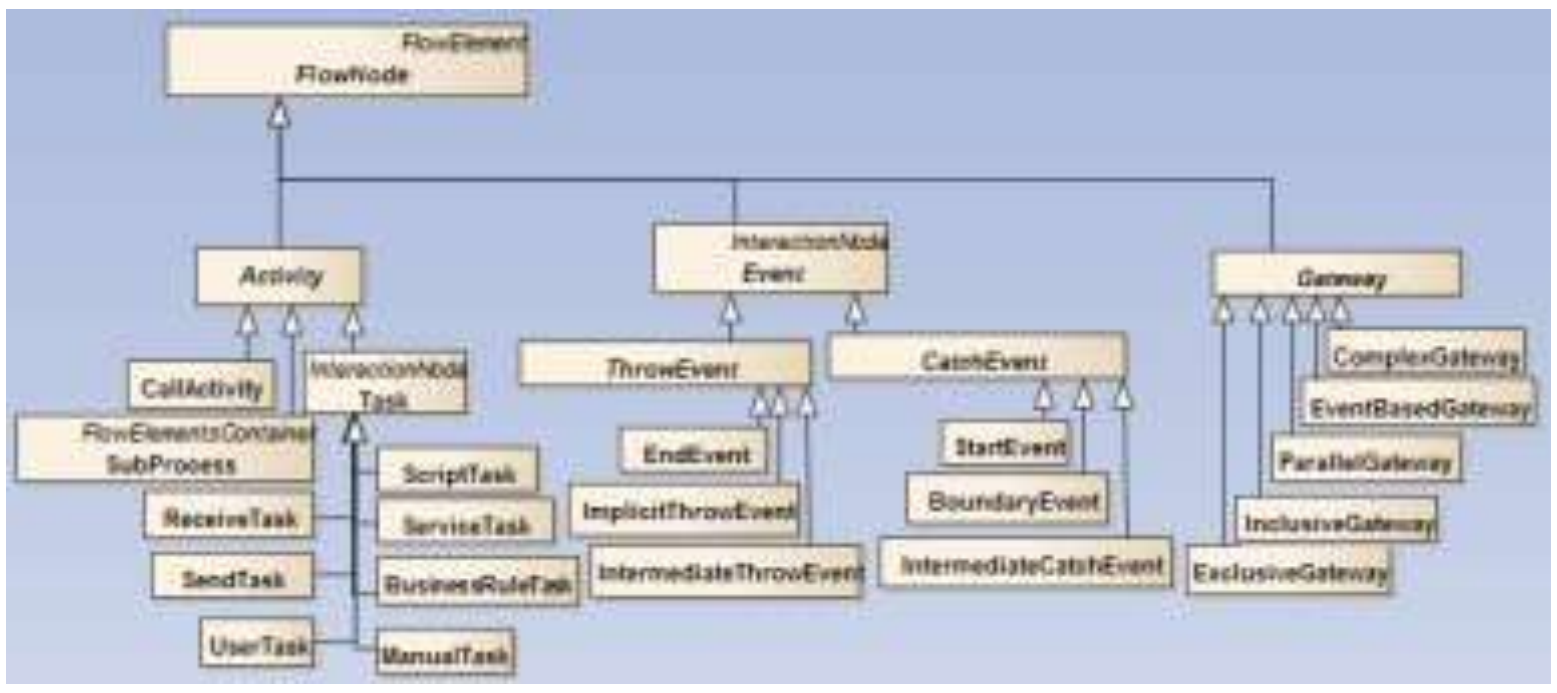

**Fig. 8.** - Main subclasses of *FlowNode*

```
abstract class FlowNode < FlowElement
operations
   -- ...
 end -- FlowNode
```

**Listing 3 – *FlowNode* class definition**

```
abstract class Activity < FlowNode, InteractionNode
      attributes
               isForCompensation : Boolean
               startQuantity : Integer
               completionQuantity : Integer
      operations
   -- ...
 end -- Activity
```

**Listing 4 – *Activity* class definition**

The definition of BPMN abstract syntax, in the USE tool notation, as well as the instantiation of the BPMN metamodel, allowed to find the following issues in the BPMN standard's documentation:

- In the specification it is considered a visual shortcut the use of a non-directional data association connected to a sequence flow (page 225). However, the metamodel only allows links among instances of subclasses *DataAssociation* and Activity (see Fig. 7). So, a *Sequence Flow* cannot be directly linked to a *DataObject* via an instance of type *DataAssociation*. Tools that implement this visual shortcut, should instantiate the same metaclasses as the regular solution (a *DataOutputAssociation* going out an activity to a *DataObject* and *DataInputAssociation* coming from the same *DataObject* instance to other activity);
- The metamodel does not allow a *Subprocess* to receive/send a message flow. This constraint introduces a limitation in the modularization of a process in sub-processes when there are interactions among participants. The elements that participate in an interaction must appear at top level in the process. Modelers tend to ignore this constraint and deliberately, or not, they violate the metamodel. In order to allow the abovementioned representation an update was introduced in the metamodel: the inheritance relationship between Task and *InteractionNode* was replaced by a new inheritance between Activity and *InteractionNode*.

An equivalent BPMN abstract syntax, albeit more concise, based on the set theory such the one presented in [69], and targeting specifically the BPMN graphical elements', was also derived concurrently to the one attained using the USE tool. For the abstract syntax based on set theory, the definition of a *process orchestration* was based on the following propositions:

- N is a set of *FlowNode* and a superset of disjoint sets of *Activities* A, *Events* E, and *Gateways* G, i.e., $N \supseteq A$, $N \supseteq E$, $N \supseteq G$, and $A \cap E \cap G = \emptyset$
- A is a superset of disjoint sets of *Task* $A_T$, *CallActivity* $A_C$, and *SubProcess* $A_S$, i.e., $A \supseteq A_T$, $A \supseteq A_C$, $A \supseteq A_S$, and $A_T \cap A_C \cap A_S = \emptyset$
- $A_T$ is a superset of disjoint sets of *ReceiveTask* $A_{TR}$, *SendTask* $A_{TS}$, *UserTask* $A_{TU}$, *ScriptTask* $A_{TP}$, *BusinessRuleTask* $A_{TB}$, *ManualTask* $A_{TM}$, i.e., $A_T \supseteq A_{TR}$, $A_T \supseteq A_{TS}$, $A_T \supseteq A_{TU}$, $A_T \supseteq A_{TP}$, $A_T \supseteq A_{TB}$, $A_T \supseteq A_{TM}$, and $A_{TR} \cap A_{TS} \cap A_{TU} \cap A_{TP} \cap A_{TB} \cap A_{TM} = \emptyset$
- $A_S$ is a superset of disjoint sets of *EventSubProcess* $A_{SE}$, *Transaction* $A_{ST}$, *EmbeddedSubProcess* $A_{SB}$
- E is a superset of disjoint sets of *ThrowEvent* $E_T$, and *CatchEvent* $E_C$, i.e., $E \supseteq E_T$, $E \supseteq E_C$, and $E_T \cap E_C = \emptyset$
- $E_T$ is a superset of disjoint sets of *EndEvent* $E_{TE}$, *IntermediateThrowEvent* $E_{TI}$, i.e., $E_T \supseteq E_{TE}$, $E_T \supseteq E_{TI}$, and $E_{TE} \cap E_{TI} = \emptyset$
- $E_C$ is a superset of disjoint sets of *StartEvent* $E_{CS}$, *BoundaryEvent* $E_{CB}$, *IntermediateCatchEvent* $E_{CI}$, i.e., $E_C \supseteq E_{CS}$, $E_C \supseteq E_{CB}$, $E_C \supseteq E_{CI}$, and $E_{CS} \cap E_{CB} \cap E_{CI} = \emptyset$
- G is a superset of disjoint sets of *ExclusiveGateway* $G_E$, *InclusiveGateway* $G_I$, *ParallelGateway* $G_P$, *EventBasedGateway* $G_V$, *ComplexGateway* $G_C$, i.e., $G \supseteq G_E$, $G \supseteq G_I$, $G \supseteq G_P$, $G \supseteq E_{CS}$, $G \supseteq E_{CB}$, and $G_E \cap G_I \cap G_P \cap G_V \cap G_C = \emptyset$
- Each $G_k$ (with $k = E, I, P, V, C$) is a superset of disjoint sets of join gateway $G_{kj}$ and fork gateway $G_{kf}$, i.e., $G_k \supseteq G_{kj}$, $G_k \supseteq G_{kf}$, and $G_{kj} \cap G_{kf} = \emptyset$
- $S \subseteq N \times N$ is a subset of *SequenceFlows* connecting *FlowNode*, and denote the process control flow
- O represents the subset of *SequenceFlows* that are part of the normal flow, with $O \subseteq S$. In this path are only include FlowNodes $F \backslash E_{CB}$
- X represents the subset of *SequenceFlows* that are part of the exception flow, with $X \subseteq S$. In this path are only include FlowNodes $E_{CB}$ such that its occurrence interrupts the execution of an activity
- $S = O \cup X$ the process control flow is the union of the normal flow and exception flow
- $C \subseteq S$ is a subset of *SequenceFlows* connecting *FlowNode*, and denote all the conditions which are part of the control flow

So, a BPMN process, described using the graphical elements of BPMN elements shown in Fig. 3, can be defined alternatively as follows:

**Definition** (*BPMN Process Orchestration*).
A BPMN process Orchestration is a tuple $P = (N, A, E, G, T, S, A_T, A_C, A_S\, A_{TR}, A_{TS}, A_{TU}, A_{TP}, A_{TB}, A_{TM}, A_{SE}, A_{ST}, A_{SB}, E_T, E_C, E_{TE}, E_{TI}, E_{CS}, E_{CB}, E_{CI}, G_E, G_I, G_P, G_V, G_C, G_{ky}$ (with $k = E, I, P, V, C$ and $y = j, f$), $S, O, X, C)$

# 5. The catalog of BPMN rules

We present herein the rules that, attached to the metamodel, can contribute for achieving better quality in the representation of process models through BPMN. Each rule will be presented either in one or all of three forms: (1) in textual form; (2) with pictures illustrating its correct usage and exemplifying its violation; and (3) in a formal form using OCL syntax. Since OCL expressions are generally too large, we opt to wrap them up with a logical operation. The details of logical operations are available in the already mentioned script made available on-line[4], which merges the metaclasses and associations definitions as well as the OCL constraints.

Some of rules are quite simple and are explicitly stated in BPMN specification. Only tools that only provide mere design aid do not enforce them. Other rules are more elaborated and distilled from the details of process design and execution referenced in several sources [7]. Practitioners have also given some contributions, namely by promoting process diagrams' modeling best-practices [70] [71].

In this document we do not elicited all the BPMN modeling rules. The reason was that plenty of those rules can be directly derived from the metamodel and therefore do not need to be explicitly stated or enforced by OCL constraints. Some examples of those rules are the following:

- A Gateway cannot have incoming or outgoing Message Flow: since a Gateway is not an *InteractionNode*, it cannot be the source or target of a Message Flow;
- A Data Store cannot be the source or a target of a Sequence Flow: A *SequenceFlow* must have as source or a target a *FlowNode*, and a Data Store is an *ItemAwareElement* and not a *FlowNode*;
- A Message Flow must have a source and a target: the cardinality of the associations *A_targetRef_messageFlow*, and *A_sourceRef_messageFlow* in the metamodel is 1 regarding the *InteractionNode* side, which makes it mandatory a Message Flow to have a source and a target.

## 5.1. Flow control well-formedness rules

**Collaboration (aka pools)**

The Collaboration diagram depicts the interactions among participants (shown as pools) through message flows that connect two pools (or the objects within the pools). Collaboration may include processes within the pools and/or Choreographies between the pools [1] (pages 109 and 111).

### 5.1.1. Only one implicit pool can exist in a collaboration

A pool is not required to contain a process, i.e., it can be a "black box". On the other hand, one and only one pool in a collaboration diagram may be presented without a boundary. In this case it is assumed that the Process has an implicit pool. If there is more than one Pool in the diagram, then all the remaining pools must have a boundary [1] (page 112).

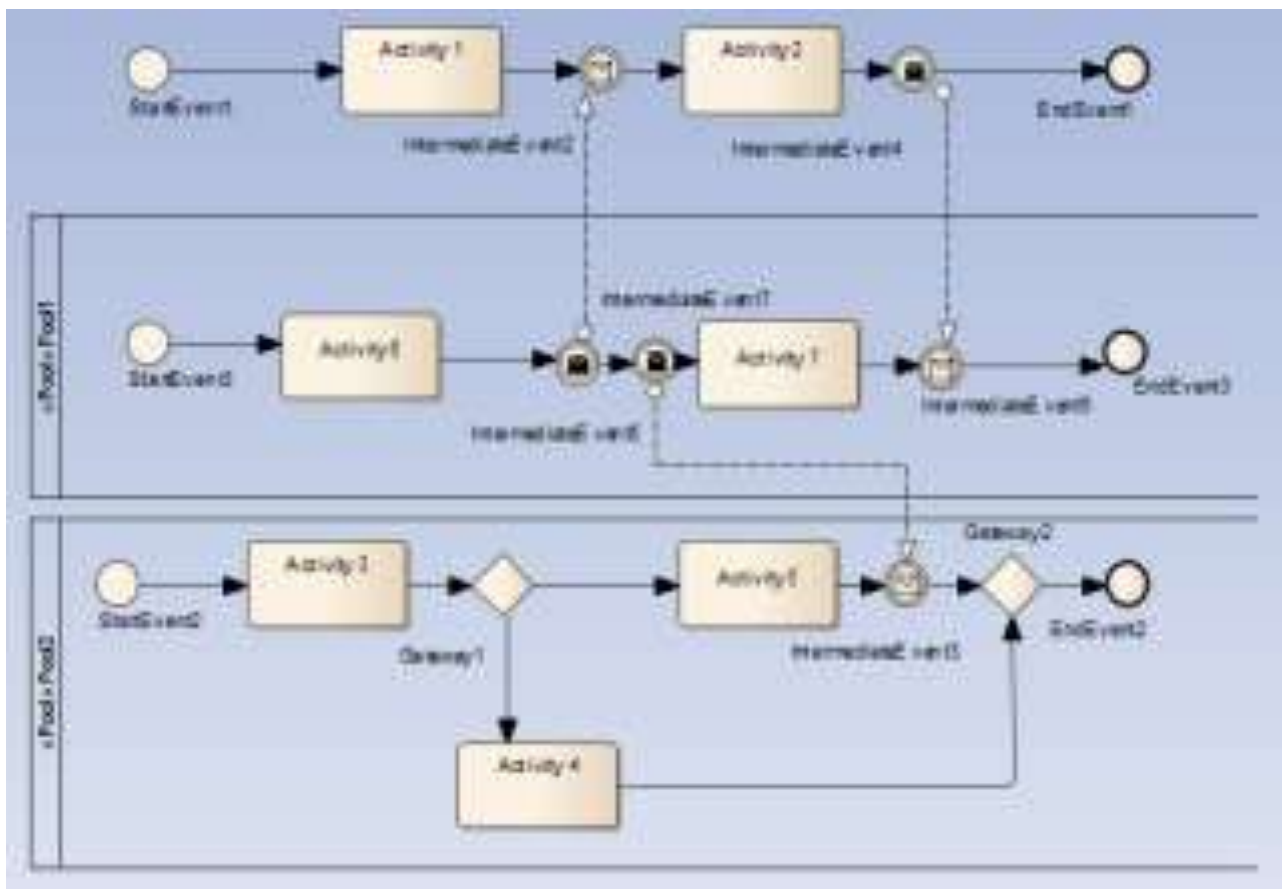

---
[4] http://sdrv.ms/16EvDjG

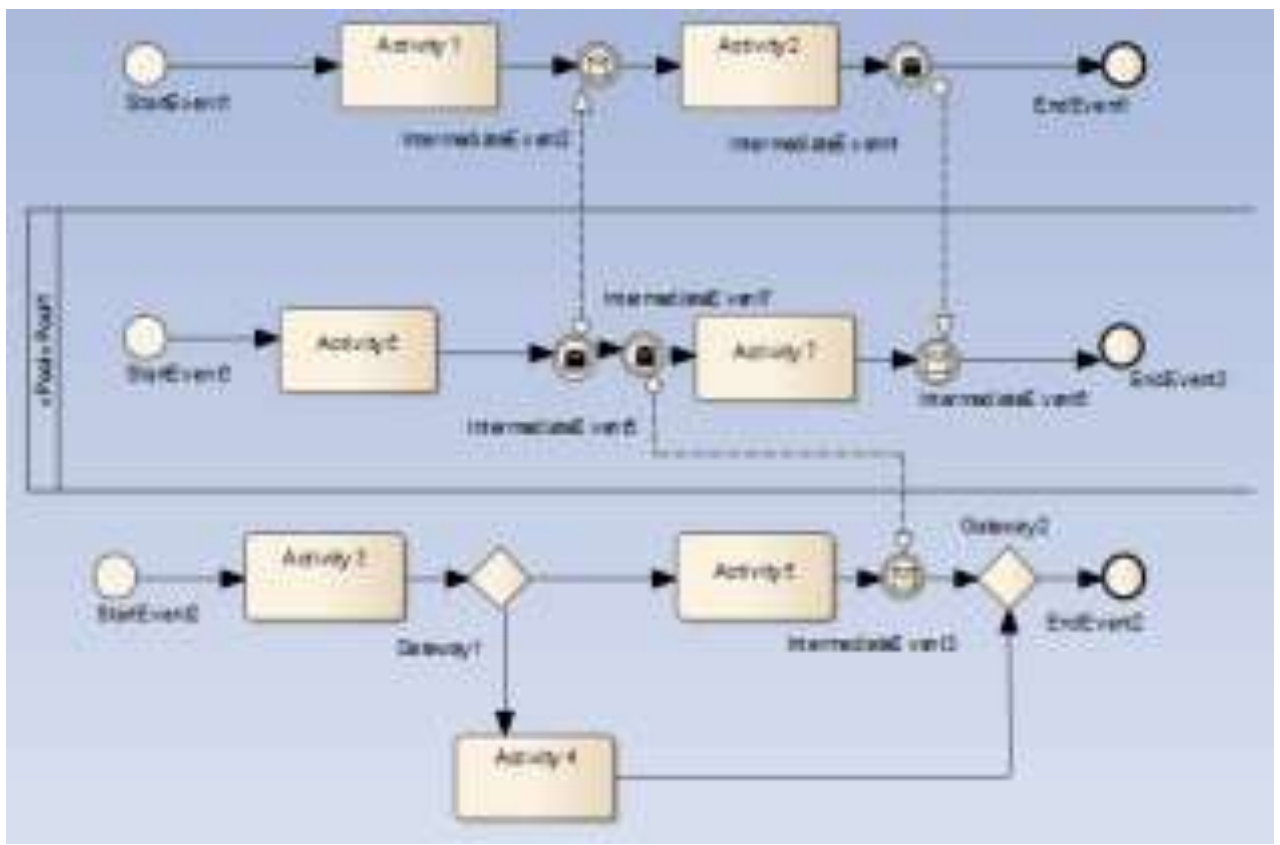

**Fig. 9.** – Only one implicit pool can exist in a collaboration

**Picture's interpretation**: Correct: Only one implicit pool may exist in the collaboration (Fig. 9 top). Wrong: Two implicit pools are not allowed in a collaboration (Fig. 9 bottom).

A well-formedness rule regarding processes and participants can be enforced by attaching the following invariant to the *Collaboration* element of the BPMN metamodel: The difference among the number of processes and participants must not surpass one. Therefore only one Process can have an implicit pool.

The well-formedness rule regarding implicit pool can be enforced by attaching the following invariant to the *Collaboration* element of the BPMN metamodel.

```
context Collaboration
  inv aCollaborationCanOnlyHaveOneImplicitProcess:
    countAllProcessesInCollaboration() - countAllPoolsInCollaboration()= 0
```

**Listing 5 – Only one implicit pool can exist in a collaboration**

**Process**

A Process describes a sequence or flow of Activities in an organization with the objective of carrying out work. In BPMN a Process is depicted as a graph of Flow Elements, which are a set of Activities, Events, Gateways, and Sequence Flows that represents what an organization does in order to accomplish specific objectives. Processes can be defined either at enterprise-wide level or low level such as the work performed by a single person [1] (page 145).

### 5.1.2. A Process is an abstraction of more detailed Process, if its normal flow is contained by the second Process's normal flow

A full detailed process, also known as *private* process, includes several features of process modeling, such as documentation elements (Artifacts), organization elements (Pools/Lanes), exception flow elements (Intermediate and Boundary Events), and data flow elements (Data Objects and Data Stores).

The *normal flow* is the description of the basic control flow of the Process. Therefore, the *normal flow* is a simplified view that provides an abstraction of the full detailed process. By collapsing flow elements (Activities, Events, and Gateways) and Sequence Flow of the *normal flow*, one can transform a private process, into a public process without changing its underlying structure.

Furthermore, one can build a more concrete perspective of the process from a more abstract one, by adding elements related to documentation, organization, exception flow, or data flow. Conversely, one can turn a process more abstract by removing detail from a more concrete perspective. Boundary Intermediate Events, in general, and Cancel, Error, Multiple, and Parallel Multiple Intermediate Events in particular, should not be part of a normal flow.

The activities that are used to communicate to the other participant(s) are the first candidates to be included in the public (abstract) process. Internal activities of a private process can be collapsed or even removed from the public process. Therefore, the public process would show the message flows as well as the order of those message flows that are needed to interact with another processes, since public processes can be modelled separately or within a collaboration to show the flow of messages between the public process activities and other participants [1] (page 23).

Given two processes A and B, it is necessary for B to be an abstraction of A that in the two top-level processes:

- B possesses lesser or equal number of flow nodes (activities, events and gateways) than A;
- B possesses lesser or equal number of sequence flows than A;
- B possesses lesser or equal number of incoming message flow than A;
- B possesses lesser or equal number outgoing message flow than A;
- B possesses greater or equal number of sub-processes than A;

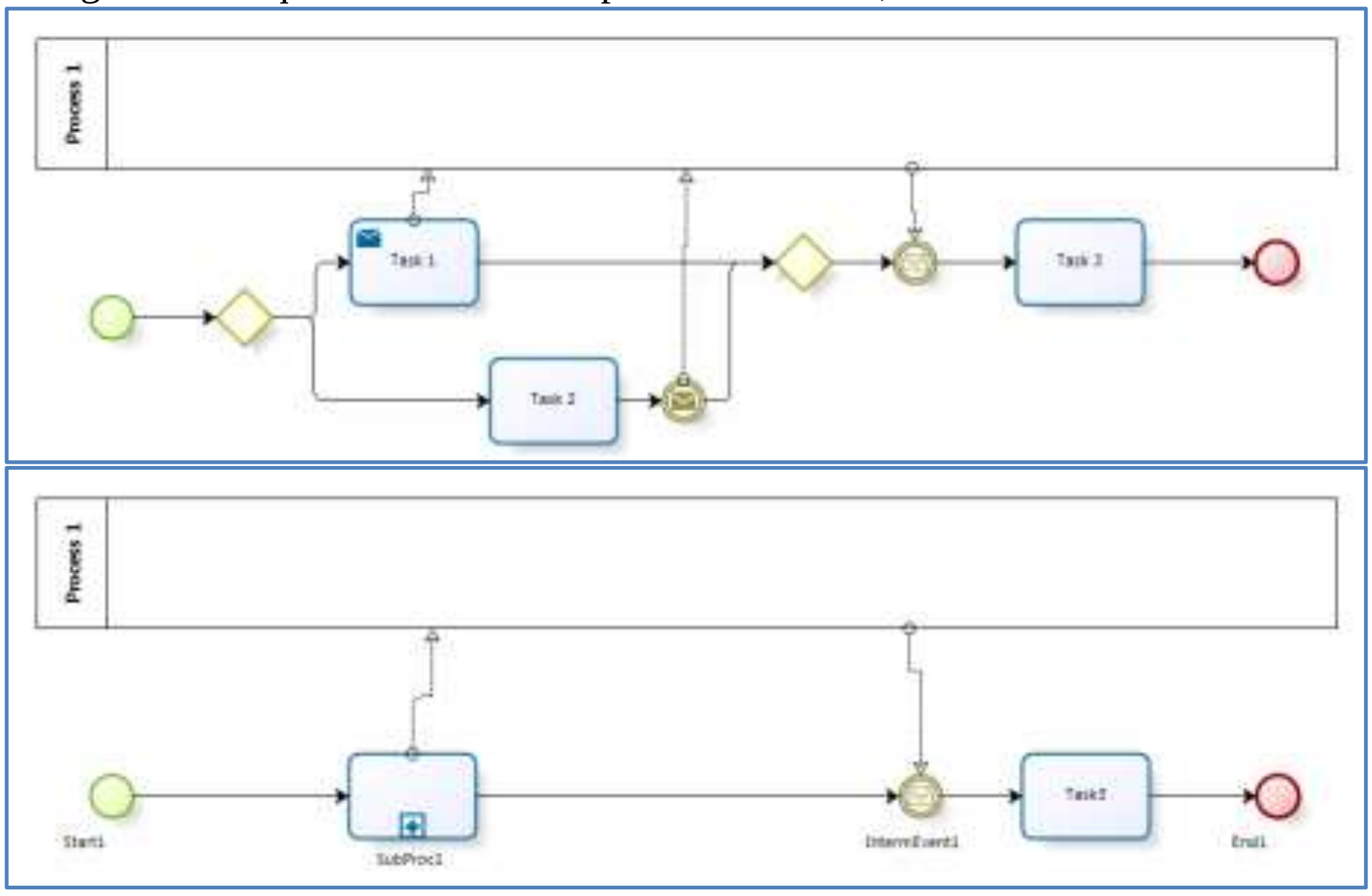

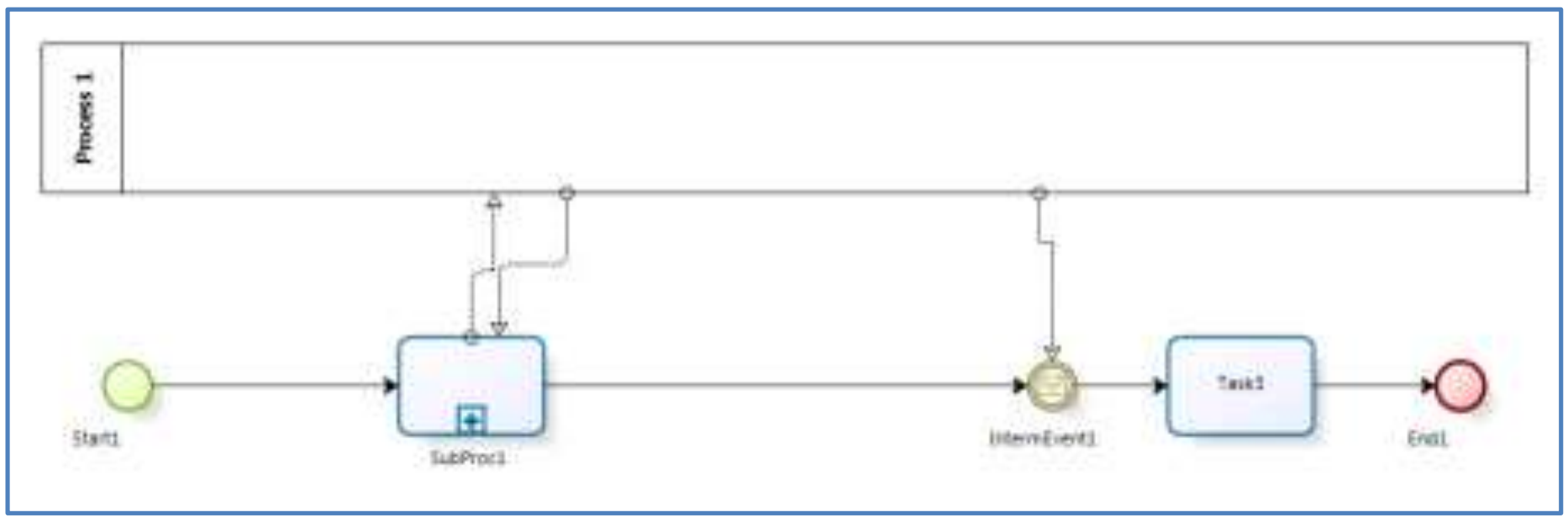

**Fig. 10.** –A Process is an abstraction of more detailed Process, if its normal flow is contained by the second Process's normal flow

**Picture's interpretation**: Correct: The private process depicted at top, can be abstracted through the public process depicted at middle; Wrong: The process depicted at bottom is not an abstraction of the public process since there are more incoming message flows than in the detailed process.

To check whether a process is a valid abstraction of another process, can be done by an OCL function attached to the Process element of the BPMN metamodel.

```
Process::isAbstraction(p : Process)  : Boolean =
 self.totalNumberContainerFlowNodes() <=
    p.totalNumberContainerFlowNodes() and
 self.totalNumberContainerSequenceFlows() <=
    p.totalNumberContainerSequenceFlows() and
 self.bpmnOutputMessageFlows()->size() <=
    p.bpmnOutputMessageFlows()->size() and
 self.bpmnInputMessageFlows()->size() <=
    p.bpmnInputMessageFlows()->size() and
 self.totalNumberContainerSubProcesses() >=
    p.totalNumberContainerSubProcesses()
```

**Listing 6 – A Process is an abstraction of more detailed Process, if its normal flow is contained by the second Process's normal flow**

### 5.1.3. A public Process may not be executable

A private process can be *executable* and *non-executable*. An executable process is a process that has been modelled for the purpose of being executed according to the semantics of a process execution language. A non-executable private process is a process that has been modelled only for documentation purposes. Therefore, information provided for execution, doesn't need to be included in a non-executable private process [1] (page 23).

Conversely, public processes are usually modeled to show only the relevant flow elements of the collaboration they made with external clients. Therefore, since the internal details of public processes are omitted from the model they cannot be executed [1] (page 147).

The well-formedness rule regarding process execution can be enforced by attaching the following invariant to the process element of the BPMN metamodel.

```
context Process
inv publicProcessIsNotExecutable:
  self.processType = ProcessType::Public
  implies
  self.isExecutable  <> true
```

**Listing 7 – A public Process may not be executable**

Moreover, if a process is an abstract version of another private process, the former process is necessarily a public process and therefore cannot be executable. Regarding Fig. 10, the process depicted at middle must be of type Public and therefore non-executable.

### 5.1.4. A Top-Level Process can only be instantiated by a restricted set of Start Events types

Any container (process or sub-process) that does not have a parent container is considered a *top-level* Process [1] (page 238). Top-level processes can have one of seven types of start events (see Fig. 4, first column): none, message, timer, conditional, signal, multiple, and parallel [1] (page 112) .

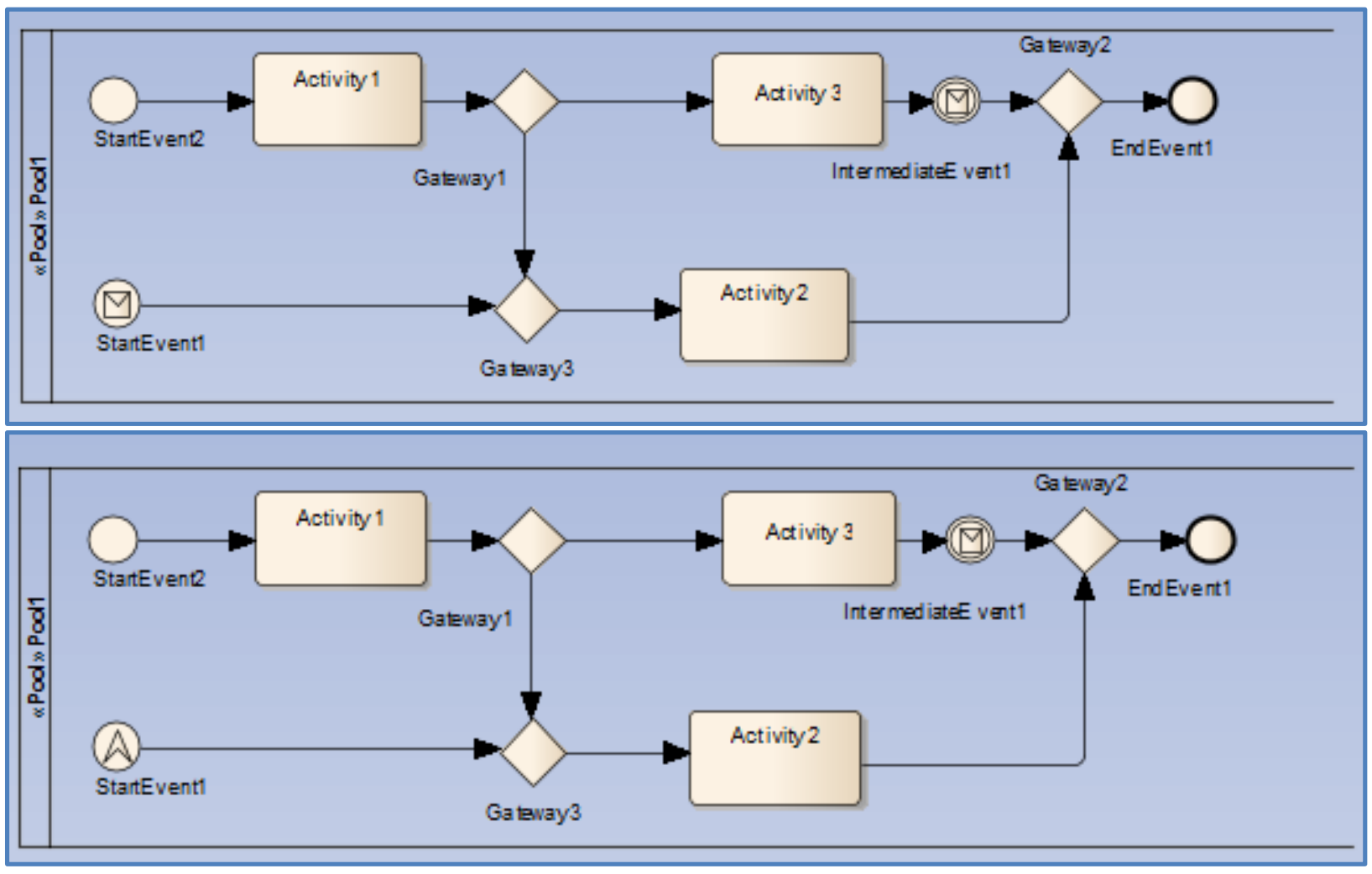

**Fig. 11.** – A Top-Level Process can only be instantiated by a restricted set of Start Events types

**Picture's interpretation**: Correct: A top-level process is instantiated by allowed types of start events (none and a message types) (top); Wrong: A top-level process is instantiated by an invalid type of start event (escalation type) (bottom).

The well-formedness rule regarding start events can be enforced by attaching the following invariant to the StartEvent element of the BPMN metamodel.

```
context StartEvent
inv topLevelProcessHasRestrictedTypeOfStartEvents:
    isContainerTopLevelProcess()
```

**Listing 8 – A Top-Level Process can only be instantiated by a restricted set of Start Events types**

### 5.1.5. A Call Activity must only call a Callable Element

A Callable Element can be called, by a Call Activity (known as Reusable Sub-Process in BPMN 1.2) [1] (page 186). The Call Activity inherits the attributes and model associations of Activity [1] (page 185). The elements that can be called by Call Activities are Process and *GlobalTask* (i.e., elements of type *CallableElements*) [1] (page 186).

When the callable element is a process the None Start Event is used for invoking the process from the call activity. All other types of start events are only applicable when the process is used as a top-level process.

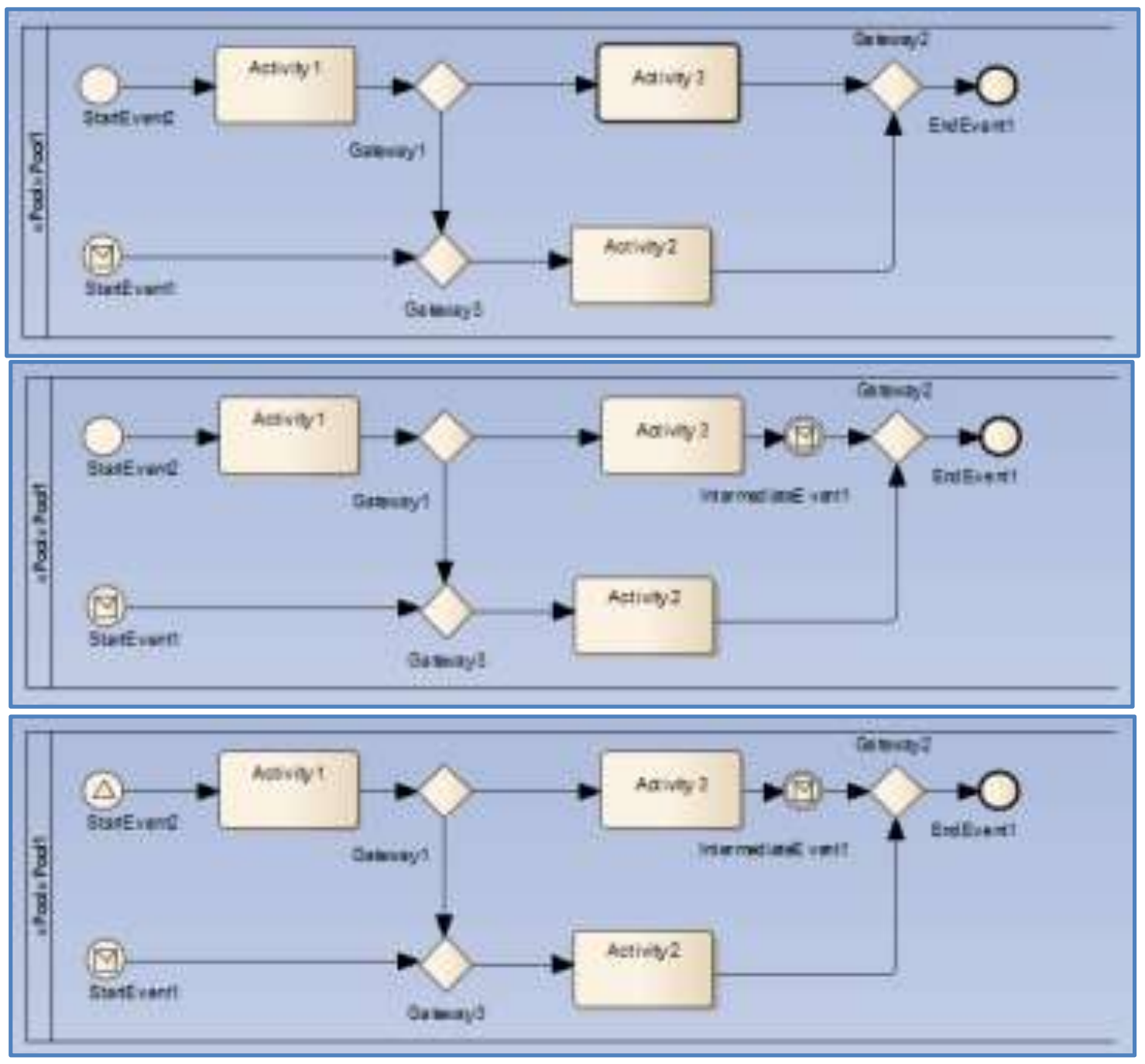

**Fig. 12.** – A Call Activity must only call a Callable Element

**Picture's interpretation**: Correct: The Call Activity Activity3 (top) can reuse the process such as the one at the middle with a none and a message type start events; Wrong: The Call Activity Activity3 (top) cannot reuse the process such as the one at the bottom because it is not a reusable process (there is no untyped start event).

The well-formedness rule regarding callable activities can be enforced by attaching the following invariant to the *CallActivity* element of the BPMN metamodel.

```
context CallActivity
inv callableElementIsReusable:
    isCallable()
```

**Listing 9 – A Call Activity must only call a Callable Element**

**SubProcess**

A Sub-Process is a compound Activity which can be used to break down another container (*Process* or *SubProcess*) and give it further detail.

### 5.1.6. The messages flows interacting with a collapsed view of a sub-process must be the same as the depicted in the detailed view of the same sub-process

In BPMN, a sub-process can be used for business process modularization, to hide complexity at upper levels of the process representation, and unveiling specific details only at the lower levels [7].

As referred in section 4, since BPMN metamodel do not allow a sub-process to directly exchange messages with other external process, we replaced the inheritance relationship between Task and *InteractionNode* in the metamodel by a new inheritance between Activity and *InteractionNode*. However, to implement this modification in the metamodel, care should be taken to avoid inconsistent models, i.e., models with messages exchanged with a collapsed *SubProcess*, not seen in the detailed view of the *SubProcess*. This will be ensured, through OCL constraints that will enforce the following:

- A message flow from a collapsed sub-process must be replicated in the child-level diagram.
- A message flow to a collapsed sub-process must be replicated in the child level diagram.
- An incoming message flow showed in the sub-process detailed view diagram is replicated in the collapsed view of the sub-process.
- An outgoing message flow showed in the sub-process detailed view diagram should is replicated in the collapsed view of the sub-process.
- The set of elements that originate the incoming messages are the same for the collapsed and detailed sub-process
- The set of elements that are the target of the outgoing messages are the same for the collapsed and detailed sub-process

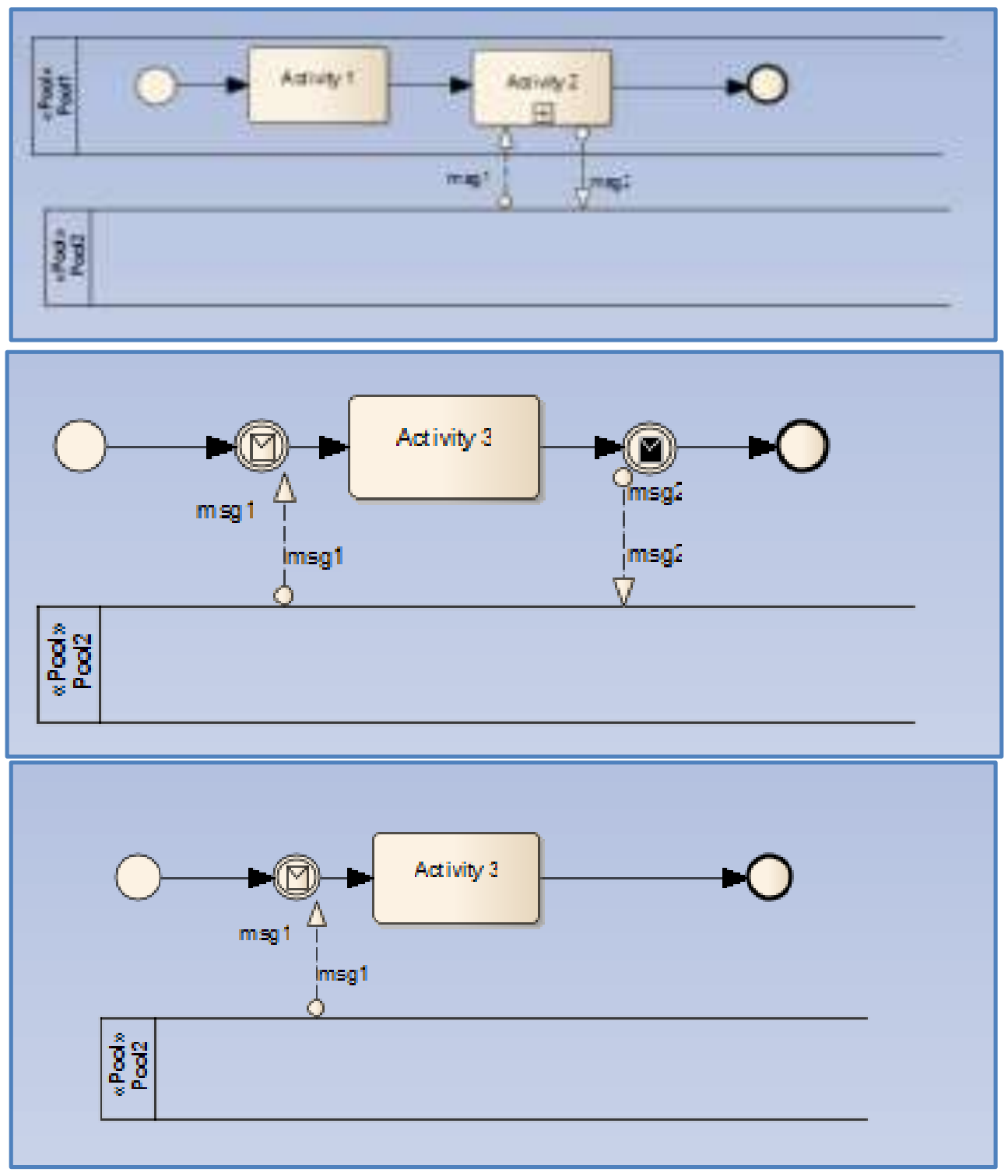

**Fig. 13.** – The messages flows interacting with a collapsed view of a sub-process must be the same as the depicted in the detailed view of the same sub-process

**Picture's interpretation**: <u>Correct</u>: The Message Flows exchanged with the collapsed sub-process, at process level (top) are the same showed in the detailed view of the sub-process (middle); <u>Wrong</u>: There are inconsistencies between sub-process's representations at detail level (bottom) compared with the collapsed view of the sub-process (top).

The well-formedness rule regarding message flow can be enforced by attaching the following invariant to the *SubProcess* element of the BPMN metamodel.

```
context SubProcess
inv messageFlowIsConsistent:
    isMessageFlowconsistent()
```

**Listing 10 – The messages flows interacting with a collapsed view of a sub-process must be the same as the depicted in the detailed view of the same sub-process**

### 5.1.7. A Sub-Process can only have one None Start Event

A *SubProcess* (known as *Embedded SubProcess* in BPMN 1.2) can only have one type of *Start Event*: the *None Start Event* [1] (page 241). Since the sub-process starts after having been triggered by the parent process, it must be clear where the sub-process begins, i.e., the *None Start Event*.

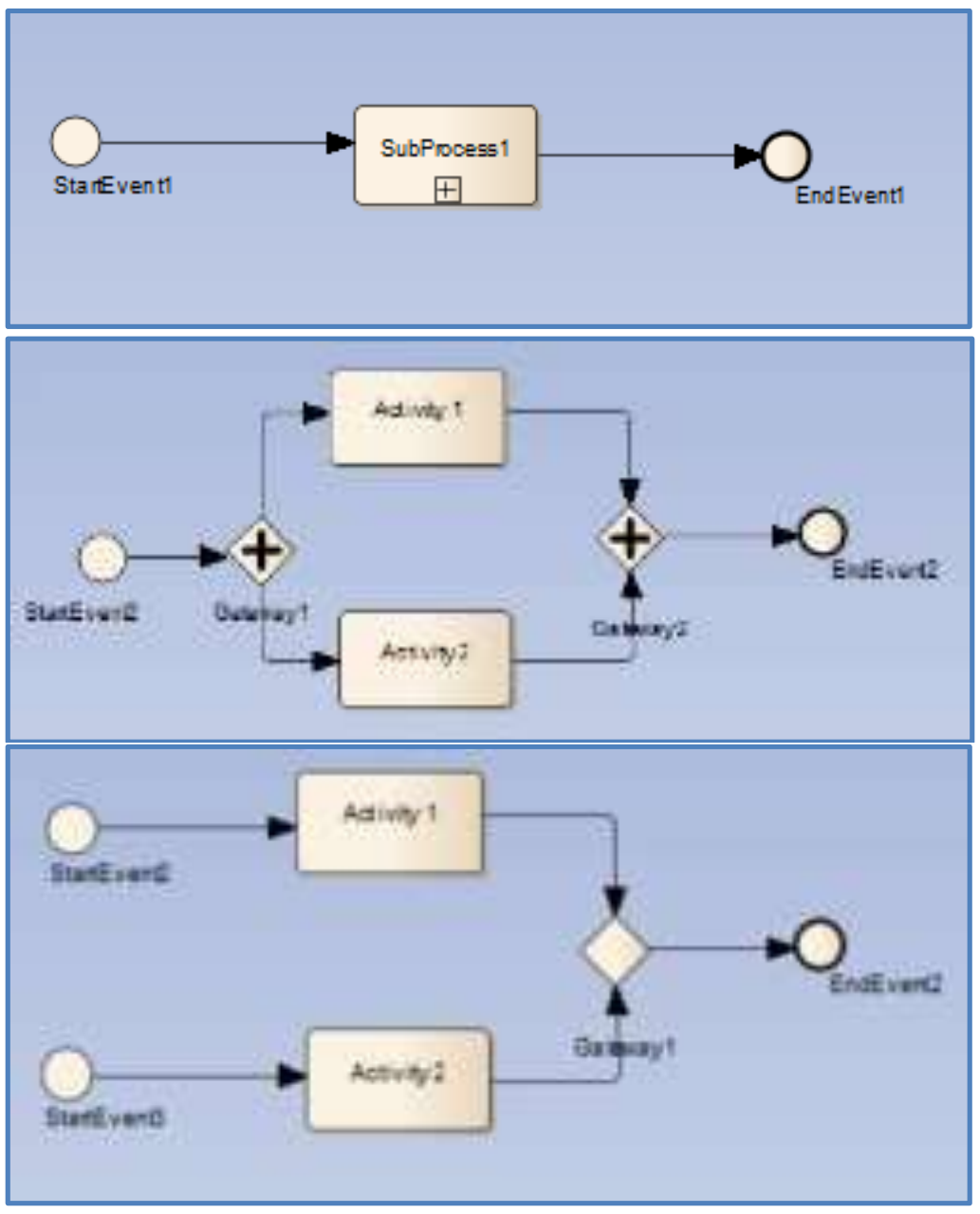

**Fig. 14.** – A *SubProcess* can only have one *None Start Event*

**Picture's interpretation**: Correct: In the middle we have only one *None Start Event* for the sub-process depicted at top; Wrong: Several *None Start Events* for the same sub-process.

The well-formedness rule regarding start events can be enforced by attaching the following invariant to the *Subrocess* element of the BPMN metamodel.

```
context SubProcess
  inv onlyOneNoneStartEventInSubProcess:
    subProcesshasOnlyOneNoneStartEvent()
```

**Listing 11 – A *SubProcess* can only have one *None Start Event***

### 5.1.8. An Event Sub-Process can only have a single Start Event and it must be typed

An Event Sub-Process must not have more than one Start Event. The allowed Start Event types are: Message, Timer, Escalation, Error, Compensation, Conditional, Signal, Multiple, and Parallel (see second column in Fig. 4). Furthermore, the non-interrupting variants of the mentioned catch events are only possible within an event sub-process, and not in a normal process (see third column in Fig. 4) [1] (page 242).

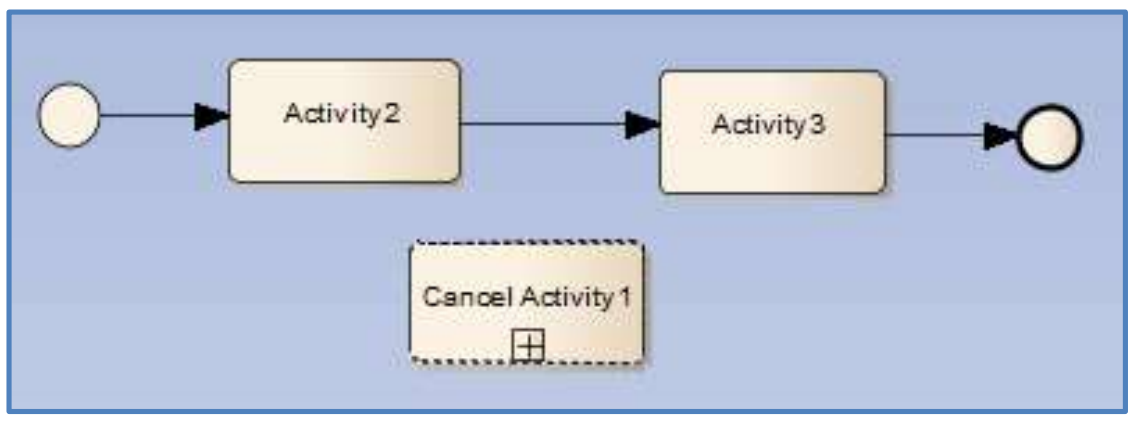

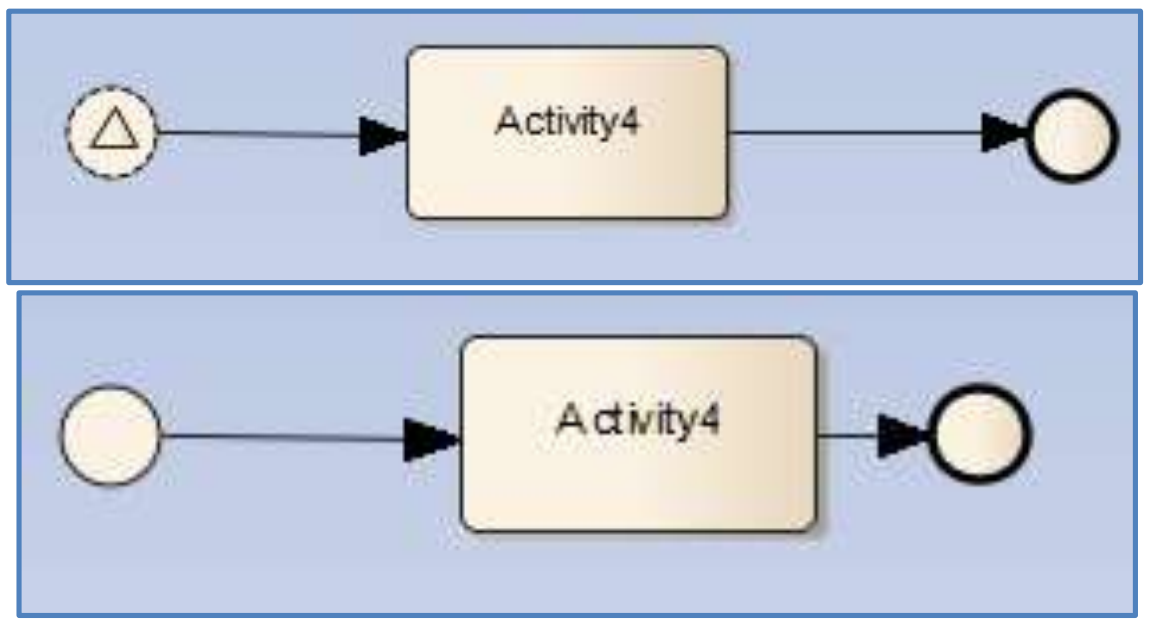

**Fig. 15.** – An *Event Sub-Process* can only have a single *Start Event* and it must be typed

**Picture's interpretation**: Correct: The picture in the middle shows one *Start Event* of allowed type in an *Event SubProcess* "Cancel Activity1" (top); Wrong: The picture in the bottom shows an untyped *Start Event* not allowed in an *Event SubProcess*.

The well-formedness rule regarding start events can be enforced by attaching the following invariants to the *Sub-Process* and *StartEvent* elements of the BPMN metamodel.

```
context SubProcess
  inv eventSubProcessTypedStartEventAllowed:
    onlyTypedStartEventAllowed()
  inv eventSubProcessesDoNotAllowUntypedStartEvent:
    self.isEventSubProcess() implies
           self.totalNumberContainerNoneStartEvents() = 0
```

**Listing 12 – An *Event Sub-Process* can only have a single *Start Event* and it must be typed**

### 5.1.9. An Event Sub-Process must not have any incoming or outgoing Sequence Flows

An *Event SubProcess* must not have incoming or outgoing *Sequence Flow*, since it is the *Start Event* that triggers the sub-process execution [1] (page 176).

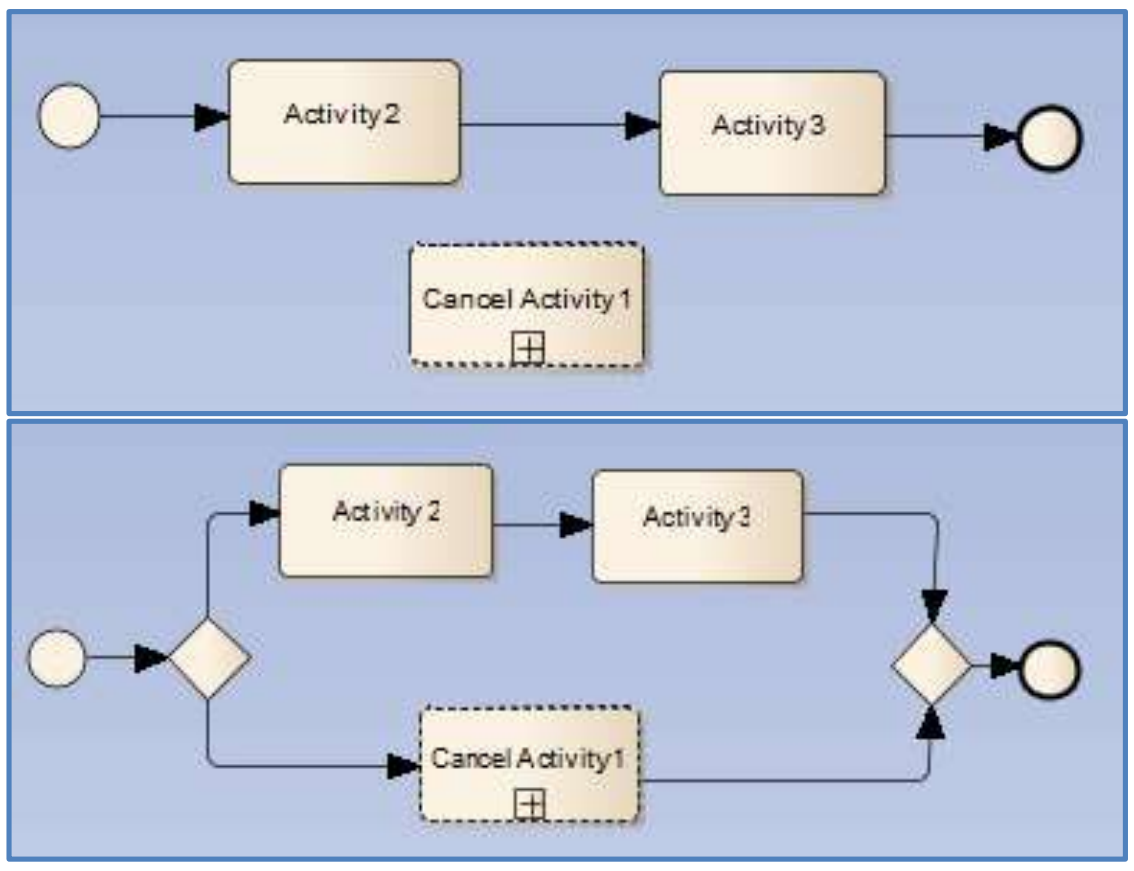

**Fig. 16.** – An Event Sub-Process must not have any incoming or outgoing Sequence Flows

**Picture's interpretation**: Correct: An event sub-process without incoming or outgoing Sequence Flows (top); Wrong: An event sub-process must not have any incoming or outgoing Sequence Flows (bottom).

The well-formedness rule regarding incoming or outgoing sequence flows can be enforced by attaching the following invariant to the SubProcess element of the BPMN metamodel.

```
context SubProcess
  inv noIncomingAndOutgoingSequenceFlow:
    hasNoIncomingOutgoingAndSequenceFlow()
```

**Listing 13 – An Event Sub-Process must not have any incoming or outgoing Sequence Flows**

### Flow Nodes

Flow nodes are the elements that create the main structure of a Process. These elements are *Activities*, *Events*, and *Gateways*.

#### 5.1.10. If a container includes Start and End Events, as general rule, all Flow Nodes must have at least one incoming or one outgoing Sequence Flow

As general rule it is not allowed to have isolated flow nodes. So flow nodes must have one incoming or one outgoing sequence flow [1] (page 99, 153, 259, 431*)*. There are some exceptions to the rule which are *Event SubProcess, Compensation Activity*, *Boundary Event* of type compensation, and the activities contained within an *Ad-Hoc Sub-Process*.

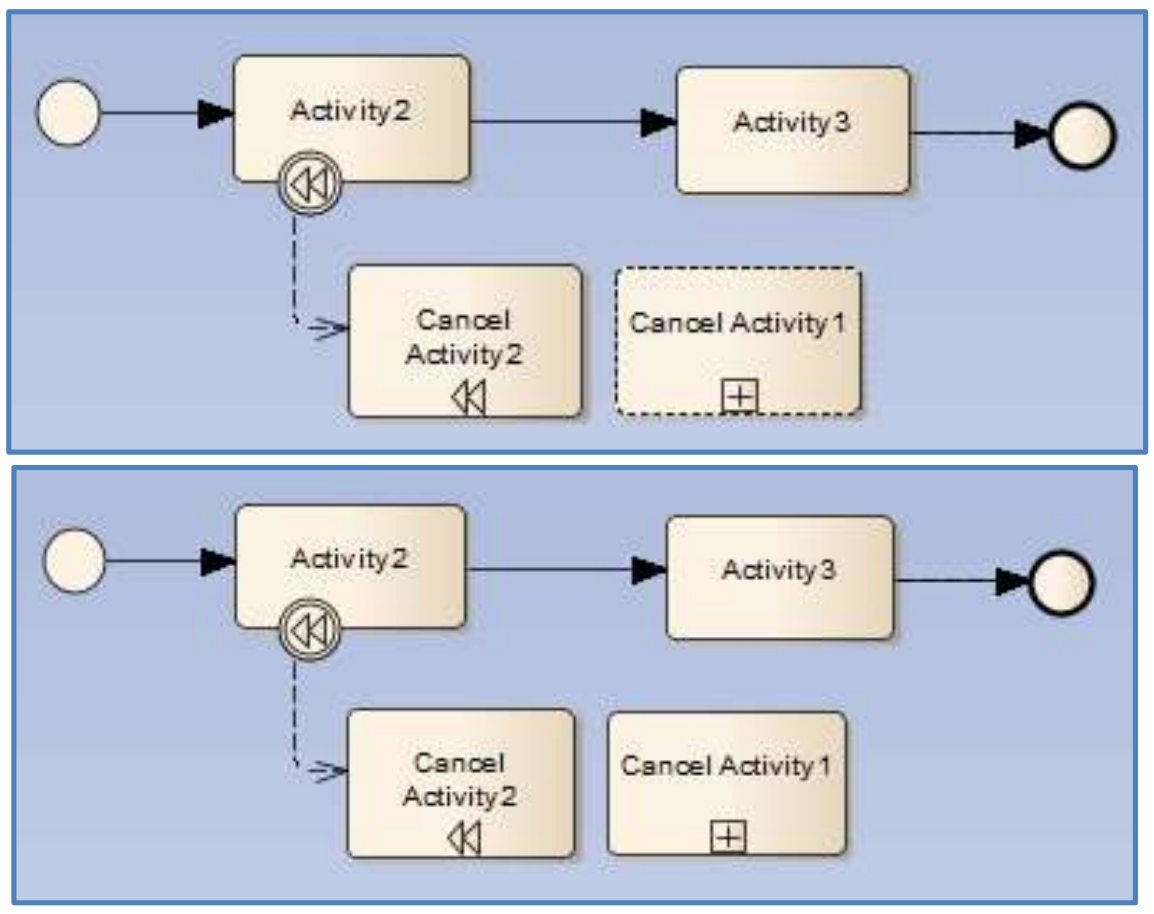

**Fig. 17.** – If a container includes Start and End Events, as general rule, all Flow Nodes must have at least one incoming or one outgoing Sequence Flow

**Picture's interpretation**: Correct: All flow nodes are connected. The allowed exceptions are an event sub-process and a compensation activity (top); Wrong: A sub-process doesn't have at least one incoming or one outgoing sequence flow (bottom).

The well-formedness rule regarding incoming or outgoing sequence flows can be enforced by attaching the following invariant to the *FlowElementsContainer* element of the BPMN metamodel. This rule is also covered by the invariant presented in section 5.1.55.

```
context FlowElementsContainer
inv mandatoryIncomingOrOutgoingSequenceFlow:
    hasIncomingOrOutgoingSequenceFlow()
```

**Listing 14 – If a container includes Start and End Events, as general rule, all Flow Nodes must have at least one incoming or one outgoing Sequence Flow**

#### 5.1.11. A Flow Node, in a container that includes start and end events, must have at least one incoming or one outgoing sequence flow

If the container includes start and end events, all flow objects must have [1] (pages 99, 153, 245, 249, 259, 431):

- an incoming sequence flow. Exceptions are: Start Events, Boundary Events, Event SubProcess , and Compensation Activities;
- an outgoing sequence flow. Exceptions are:  End Events, Event SubProcess, and Compensation Activities.

The well-formedness rule regarding incoming or outgoing sequence flows can be enforced by attaching the following invariant to the FlowNode element of the BPMN metamodel.

```
context FlowNode
inv mandatoryIncomingAndOutgoingSequenceFlow:
    hasIncomingAndOutgoingSequenceFlow()
```

**Listing 15 – A Flow Node, in a container that includes start and end events, must have at least one incoming or one outgoing sequence flow**

**Events**

An Event is something that affects the flow of the Process and usually has a *trigger* or a *result.*

### 5.1.12. Only some predefined types of *Start, Intermediate* and *End Events* are allowed in specific contexts

The allowed triggers and results of *Start*, *Intermediate* and *End Events* in specific contexts are explicitly defined in [1] (page 261). Fig. 4 depicts all possible cases, which are enforced by the following invariants.

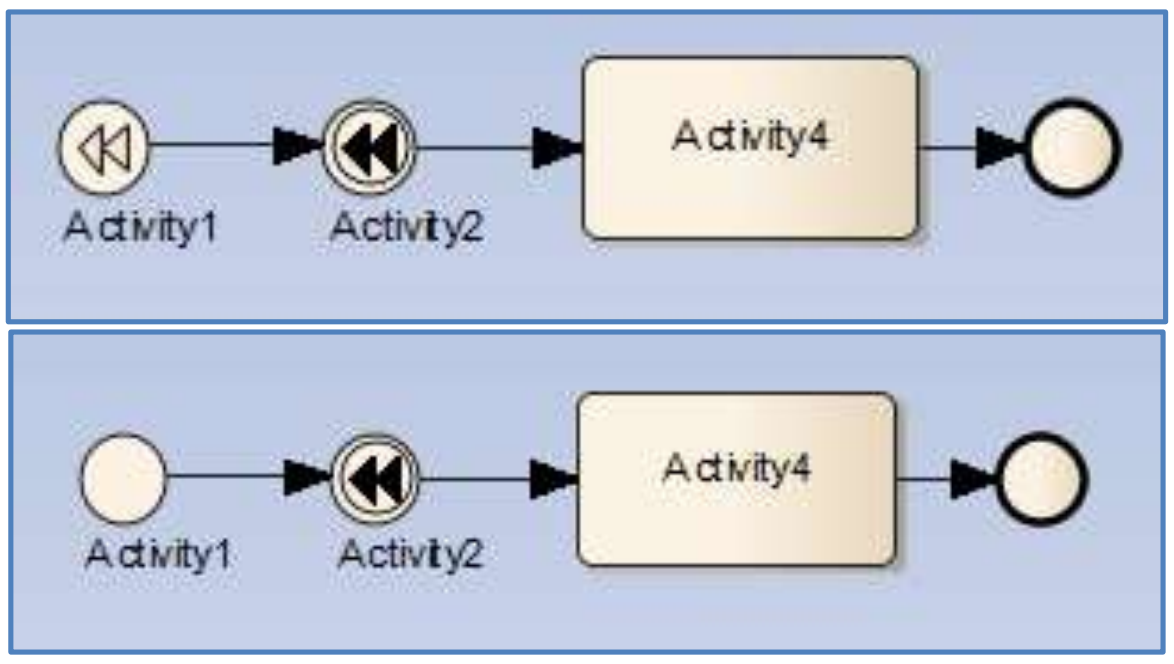

**Fig. 18.** – Only some predefined types of *Start*, *Intermediate* and *End Events* are allowed in specific contexts (I)

**Picture's interpretation**: Correct: The picture at the top shows one interrupting *Start Event* of allowed type in an *Event SubProcess*; Wrong: The picture at the bottom shows an untyped interrupting *Start Event* not allowed in an *Event SubProcess.*

The well-formedness rule regarding events can be enforced by attaching the following invariants to the BPMN metamodel.

```
context StartEvent
inv eventSubProcessAllowedEventType:
      (self.container.oclIsKindOf(SubProcess)
        and self.container.oclAsType(SubProcess).isEventSubProcess())
      implies
      ((self.isInterruptingEvent() implies
      (not (self.isCancelEvent() or
            self.isNoneEvent() or
            self.isLinkEvent() or
            self.isTerminateEvent())))
      and
      (self.isNonInterruptingEvent() implies
      (not (self.isCancelEvent() or
            self.isNoneEvent() or
            self.isLinkEvent() or
            self.isTerminateEvent() or
            self.isErrorEvent() or
            self.isCompensateEvent())))
```

**Listing 16 – Only some predefined types of *Start, Intermediate* and *End Events* are allowed in specific contexts (I)**

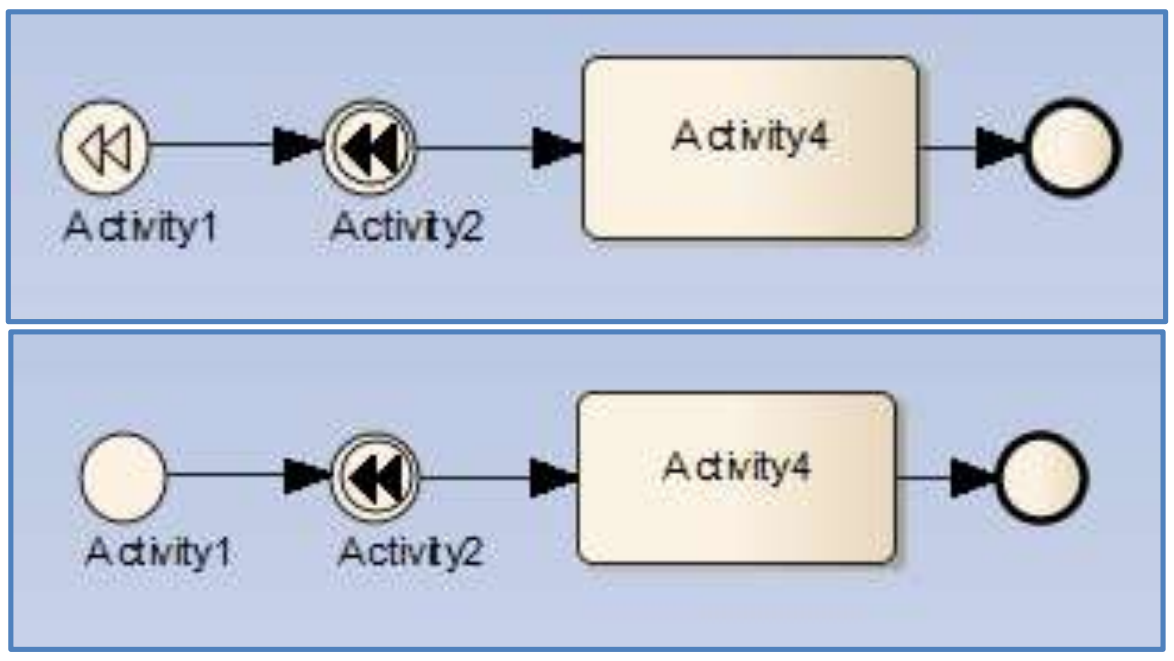

**Fig. 19.** – Only some predefined types of Start, Intermediate and End Events are allowed in specific contexts (II)

**Picture's interpretation**: Correct: The picture at the top shows one interrupting *Start Event* of allowed type in an *Event SubProcess*; Wrong: The picture at the bottom shows an untyped interrupting *Start Event* not allowed in an *Event SubProcess.*

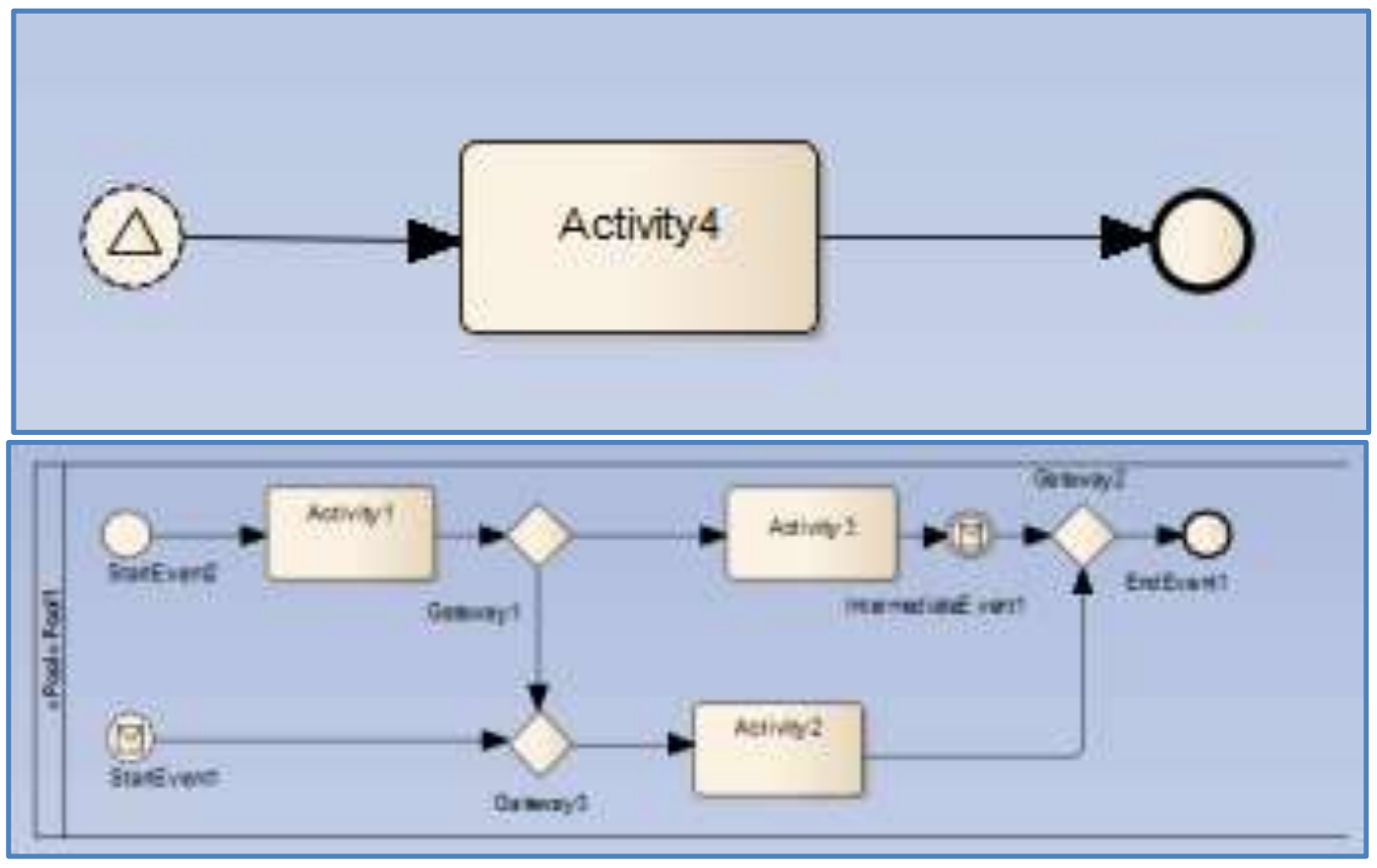

**Fig. 20.** – Only some predefined types of Start, Intermediate and End Events are allowed in specific contexts (III)

**Picture's interpretation**: Correct: The picture at the top shows one non-interrupting *Start Event* inside an *Event SubProcess*; Wrong: The picture at the bottom shows a non-interrupting *Start Event* not allowed in a top-level process.

```
context StartEvent
  inv nonInterruptingStartEvTypesRestrictedToEventSubProcess:
    isContainerEventSubProcess()
```

**Listing 17 – Only some predefined types of *Start, Intermediate* and *End Events* are allowed in specific contexts (II)**

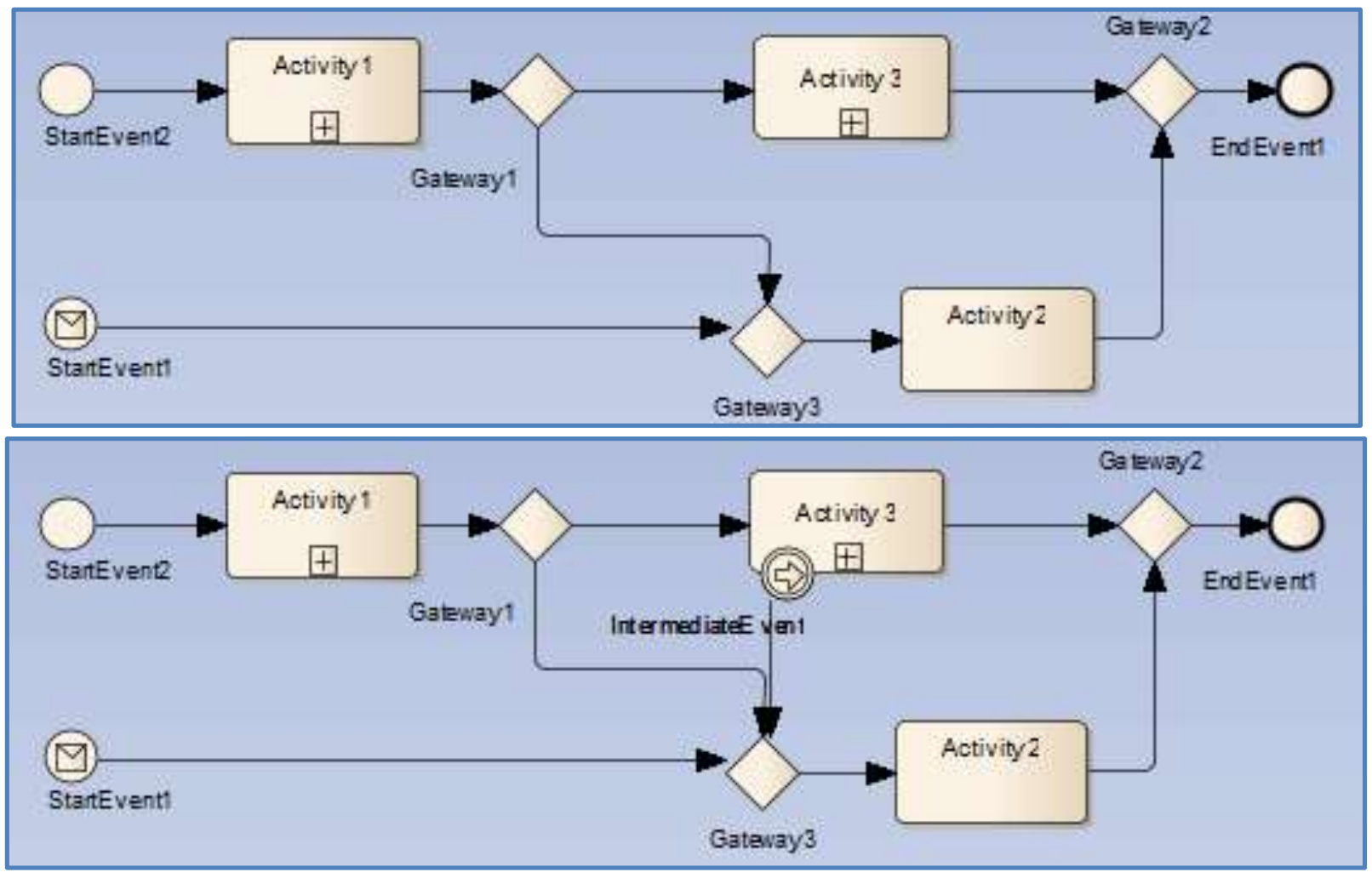

**Fig. 21.** – Only some predefined types of *Start*, *Intermediate* and *End Events* are allowed in specific contexts (IV)

**Picture's interpretation**: Correct: The picture at the top shows allowed event types in a top-level process; Wrong: The picture at the bottom shows an invalid type of a *Boundary Event*.

```
context BoundaryEvent
inv boundaryEventAllowedEventType:
      ((self.isInterruptingEvent()) implies
      (not (self.isNoneEvent() or
            self.isLinkEvent() or
            self.isTerminateEvent())))
      and
      ((self.isNonInterruptingEvent()) implies
      (not (self.isCancelEvent() or
            self.isNoneEvent() or
            self.isLinkEvent() or
            self.isTerminateEvent() or
            self.isErrorEvent() or
            self.isCompensateEvent())))
```

**Listing 18 – Only some predefined types of *Start, Intermediate* and *End Events* are allowed in specific contexts (III)**

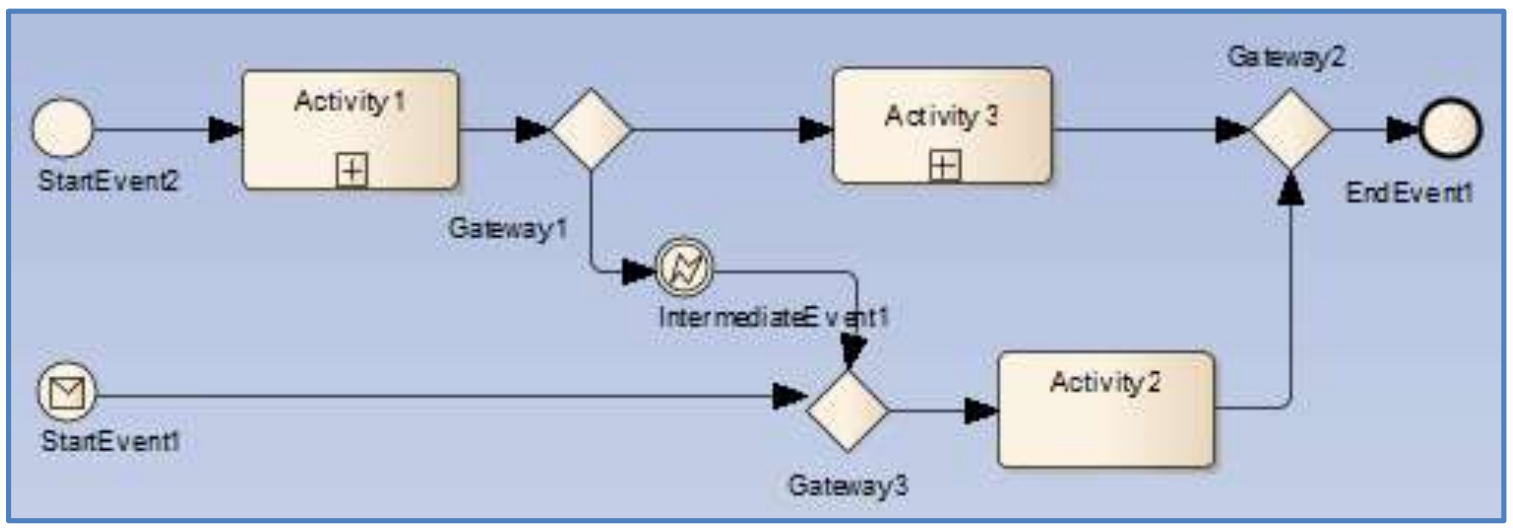

**Fig. 22.** – Only some predefined types of *Start*, *Intermediate* and *End Events* are allowed in specific contexts (V)

**Picture's interpretation**: Wrong: The picture shows an invalid type of an Intermediate Catch Event.

```
context IntermediateCatchEvent
inv intermediateCatchEventAllowedEventType:
      (not (self.isNoneEvent() or
            self.isErrorEvent()or
            self.isEscalationEvent()or
            self.isCancelEvent()or
            self.isTerminateEvent()or
            self.isCompensateEvent()))
```

**Listing 19 – Only some predefined types of *Start, Intermediate* and *End Events* are allowed in specific contexts (IV)**

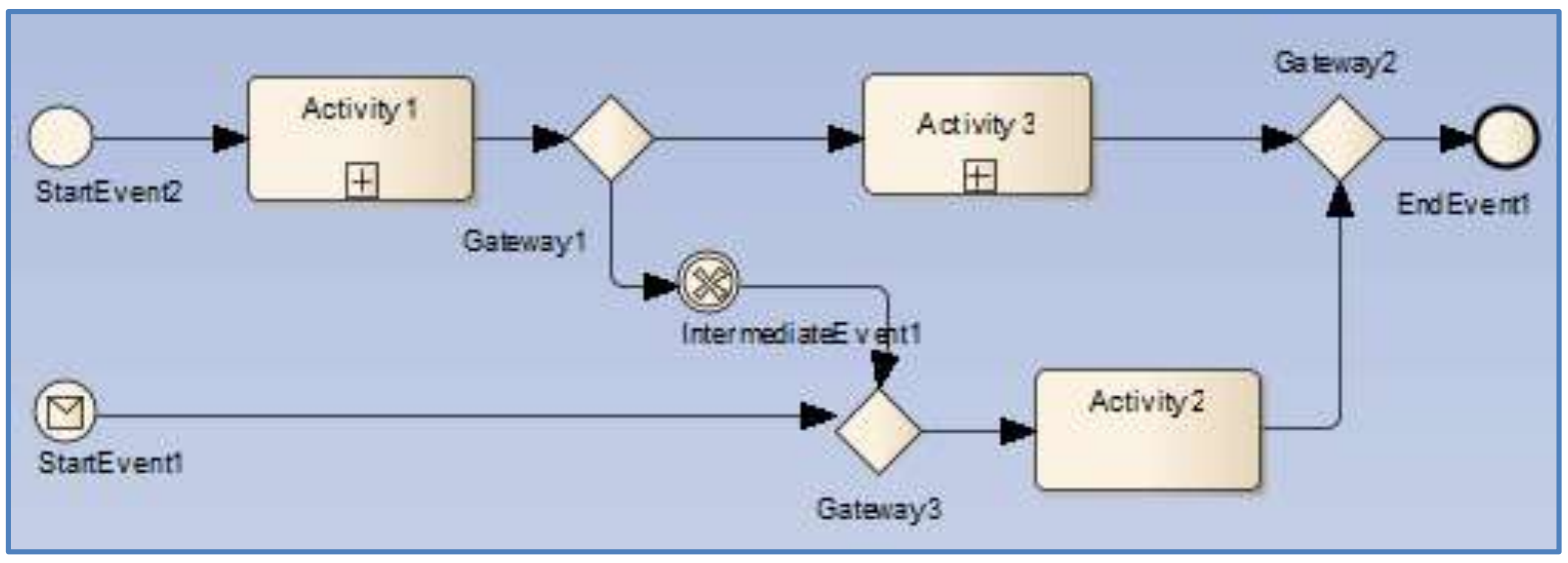

**Fig. 23.** – Only some predefined types of *Start*, *Intermediate* and *End Events* are allowed in specific contexts (VI)

**Picture's interpretation**: Wrong: The picture shows an invalid type of an *Intermediate Throw Event*.

```
context IntermediateThrowEvent
inv intermediateThrowEventAllowedEventType:
      (not (self.isCancelEvent()or
            self.isTimerEvent()or
            self.isErrorEvent()or
            self.isTerminateEvent()or
            self.isParallelMultipleEvent()or
            self.isConditionalEvent()))
```

**Listing 20 – Only some predefined types of *Start, Intermediate* and *End Events* are allowed in specific contexts (V)**

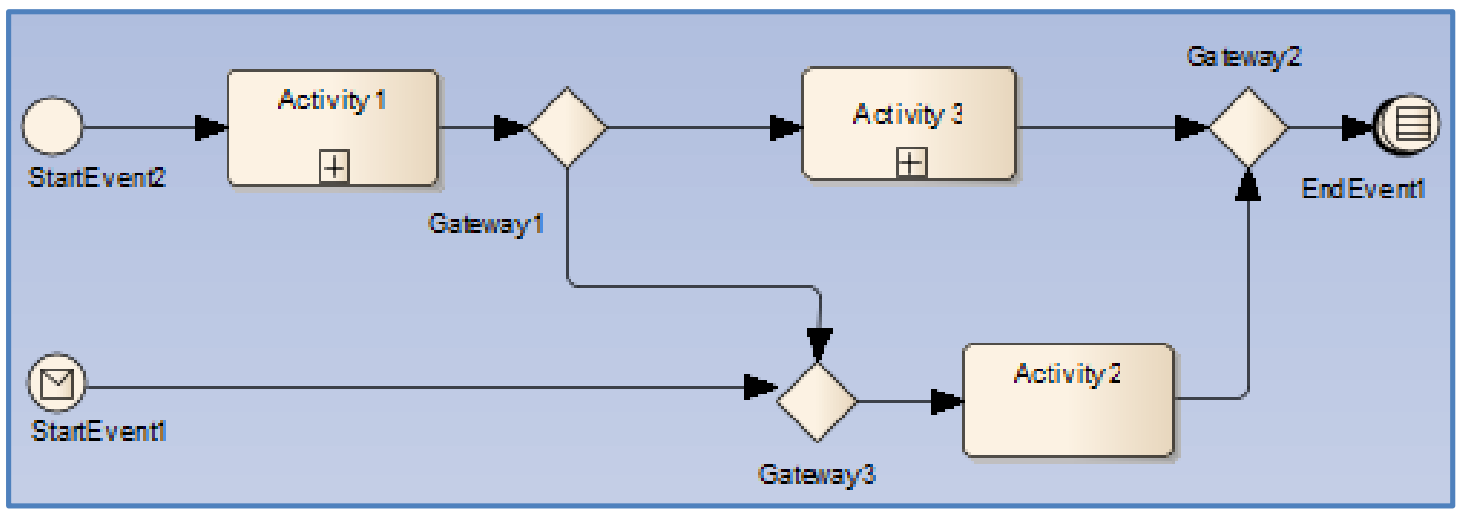

**Fig. 24.** – Only some predefined types of *Start*, *Intermediate* and *End Events* are allowed in specific contexts (VII)

**Picture's interpretation**: Wrong: The picture shows an invalid type of an *End Event*.

```
context EndEvent
inv endEventAllowedEventType:
      not (self.isTimerEvent()or
            self.isConditionalEvent()or
            self.isLinkEvent()or
            self.isParallelMultipleEvent())
```

**Listing 21 – Only some predefined types of *Start, Intermediate* and *End Events* are allowed in specific contexts (VI)**

### 5.1.13. Incoming *Sequence Flow* not allowed in a *Start Event*

The Start Event indicates where a particular process will start. In terms of sequence flows, the Start Event starts the flow of the process, and thus, will not have any incoming sequence flows [1] (page 245).

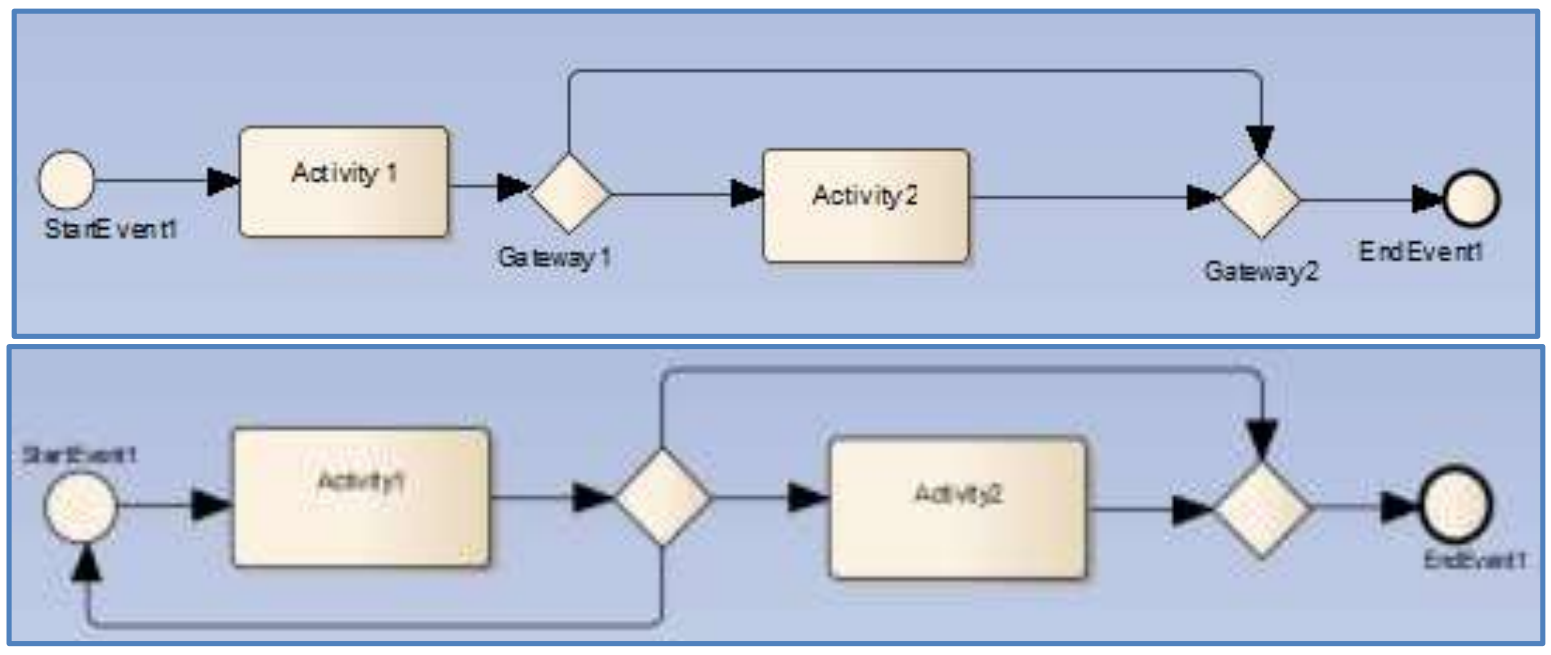

**Fig. 25.** – Incoming *Sequence Flow* not allowed in a *Start Event*

**Picture's interpretation**: Correct: Start event has no incoming sequence flows (top); Wrong: Start event has an incoming sequence flow (bottom).

The well-formedness rule regarding start events can be enforced by attaching the following invariant to the *FlowElementsContainer* and *StartEvent* elements of the BPMN metamodel.

```
context FlowElementsContainer
inv startEventsHaveNoIncomingSequenceFlow:
  self.totalContainerStartEvents()
   ->forAll(numberInputSequenceFlows() = 0)

context StartEvent
  inv cannotHaveIncomingButHaveOutgoingSequenceFlow:
    self.inputSequenceFlows()->isEmpty() and
      self.outputSequenceFlows()->size() > 0
```

**Listing 22 – Incoming *Sequence Flow* not allowed in a *Start Event***

### 5.1.14. Outgoing Sequence Flow not allowed in an End Event

The end event indicates where a process will end. In terms of sequence flows, the end event ends the flow of the process, and thus, will not have any outgoing sequence flows [1] (page 249).

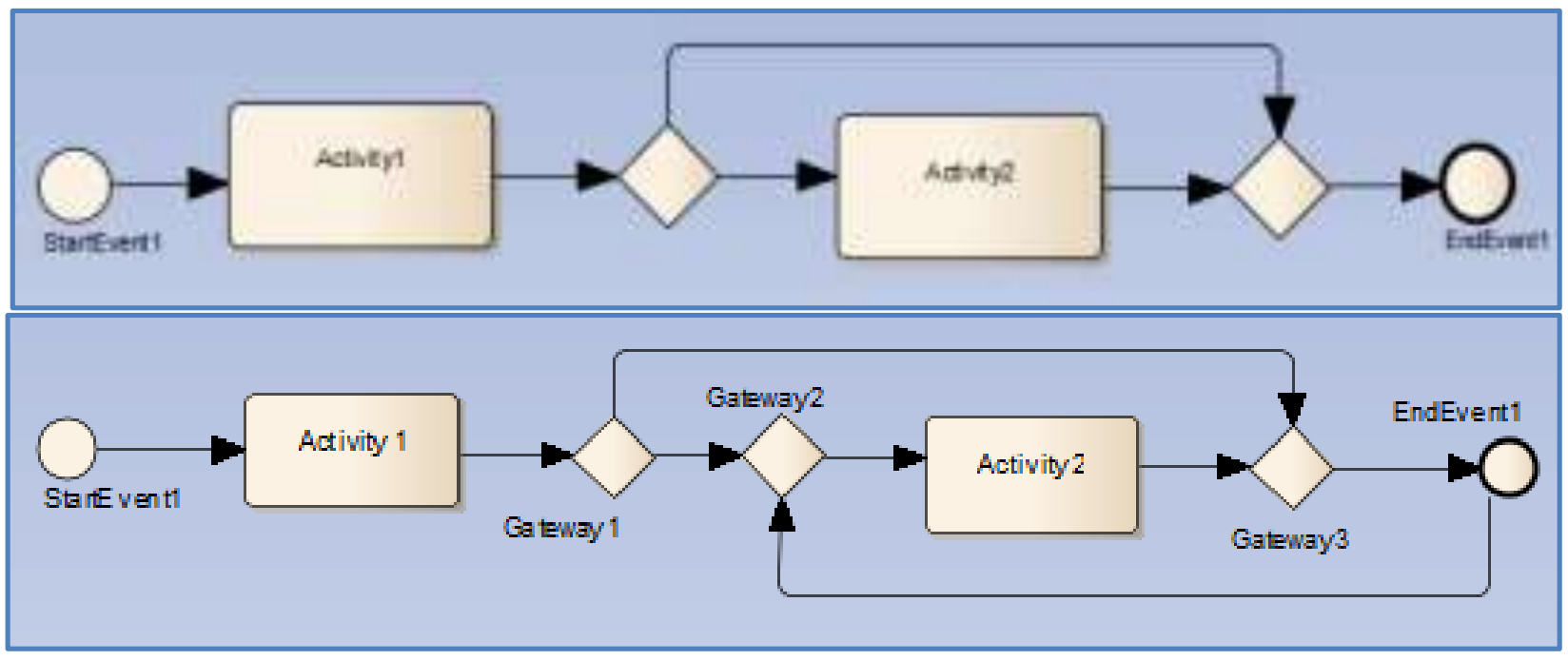

**Fig. 26.** – Outgoing *Sequence Flow* not allowed in an *End Event*

**Picture's interpretation**: Correct: *End Event* has no outgoing sequence flows (top); Wrong: *End Event* has an outgoing sequence flow (bottom).

The well-formedness rule regarding end events can be enforced by attaching the following invariant to the *FlowElementsContainer* element of the BPMN metamodel.

```
context FlowElementsContainer
  inv endEventsHaveNoOutgoingSequenceFlow:
  self.totalContainerEndEvents()
    ->forAll(numberOutputSequenceFlows() = 0)
```

**Listing 23 – Outgoing *Sequence Flow* not allowed in an *End Event***

#### 5.1.15. A Catch Event with incoming Message Flow must have Message or Multiple type

A *Start Event* or a catching *Intermediate Event* with incoming *Message Flow* must be of type *Message* of *Multiple* [1] (page 44, 271).

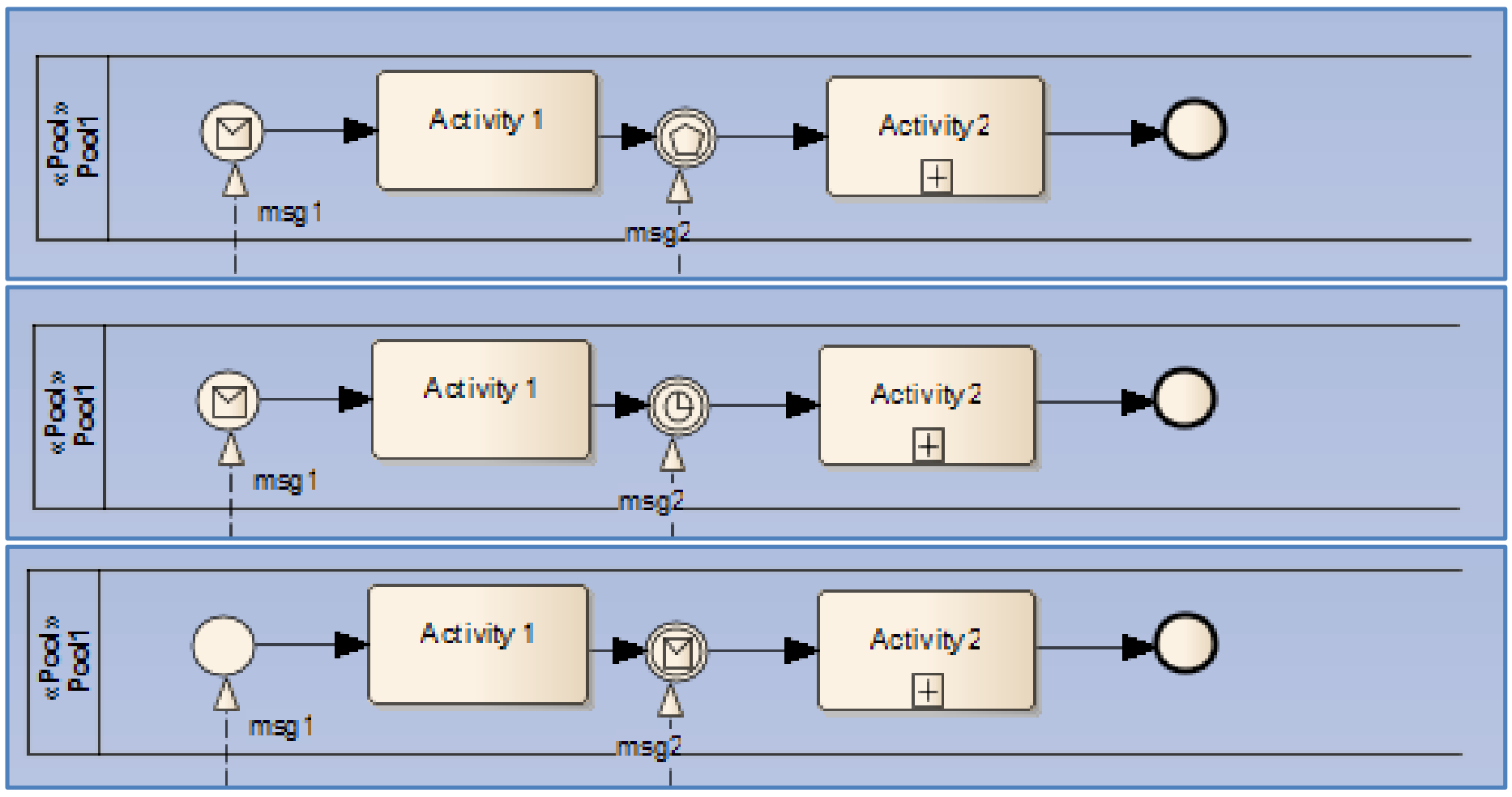

**Fig. 27.** – A *Catch Event* with incoming *Message Flow* must have *Message* or *Multiple* type

**Picture's interpretation**: <u>Correct</u>: *Catch Events* with incoming *Message Flow* are *Message* or *Multiple* types (top); <u>Wrong</u>: *Catch Events* with incoming *Message Flow* cannot be of *Timer* type (middle) or untyped (bottom).

The well-formedness rule regarding Start Events or catching Intermediate Events can be enforced by attaching the following invariant to the *CatchEvent* element of the BPMN metamodel.

```
context CatchEvent
  inv incomingMessageFlowHasMessageMultipleType:
    self.numberInputMessageFlows() > 0
    implies
    (self.isMessageEvent() or self.isMultipleEvent())
```

**Listing 24 – A *Catch Event* with incoming *Message Flow* must have *Message* or *Multiple* type**

#### 5.1.16. A *Catch Event* must have *Multiple* type if there are more than one incoming *Message Flow*

A *Start Event* or a catching *Intermediate Event* with more than one incoming *Message Flow* must be of type *Multiple* [1] (page 241, 253).

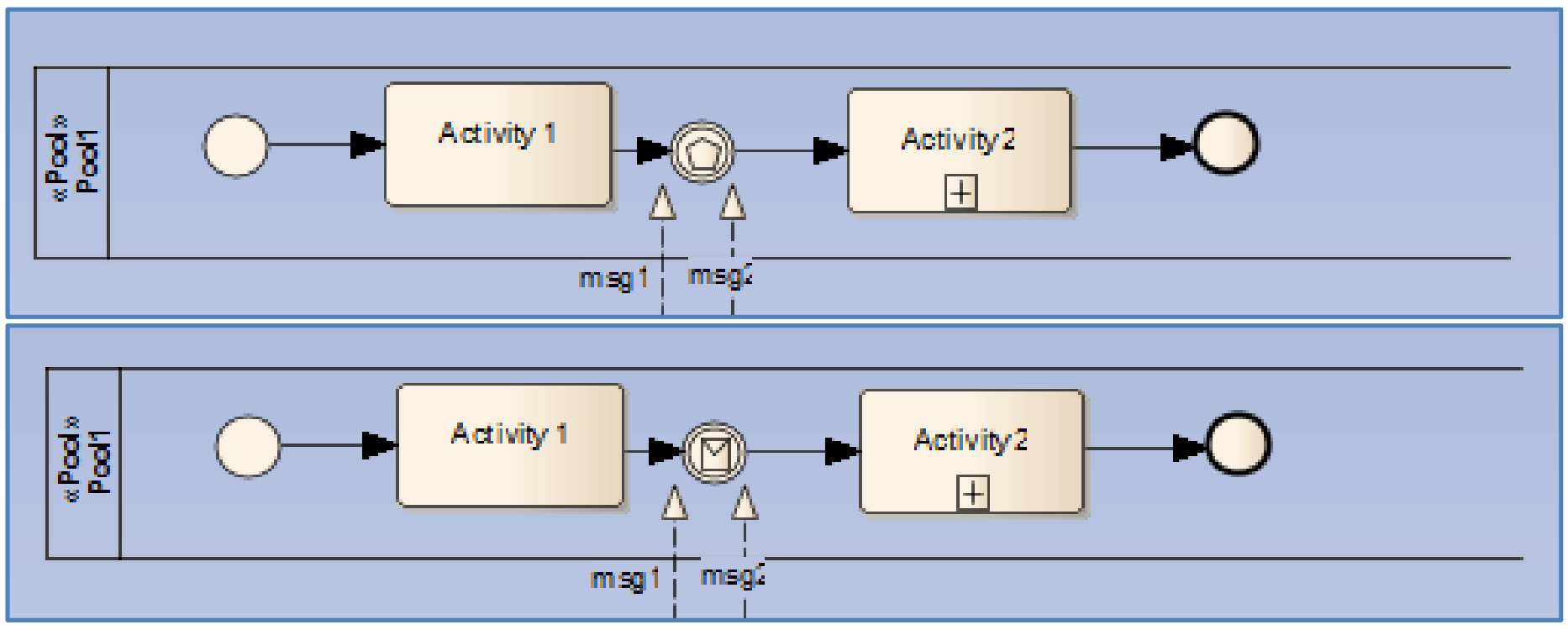

**Fig. 28.** –A *Catch Event* must have *Multiple* type if there are more than one incoming *Message Flow*

**Picture's interpretation**: <u>Correct</u>: *Catch Events* with several incoming *Message Flow* are of *Multiple* type (top); <u>Wrong</u>: *Catch Events* with several incoming *Message Flow* must not be of type *Message* (bottom).

The well-formedness rule regarding *Start Events* or *Catching Intermediate* events can be enforced by attaching the following invariant to the *CatchEvent* element of the BPMN metamodel.

```
context CatchEvent
inv multipleIncomingMesgFlowHasMultipleType:
    self.numberInputMessageFlows() > 1
    implies
    self.isMultipleEvent()
```

**Listing 25 – A *Catch Event* must have *Multiple* type if there are more than one incoming *Message Flow***

### 5.1.17. Outgoing *Message Flow* not allowed in a *Catch Event*

A *Start Event* or a catching *Intermediate Event* cannot have outgoing message flows [1] (page 251).

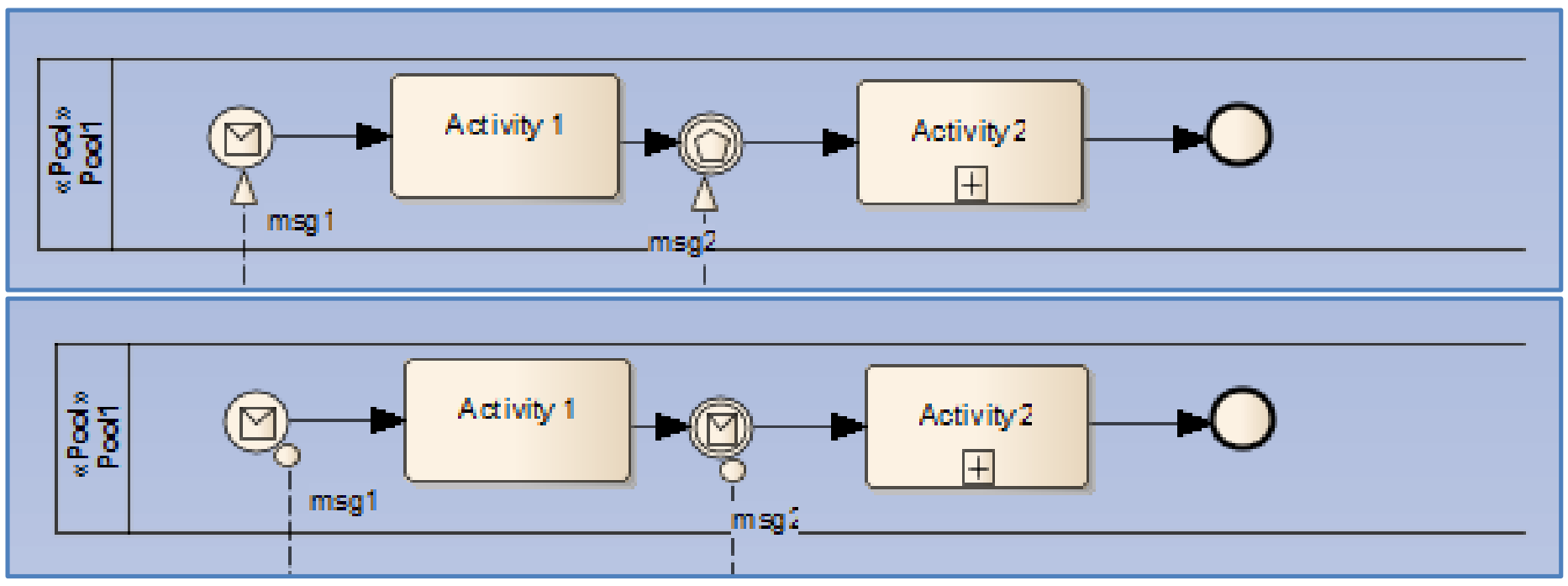

**Fig. 29.** – Outgoing *Message Flow* not allowed in a *Catch Event*

**Picture's interpretation**: Correct: *Catch Events* with incoming *Message Flow* (top); Wrong: *Catch Events* with outgoing *Message Flow* (bottom).

The well-formedness rule regarding start events or catching intermediate events can be enforced by attaching the following invariant to the *CatchEvent* element of the BPMN metamodel.

```
context CatchEvent
inv outgoingMessageFlowNotAllowedMessageType:
    (self.isMessageEvent() or self.isMultipleEvent())
    implies
    self.numberOutputMessageFlows() = 0
```

**Listing 26 – Outgoing *Message Flow* not allowed in a *Catch Event***

### 5.1.18. A *Throw Event* with outgoing *Message Flow* must have *Message* or *Multiple* types.

A *End Event* or a throwing *Intermediate Event* with outgoing message flow must be of type *Message* of *Multiple* [1] (page 249, 251). An *End Event* with outgoing *Message Flow* must have "Message" result.

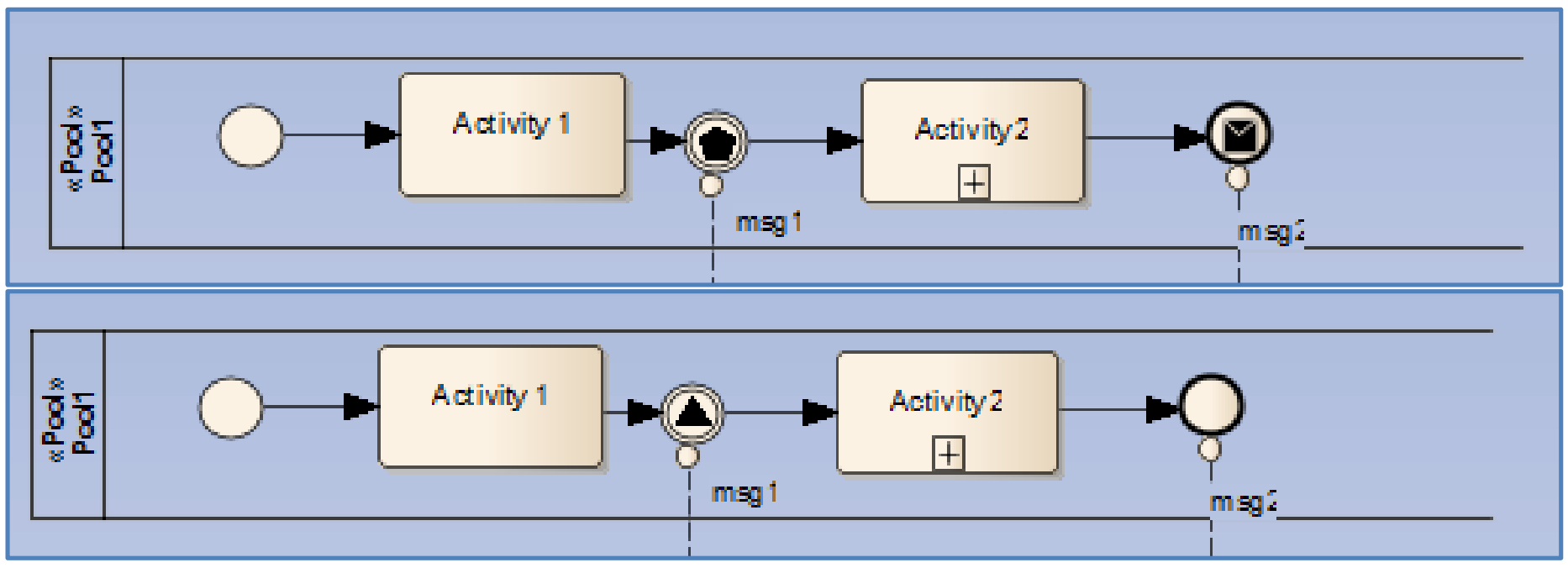

**Fig. 30.** – A *Throw Event* with outgoing *Message Flow* must have *Message* or *Multiple* types

**Picture's interpretation**: Correct: *Throw Events* with outgoing *Message Flow* are *Message* or *Multiple* types (top); Wrong: *Throw Events* with outgoing *Message Flow* are neither *None* or *Signal* types (bottom).

The well-formedness rule regarding end events or throwing intermediate events can be enforced by attaching the following invariant to the *ThrowEvent* element of the BPMN metamodel.

```
context ThrowEvent
inv outgoingMessageFlowHasMessageMultipleType:
    self.numberOutputMessageFlows() > 0
    implies
    (self.isMessageEvent() or self.isMultipleEvent())
```

**Listing 27 – A *Throw Event* with outgoing *Message Flow* must have *Message* or *Multiple* types**

### 5.1.19. A *Throw Event* must have *Multiple* type if there are more than one outgoing *Message Flow*

A *End Event* or a throwing *Intermediate Event* with more than one outgoing *Message Flow* must be of type *Multiple* [1] (page 249, 251).

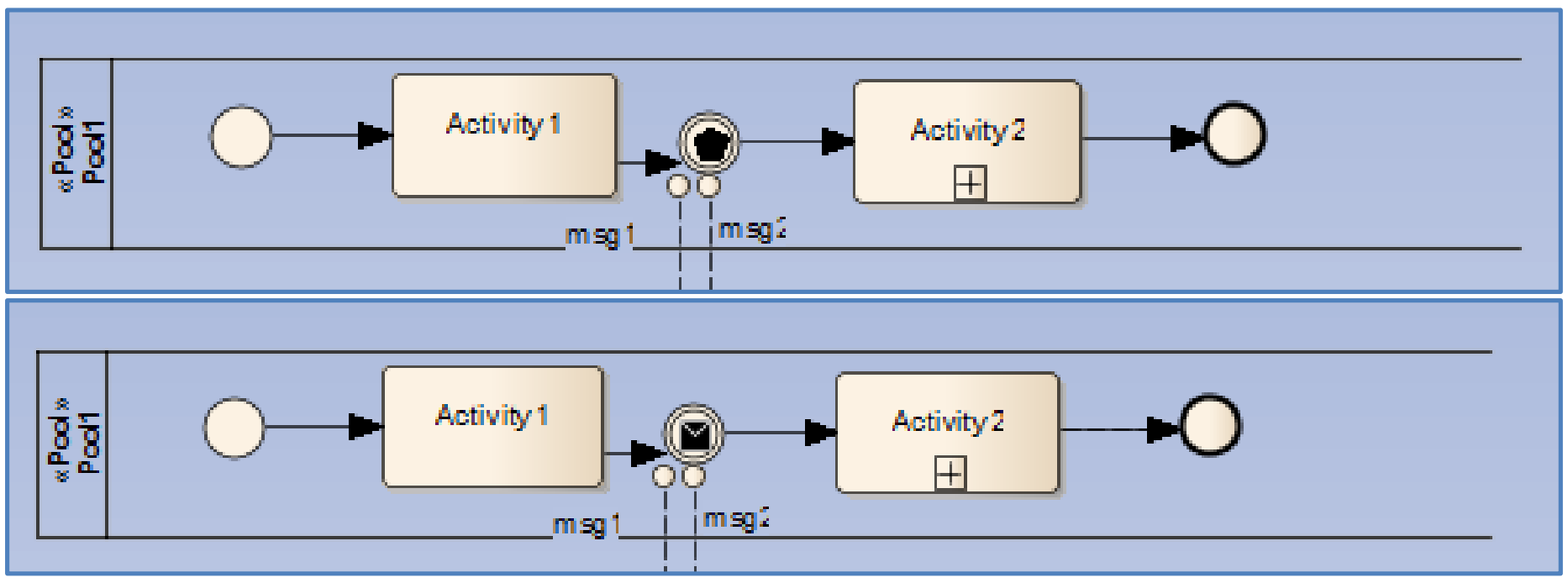

**Fig. 31.** –A *Throw Event* must have *Multiple* type if there are more than one outgoing *Message Flow*

**Picture's interpretation**: Correct: *Throw Events* with several outgoing *Message Flow* are of *Multiple* type (top); Wrong: *Throw Events* with several outgoing *Message Flow* must not be of type *Message* (bottom).

The well-formedness rule regarding *End Events* or throwing *Intermediate Events* can be enforced by attaching the following invariant to the *ThrowEvent* element of the BPMN metamodel.

```
context ThrowEvent
inv multipleOutgoingMesgFlowHasMultipleType:
    self.numberOutputMessageFlows() > 1
    implies
    self.isMultipleEvent()
```

**Listing 28 – A *Throw Event* must have *Multiple* type if there are more than one outgoing *Message Flow***

### 5.1.20. Incoming *Message Flow* not allowed in a *Throw Event*

A *End Event* or a throwing *Intermediate Event* cannot have incoming message flows [1] (page 249, 251).

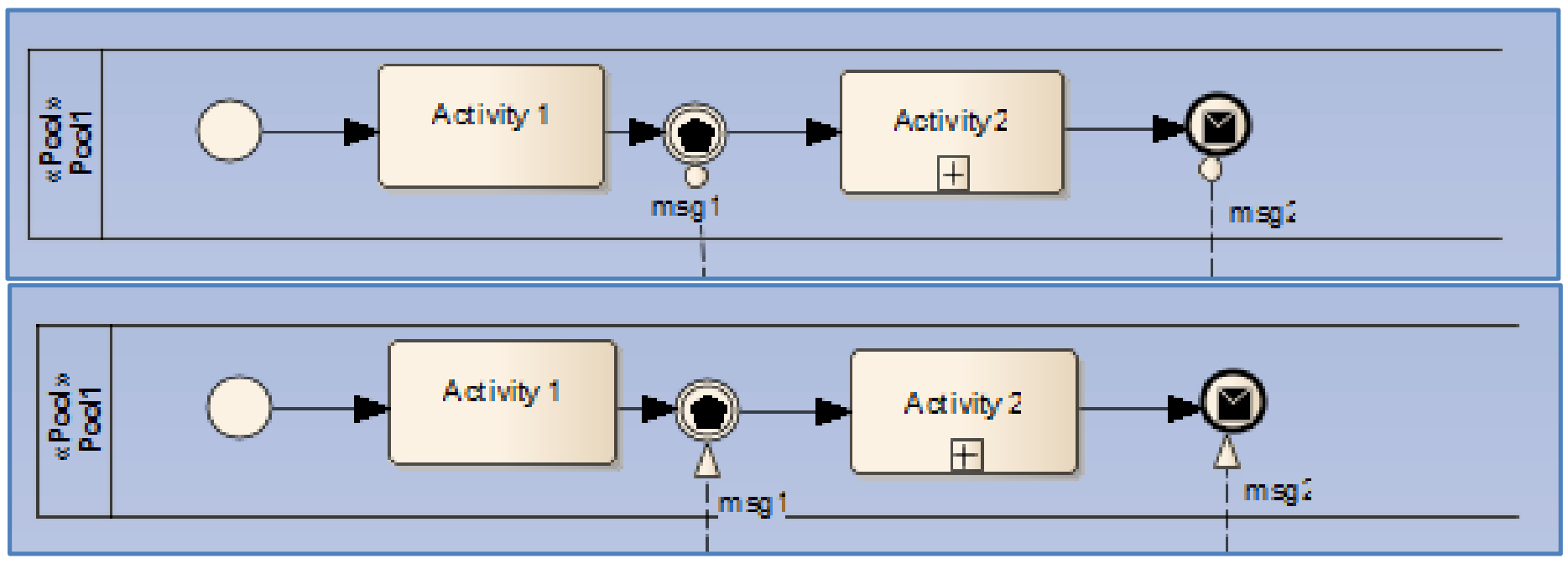

**Fig. 32.** –Incoming *Message Flow* not allowed in a *Throw Event*

**Picture's interpretation**: Correct: *Throw Events* with outgoing *Message Flow* (top); Wrong: *Throw Events* with incoming *Message Flow* (bottom).

The well-formedness rule regarding *End Events* or throwing *Intermediate Events* can be enforced by attaching the following invariant to the *ThrowEvent* element of the BPMN metamodel.

```
context ThrowEvent
inv incomingMessageFlowNotAllowedMessageType:
    (self.isMessageEvent() or self.isMultipleEvent())
    implies
```

```
self.numberInputMessageFlows() = 0
```

**Listing 29 – Incoming *Message Flow* not allowed in a *Throw Event***

### 5.1.21. A *Catch Link Event* must have an incoming *Sequence Flow*. A *Throw Link Event* must have outgoing *Sequence Flow*

A catching (throwing) link intermediate event is not allowed to have incoming (outgoing) sequence flows [1] (page 267-270).

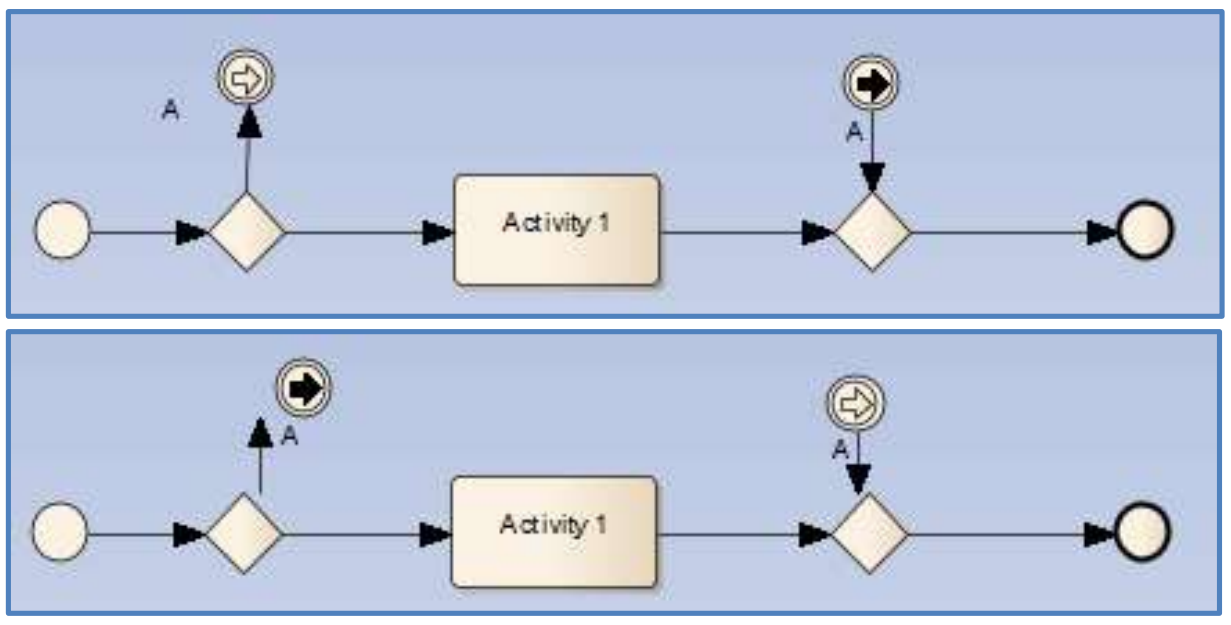

**Fig. 33.** – A *Catch Link Event* must have an incoming *Sequence Flow*. A *Throw Link Event* must have outgoing *Sequence Flow*

**Picture's interpretation**: Correct: The *Throw Link Event* has an incoming *Sequence Flow* and the *Catch Link Event* has an outgoing *Sequence Flow* (top); Wrong: The *Throw Link Event* has an outgoing *Sequence Flow* and the *Catch Link Event* has an incoming *Sequence Flow* (bottom).

The well-formedness rule regarding link events can be enforced by attaching the following invariants to the BPMN metamodel.

```
context CatchEvent
  inv catchingLinkEventHasOnlyOutgoingSequenceFlow:
    self.isLinkEvent() implies
    self.hasOutputSequenceFlows() and
    self.withoutInputSequenceFlows()
```

**Listing 30 – A *Catch Link Event* must have an incoming *Sequence Flow***

```
context ThrowEvent
  inv throwingLinkEventHasOnlyIncomingSequenceFlow:
    self.isLinkEvent() implies
    self.hasInputSequenceFlows() and
    self.withoutOutputSequenceFlows()
```

**Listing 31 – A *Throw Link Event* must have outgoing *Sequence Flow***

### 5.1.22. Multiple *Throw Link Events* allowed, but only one *Catch Link Event*

There can be multiple source *Link Events*, but only one can be target *Link Event* (page 267-270). There would be a separate virtual *Sequence Flow* for each of the Source *Link Events*.

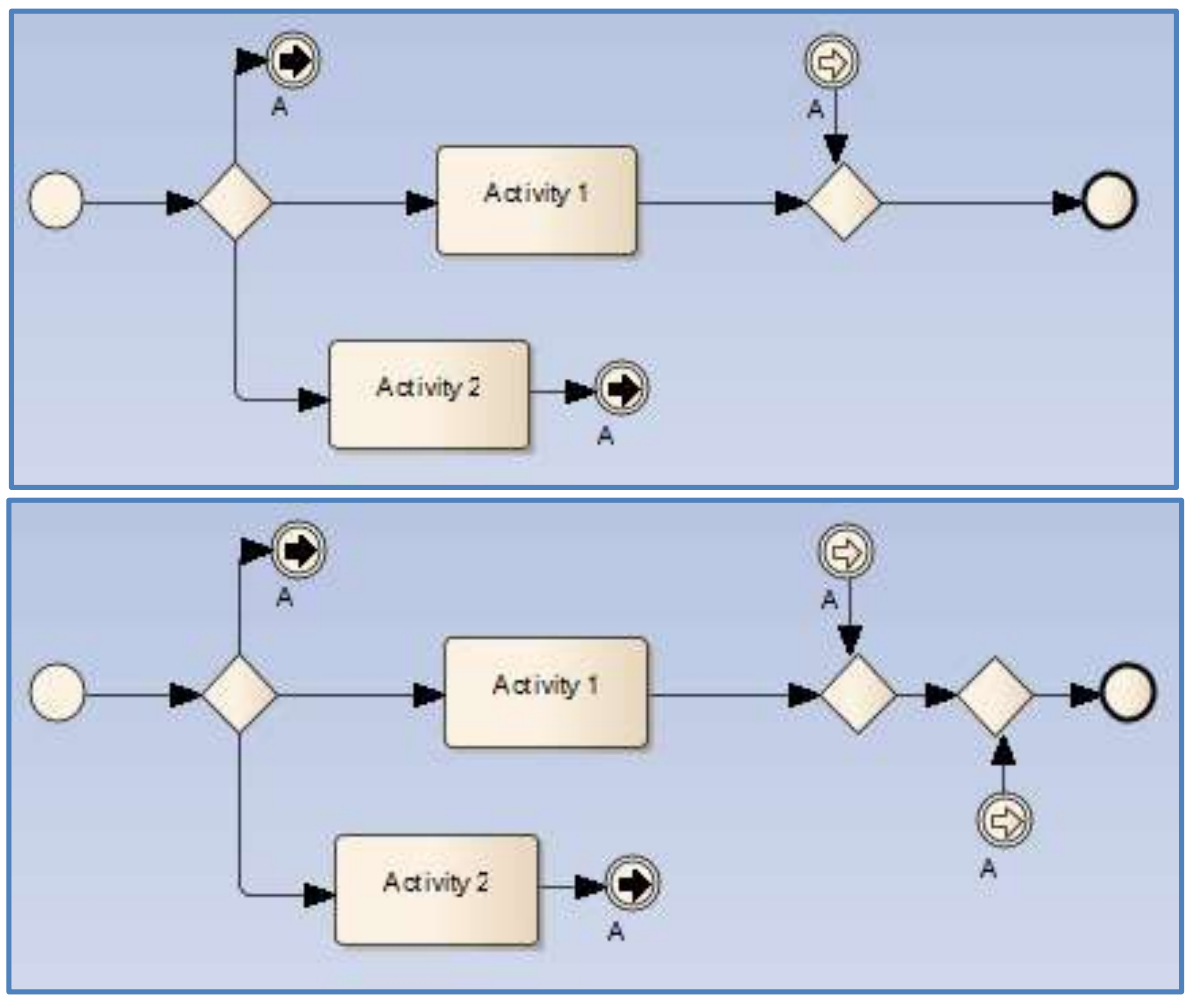

**Fig. 34.** –Multiple *Throw Link Events* allowed, but only one *Catch Link Event*

**Picture's interpretation**: Correct: The number of *Throw Link Event* with the same name can be several, however there must be a single *Catch Link Event* for each name (top); Wrong: There are several *Catch Link* Events with the same name (bottom).

The well-formedness rule regarding link events can be enforced by attaching the following invariant to the *FlowElementsContainer* element of the BPMN metamodel.

```
context FlowElementsContainer
  inv nameOfCatchLinkEventsMustBeUnique:
    nameOfCatchLinkEventsIsUnique()
```

**Listing 32 – Multiple *Throw Link Events* allowed, but only one *Catch Link Event***

### 5.1.23. Source and target *Link Events* must have names

*Link Events* are used in pairs. In order to match throw and catch *Link Events*, their names must be defined [1] (page 268).

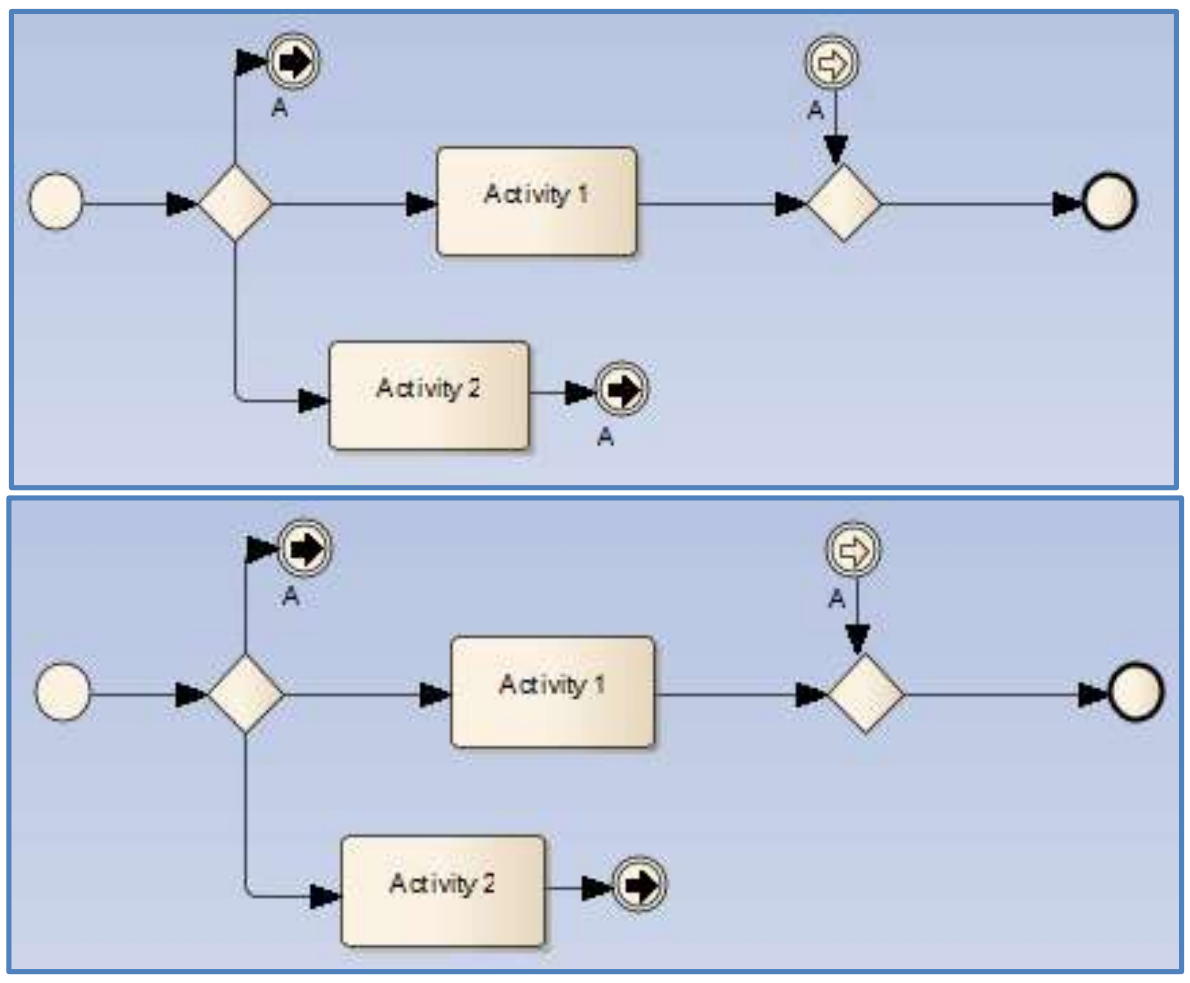

**Fig. 35.** – Source and target *Link Events* must have names

**Picture's interpretation**: Correct: The names of the *Throw Link Events* and *Catch Link Events* are defined; Wrong: There is a *Throw Link Event* without a name.

The well-formedness rule regarding name of link events can be enforced by attaching the following invariants to the *ThrowEvent* and *CatchEvent* elements of the BPMN metamodel.

```
context ThrowEvent
  inv nameOfThrowLinkEventsMustBeDefined:
  isNameOfThrowLinkEventsDefined()
context CatchEvent
  inv nameOfCatchLinkEventsMustBeDefined:
```

```
isNameOfCatchLinkEventsDefined()
```

**Listing 33 – Source and target *Link Events* must have names**

### 5.1.24. *Throw* and *Catch Link Events* names must match in the same container

*Link Events* are only used as Intermediate Events and must exist within a single Process level. So, the name of a throw *Link Event* must match at most one target *Link Event* in the same container (*Process* or *SubProcess*) (page 259).

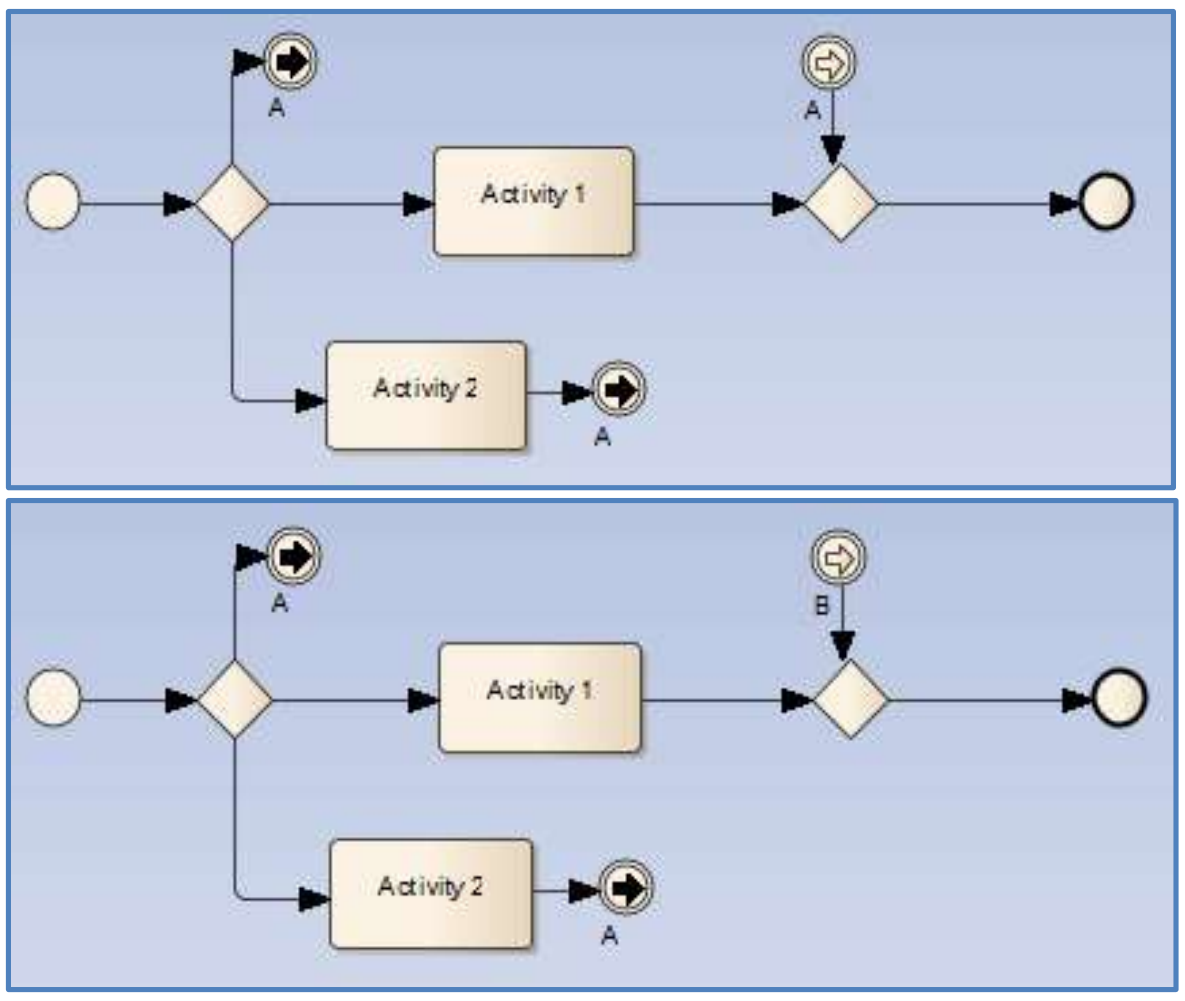

**Fig. 36.** – *Throw* and *Catch Link Events* names must match in the same container

**Picture's interpretation**: Correct: The names of the *Throw* and the *Catch Link Event* are the same (top); Wrong: The names of the *Throw* and the *Catch Link Event* do not match (bottom).

The well-formedness rule regarding link events can be enforced by attaching the following invariant to the *FlowElementsContainer* element of the BPMN metamodel.

```
context FlowElementsContainer
  inv namesOfCatchAndThrowLinkEventMustMatch:
  matchInNamesOfCatchAndThrowLinkEvent()
```

**Listing 34 – *Throw* and *Catch Link Events* names must match in the same container**

### 5.1.25. *Intermediate Events* used within normal flow require incoming and outgoing *Sequence Flows*

An *Intermediate Event* must be a source for a *Sequence Flow* and must be also a target of a *Sequence Flow*, if used in normal flow and the type is *None*, *Message*, *Timer*, *Compensation*, *Conditional*, *Signal*, *Multiple*, or *ParallelMultiple*.

So, an *Intermediate Event* used within normal flow, must have at least one incoming and one outgoing sequence flows [1] (page 259). There are some exceptions to this rule:

- *Target Link Event* cannot have incoming *Sequence Flow*;
- *Source Link Event* cannot have outgoing *Sequence Flow*;

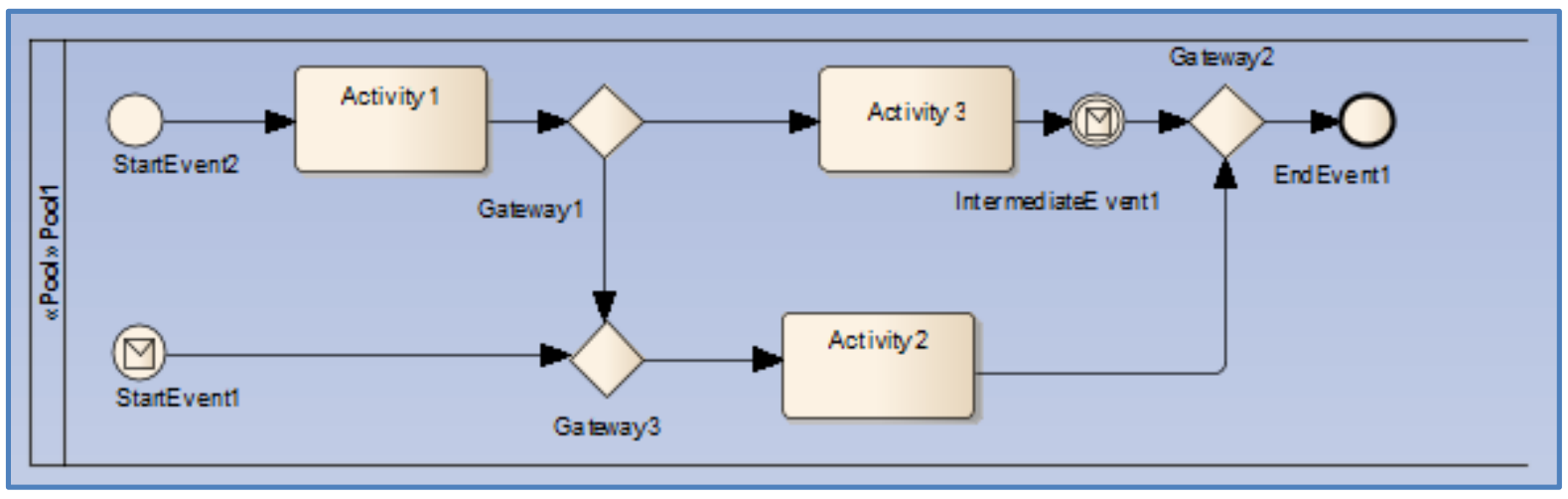

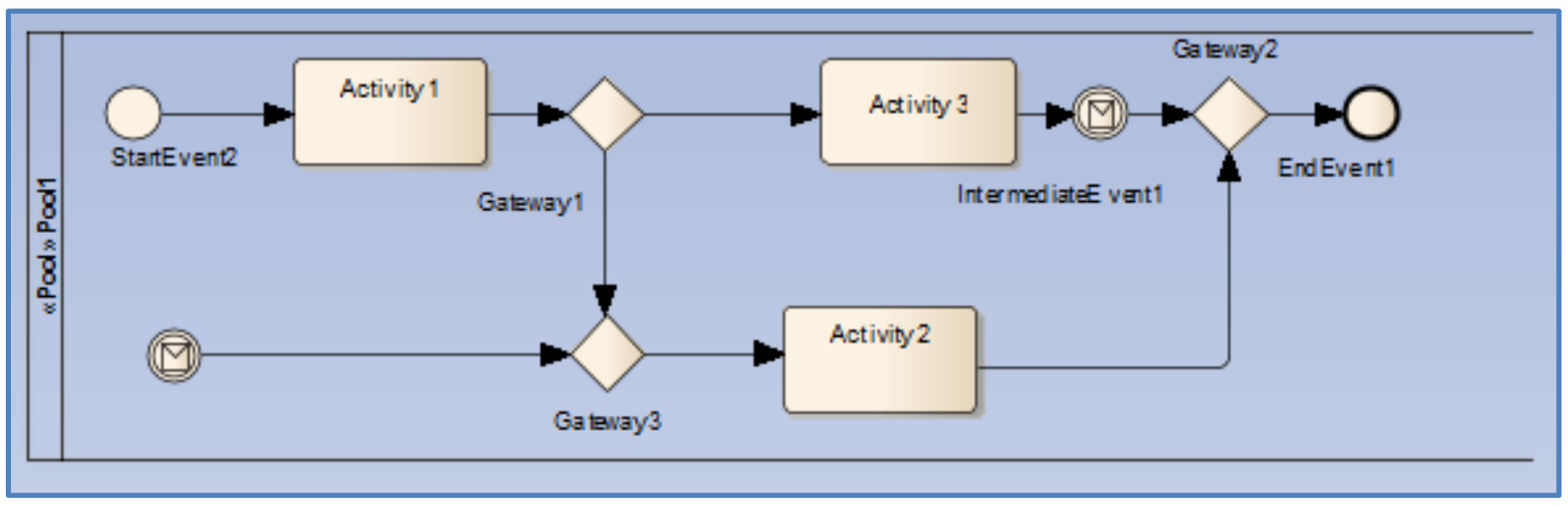

**Fig. 37.** – *Intermediate Events* used within normal flow require incoming and outgoing *Sequence Flows*

**Picture's interpretation**: <u>Correct</u>: The *Intermediate Event* is the target and source of a *Sequence Flow*; <u>Wrong</u>: One of the *Intermediate Event* is not the target of a *Sequence Flow*.

The well-formedness rule regarding intermediate events can be enforced by attaching the following invariant to the *FlowElementsContainer* element of the BPMN metamodel.

```
context FlowElementsContainer
  inv intermediateEventsMustHaveIncAndOutSeqFlow:
    intermediateEventsHaveIncAndOutSeqFlows()
```

**Listing 35 – *Intermediate Events* used within normal flow require incoming and outgoing *Sequence Flows***

### 5.1.26. *Activities* or *Gateways* without incoming *Sequence Flow* do not allow explicit *Start Events*

When a container (*Process* or *SubProcess*) is modeled with Activities or Gateways (other than *Compensations Activity* and *Event SubProcess*), without incoming sequence flow, it means that it has an implicit *Start Event* [1] (page 153, 430). Therefore explicit *Start Events* are not allowed [1] (page 239).

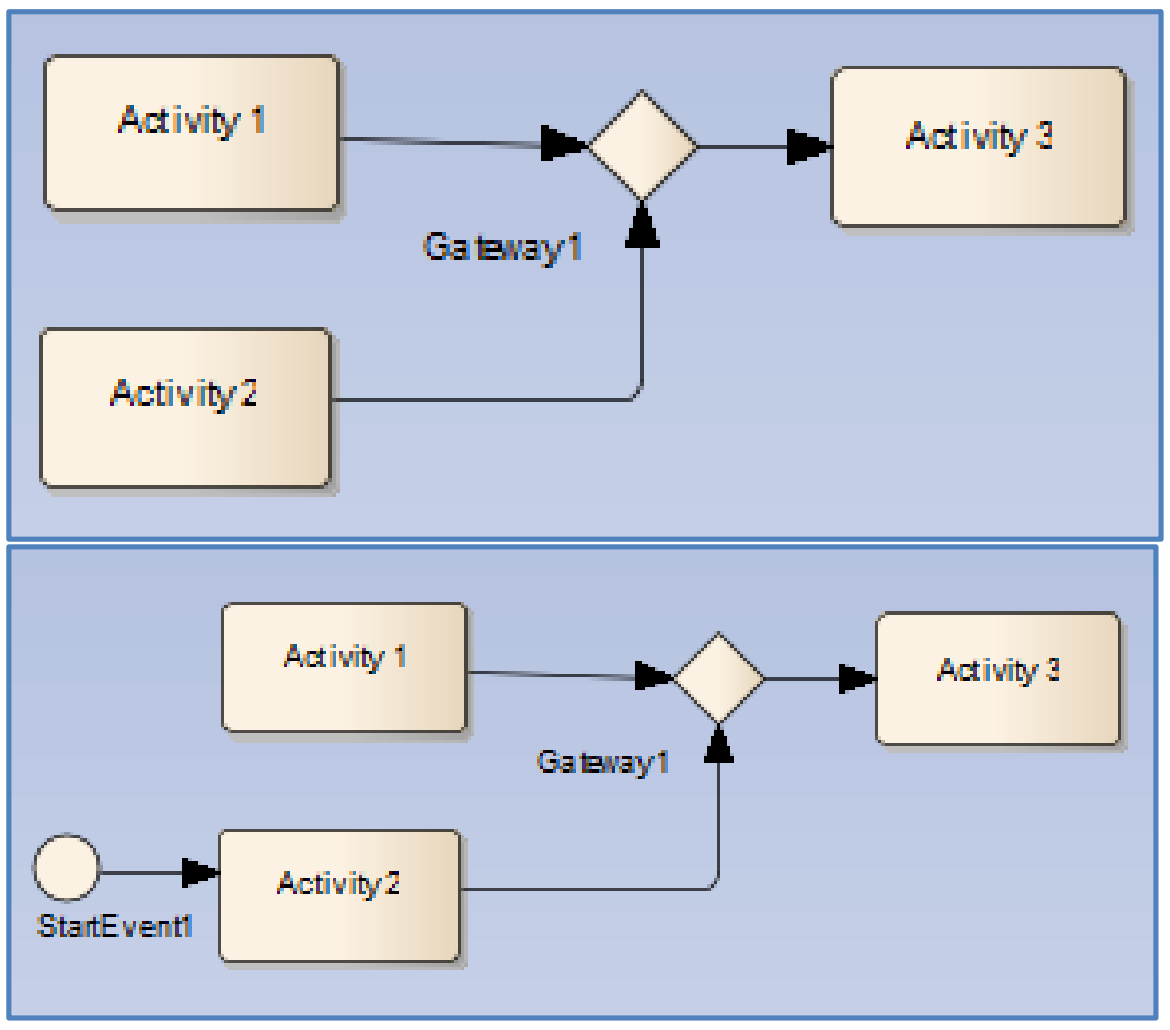

**Fig. 38.** – *Activities* or *Gateways* without incoming *Sequence Flow* do not allow explicit *Start Events*

**Picture's interpretation**: <u>Correct</u>: Implicit Start event from activities without incoming sequence flow; <u>Wrong</u>: Explicit Start Event is not allowed when there is an implicit one.

The well-formedness rule regarding flow nodes can be enforced by attaching the following invariant to the *FlowElementsContainer* element of BPMN metamodel.

```
context FlowElementsContainer
 inv explicitNodeWithoutIncSeqFlowRequiresNoStartEv:
   existsOnlyImplicitStartEvent()
```

**Listing 36 – *Activities* or *Gateways* without incoming *Sequence Flow* do not allow explicit *Start Events***

### 5.1.27. Explicit *Start/End Events* do not allow *Activities* or *Gateways* without incoming/outgoing *Sequence Flow*

*Start Event* and *End event* are optional [1] (page 238). However, if there is at least one explicit *Start/End Event* in a container (*Process* or *SubProcess*), there must not be other flow nodes such as *Activity* and *Gateway*, without incoming/outgoing sequence flow [1] (page 153, 289, 430). There are some exceptions: *Compensation Activity* an *Event SubProcess* do not have incoming and outgoing *Sequence Flows*.

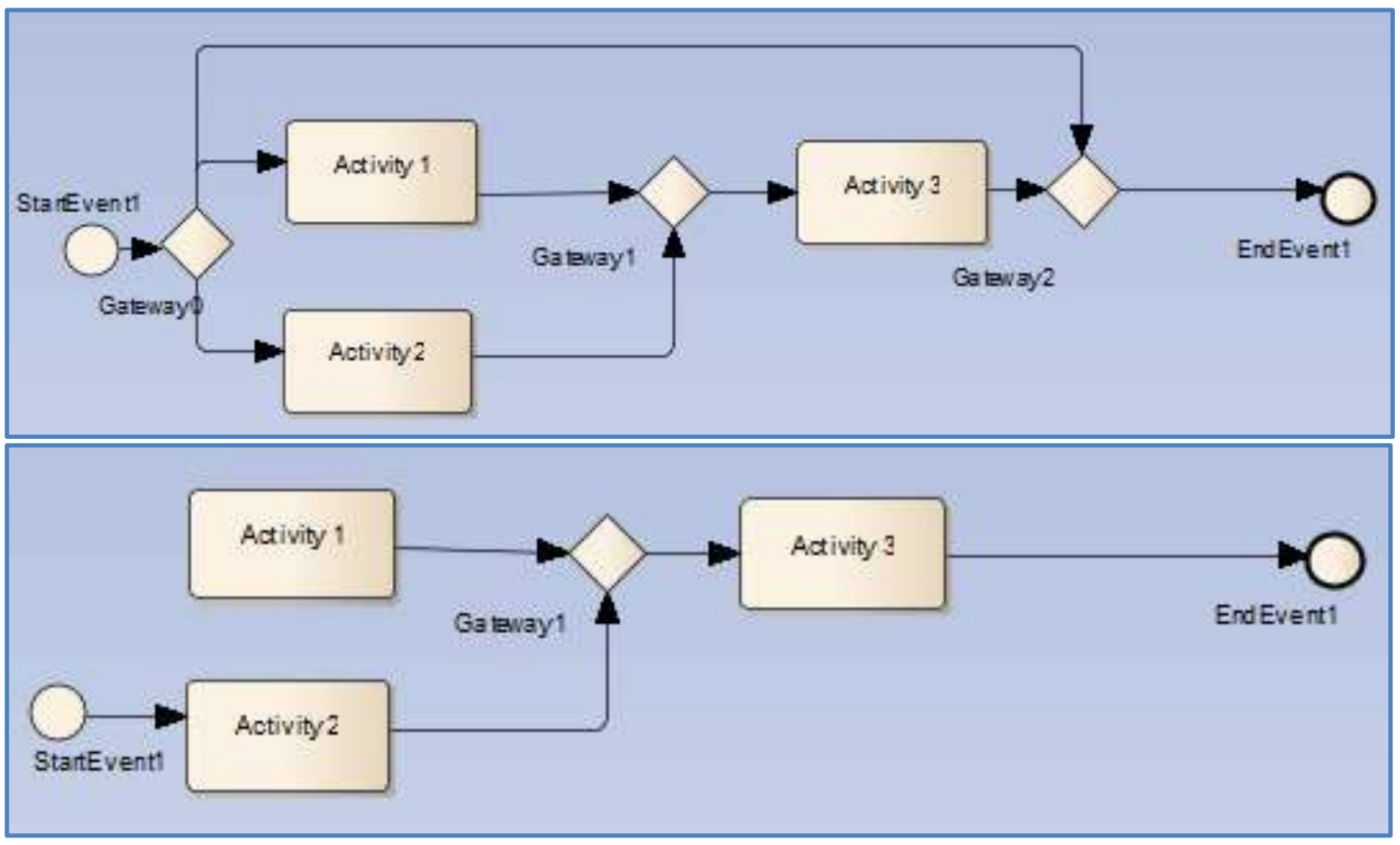

**Fig. 39.** – Explicit *Start/End Events* do not allow *Activities* or *Gateways* without incoming/outgoing *Sequence*

**Picture's interpretation**: Correct: *Start* event and without activities or gateways without incoming sequence flow (top); Wrong: *Start* event and activity without incoming sequence flow (bottom).

The well-formedness rule regarding flow nodes can be enforced by attaching the following invariant to the *FlowElementsContainer* element of BPMN metamodel.

```
context FlowElementsContainer
 inv explicitStartEvRequiresNoActivOrGatWithoutInSeqFlow:
  explicitStartWithoutNodesWithIncomingSequenceFlow()
```

**Listing 37 – Explicit *Start/End Events* do not allow *Activities* or *Gateways* without incoming/outgoing *Sequence Flow***

### 5.1.28. Implicit *Start Events* require implicit *End Events*, and vice versa

Explicit *Start Events* and *End Events* can be omitted. When implicit *Start* (*End*) *Event* is use, it requires also the use of implicit *End* (*Start*) *Event* [1] (page 239, 247). In this case, all *Activities* and *Gateways* without incoming sequence flows have implicit start events which have the same behavior as *None Start Events*. Furthermore, all *Activities* and *Gateways* without outgoing sequence flows have also implicit end events which have the same behavior as *None End Events*.

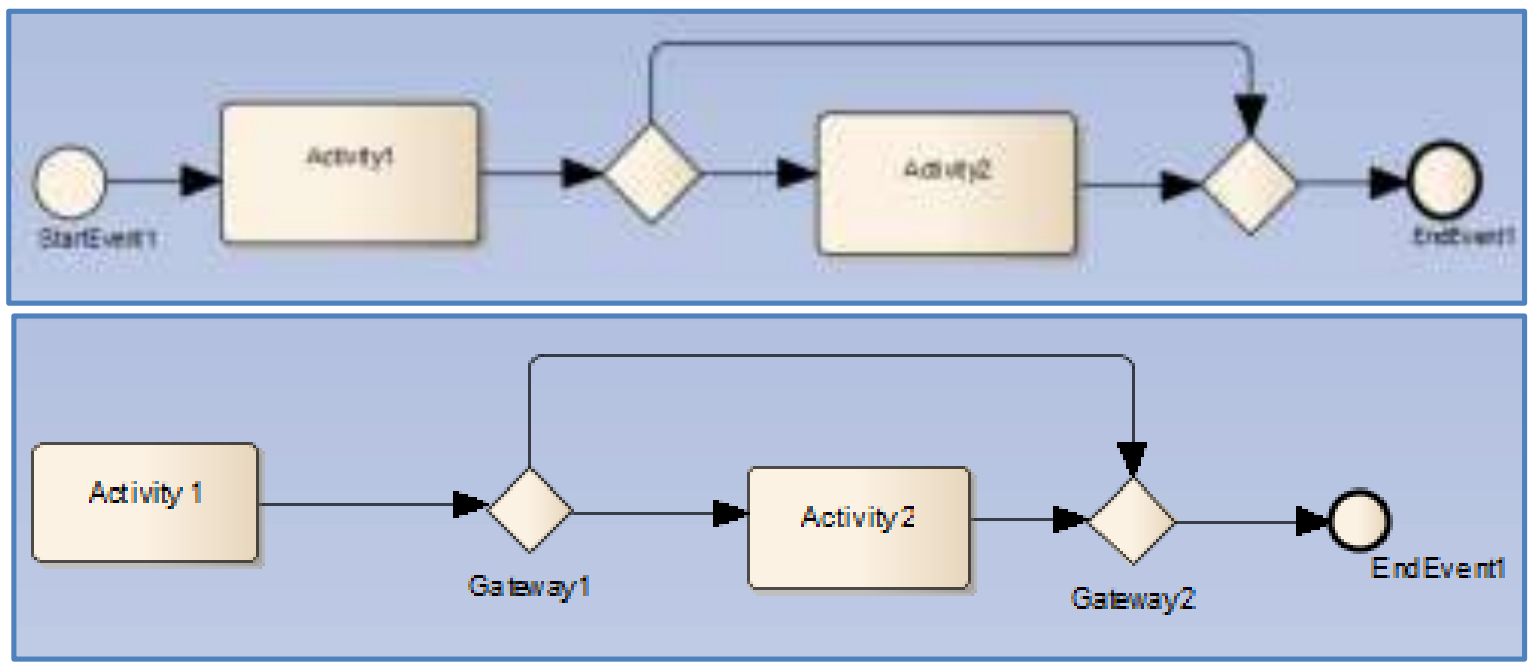

**Fig. 40.** – Implicit *Start Events* require implicit *End Events*, and vice versa

**Picture's interpretation**: Correct: Explicit *Start* and *End Event* (top); Wrong: *Implicit Start Event* and explicit *End Event* (bottom).

The well-formedness rule *regarding* flow nodes can be enforced by attaching the following invariant to the *FlowElementsContainer* element of the BPMN metamodel.

```
context FlowElementsContainer
 inv implicitStartEventsRequiresImplicitEndEvents:
  noStartOrEndEvents()
```

**Listing 38 – Implicit *Start Events* require implicit *End Events*, and vice versa**

#### 5.1.29. Non-interrupting *Start Events* are only allowed in *Event SubProcess*

When using interrupting *Start Events* in an *Event SubProcess*, the occurrence of the *Start Event* results in an interruption of the containing process. If however it is desirable, in spite of the *Start Event* occurrence, to proceed with the containing process, this can be ensured by using non-interrupting *Start Events*. However non-interrupting start events are only allowed inside an *Event SubProcess* [1] (page 242).

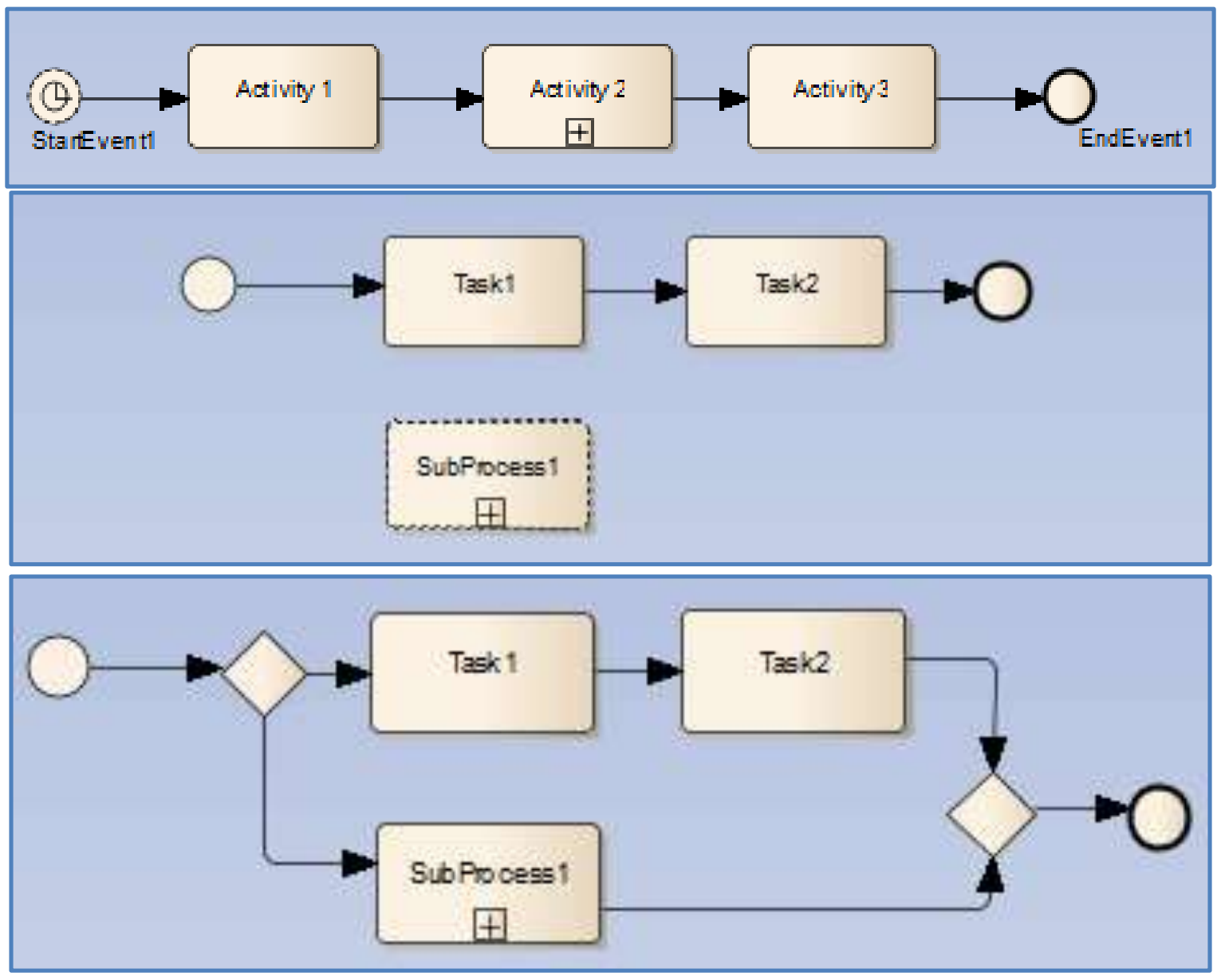

**Fig. 41.** – Non-interrupting *Start Events* are only allowed in *Event SubProcess*

**Picture's interpretation**: Correct: *Non-interrupting Start Event* (top) only allowed inside an event sub-process (middle); Wrong: Non-interrupting *Start Event* (top) not allowed in an embedded sub-process (bottom).

A well-formedness rule can be enforced by attaching the following invariant to the *FlowElementsContainer* element of the BPMN metamodel. This rule is also covered by the invariant presented in section 5.1.12.

```
context FlowElementsContainer
  inv nonInterruptingStartEventsHostedOnlyByEventSubProcess:
    nonInterruptingStartEventsOnlyInEventSubProcess()
```

**Listing 39 – Non-interrupting *Start Events* are only allowed in *Event SubProcess***

#### 5.1.30. *Error Intermediate Events* can only be attached to *Activity* boundaries

*Error Intermediate Events* can only be attached to *Activity* boundaries. They are catching events and can only be attached either to the boundary of *Tasks* or *SubProcesses* [1] (page 261).

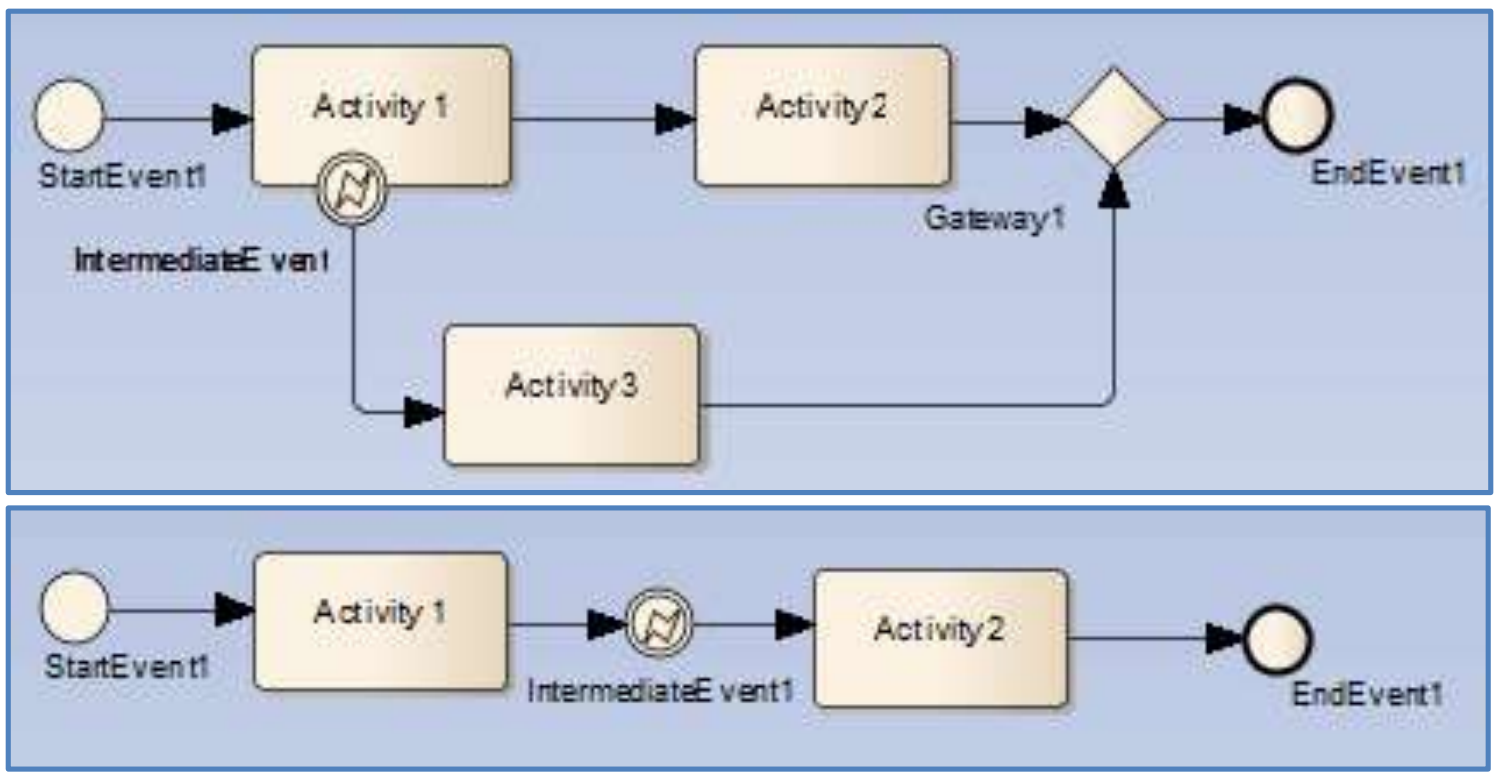

**Fig. 42.** – *Error Intermediate Events* can only be attached to *Activity* boundaries

**Picture's interpretation**: Correct: *Interrupting Error Event* attached to an activity (top); Wrong: *Interrupting Error Event* not attached to an activity (bottom).

A well-formedness rule can be enforced by attaching the following invariant to the *CatchEvent* element of the BPMN metamodel.

```
context CatchEvent
inv catchingErrorEventCanOnlyBeAttachedToActivityBoundary:
     isCatchingErrorAttachedToActivityBoundary()
```

**Listing 40 – *Error Intermediate Events* can only be attached to *Activity* boundaries**

### 5.1.31. *Catch Error Event* must trigger an exception flow

*Error Intermediate Events* cannot be used within normal *Sequence Flows*. This catching event should trigger an exception flow [1] (page 255, 277).

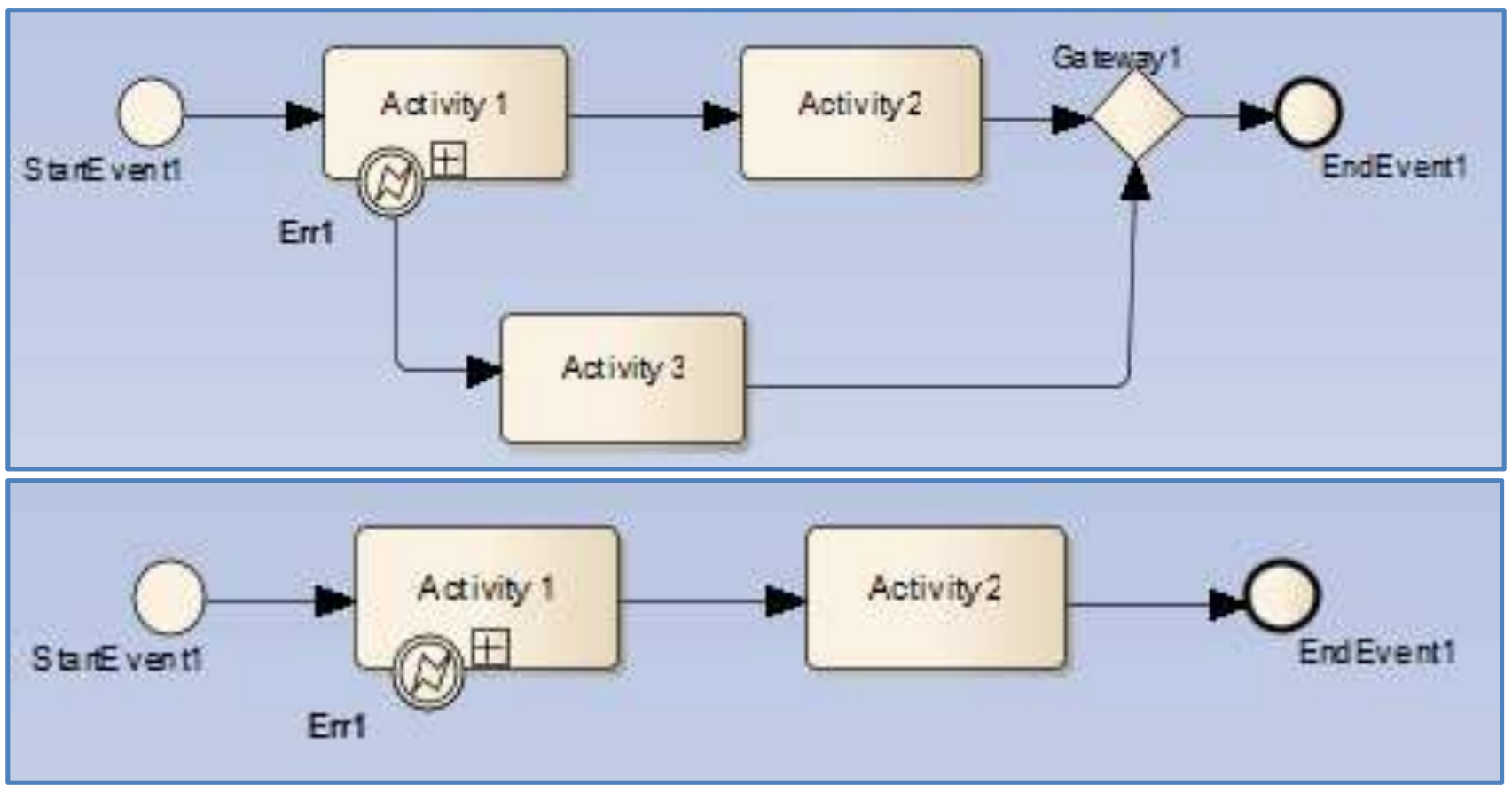

**Fig. 43.** – *Catch Error Event* must trigger an exception flow

**Picture's interpretation**: Correct: *Error Event* triggering an exception flow (top); Wrong: *Error Event* without an exception flow (bottom).

A well-formedness rule can be enforced by attaching the following invariant to the *CatchEvent* element of the BPMN metamodel. This rule is also covered by the invariant presented in section 5.1.55.

```
context CatchEvent
inv catchErrorEventTriggerExceptionFlow:
  self.isErrorEvent() implies
    (self.oclAsType(BoundaryEvent).attachedToRef.isDefined() and
      self.outgoing_a.targetRef->notEmpty())
```

**Listing 41 – *Catch Error Event* must trigger an exception flow**

### 5.1.32. A *Throw Error Event* must have an unnamed *Catch Error Event* or a *Catch Error Event* with the same name

A throwing error is caught by an intermediate error event with the same name or unnamed attached to the sub-process border [1] (page 255) [70].

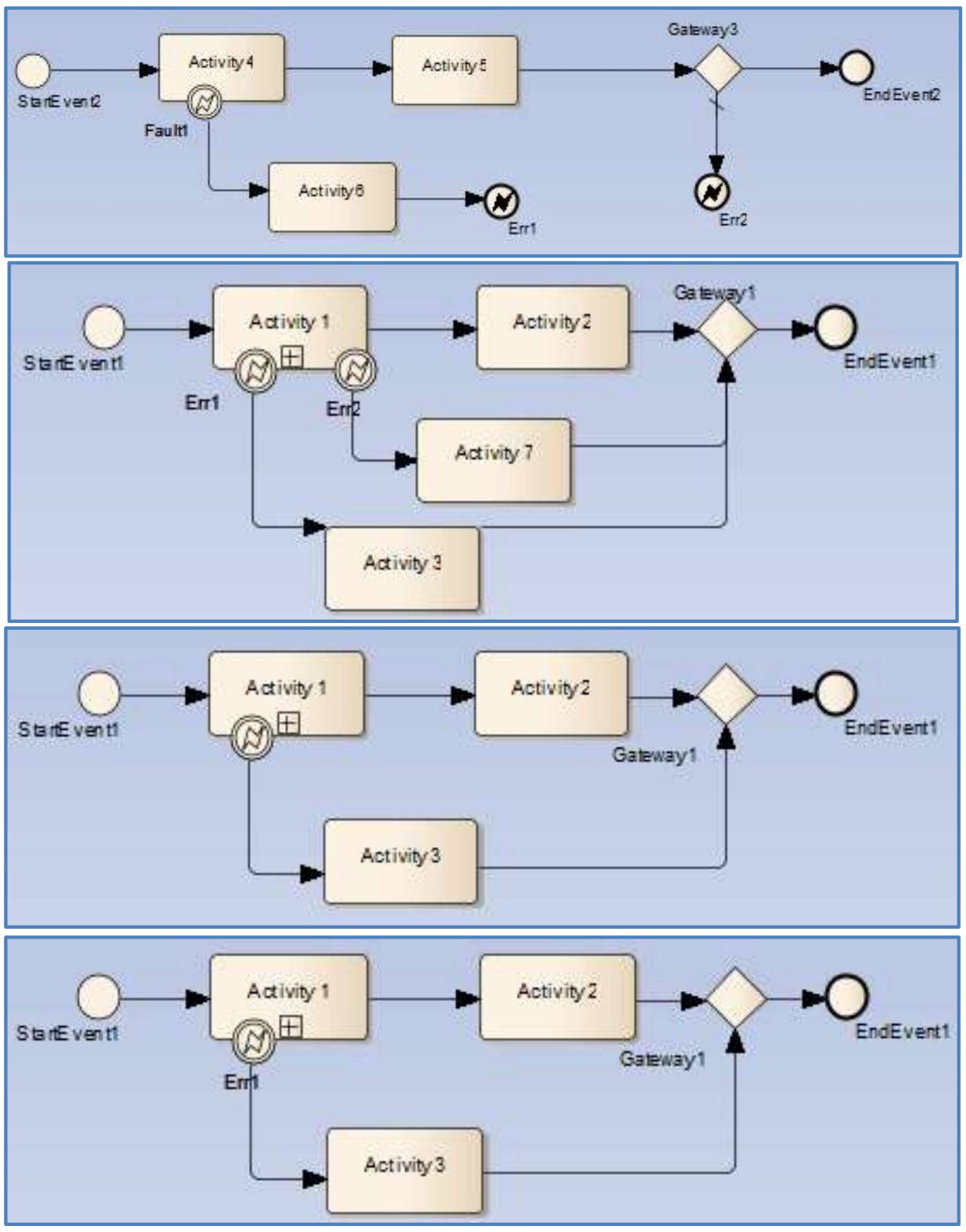

**Fig. 44.** – A *Throw Error Event* must have an unnamed *Catch Error Event* or a *Catch Error Event* with the same name

**Picture's interpretation**: <u>Correct</u>: The two names of the *End Error Events* (top diagram) match the names of the catch *Error Events* (second diagram) or have an unnamed *Error Event* (third diagram); <u>Wrong</u>: The "Err2" name of the *End Error Event* (top) does not match the name of the catch *Error Event* (bottom diagram).

A well-formedness rule can be enforced by attaching the following invariant to the *ThrowEvent* element of the BPMN metamodel.

```
context ThrowEvent
inv throwingErrorEventHasACatchErrorEventWithSameNameXORUnnamed:
    throwErrorHasCatchEvent()
```

**Listing 42 – A *Throw Error Event* must have an unnamed *Catch Error Event* or a *Catch Error Event* with the same name**

### 5.1.33. A catching *Error Event* name must match the name of a thrown *Error Event* or be unnamed

A labeled catching error must have a thrown error with the same label [1] (page 255) [70]. The catching error reacts to thrown errors with the same label.

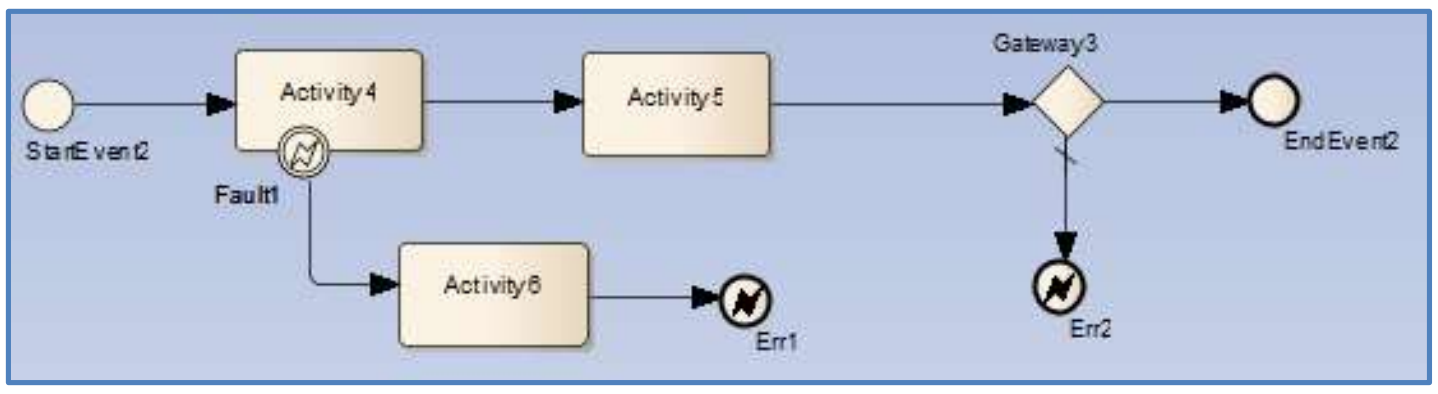

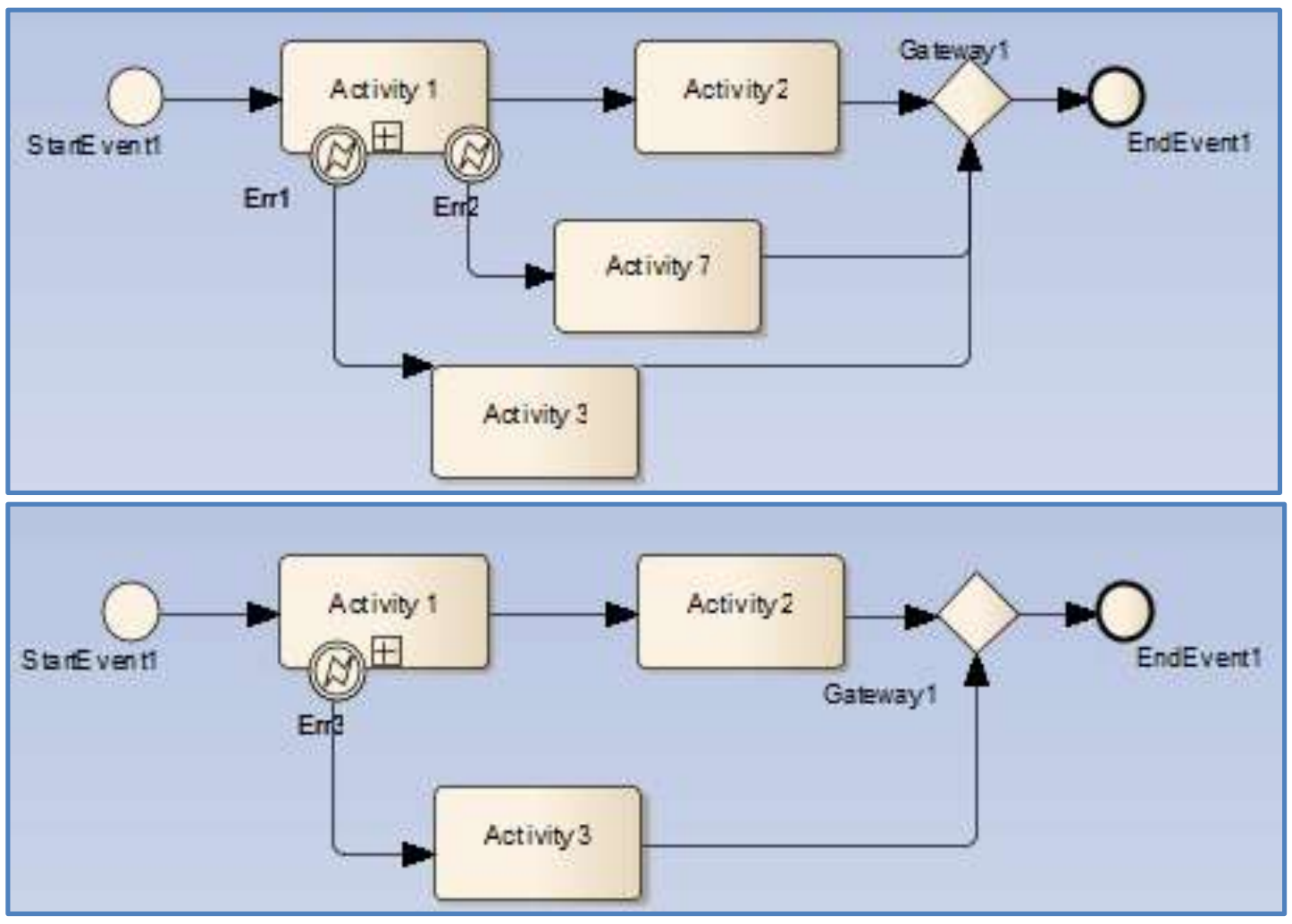

**Fig. 45.** – A catching *Error Event* name must match the name of a thrown *Error Event* or be unnamed

**Picture's interpretation**: <u>Correct</u>: The two names of the *End Error Events* (top) are matched by the names of the two *Catch Error Events* (middle); <u>Wrong</u>: The name of the *Catch End Error Event* "Err3" (bottom) does not match the name of the *End Error Events* (top).

A well-formedness rule can be enforced by attaching the following invariant to the *CatchEvent* element of the BPMN metamodel. This rule is also covered by the invariant presented in section 5.1.32.

```
context CatchEvent
  inv namedCatchingErrorEventHasThrowEventWithSameName:
    catchingErrorEventNameMatchThrowEventName()
```

**Listing 43 – A catching *Error Event* name must match the name of a thrown *Error Event* or be unnamed**

#### 5.1.34. Unnamed and named *Catch Error Events* must not be mixed

If there are several catch error events attached to a sub-process boundary, they should be always labeled. Since an unlabeled error intermediate event always catches any type of error, it also reacts to errors caused by labeled events. Therefore, the resulting implicit parallel flows would be difficult to understand [1] (page 255) [70].

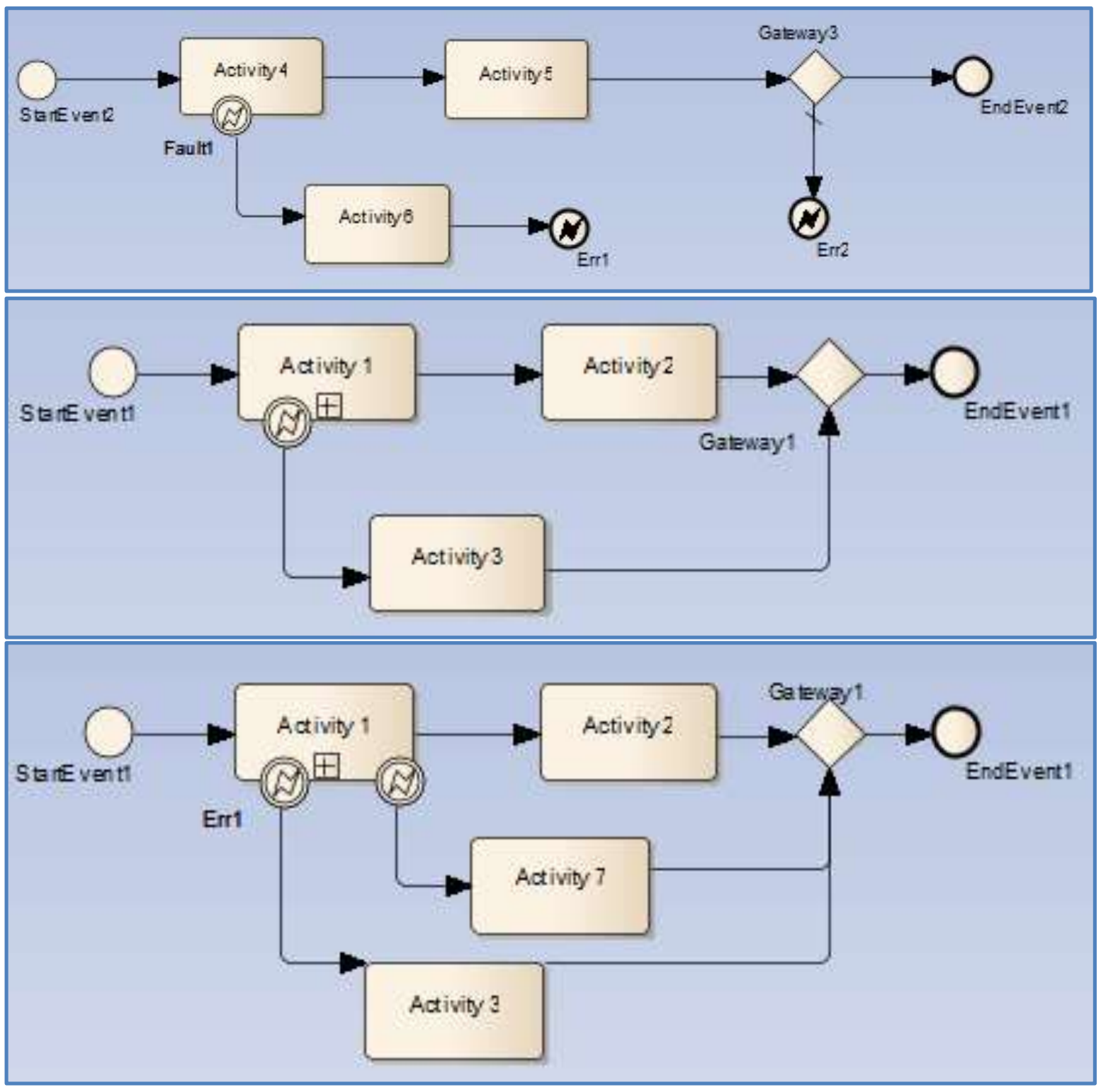

**Fig. 46.** – Unnamed and named *Catch Error Events* must not be mixed

**Picture's interpretation**: <u>Correct</u>: Single unlabeled *Catch Error Event* (middle) that can react to *Throwing Error Event* "Err1" and "Err2" (top); <u>Wrong</u>: Unlabeled *Catch Error Event* mixed with a labeled *Catch Error Event* "Err1" (bottom).

A well-formedness rule can be enforced by attaching the following invariant to the *CatchEvent* element of the BPMN metamodel.

```
context CatchEvent
inv catchingErrorMustNotMixNamedAndUnnamedEvents:
  noMixNamedAndUnnamedcatchingErrorEvents()
```

**Listing 44 – Unnamed and named *Catch Error Events* must not be mixed**

### 5.1.35. A *Throwing Error Event* must be an *End Event*

*Throwing Error Events* are always *End Events*, since they completely abort the surrounding sub-process [1] (page 261).

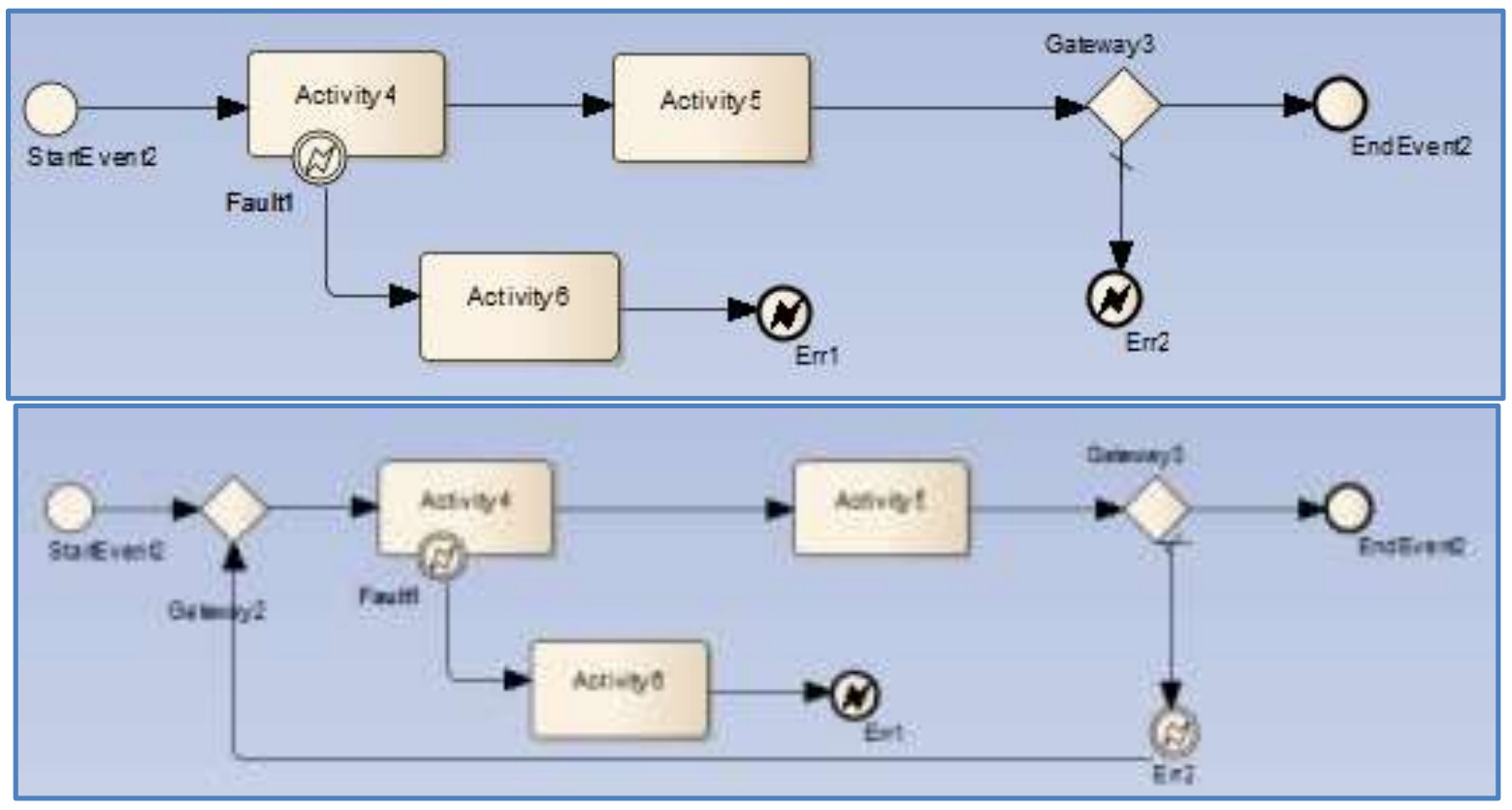

**Fig. 47.** – A *Throwing Error Event* must be an *End Event*

**Picture's interpretation**: <u>Correct</u>: *Throw Error Events* are *End Events*; <u>Wrong</u>: A *Throw Error Event* cannot be an *Intermediate Event*.

A well-formedness rule can be enforced by attaching the following invariant to the *ThrowEvent* element of the BPMN metamodel. This rule was already covered by the invariant presented in section 5.1.12.

```
context ThrowEvent
  inv errorEventIsAlwaysEndEvent:
    self.isErrorEvent() implies
      self.oclIsTypeOf(EndEvent)
```

**Listing 45 – A *Throwing Error Event* must be an *End Event***

### 5.1.36. *Catch Escalation Events* can only be attached to activity boundaries

Interrupting or non-interrupting *Catch Escalation* intermediate events must be attached to activity boundaries. They can be attached either to the boundary of tasks or sub-processes [1] (page 261).

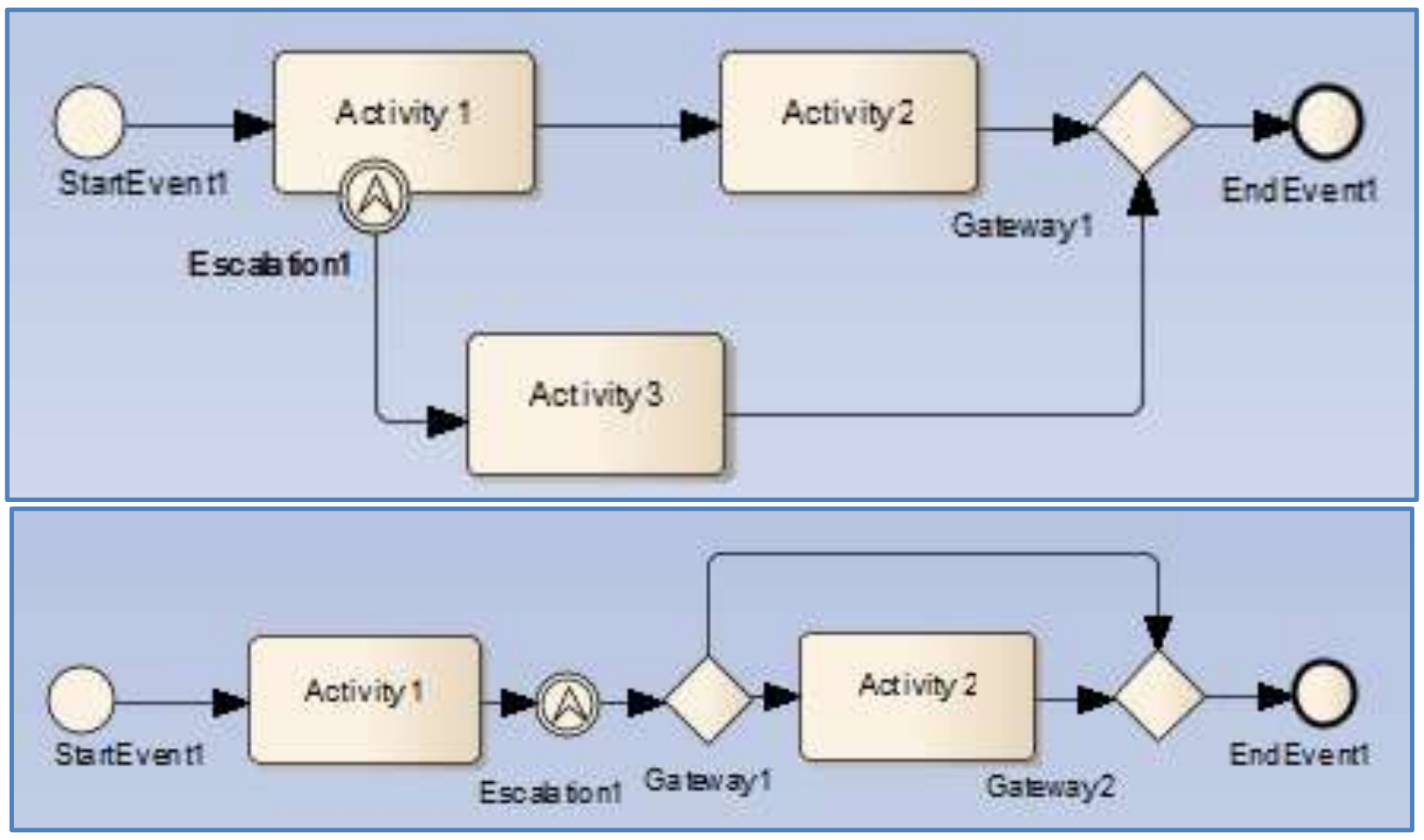

**Fig. 48.** – *Catch Escalation Events* can only be attached to activity boundaries

**Picture's interpretation**: <u>Correct</u>: *Interrupting Catch Escalation Event* attached to an activity; <u>Wrong</u>: *Catch Escalation Event* as intermediate event.

A well-formedness rule can be enforced by attaching the following invariant to the *CatchEvent* element of the BPMN metamodel.

```
context CatchEvent
inv catchEscalationEventIsBoundaryEvent:
     isCatchingEscalationAttachedToActivityBoundary()
```

**Listing 46 – *Catch Escalation Events* can only be attached to activity boundaries**

### 5.1.37. *Catch Escalation Event* must trigger an exception flow

*Catch Escalation Events* cannot be used within normal sequence flows. This catching event should trigger an exception flow [1] (page 277).

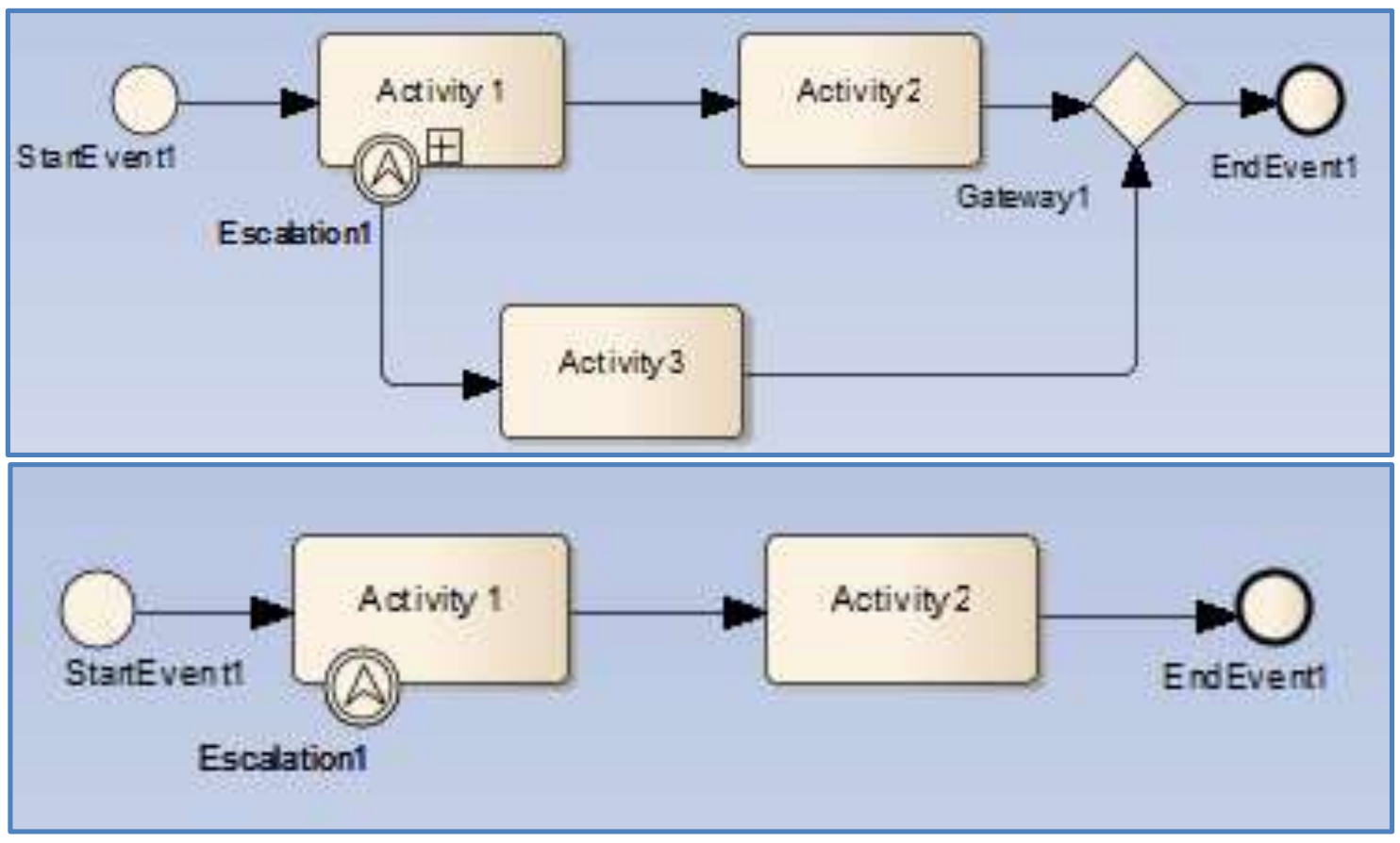

**Fig. 49.** – *Catch Escalation Event* must trigger an exception flow

**Picture's interpretation**: <u>Correct</u>: *Catch Escalation Event* triggering an exception flow; <u>Wrong</u>: *Catch Escalation Event* without an exception flow.

A well-formedness rule can be enforced by attaching the following invariant to the *CatchEvent* element of the BPMN metamodel. This rule is also covered by the invariant presented in section 5.1.55.

```
context CatchEvent
inv catchEscalationEventTriggerExceptionFlow:
  self. isEscalationEvent() implies
    (self.oclAsType(BoundaryEvent).attachedToRef.isDefined() and
      self.outgoing_a.targetRef->notEmpty())
```

**Listing 47 – *Catch Escalation Event* must trigger an exception flow**

### 5.1.38. A *Throw Escalation Event* must have an unnamed *Catch Escalation Event* or a *Catch Escalation Event* with the same name

A labeled catching *Escalation Event* must have a thrown *Escalation Event* with the same label [1] (page 248) [70]. The catching *Escalation Event* reacts to thrown *Escalation Event* with the same label.

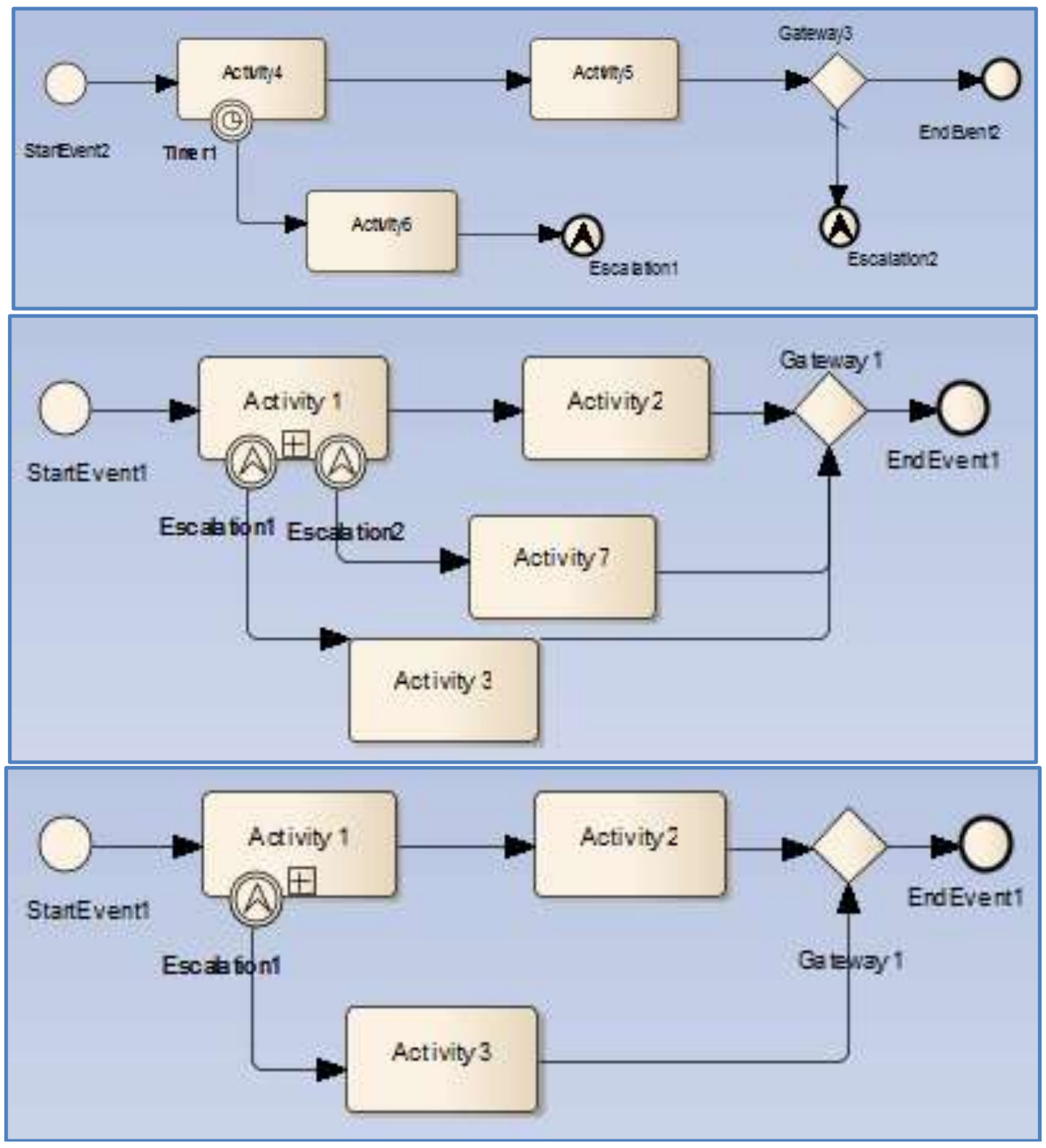

**Fig. 50.** – A *Throw Escalation Event* must have an unnamed *Catch Escalation Event* or a *Catch Escalation Event* with the same name

**Picture's interpretation**: Correct: The two names of the *End Escalation Events* (top) match the names of the Interrupting *Catch Escalation Events* (middle); Wrong: The "Escalation2" name of the *End Escalation Event* (top) does not match the name of any *Catch Escalation Event* (bottom).

A well-formedness rule can be enforced by attaching the following invariant to the *CatchEvent* element of the BPMN metamodel.

```
context ThrowEvent
  inv throwingEscalationEndEventIsInterrupting:
    isThrowingEscalationEndEventInterrupting()
```

**Listing 48 – A *Throw Escalation Event* must have an unnamed *Catch Escalation Event* or a *Catch Escalation Event* with the same name**

### 5.1.39. Named and unnamed *Interrupting Catch Escalation Events* must not be mixed

If there are several *Catch Interrupting Escalation Events* attached to a sub-process boundary, they should be always labeled. Since an unlabeled escalation intermediate event always catches any type of escalation, it also reacts to escalations caused by labeled events. Therefore, the resulting flows would be difficult to understand [1] (page 248) [70].

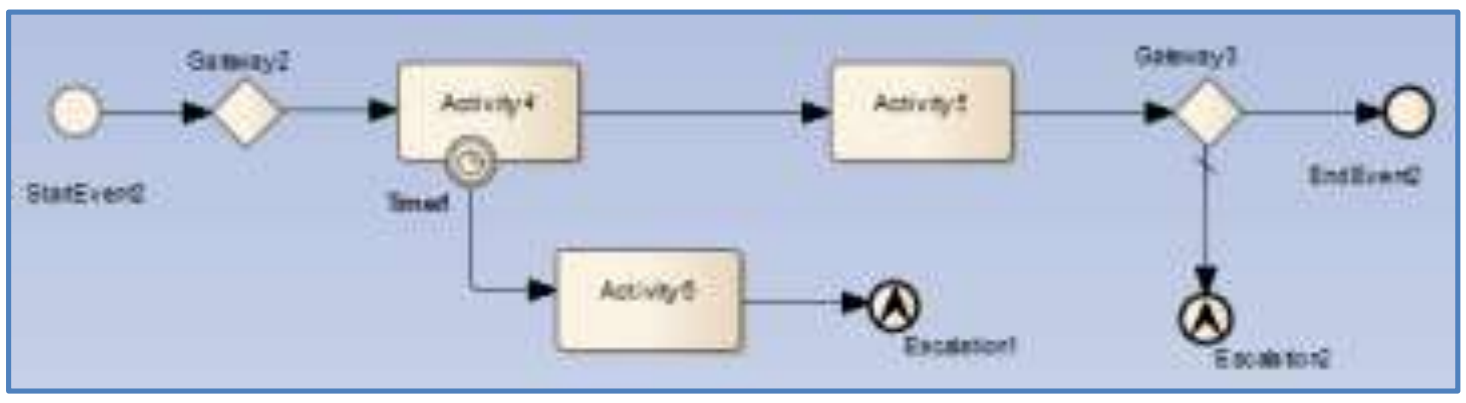

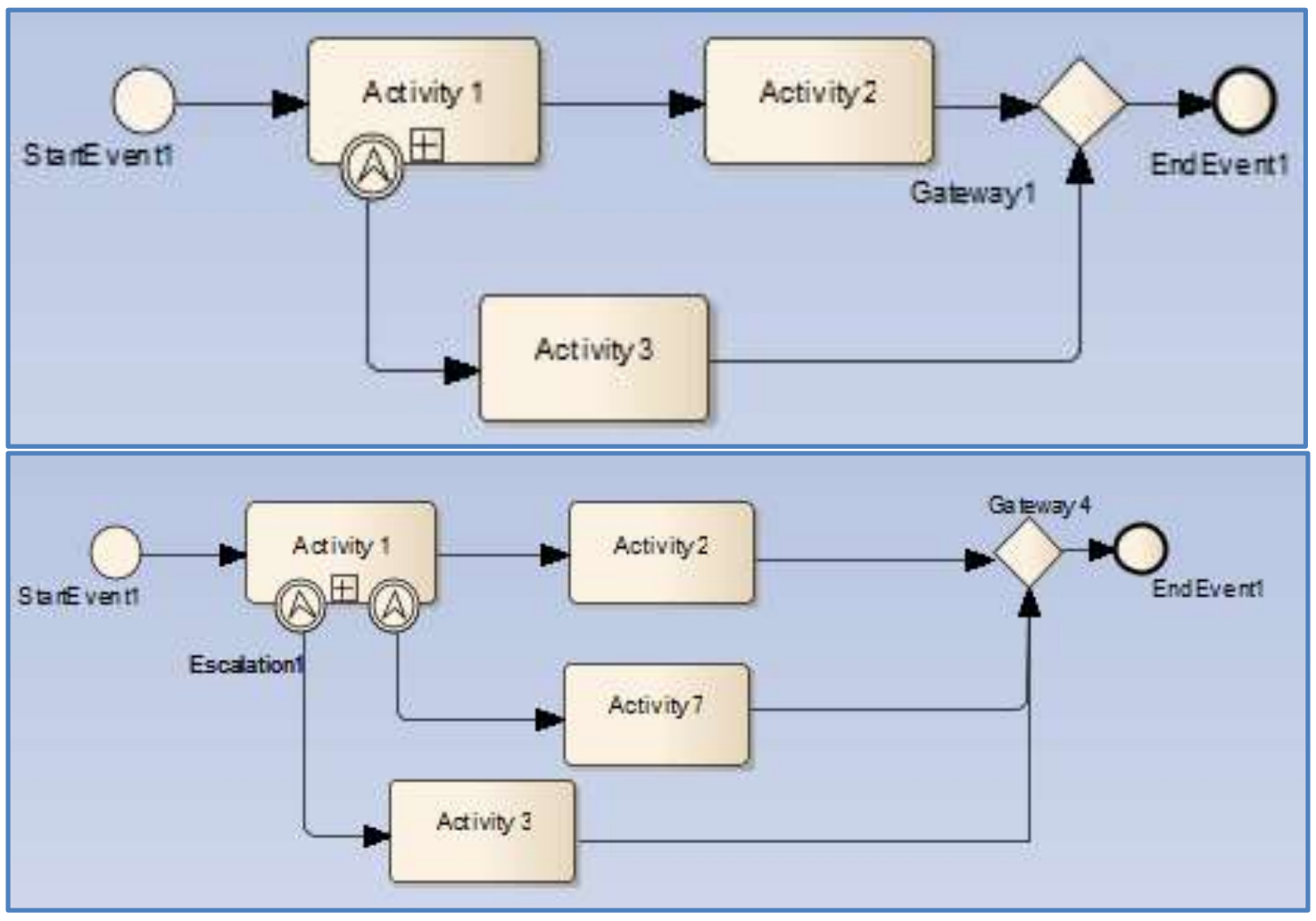

**Fig. 51.** – Named and unnamed *Interrupting Catch Escalation Events* must not be mixed

**Picture's interpretation**: Correct: Single unlabeled Interrupting *Catch Escalation Event* (middle) that can react to *Throwing Escalation End Event* "Escalation1" and "Escalation2" (top); Wrong: Unlabeled *Interrupting Catch Escalation Event* mixed with a labeled *Interrupting Catch Escalation Event* "Escalation1" (bottom).

A well-formedness rule can be enforced by attaching the following invariant to the *CatchEvent* element of the BPMN metamodel.

```
context CatchEvent
  inv interrCatchingEscalationMustNotMixNamedAndUnnamedEvents:
    noMixNamedAndUnnamedCatchingInterEscalationEvents()
```

**Listing 49 – Named and unnamed *Interrupting Catch Escalation Events* must not be mixed**

### 5.1.40. Unnamed and named *Catch non-Interrupting Escalation Events* must not be mixed

If there is several *Catch non-Interrupting Escalation Events* attached to a sub-process boundary, they should be always labeled. Since an unlabeled escalation intermediate event always catches any type of escalation, it also reacts to escalations caused by labeled events. Therefore, the resulting implicit parallel flows would be difficult to understand [1] (page 248) [70].

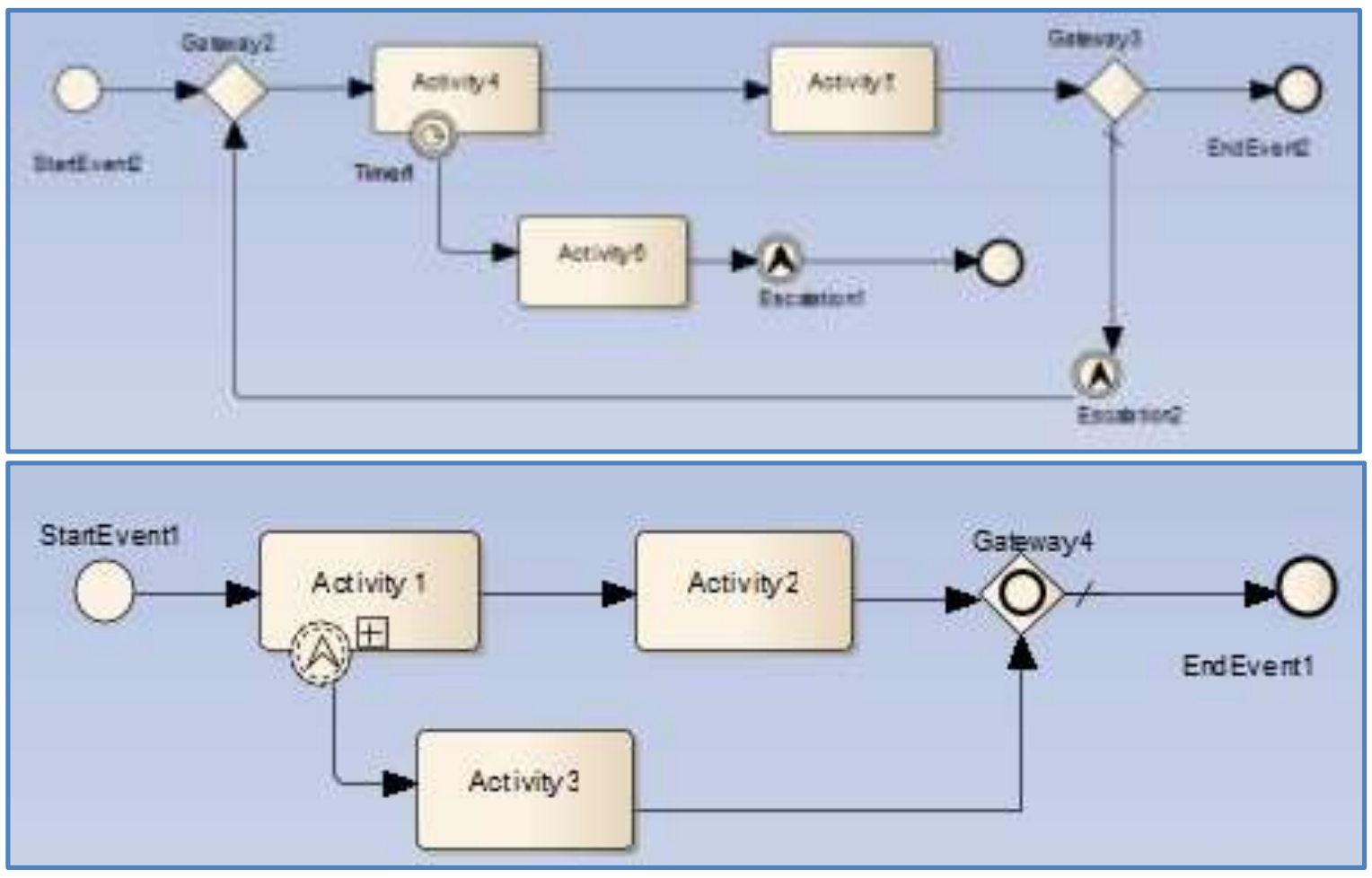

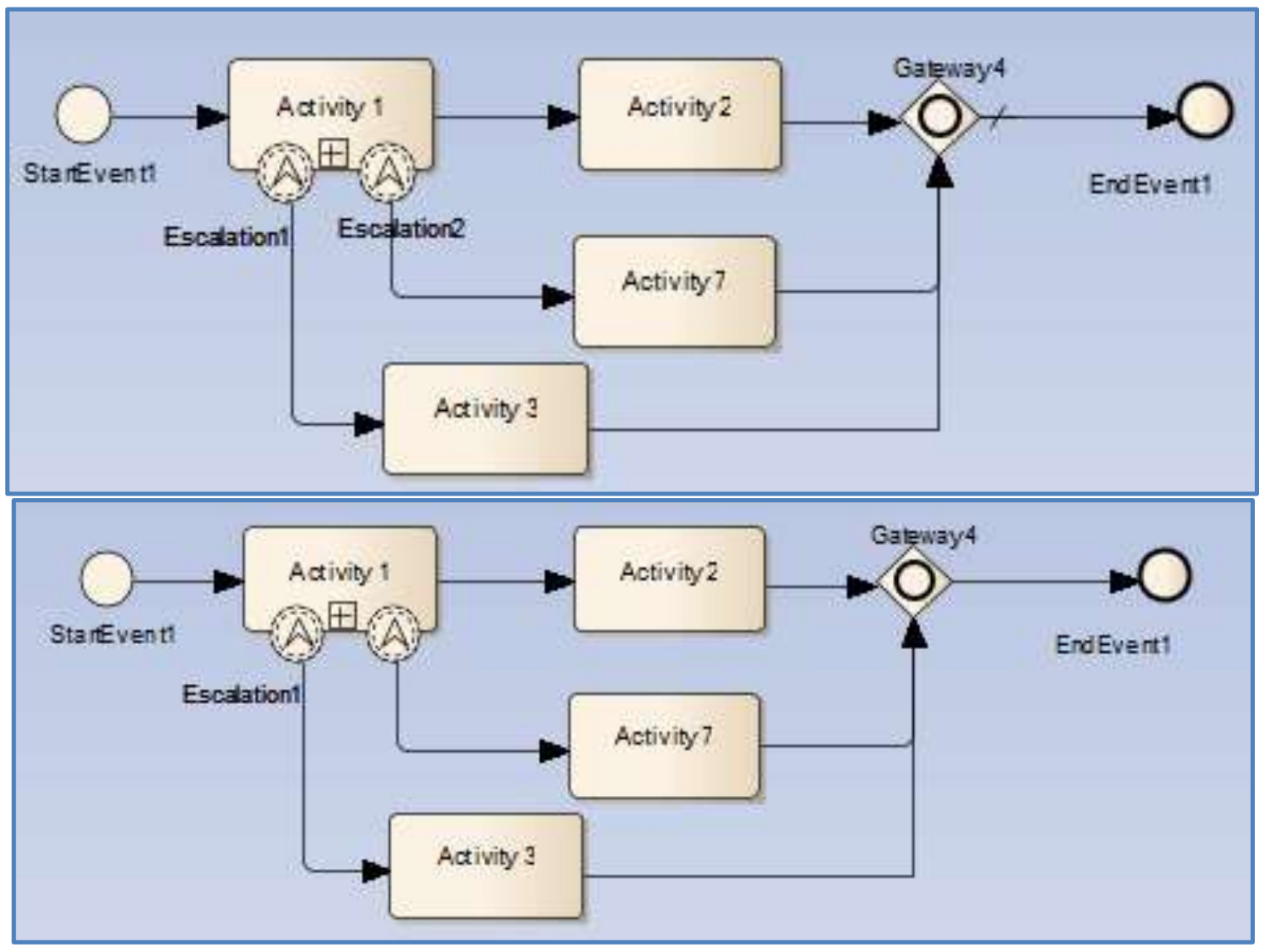

**Fig. 52.** – Unnamed and named *Catch non-Interrupting Escalation Events* must not be mixed

**Picture's interpretation**: Correct: Single unlabeled *non-Interrupting Catch Escalation Event* (second) or *Catch Escalation Intermediate Events* "Escalation1" and "Escalation2" (third) that can react to *Throwing Escalation Intermediate Event* "Escalation1" and "Escalation2" (top); Wrong: Unlabeled non-Interrupting *Catch Escalation Event* mixed with a labeled *non-Interrupting Catch Escalation Event* "Escalation1" (bottom)

A well-formedness rule can be enforced by attaching the following invariant to the *CatchEvent* element of the BPMN metamodel.

```
context CatchEvent
inv nonInterrCatchingEscalationMustNotMixNamedAndUnnamedEvents:
  noMixNamedAndUnnamedCatchingNonInterEscalationEvents()
```

**Listing 50 – Unnamed and named *Catch non-Interrupting Escalation Events* must not be mixed**

### 5.1.41. A *Throwing Escalation End Event* must be caught by an Interrupting *Escalation Catch Event*

If an escalation is required to abort the entire activity, the attached intermediate event is an interrupting event with the same name or unnamed. In this case the throwing escalation event in the sub-process needs to be an end event. So, an *Escalation End Event* needs a *Catch Escalation Event* to react to it [1] (page 248) [70].

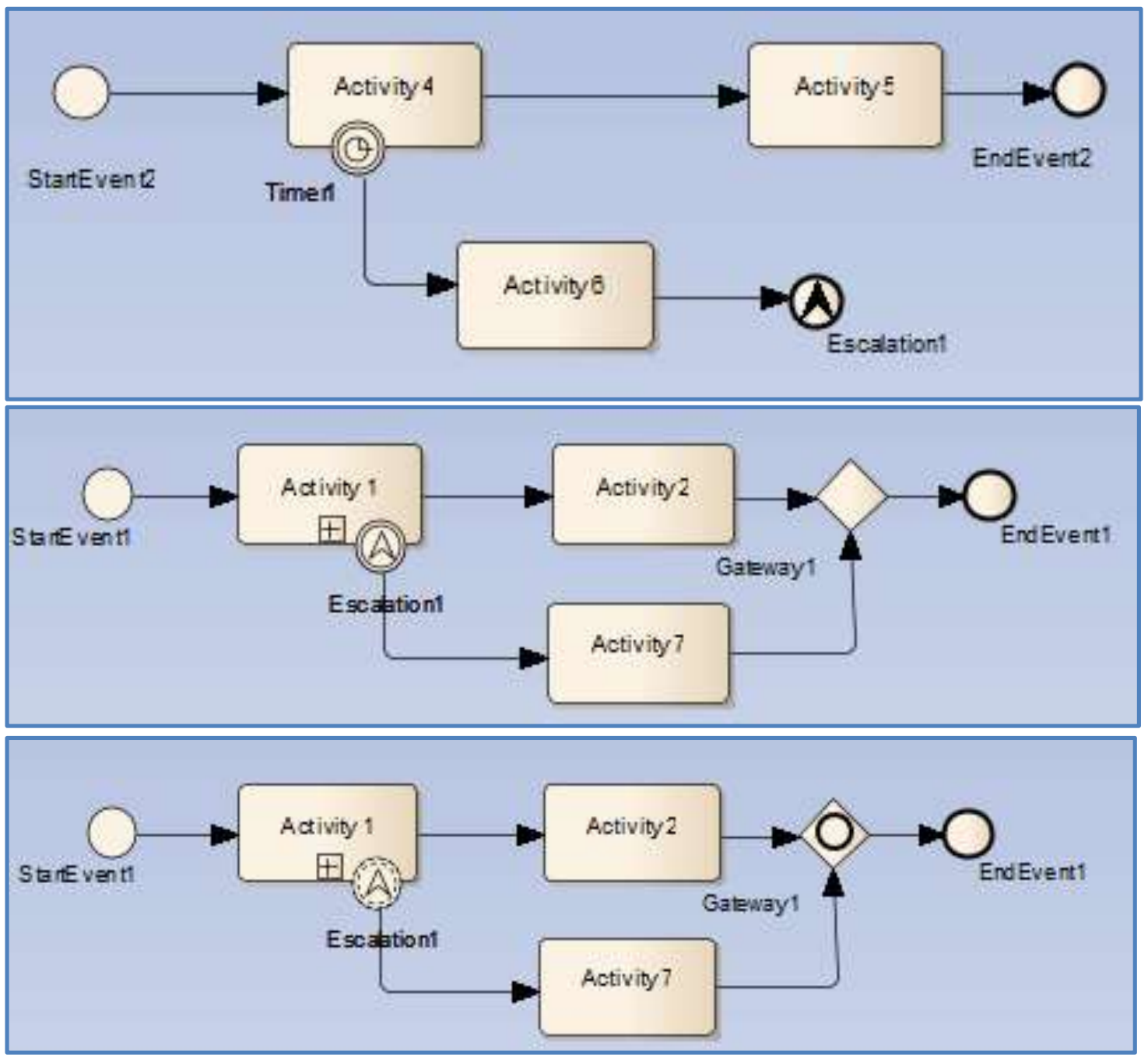

**Fig. 53.** – A *Throwing Escalation End Event* must be caught by an Interrupting *Escalation Catch Event*

**Picture's interpretation**: <u>Correct</u>: A *Throw Escalation End Event* "Escalation1" (top) is caught by an *Interrupting Escalation Event* "Escalation1 (middle); <u>Wrong</u>: A *Throw Escalation End Event* "Escalation1" (top) is caught by a *non-Interrupting Escalation Event* "Escalation1" (bottom).

A well-formedness rule can be enforced by the following invariant to the *ThrowEvent* element of the BPMN metamodel, already considered in section 5.1.38.

```
context ThrowEvent
inv throwingEscalationEndEventIsInterrupting:
    isThrowingEscalationEndEventInterrupting()
```

**Listing 51 – A *Throwing Escalation End Event* must be caught by an Interrupting *Escalation Catch Event***

### 5.1.42. A *Throwing Escalation Intermediate Event* must be caught by a *non-Interrupting Escalation Catch Event*

If an intermediate escalation event does not abort its current activity, the attached boundary event of the activity, must be either a non-interrupting event with the same name of the intermediate escalation event, or unnamed. In this case the throwing escalation event in the sub-process needs to be an intermediate event. So, an *Escalation Intermediate Event* needs a *non-interrupting Catch Escalation Event* to capture it [1] (page 248) [70].

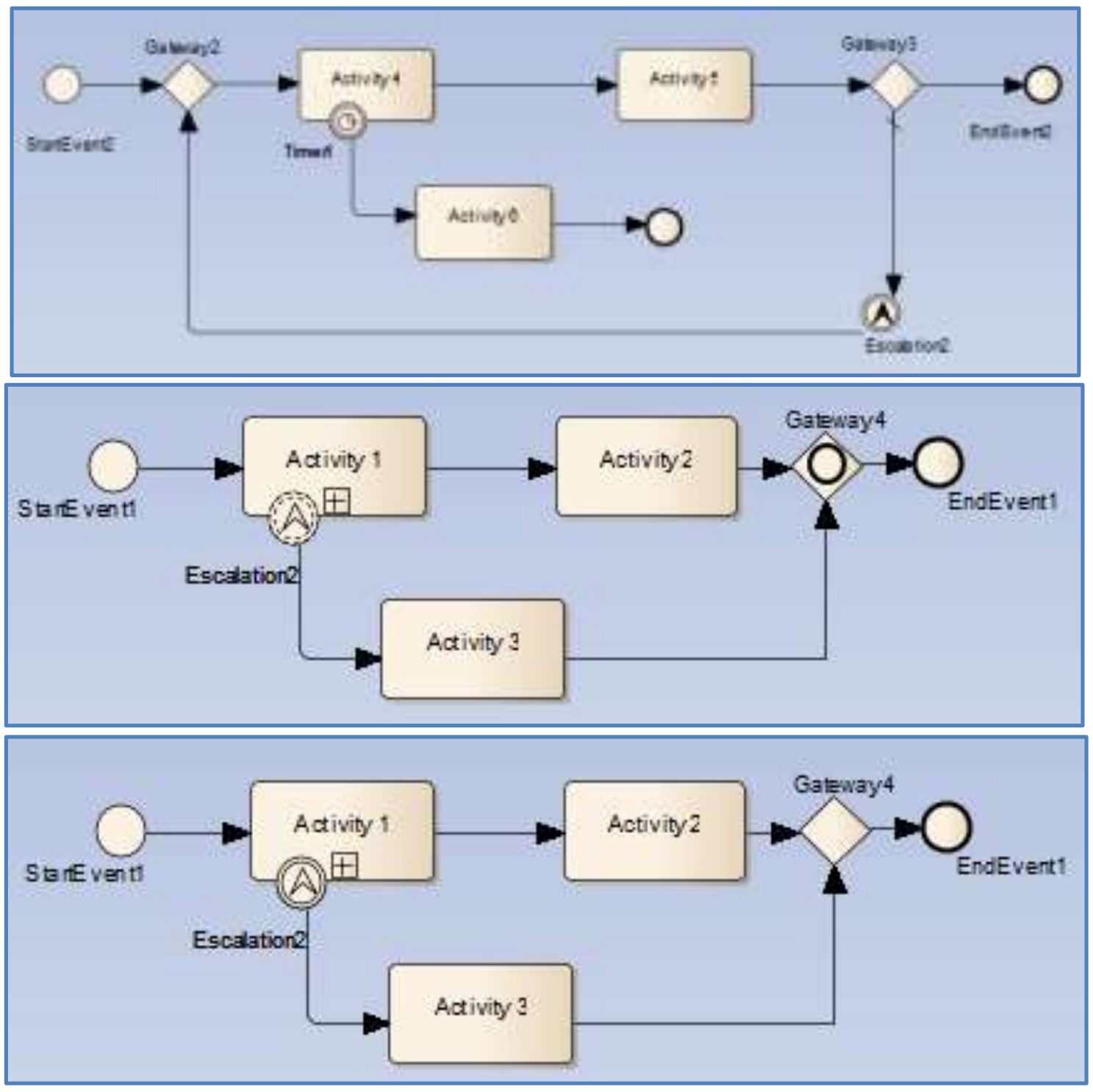

**Fig. 54.** – A *Throwing Escalation Intermediate Event* must be caught by a *non-Interrupting Escalation Catch Event*

**Picture's interpretation**: <u>Correct</u>: A *Throw Escalation Intermediate Event* "Escalation2" (top) is caught by a *non-Interrupting Escalation Event* "Escalation2" (middle); <u>Wrong</u>: A *Throw Escalation Intermediate Event* "Escalation2" is caught by an *Interrupting Escalation Event* "Escalation2" (bottom).

A well-formedness rule can be enforced by attaching the following invariant to the *ThrowEvent* element of the BPMN metamodel.

```
context ThrowEvent
inv throwingEscalationIntermediateEventIsNonInterrupting:
    isThrowingEscalationIntermediateEventNonInterrupting()
```

**Listing 52 – A Throwing Escalation Intermediate Event must be caught by a non-Interrupting Escalation Catch Event**

### 5.1.43. A *Throw Escalation Event* must have an unnamed *Catch Escalation Event* or a *Catch Escalation Event* with the same name

A labeled catching *Escalation Event* must have a thrown *Escalation Event* with the same label [1] (page 248) [70]. The catching *Escalation Event* reacts to thrown *Escalation Event* with the same label.

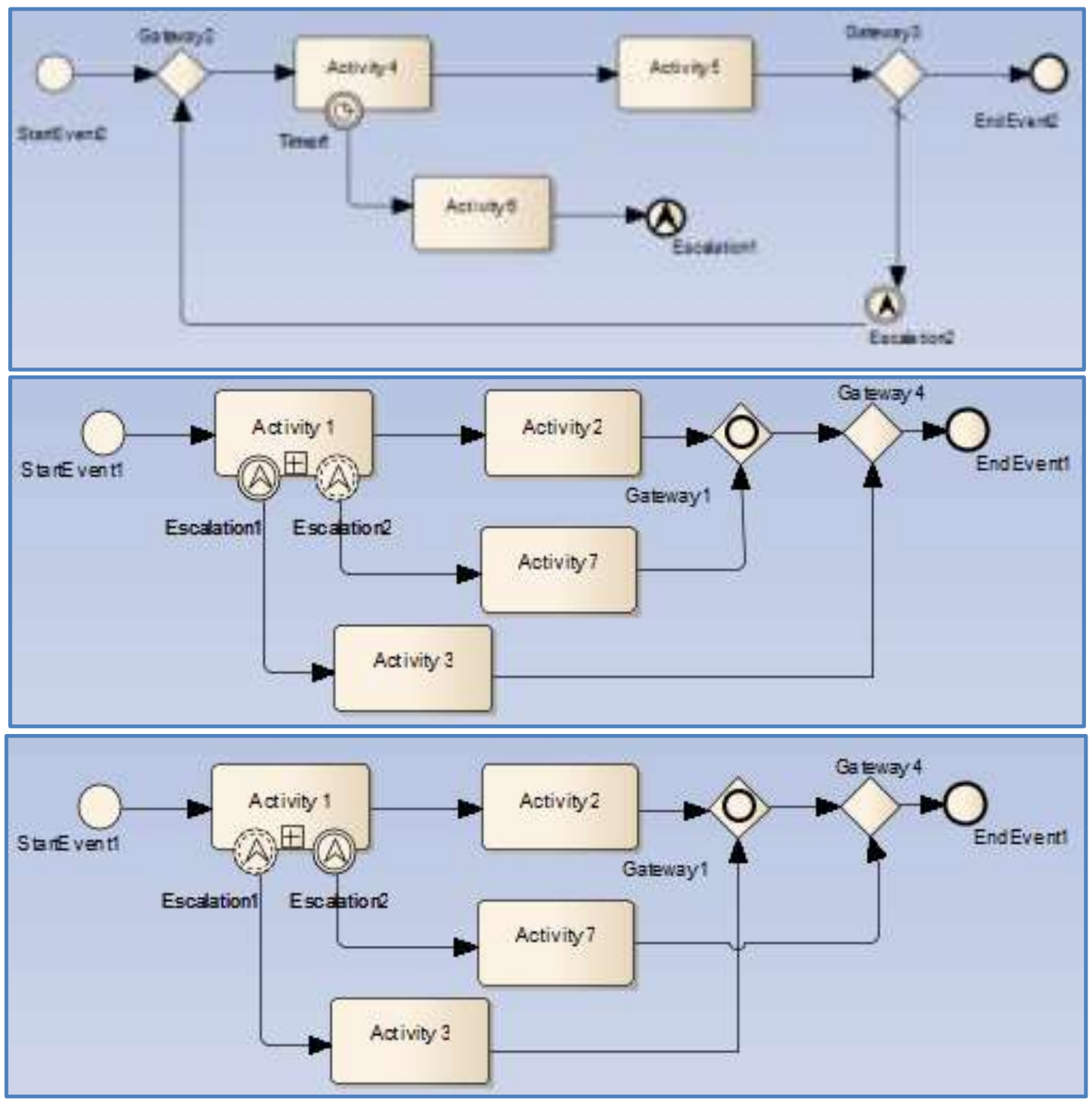

**Fig. 55.** – A *Throw Escalation Event* must have an unnamed *Catch Escalation Event* or a *Catch Escalation Event* with the same name

**Picture's interpretation**: <u>Correct</u>: The two names of the *End* and *Intermediate Escalation Events* (top) match the names of the *Interrupting* and *non-interrupting Catch Escalation Events* (middle); <u>Wrong</u>: The two names of the *End* and *Intermediate Escalation Events* (top) do not match the names of the Int*errupting and non-interrupting Catch Escalation Events* (bottom).

A well-formedness rule can be enforced by attaching the following invariant to the *CatchEvent* element of the BPMN metamodel. This rule is already covered by invariants of sections 5.1.41 and 5.1.42.

```
context CatchEvent
inv namedCatchingEscalationEventHasThrowEventWithSameName:
  catchingEscalationEventNameMatchThrowEventName()
```

**Listing 53 – A *Throw Escalation Event* must have an unnamed *Catch Escalation Event* or a *Catch Escalation Event* with the same name**

### 5.1.44. A *Throw Signal Event* must be caught by a *Catch Signal Event* with the same name or unnamed

If a *Signal Event* is used to trigger an action, a *Catch Signal Event* with the same name or unnamed must also be used to capture it. The catching *Signal Event* can filter out signals that have the same name (or it can be set to react to any signal). The *Throw Signal Event* can reside in the same process of the *Catch Signal Event* or in other Process within the same collaboration [1] (page 253) [70] [71].

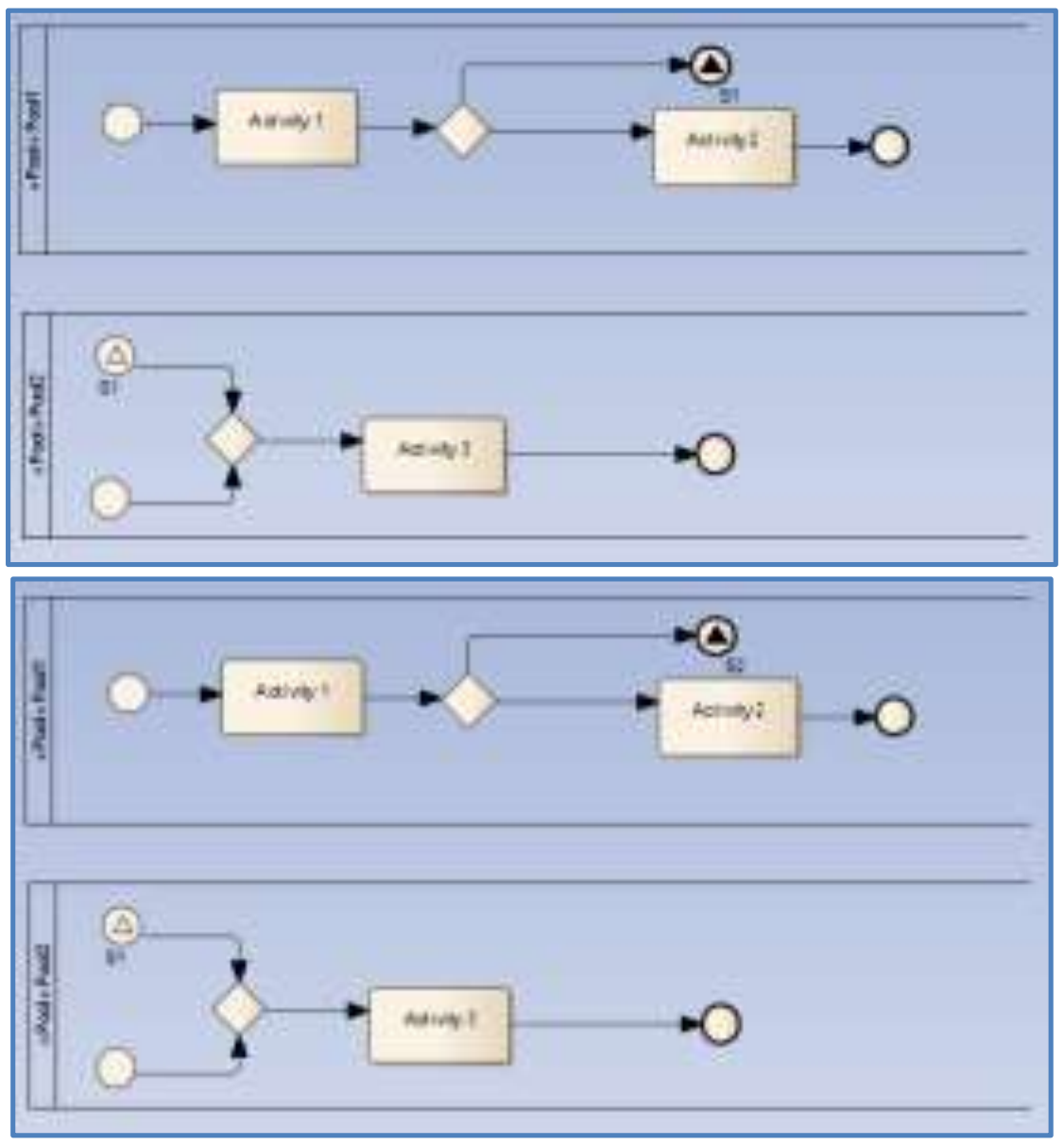

**Fig. 56.** – A *Throw Signal Event* must be caught by a *Catch Signal Event* with the same name or unnamed

**Picture's interpretation**: Correct: A *Throw Signal End Event* "S1" is caught by a *Catch Signal Event* "S1" in other process (top); Wrong: A *Throw Signal End Event* "S2" is not caught by any *Catch Signal Event* in the same or other process of collaboration (bottom).

A well-formedness rule can be enforced by attaching the following invariant to the *ThrowEvent* element of the BPMN metamodel.

```
context ThrowEvent
  inv existsCatchSignalEvent:
    hasCatchSignalEvent()
```

**Listing 54 – A *Throw Signal Event* must be caught by a *Catch Signal Event* with the same name or unnamed**

### 5.1.45. A named *Catch Signal Event* captures *Signal Throw Event* with the same name

A *Catch Signal Event* is used to capture a *Throw Signal Event* with the same name. An unnamed *Catch Signal Event* can capture any throwing *Signal Event* independently of its name. The *Catch Signal Event* can reside in the same process of the *Throw Signal Event* or in other process within the same collaboration [1] (page 253).

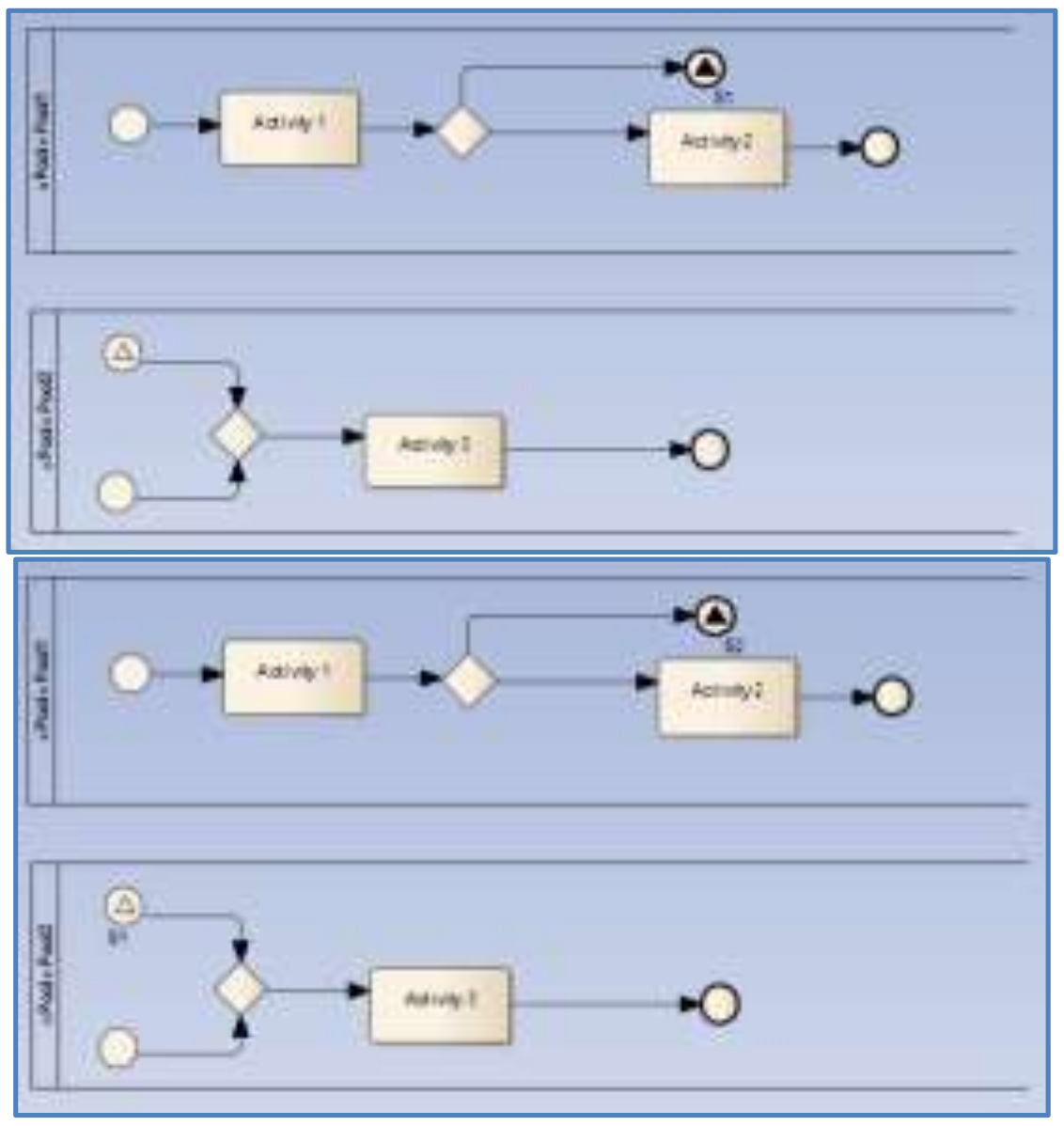

**Fig. 57.** – A named *Catch Signal Event* captures *Signal Throw Event* with the same name

**Picture's interpretation**: Correct: A *Throw Signal End Event* "S1" is caught by an unnamed *Catch Signal Event* in other process (top); Wrong: A *Catch Signal End Event* "S1" do not captures any *Throw Signal Event* in the same or other process of collaboration (bottom).

A well-formedness rule can be enforced by attaching the following invariant to the *ThrowEvent* element of the BPMN metamodel. This rule is already covered by the invariant of section 5.1.44.

```
context Process
inv catchingSignalEventHasThrowEventWithSameName:
    nameOfCatchingSignalEventMatchNameOfThrowingEvent()
```

**Listing 55 – A named *Catch Signal Event* captures *Signal Throw Event* with the same name**

#### 5.1.46. A catching *Cancel Intermediate Event* can only be used attached to the boundary of a *Transaction Sub-Process*

The *Catch Cancel Event* is triggered when a throwing *Cancel End Event* is reached inside the transaction, resulting in the abortion of the transaction. The required compensations are then carried out and the exception flow at the attached cancel intermediate event is triggered [1] (page 263). A transaction cannot have a normal intermediate event (e.g. a message) attached to its boundary, because it would not perform the compensations required by the transaction before trigger the exception flow.

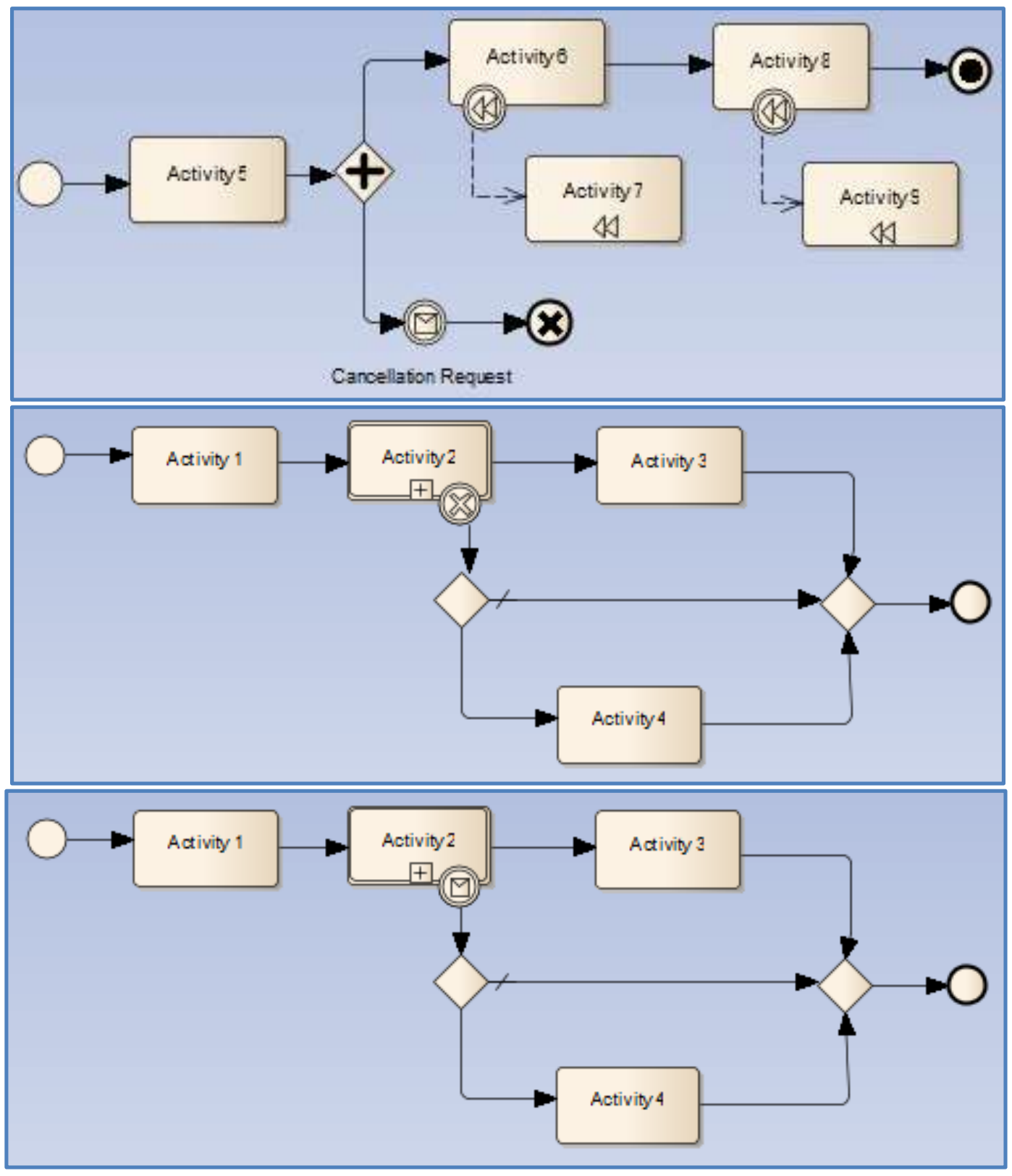

**Fig. 58.** – A catching *Cancel Intermediate Event* can only be used attached to the boundary of a *Transaction Sub-Process*

**Picture's interpretation**: Correct: A *Throw Cancel End Event* within a Transaction SubProcess (top) is caught by a *Catch Cancel Event* (middle); Wrong: An *Interrupting Catch Message Event* attached to a *Transaction SubProcess* (bottom).

A well-formedness rule can be enforced by attaching the following invariant to the *BoundaryEvent* element of the BPMN metamodel.

```
context BoundaryEvent
  inv transactionCanOnlyHaveCancelEvent:
    self.isAttachedActivityTransaction()
    implies
    self.isCancelEvent()
```

**Listing 56 – A catching *Cancel Intermediate Event* can only be used attached to the boundary of a *Transaction Sub-Process***

### 5.1.47. The *Cancel End Event* must be contained within the *Transaction SubProcess* or within a lower-level child *Transaction SubProcess*

To cancel the *Transaction SubProcess*, the *Cancel End Event* must be contained within the *SubProcess* or within a lower-level child *SubProcess* [71]. The *Cancel End Event* must only be used within a *Transaction SubProcess* and, thus, may not be used in any other type of *SubProcess* or *Process* [1] (page 263, 280).

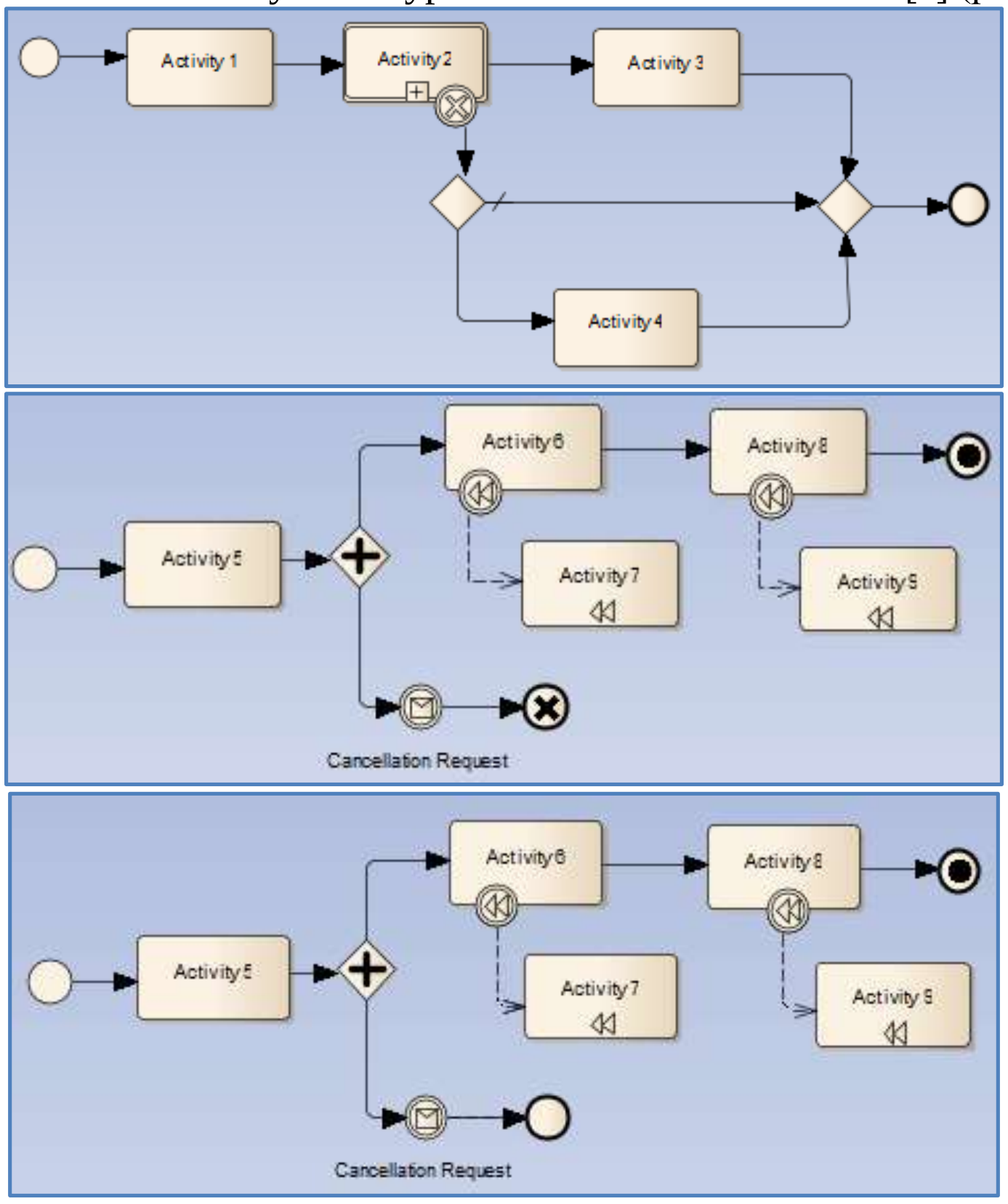

**Fig. 59.** – The *Cancel End Event* must be contained within the *Transaction SubProcess* or within a lower-level child *Transaction SubProcess*

**Picture's interpretation**: Correct: A *Throw Cancel End Event* (middle) exists within the *Transaction SubProcess* (top); Wrong: There is no *Throw Cancel End Event* within the *Transaction SubProcess* (bottom);

A well-formedness rule can be enforced by attaching the following invariant to the *Transaction* element of the BPMN metamodel.

```
context Transaction
  inv hasCancelEndEvent:
    self.boundaryEventRefs->
      select(oclIsTypeOf(BoundaryEvent) and
          isCancelEvent())->asSet()->size()>0
    implies
    self.bpmnAllElements(self)->flatten
      ->select(oclIsKindOf(EndEvent)
      and oclAsType(EndEvent).isCancelEvent())
      ->asSet()->size() > 0
```

**Listing 57 – The *Cancel End Event* must be contained within the *Transaction SubProcess* or within a lower-level child *Transaction SubProcess***

### 5.1.48. A *Terminate End Event* must exist in a *Transaction* if there are several types of *End Events*

At least one terminate end event must be used to finish a sub-process transaction, if there are several paths with different types of *End Events*. A Terminate *End Event* finishes successfully an entire transaction, including other paths, and the process can continue the normal flow [1] (page 248, 431, 443) [70].

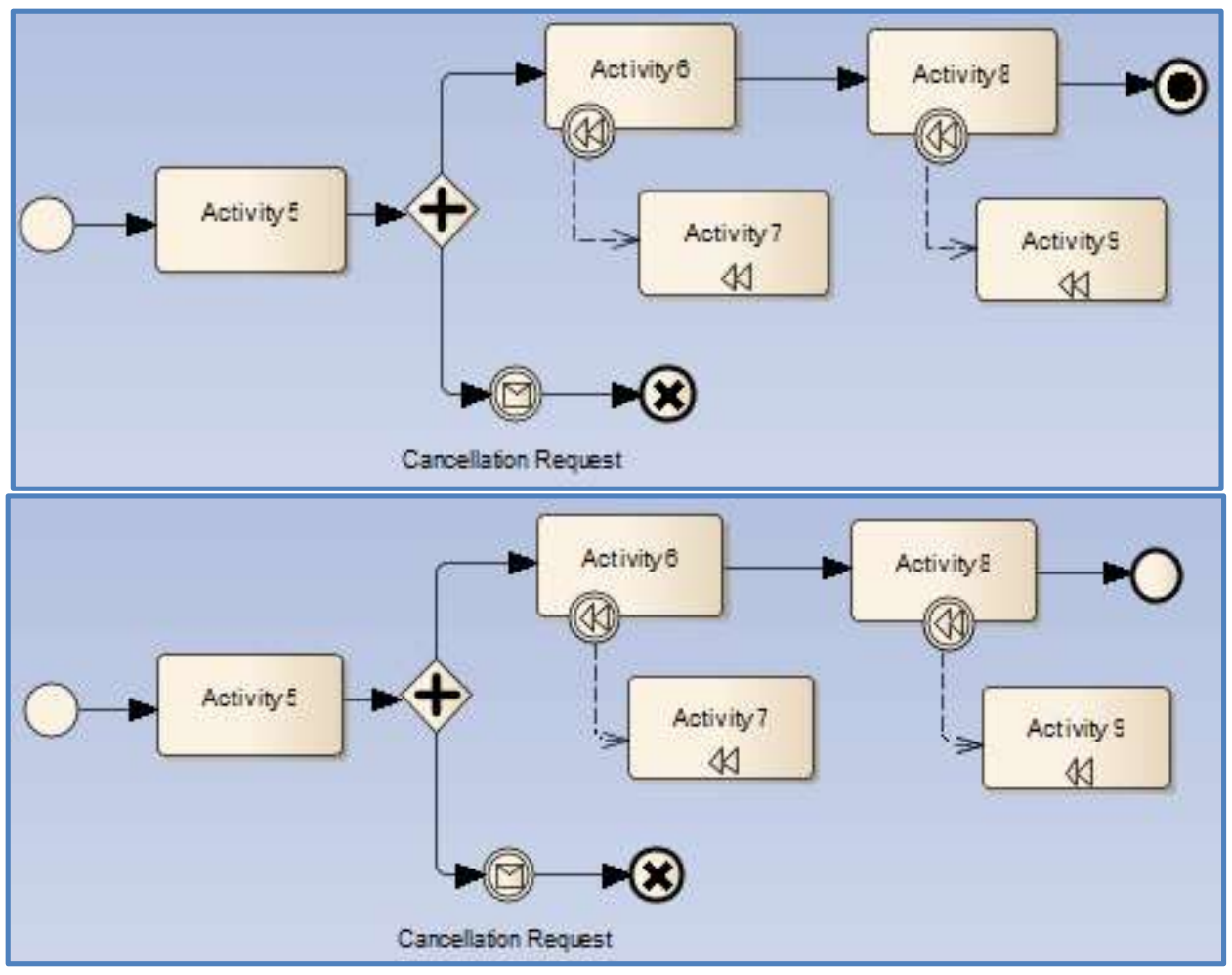

**Fig. 60.** – A *Terminate End Event* must exist in a *Transaction* if there are several types of *End Events*

**Picture's interpretation**: Correct: A *Terminate End Event* is used to finish the entire *Transaction* (top); Wrong: A *None End Event* must not be used to terminate the entire *Transaction* (bottom).

A well-formedness rule can be enforced by attaching the following invariant to the *Transaction* element of the BPMN metamodel.

```
context Transaction
  inv multiPathsRequireTerminateEndEvent:
    (totalNumberContainerEndEvents() > 1)
    implies
    (totalContainerEndEvents()
        ->select(isTerminateEvent())->size() > 0)
```

**Listing 58 – A *Terminate End Event* must exist in a *Transaction* if there are several types of *End Events***

### 5.1.49. A *Compensation End* or *Intermediate Event* can only be used in a *Sub-Process* which is not a *Transaction*

A *Throwing Compensation End* or *Intermediate Events* can be used when compensations are modeled in a *Sub-Process* which are not a *Transaction* [1] (page 248) [70]. In *Transactions* compensation activities are automatically triggered when the transaction is aborted.

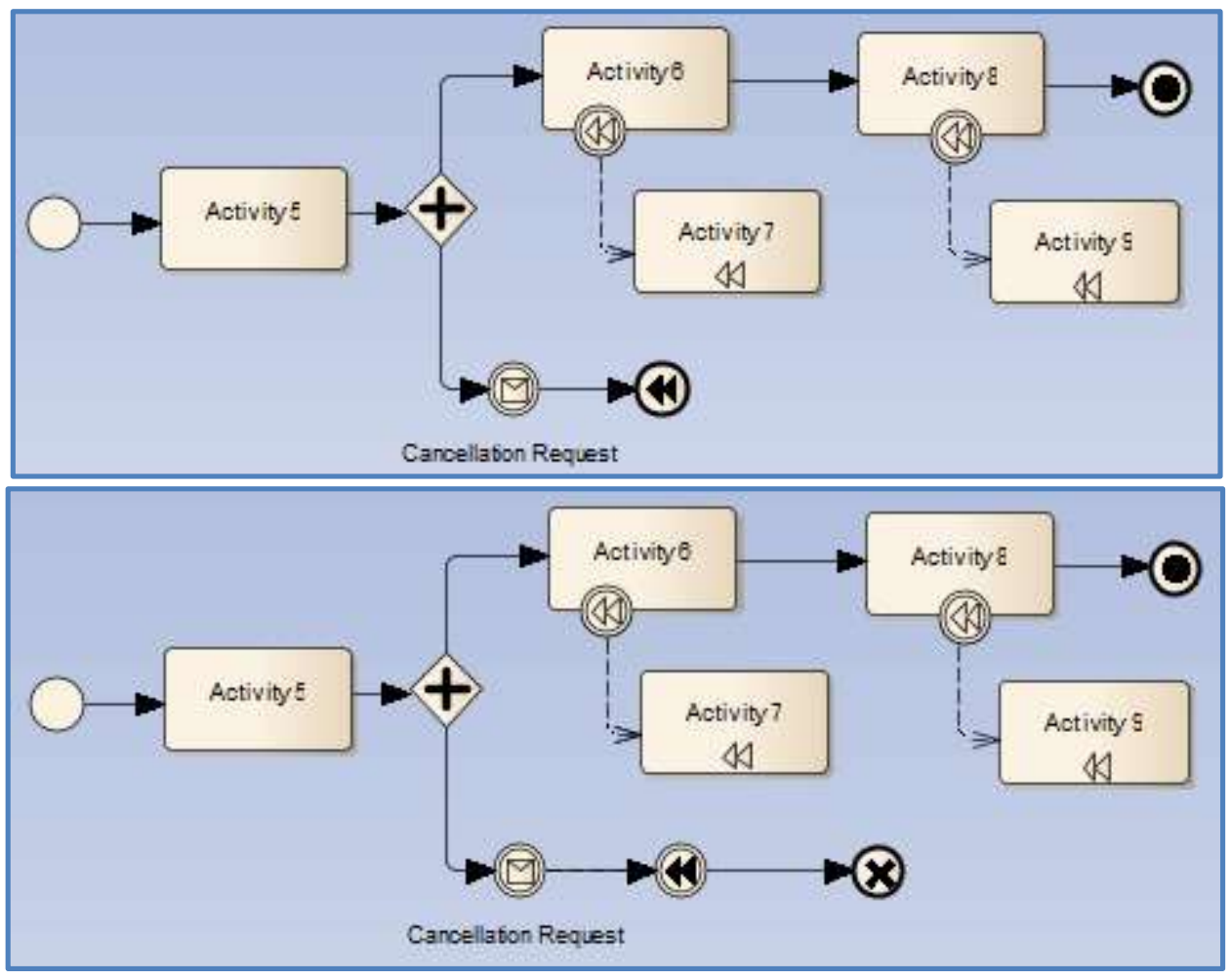

**Fig. 61.** – A *Compensation End* or *Intermediate Event* can only be used in a *Sub-Process* which is not a *Transaction*

**Picture's interpretation**: Correct: A *Compensation End Event* is used in a regular sub-process (top); Wrong: A *Compensation Intermediate Event* cannot be used in a sub-process which is a *Transaction* (bottom).

A well-formedness rule can be enforced by attaching the following invariant to the *Transaction* element of the BPMN metamodel.

```
context Transaction
  inv trowingCompensatingEventsNotAllowed:
    totalContainerThrowEvents()
        ->select(isCompensateEvent())->size() = 0
```

**Listing 59 – A *Compensation End* or *Intermediate Event* can only be used in a *Sub-Process* which is not a *Transaction***

### 5.1.50. *Embedded SubProcess* can have *Compensation Activities* explicitly called or via *Event SubProcess*

When activities that are part of embedded sub-process have compensation activities, these activities can be called explicitly with throwing compensation event [1] (page 252).

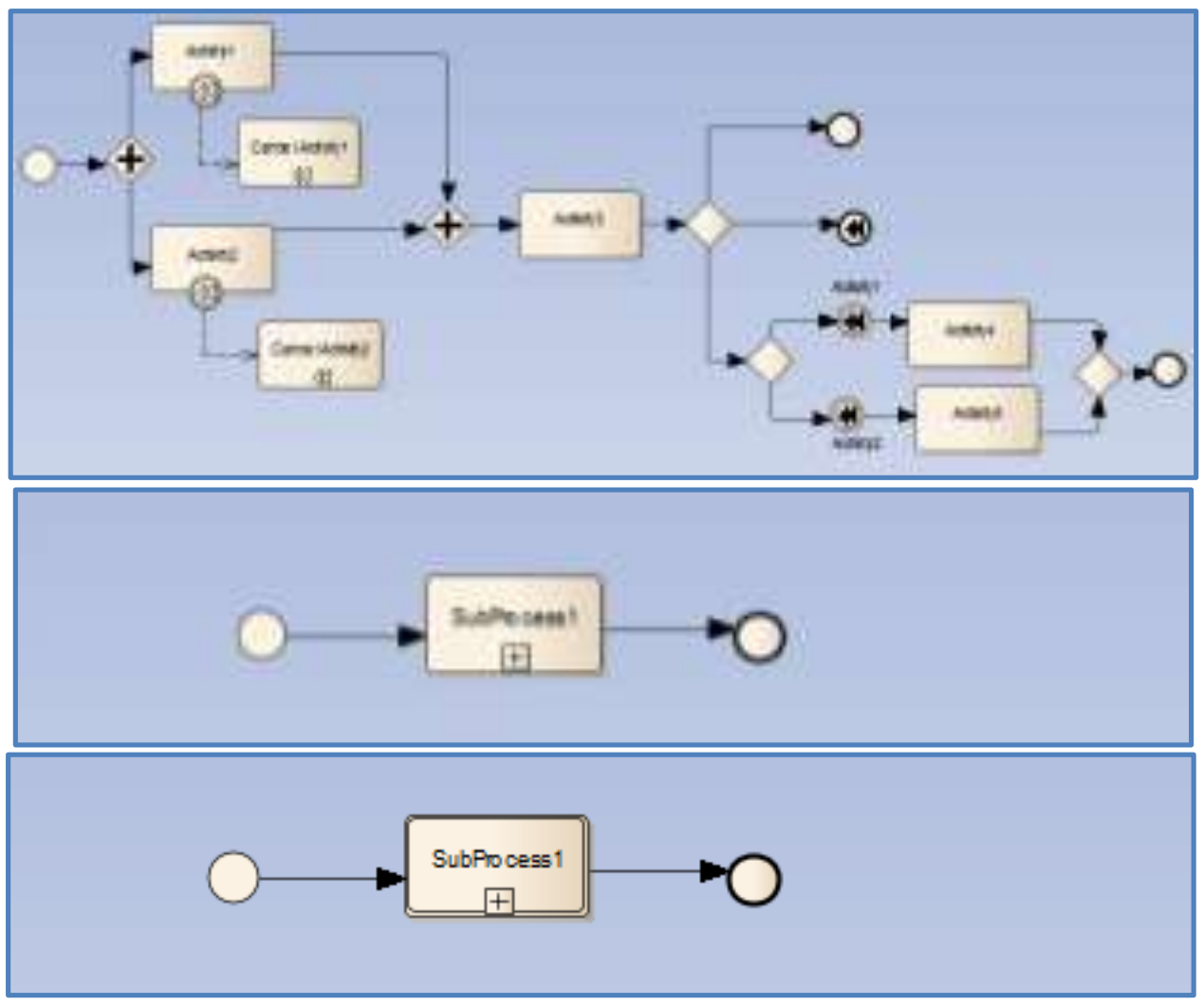

**Fig. 62.** – *Embedded SubProcess* can have *Compensation Activities* explicitly called or via *Event SubProcess* (I)

**Picture's interpretation**: Correct: *Compensation Activities* match *Compensation Intermediate* and *End Events* (top) as part of an embedded sub-process (middle); Wrong: A *Transaction* cannot explicitly throw *Compensation Events* (bottom).

Compensations of a sub-process can also be handled using *Event Sub-Process*. In this case *Event Sub-Process* must have *Compensation Start Event*. An *Event Sub-Process* is started when a throwing *Compensation Event* of the same name occurs after that process. In contrast to other event sub-process, they are not triggered by an event of the surrounding process. Thus, the event sub-process will only be performed after the surrounding process has been finished, which results in no distinction between a non-Interrupting and an Interrupting *Compensation Start Event* [1] (page 252) [70].

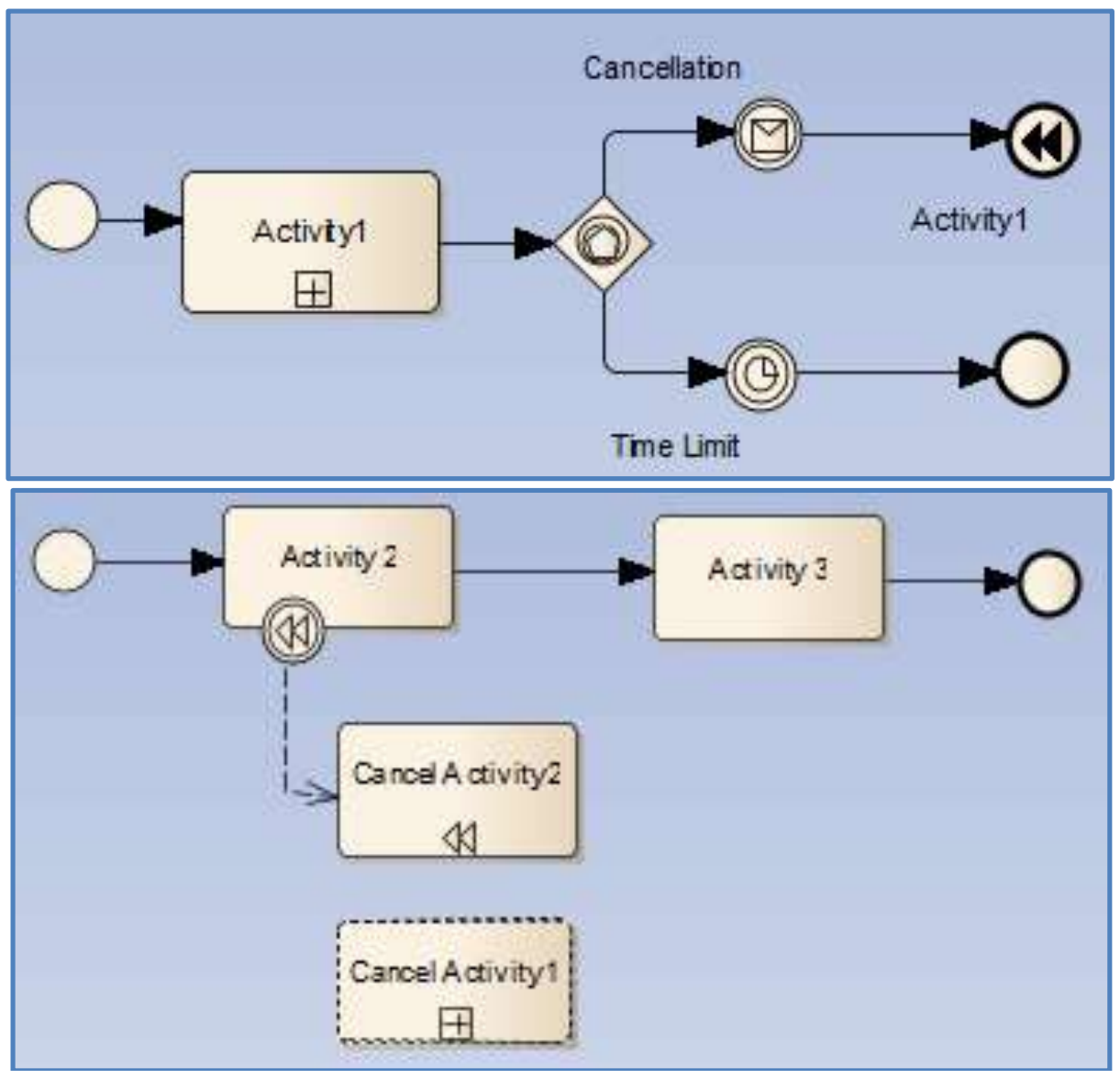

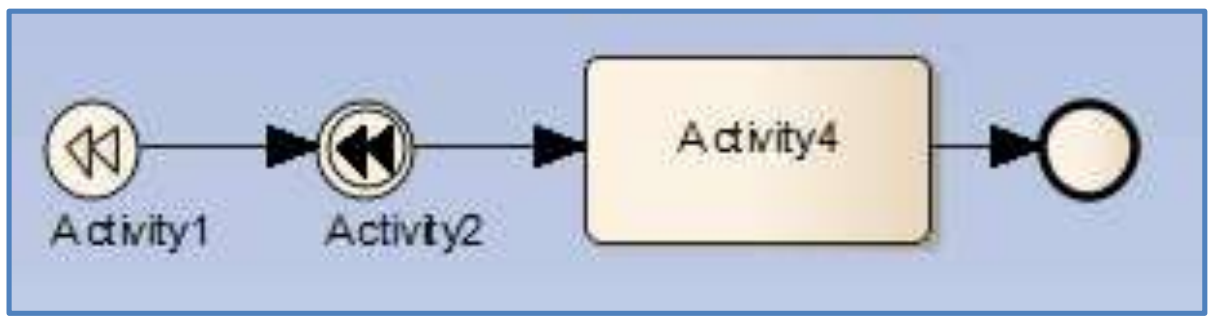

**Fig. 63.** – *Embedded SubProcess* can have *Compensation Activities* explicitly called or via *Event SubProcess* (II)

**Picture's interpretation**: Correct: *Compensation End Event* "Activity1" (top) throws an event that can be cached by the *Start Event* "Activity1" of the *Event sub-process* "Cancel Activity1" (expanded at bottom). This *Event sub-process* has an *Intermediate Throw Compensation Event* "Activity2" that starts the compensation of "Activity2" through *Compensation Activity* "Cancel Activity1" (middle).

A well-formedness rule can be enforced by attaching the following invariant to the *FlowElementsContainer* element of the BPMN metamodel.

```
context FlowElementsContainer
  inv compensationActivityMustBeCalled:
    isCompensationActivityPartOfSubProcess()
```

**Listing 60 – *Embedded SubProcess* can have *Compensation Activities* explicitly called or via *Event SubProcess***

#### 5.1.51. The name of the throwing *Intermediate Compensation Event* must match to the name of cancelled *Activities*

The name of a throwing *Intermediate Compensation Event* must match to the name of one of the activities that can be cancelled. The name of the *Compensation End Events* or *Compensation Intermediate Events* in the same process, must match Activities that can be cancelled, since they have *Compensation Activities* assigned, which are called when those compensation events are thrown [1] (page 248, 252) [70].

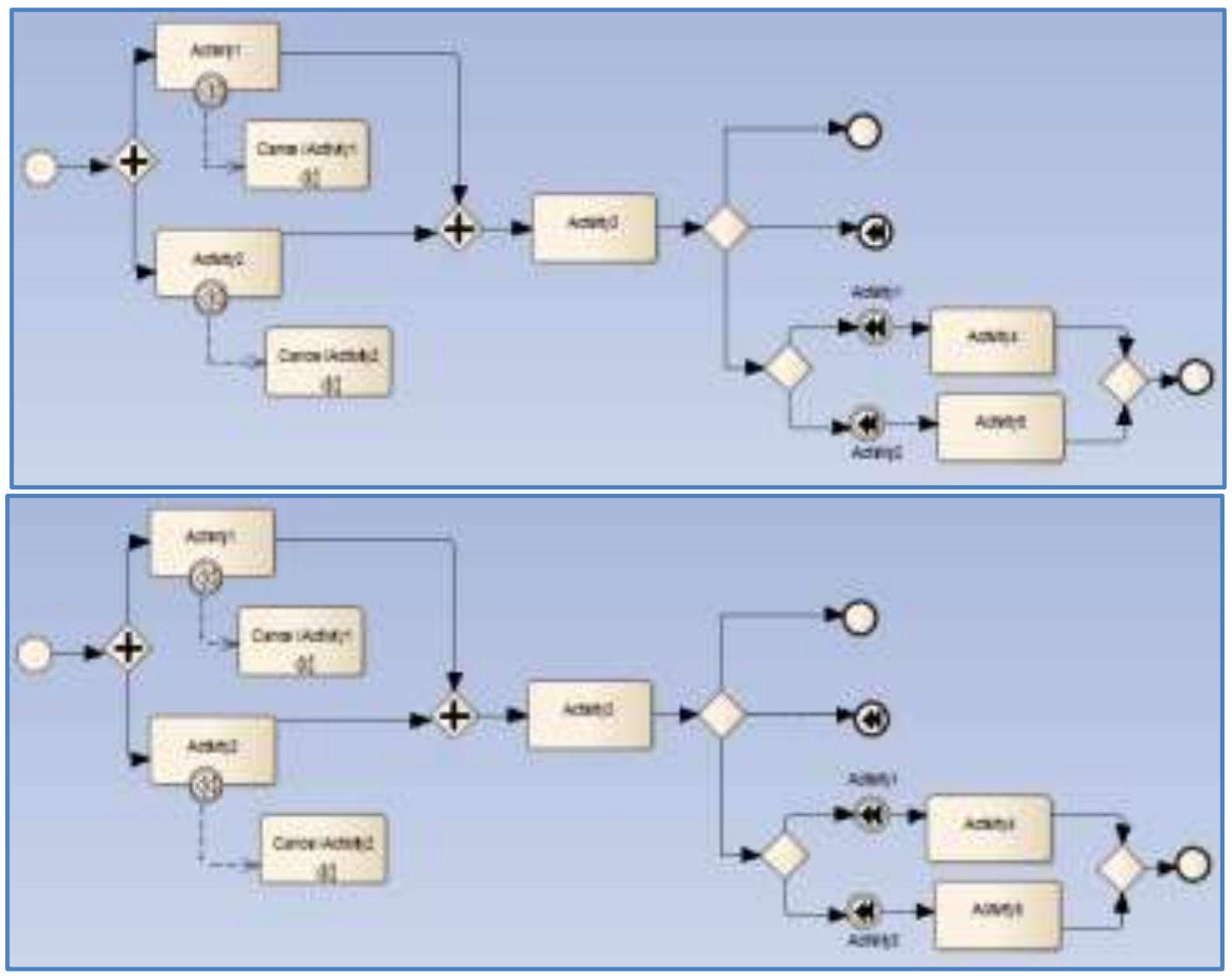

**Fig. 64.** – The name of the throwing *Intermediate Compensation Event* must match to the name of cancelled *Activities*

**Picture's interpretation**: Correct: *Compensation Throw Events* are unnamed or their name match one of *Activities* with *Compensation Activities* (top); Wrong: *Compensation Intermediate Throw Event* "Activity3" does not match any *Activity* with *Compensation Activity* (bottom).

A well-formedness rule can be enforced by attaching the following invariant to the *FlowElementsContainer* element of the BPMN metamodel.

```
context FlowElementsContainer
  inv throwCompensationEventMatchOneOfActivToBeCancelled:
   nameOfThrowCompensationEventMatchNameOfCancelledActiv()
```

**Listing 61 – The name of the throwing *Intermediate Compensation Event* must match to the name of cancelled *Activities***

### 5.1.52. An exception flow originated by *Interrupting Catch Events* can merge the normal flow through an *Exclusive Gateway*

When an activity is carried out, an exception path originated from the *Interrupting Catch Event* can be taken due to early abortion of the activity. If later the exception path joins the normal sequence flow of the *Process*, it requires an *Exclusive Gateway* [1] (page 258, 259) [70].

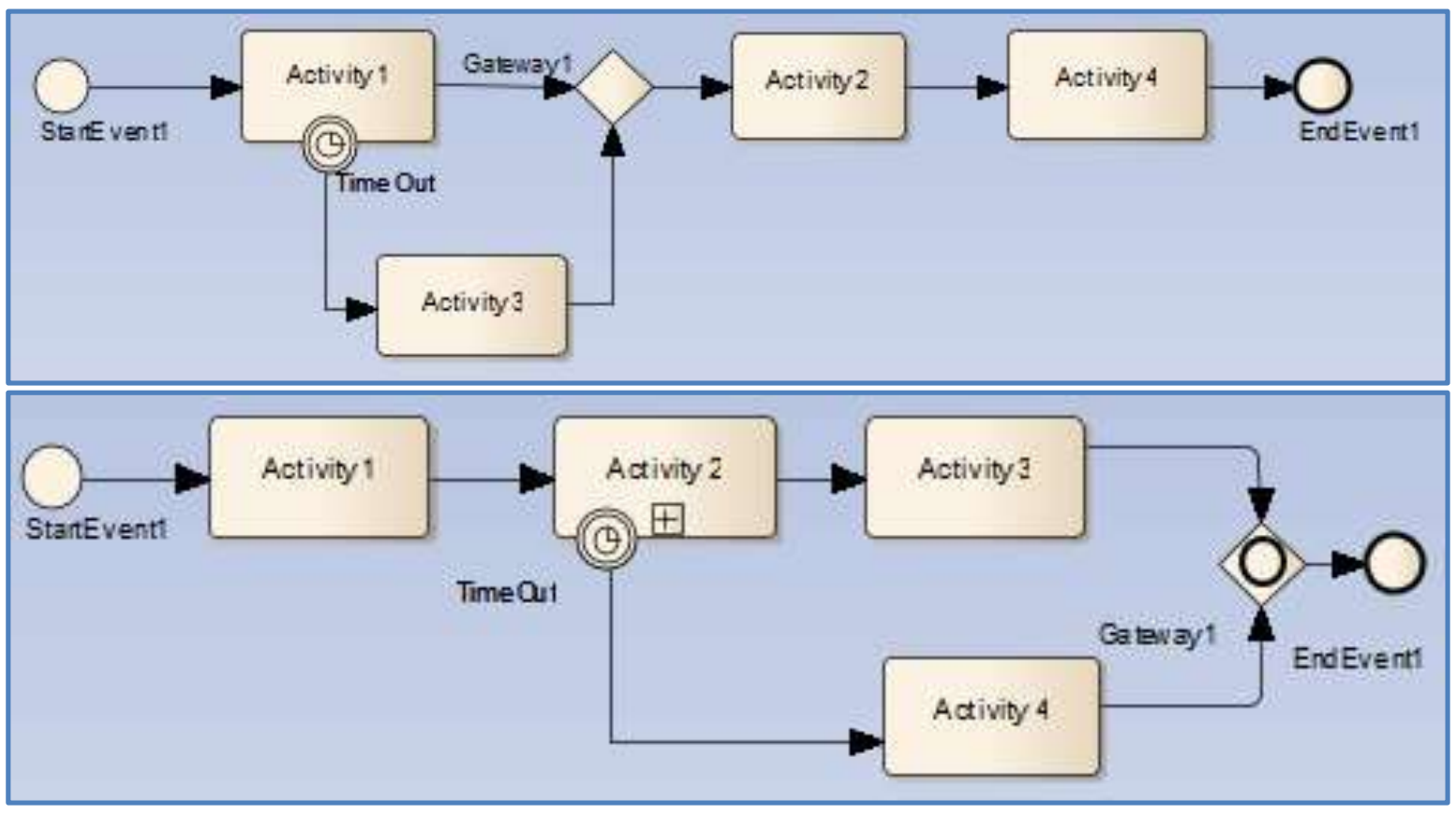

**Fig. 65.** – An exception flow originated by *Interrupting Catch Events* can merge the normal flow through an *Exclusive Gateway*

**Picture's interpretation**: <u>Correct</u>: If the event "TimeOut" occurs, "Activity1" will be aborted, and the downwards outgoing exception flow will be taken. Later it will join the normal flow of the process via an *Exclusive Gateway* (top); <u>Wrong</u>: The exception flow joins the normal flow through an *Inclusive Gateway* (bottom).

A well-formedness rule can be enforced by attaching the following invariant to the *BoundaryEvent* element of the BPMN metamodel.

```
context BoundaryEvent
  inv interruptingEventPathMergedByAnExclusiveGateway:
    pathMergedByAnExclusiveGateway()
```

**Listing 62 – An exception flow originated by *Interrupting Catch Events* can merge the normal flow through an *Exclusive Gateway***

### 5.1.53. An exception flow originated by *Non-Interrupting Catch Events* can merge the normal flow through an Inclusive *Gateway*

An attached non-interrupting event creates a parallel path if it occurs during the execution of the respective activity. If both paths (normal and exception) merge, an inclusive gateway would be required [1] (page 258, 259) [70]. Neither a parallel gateway nor exclusive gateway is suited for this scenario.

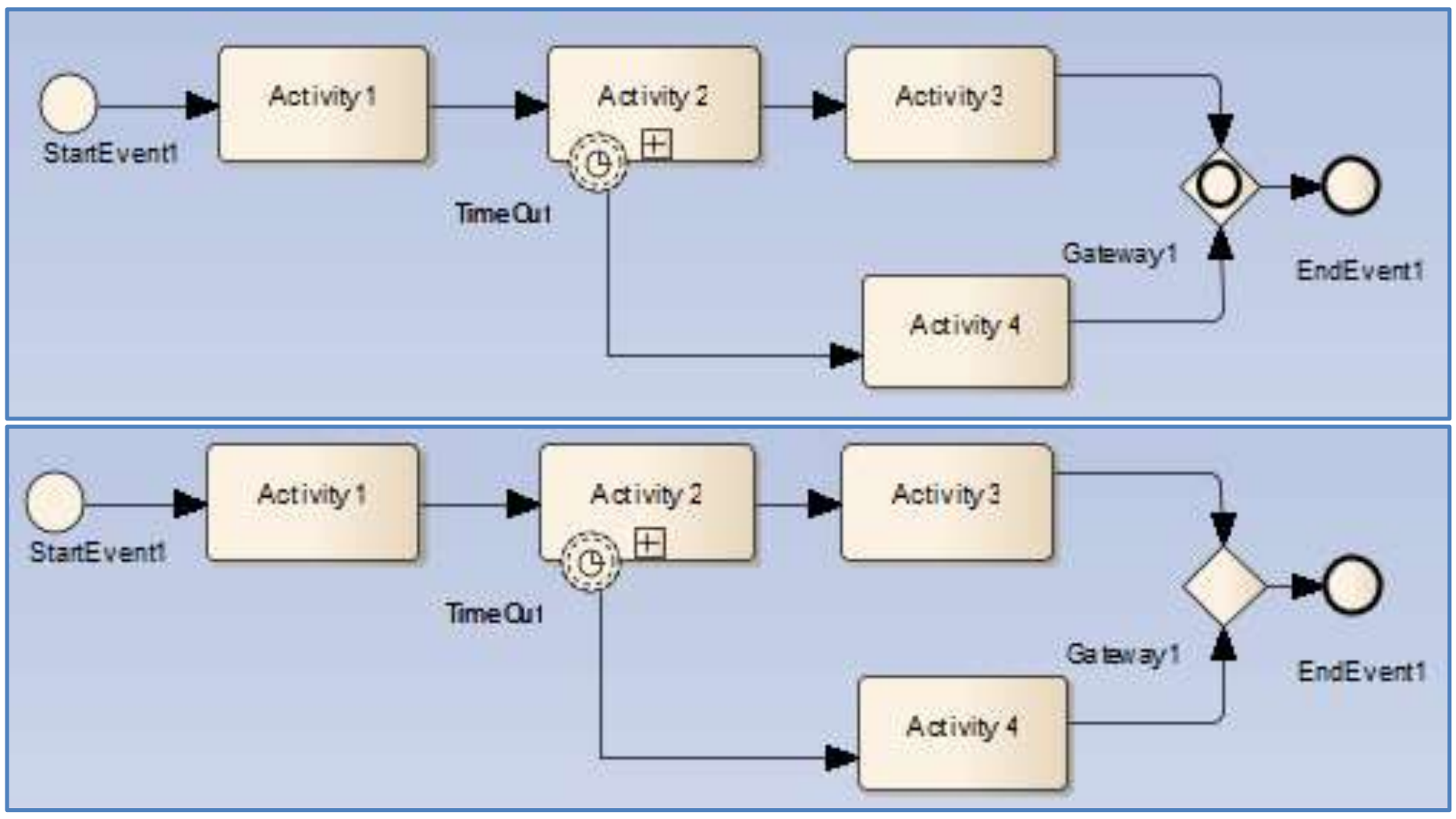

**Fig. 66.** – An exception flow originated by *Non-Interrupting Catch Events* can merge the normal flow through an Inclusive *Gateway*

**Picture's interpretation**: Correct: If the event "TimeOut" occurs, "Activity2" will continue, and besides the normal flow, an exception flow will be taken. Later they join via an Inclusive *Gateway* (top); Wrong: The exception flow joins the normal flow through an *Exclusive Gateway* (bottom).

A well-formedness rule can be enforced by attaching the following invariant to the *BoundaryEvent* element of the BPMN metamodel.

```
context BoundaryEvent
  inv nonInterruptingEventPathMergedByAnInclusiveGateway:
    pathMergedByAnInclusiveGateway()
```

**Listing 63 – An exception flow originated by *Non-Interrupting Catch Events* can merge the normal flow through an Inclusive *Gateway***

### 5.1.54. A *Condition Expression* must not be used if the source of the *Sequence Flow* is an *Event*

A *Sequence Flow* can have a *Condition Expression* that is evaluated at runtime to determine whether or not the *Sequence Flow* will be used. *Conditional* outgoing *Sequence Flows* cannot have as source *Events* [1] (page 35, 97).

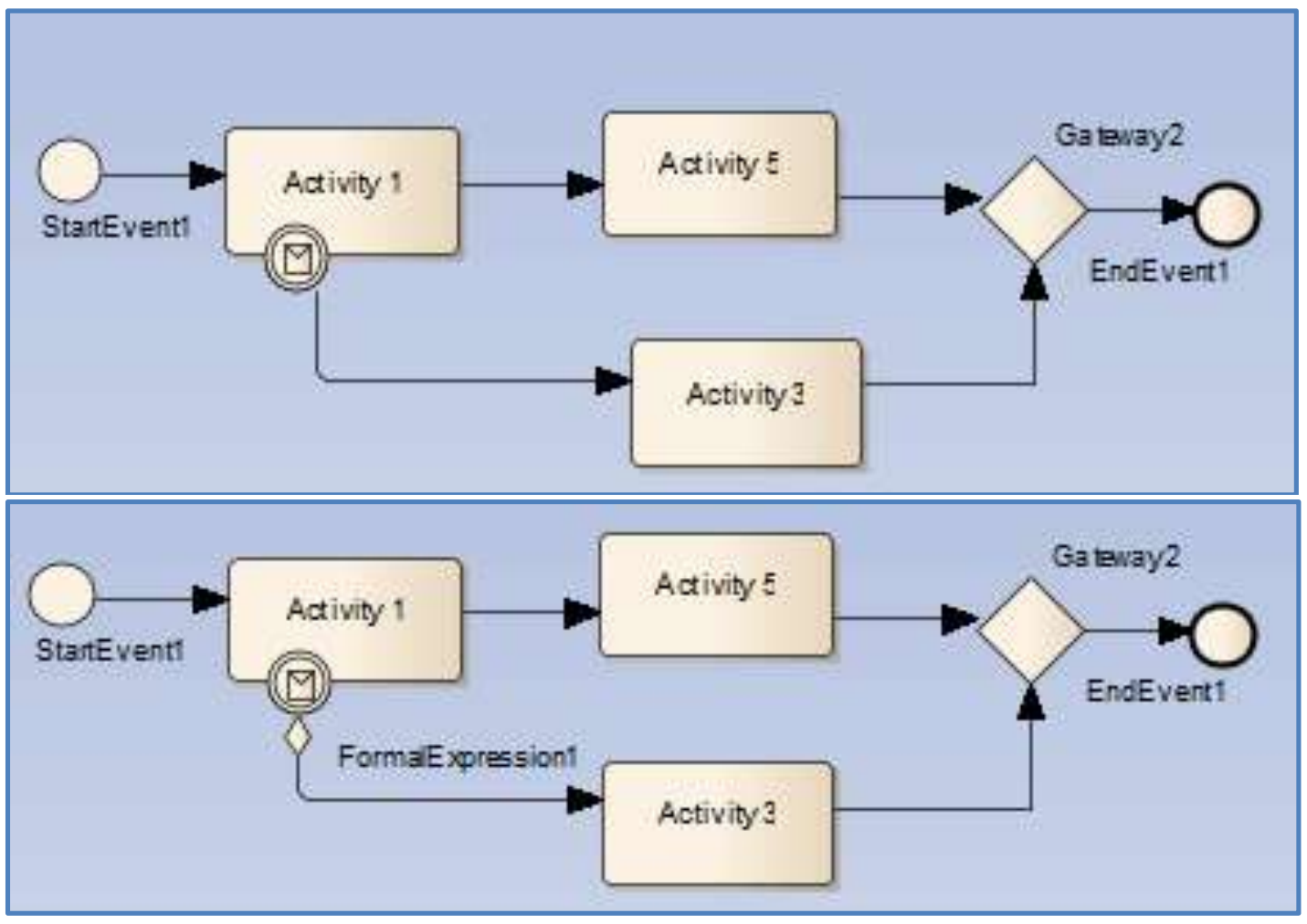

**Fig. 67.** – A *Condition Expression* must not be used if the source of the *Sequence Flow* is an *Event*

**Picture's interpretation**: Correct: The *Sequence Flow* that is source of the *Boundary Event* is unconditional (top); Wrong: The source of a conditional outgoing sequence flow is a *Boundary Event* (bottom);

A well-formedness rule can be enforced by attaching the following invariant to the *SequenceFlow* element of the BPMN metamodel.

```
context SequenceFlow
  inv sourceMustNotBeEvent:
    self.sourceRef.oclIsKindOf(Event)
    implies
    conditionExpression.isUndefined()
```

**Listing 64 – A *Condition Expression* must not be used if the source of the *Sequence Flow* is an *Event***

### 5.1.55. A *Boundary Event* must have exactly one outgoing *Sequence Flow* (unless it has the *Compensation* type)

A *Boundary Event* is attached to an Activity and an outgoing exception flow comes out from it, through a *Sequence Flow*. Exactly one *Sequence Flow* is allowed from a *Boundary Event* except in the case that it is of type *Compensation*. In this particular case an Association can replace or not the *Sequence Flow* [1] (page 259, 440, 441).

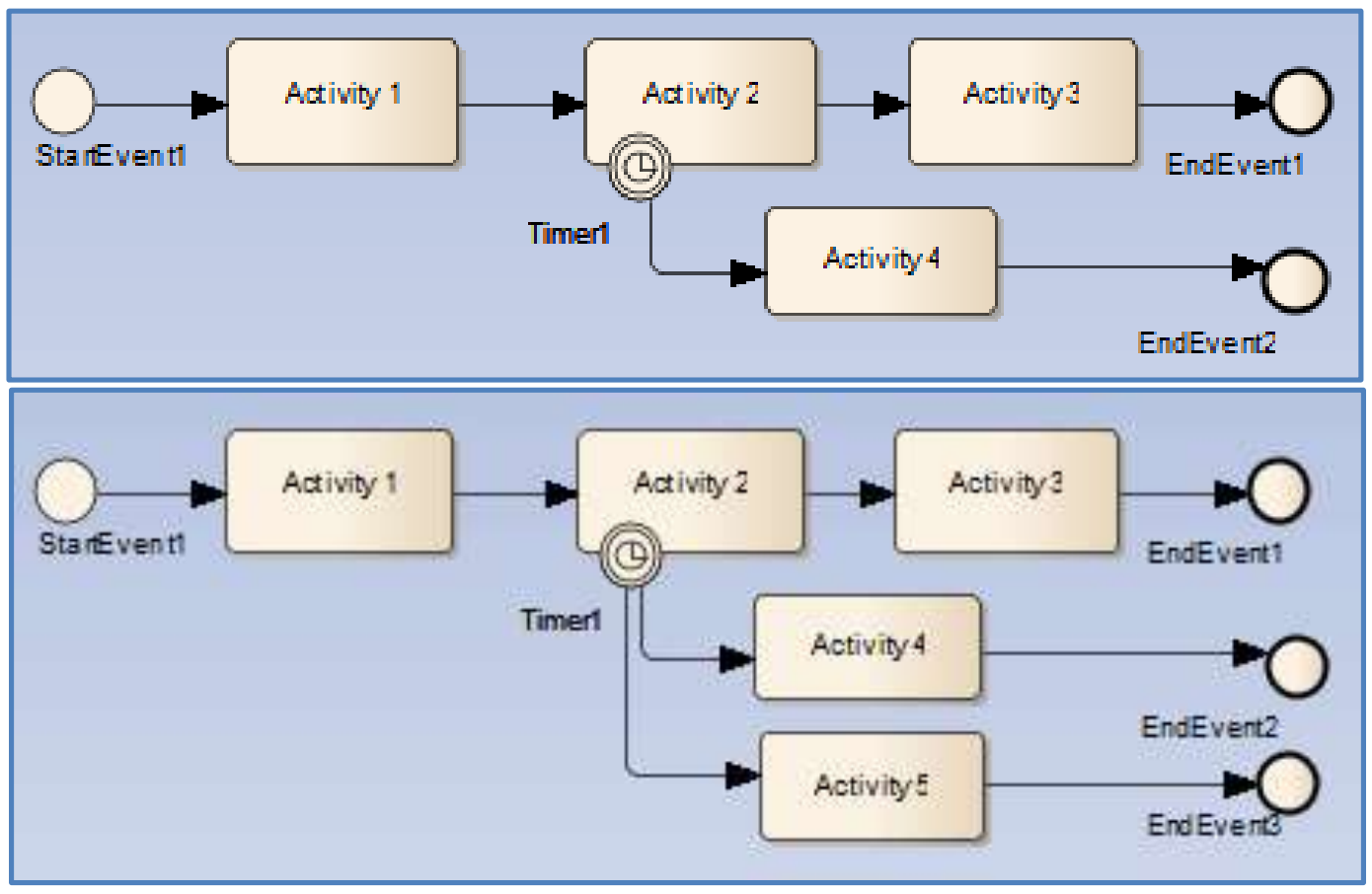

**Fig. 68.** – A *Boundary Event* must have exactly one outgoing *Sequence Flow* (unless it has the *Compensation* type)

**Picture's interpretation**: Correct: Only one sequence flow has as source a *Boundary Event* (top); Wrong: More than one sequence flow has as source a *Boundary Event* (bottom).

A well-formedness rule can be enforced by attaching the following invariant to the *BoundaryEvent* element of the BPMN metamodel.

```
context BoundaryEvent
  inv oneOutgoingSequenceFlow:
    not self.isCompensateEvent()
   implies
    self.outgoing_a->size() = 1
```

**Listing 65 – A *Boundary Event* must have exactly one outgoing *Sequence Flow* (unless it has the *Compensation* type)**

### 5.1.56. A *Boundary Event* must not have incoming *Sequence Flow*

A *Boundary Event* is attached to an *Activity* and so it cannot have incoming *Sequence Flow* [1] (page 259).

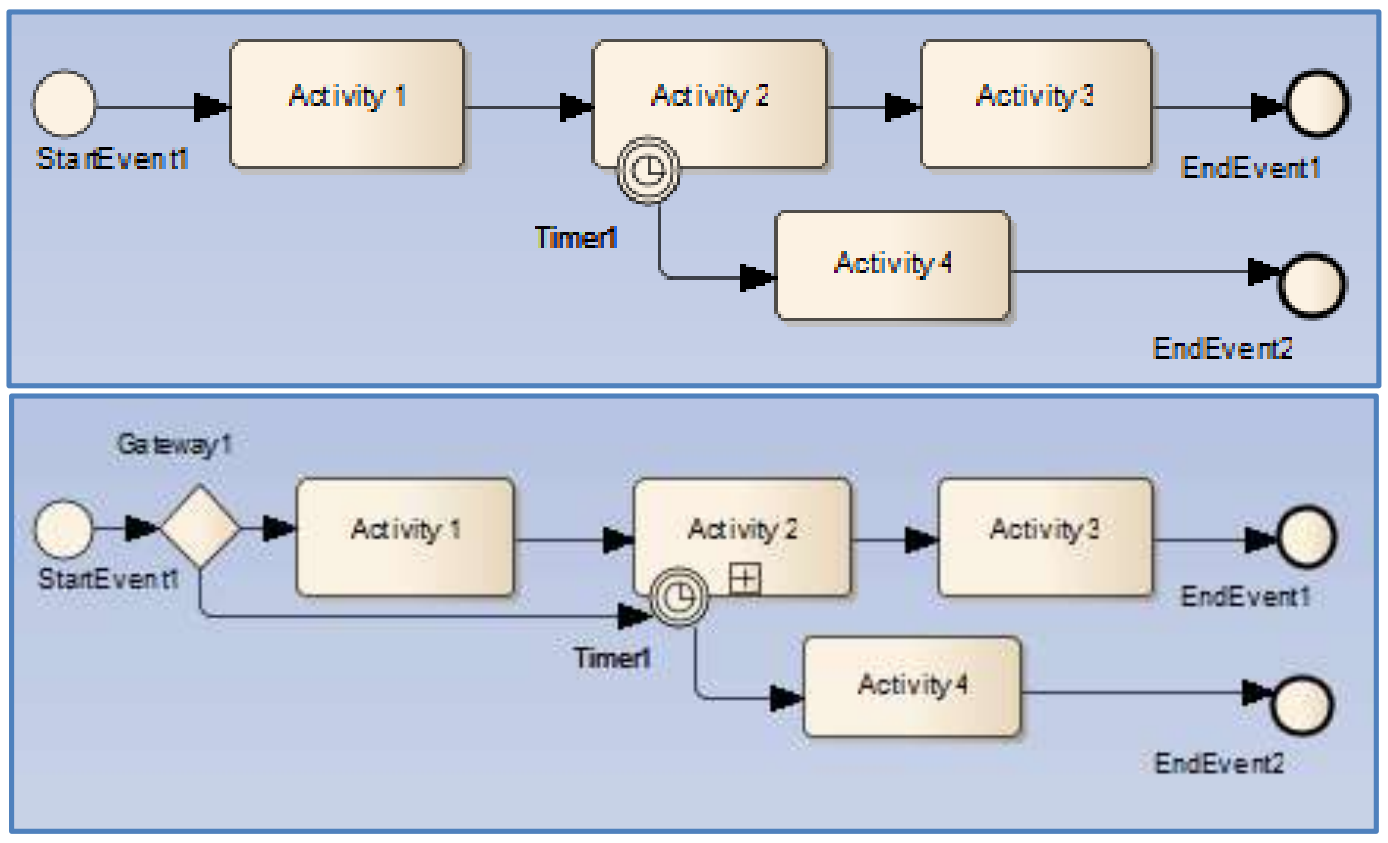

**Fig. 69.** – A *Boundary Event* must not have incoming *Sequence Flow*

**Picture's interpretation**: Correct: *Boundary Event* does not have incoming *Sequence Flow* (top); Wrong: *Boundary Event* has incoming *Sequence Flow* (bottom).

A well-formedness rule can be enforced by attaching the following invariant to the *BoundaryEvent* element of the BPMN metamodel.

```
context BoundaryEvent
  inv noIncomingSequenceFlow:
    self.incoming_a->size() = 0
```

**Listing 66 – A *Boundary Event* must not have incoming *Sequence Flow***

### 5.1.57. An *Intermediate Event* must have at least one incoming and outgoing *Sequence Flow*

An *Intermediate Event* must have [1] (page 259):

- one or more incoming and outgoing Sequence Flow if it is of type *None*, *Escalation*, *Compensation*, *Message*, *Timer*, *Conditional*, *Signal*, *Multiple* or *Parallel Multiple*.
- no incoming *Sequence Flow* for types *Cancel* and *Error*.

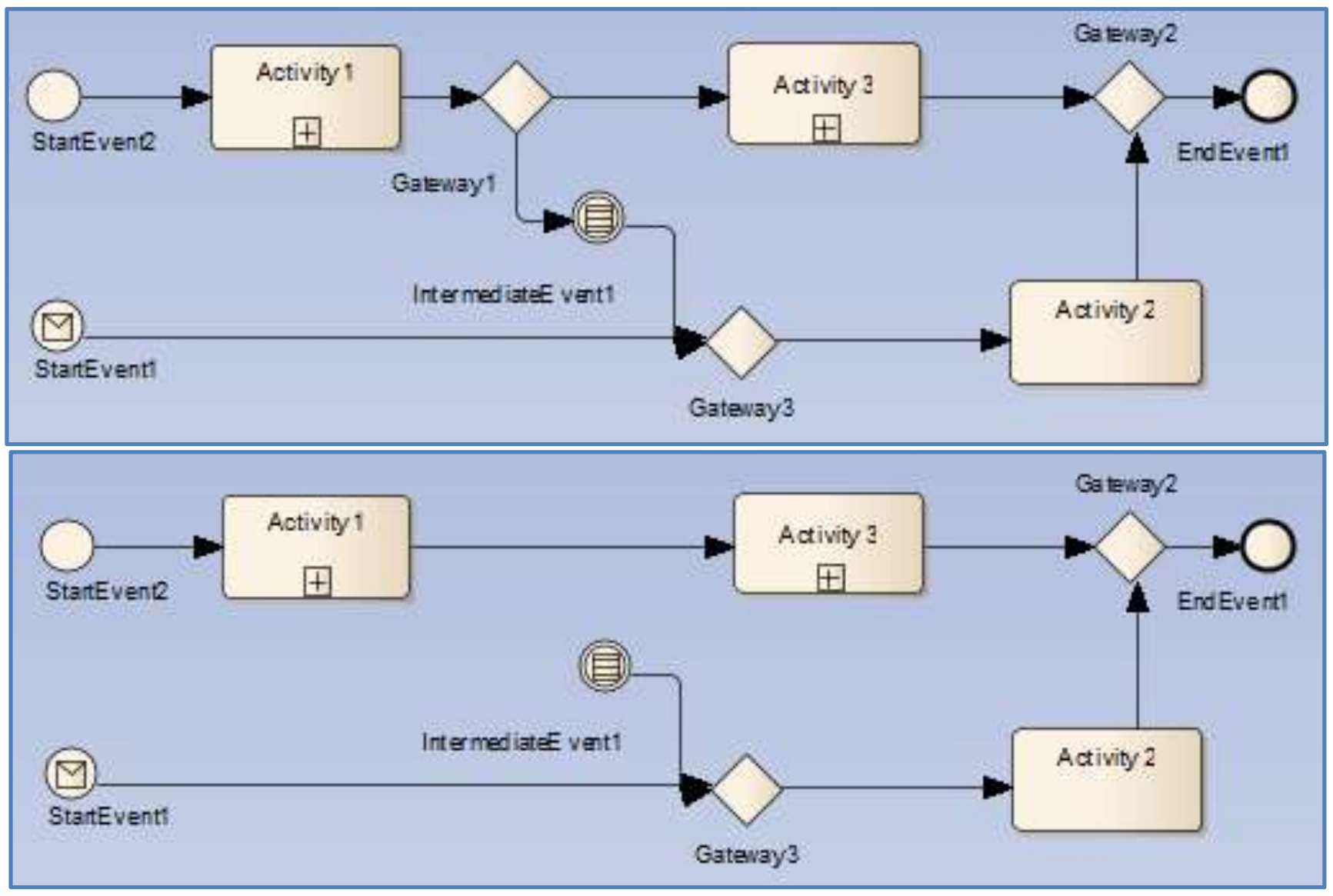

**Fig. 70.** – An *Intermediate Event* must have at least one incoming and outgoing *Sequence Flow*

**Picture's interpretation**: Correct: *Intermediate Catch Event* has incoming and outgoing *Sequence Flow* (top); Wrong: *Intermediate Catch Event* does not have incoming *Sequence Flow* (bottom);

The well-formedness rules can be enforced by attaching the following invariants to the *IntermediateThrowEvent* and *IntermediateCatchEvent* elements of the BPMN metamodel. This rules are already covered by invariant in section 5.1.25.

```
context IntermediateThrowEvent
  inv mandatoryIncomOutSequenceFlowForThrowTypedEvents:
    (self.isNoneEvent() or
    self.isEscalationEvent() or
    self.isCompensateEvent() or
    self.isMessageEvent() or
    self.isTimerEvent() or
    self.isConditionalEvent() or
    self.isSignalEvent()
    self.isMultipleEvent()
    self.isParallelMultipleEvent())
   implies
    self.incoming_a->size() > 0 and
    self.outgoing_a->size() > 0

context IntermediateCatchEvent
  inv mandatoryIncomOutSequenceFlowForCatchTypedEvents:
    (self.isNoneEvent() or
    self.isEscalationEvent() or
    self.isCompensateEvent() or
    self.isMessageEvent() or
    self.isTimerEvent() or
    self.isConditionalEvent() or
    self.isSignalEvent()
    self.isMultipleEvent()
    self.isParallelMultipleEvent())
   implies
    self.incoming_a->size() > 0 and
    self.outgoing_a->size() > 0
```

**Listing 67 – An *Intermediate Event* must have at least one incoming and outgoing *Sequence Flow***

**Gateway**

A *Gateway* represents points of control for the paths within the Process. It controls the diverging and converging of *Sequence Flow* within a container.

### 5.1.58. A *Parallel Gateway* joins only non-exclusive *Sequence Flows*

This invariant states that for merging parallel sequence flows, originated from previous split with parallel gateways, a merging parallel gateway should be used [1] (page 293-295) [70]. So, for merging non-exclusive *Sequence Flows*, a merging parallel gateway could be used. Typical scenarios where this can occur are:

- A *Parallel Gateway* joins non-exclusive tokens previously derived from an *Event Based Parallel Gateway*;
- A join *Parallel Gateway* merge non-exclusive *Sequence Flows* originated from an *Inclusive Gateway*;
- Multiple outgoing *Sequence Flows* of an activity correspond to several non-exclusive sequence flows, and they can also be merged by a parallel gateway.

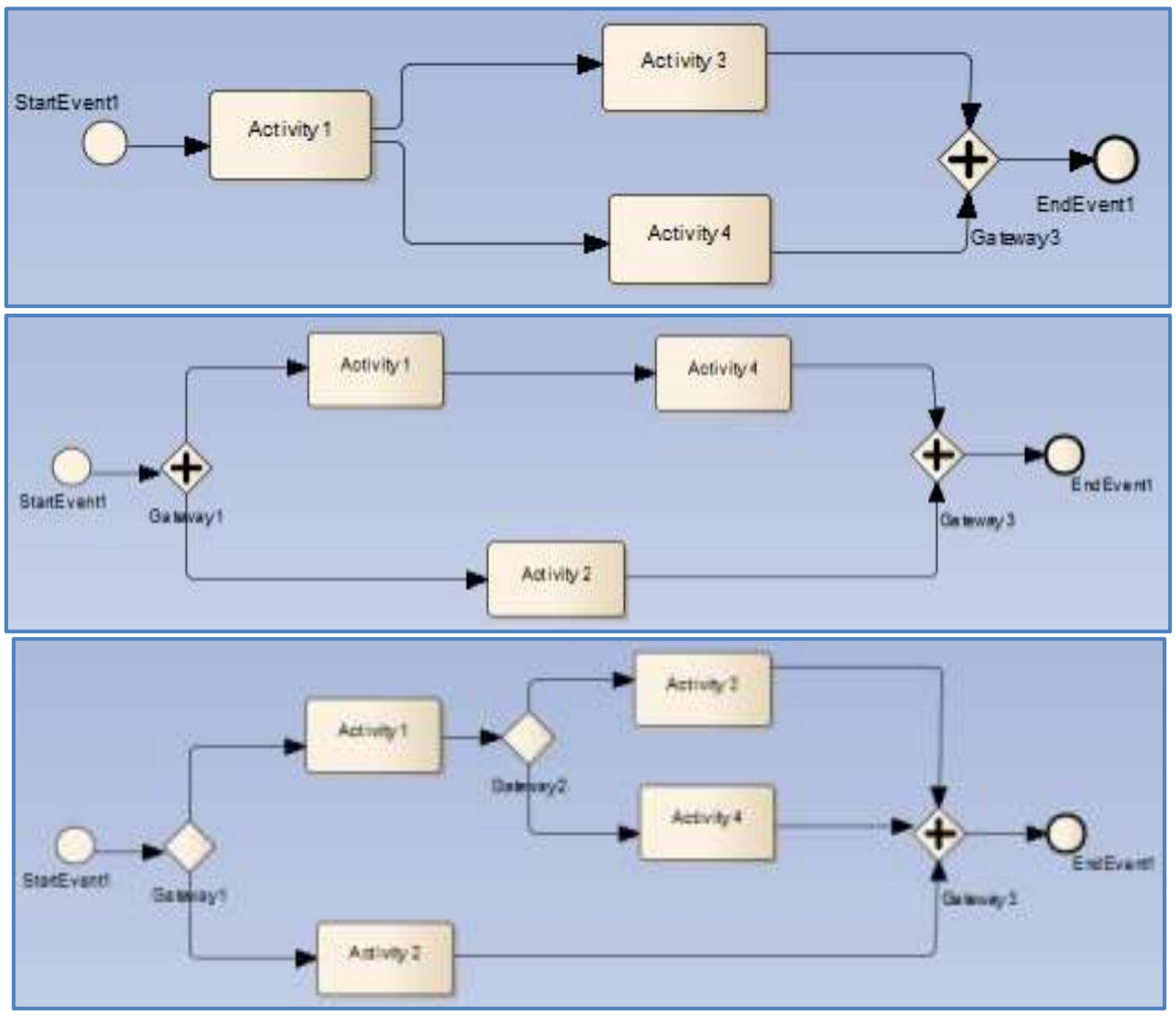

**Fig. 71.** – A *Parallel Gateway* joins only non-exclusive *Sequence Flows* (I)

**Picture's interpretation**: Correct: A parallel gateway (Gateway3) must be used to join non-exclusive sequence flows previously split apart from an activity (top diagram) and parallel gateway (Gateway1) (middle diagram); Wrong: An exclusive gateway (Gateway2 in third diagram and Gateway1 (in fourth, fifth and sixth diagrams) precedes a parallel gateway (Gateway3) that must handle only non-exclusive sequence flows (bottom).

If a Parallel Gateway is expecting a token from all of its incoming Sequence Flow and one never arrives, the Process will *deadlock*. A deadlock is a situation where the flow of the Process cannot continue because a requirement of the model is not satisfied [71]. The diagrams below depict some situations where deadlock will occurs.

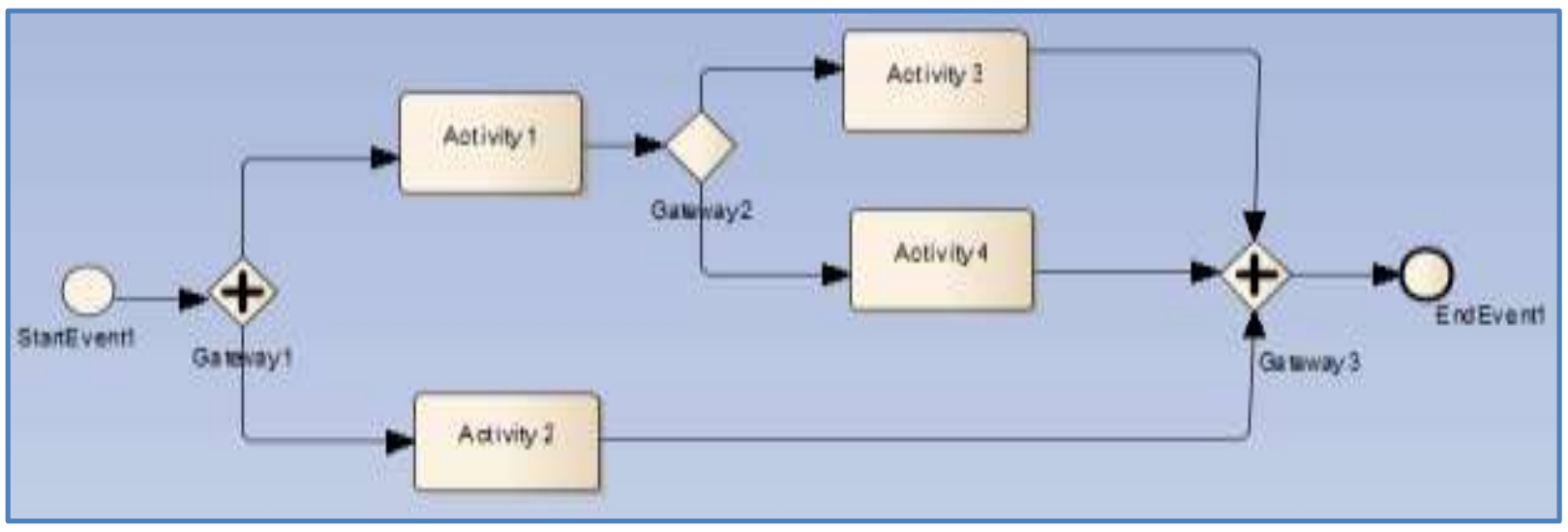

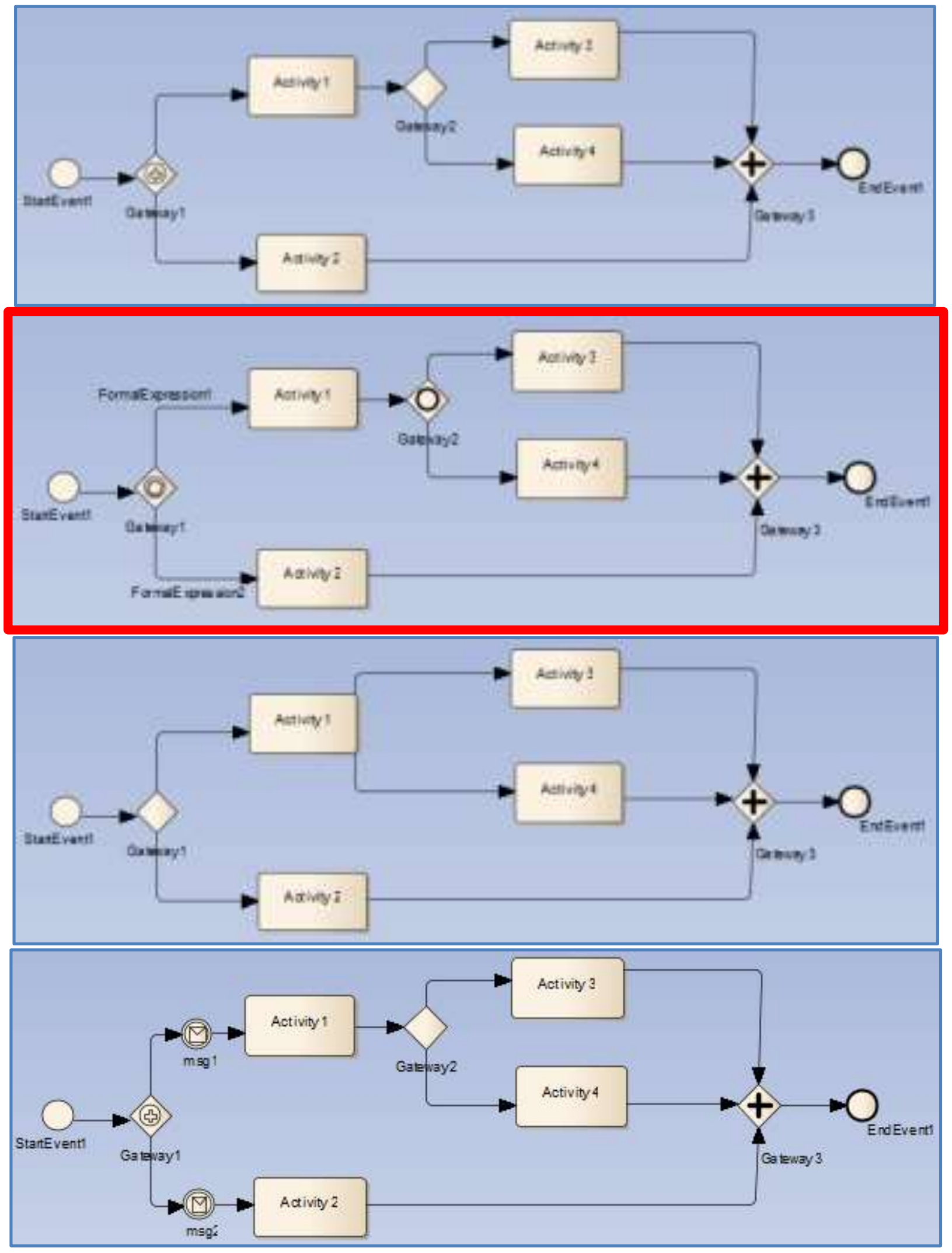

**Fig. 72.** – A *Parallel Gateway* joins only non-exclusive *Sequence Flows* (II)

**Picture's interpretation**: Wrong: **(1)** A Parallel Gateway ("Gateway3") is used to join non-exclusive Sequence Flows previously split from a Parallel Gateway ("Gateway1") and exclusive Sequence Flows from an Exclusive Gateway ("Gateway2") (top); **(2)** A Parallel Gateway ("Gateway3") is used to join non-exclusive Sequence Flows previously split from an Event Based Parallel Gateway ("Gateway1") and exclusive Sequence Flows from an Exclusive Gateway ("Gateway2") (2nd diagram); **(3)** A Parallel Gateway ("Gateway3") is used to join non-exclusive Sequence Flows resulting from split from an Inclusive Gateway ("Gateway2") and exclusive Sequence Flows from an Event Based Exclusive Gateway ("Gateway1") (3rd diagram); **(4)** A Parallel Gateway ("Gateway3") is used to join non-exclusive Sequence Flows resulting from an Activity ("Activity1") and exclusive Sequence Flows from an Exclusive Gateway ("Gateway1") (4rd diagram); **(5)** A Parallel Event Based Gateway ("Gateway3") is used to join exclusive tokens coming from an Exclusive Gateway ("Gateway1"). Furthermore, there are two parallel branches with catch events expecting messages that could never arrive (bottom).

The well-formedness rule regarding parallel gateways can be enforced by attaching the following invariant to the *Gateway* element of the BPMN metamodel.

```
context Gateway
  inv mergingParallelGwayIsPrecededBySplitWithParallelGway:
    (self.isJoin() and self.oclIsTypeOf(ParallelGateway))
      implies
    existsPrecedentSplitElementNonExclusive()
```

**Listing 68 – A *Parallel Gateway* joins only non-exclusive *Sequence Flows***

An *Inclusive Gateway* can be used for merging of *Sequence Flows* initiated from splitting *Exclusive* or *Parallel Gateways*. The use of an *Inclusive Gateway* for merging sequence flows that were created by a combination of various types of gateways, can produce a more compact representation. However this is not a recommended modeling practice [70].

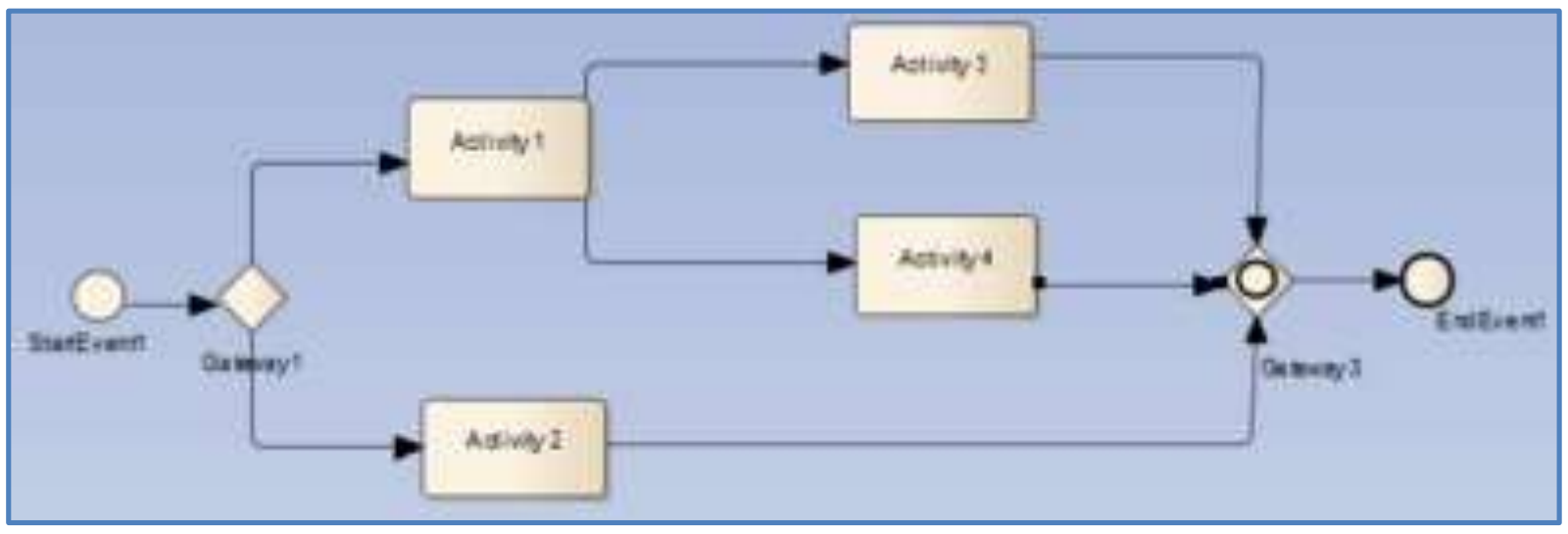

**Fig. 73.** – A *Parallel Gateway* joins only non-exclusive *Sequence Flows* (III)

**Picture's interpretation**: Not Recommended: An *Inclusive Gateway* ("Gateway3") must not be used to join exclusive tokens previously coming from an *Exclusive Gateway* ("Gateway1").

The best-practice rule regarding inclusive gateways can be enforced by attaching the following invariant to the *Gateway* element of the BPMN metamodel.

```
context Gateway
  inv mergingInclusiveGwayIsPrecededBySplitWithSeveralGwayTypes:
    (self.isJoin() and self.oclIsTypeOf(InclusiveGateway))
      implies
    existsPrecedentSeveralSplitElementTypes()
```

**Listing 69 – A *Parallel Gateway* joins only non-exclusive *Sequence Flows***

### 5.1.59. A join *Exclusive Gateway* must merge only exclusive *Sequence Flows*

This invariant states that the sequence flows arriving at a merging exclusive gateway must be exclusive [1] (page 290-292) [70]. So, if an exclusive gateway is used for merging, one must ensure that only one entrance receives a token. This can be achieved by checking whether exactly one sequence flow will be selected in every case, such as in the following situations:

- modeling the preceding splits only with *Exclusive Gateways*;
- modeling the preceding splits with previously mentioned elements or/and *Event Based Exclusive Gateway*. When an *Event Based Exclusive Gateway* is used without intermediate events or *Receive Activities*, an *Expression* must be attached to the outgoing sequence flows of the *Event Based Exclusive Gateway*;
- modeling the preceding splits with previously mentioned elements and/or *Activities* with conditional sequence flows.
- modeling a sequence flow loop.

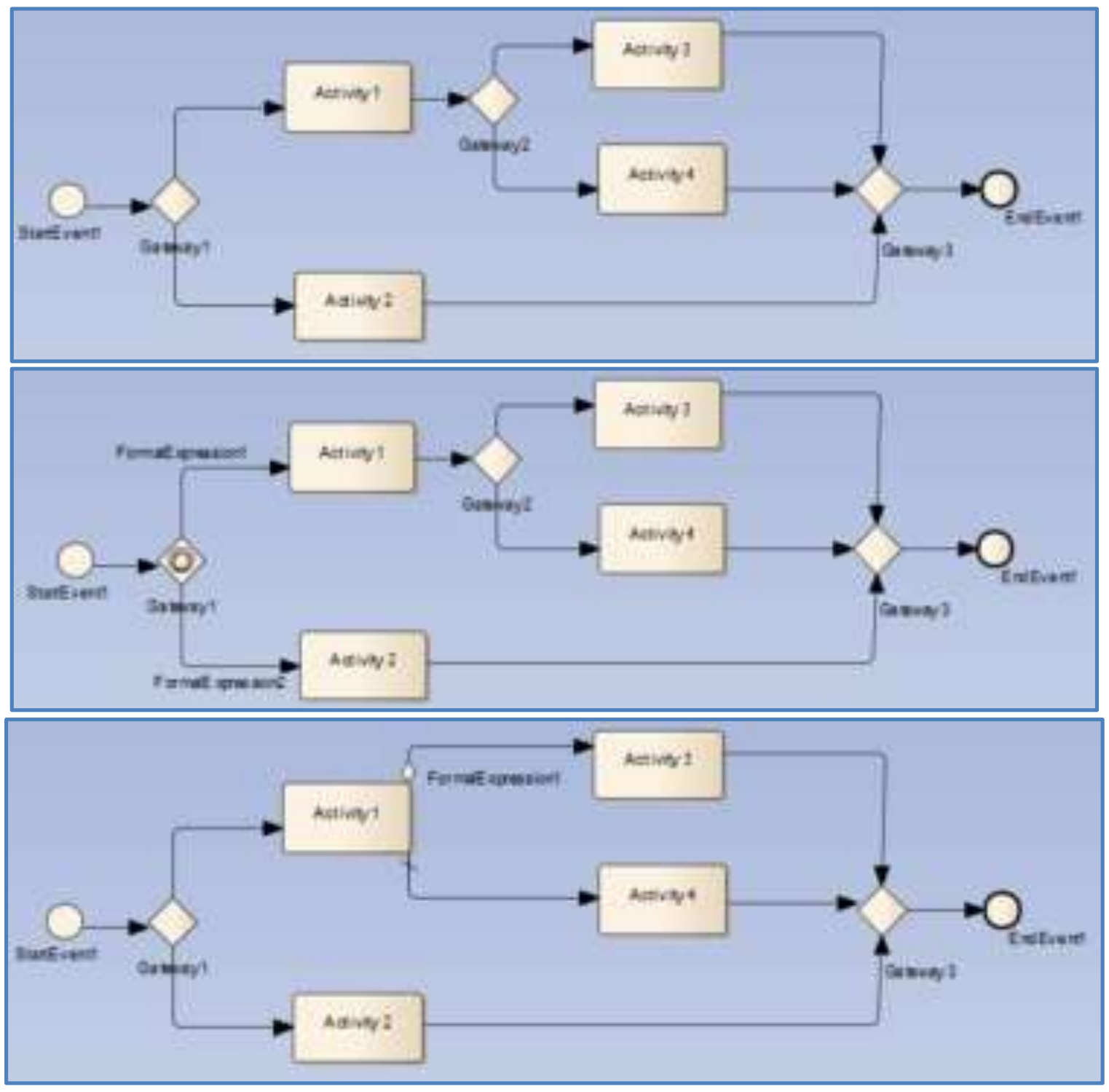

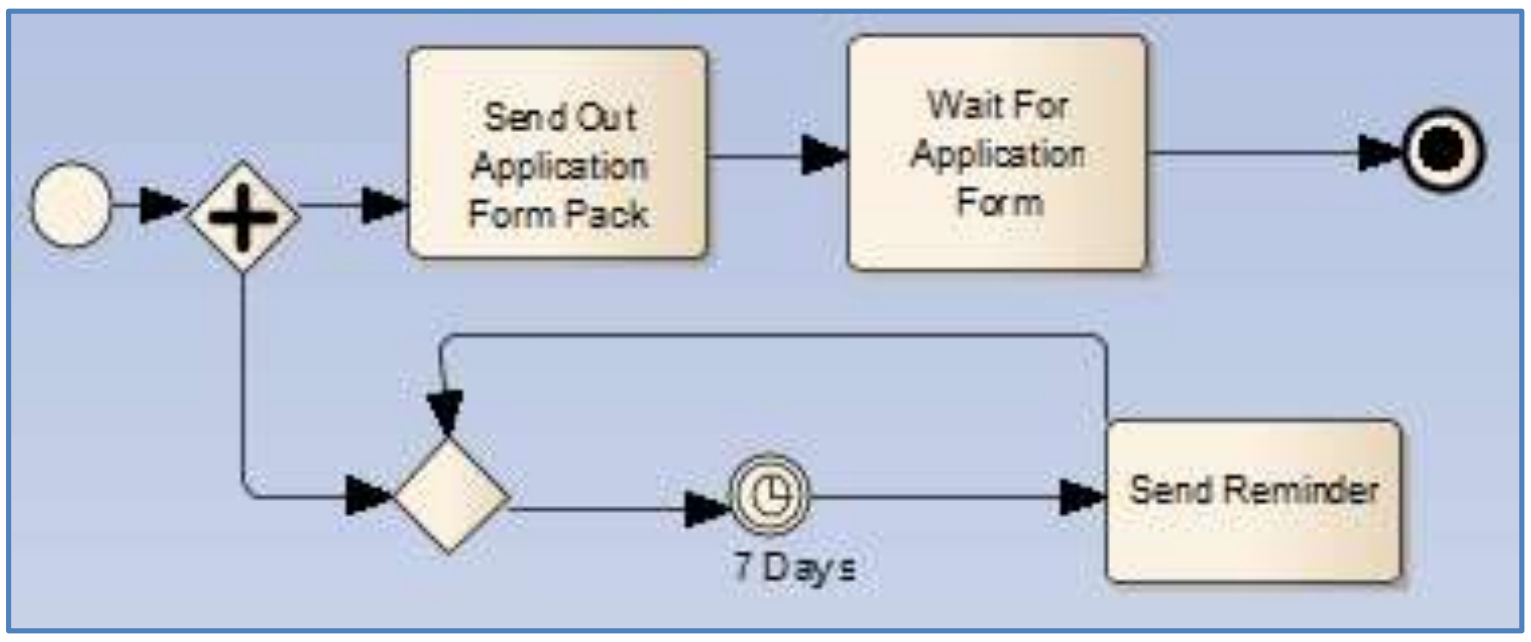

**Fig. 74.** – A join *Exclusive Gateway* must merge only exclusive *Sequence Flows* (I)

**Picture's interpretation**: <u>Correct</u>: **(1)** A merging exclusive gateway (Gateway3) receives only exclusive sequence flows, since it is preceded by splits with exclusive sequence flows (Gateway1 and Gateway2) (top); **(2)** A merging exclusive gateway (Gateway3) receives only exclusive sequence flows, since it is preceded by splits from an Event Based Exclusive Gateway with Expression attached to outgoing Sequence Flows (Gateway1) and an Exclusive Gateway (Gateway2) (2nd from top); **(3)** A merging exclusive gateway (Gateway3) receives only exclusive sequence flows, since it is preceded by an Activity (Activity1) that splits conditional sequence flows and an Exclusive Gateway (Gateway1) (3rd from top); **(4)** A merging exclusive gateway as result of a sequence flow loop (bottom).

Likewise the abovementioned situations, invalid models can also arise when the merging *Exclusive Gateway* does not matched with the appropriate split gateway such as in the following cases:

- A splitting *Parallel Gateway* or *Event-Based Parallel Gateway* precedes a merging *Exclusive Gateway*, thus non-exclusive tokens are forward, resulting in more than one non-exclusive *Sequence Flow* arriving to the *Exclusive Gateway* in the same process instance, even though the Exclusive Gateway could only deal with one of them.
- A splitting *Inclusive Gateway* precedes a merging *Exclusive Gateway*, and thus non-exclusive *Sequence Flows* could be sent by the *Inclusive Gateway* resulting in more than one token arriving the *Exclusive Gateway* which cannot be dealt by it;
- Parallel sequence flows generated by a splitting *Activity* cannot be processed by a merging *Exclusive Gateway*.

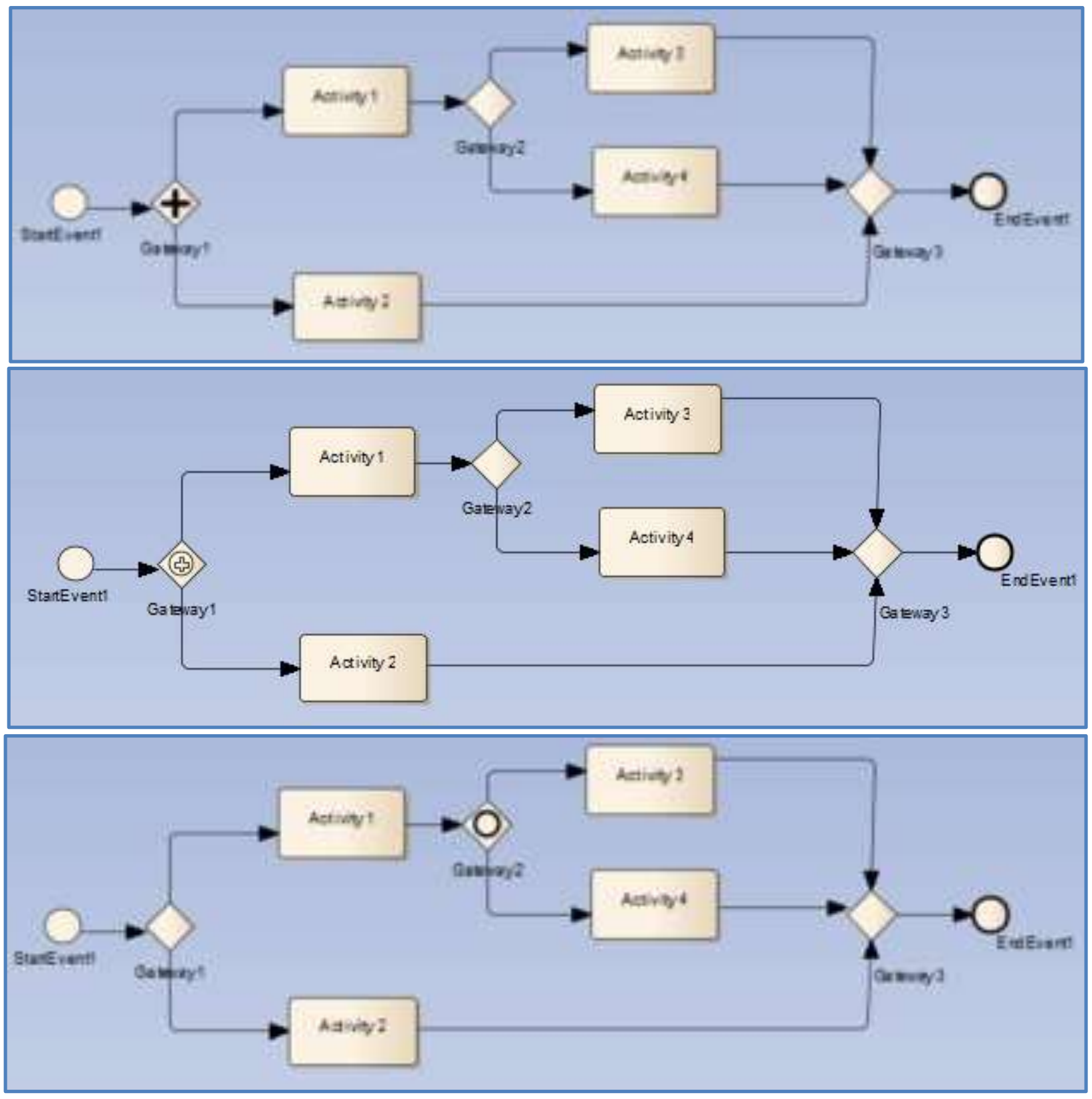

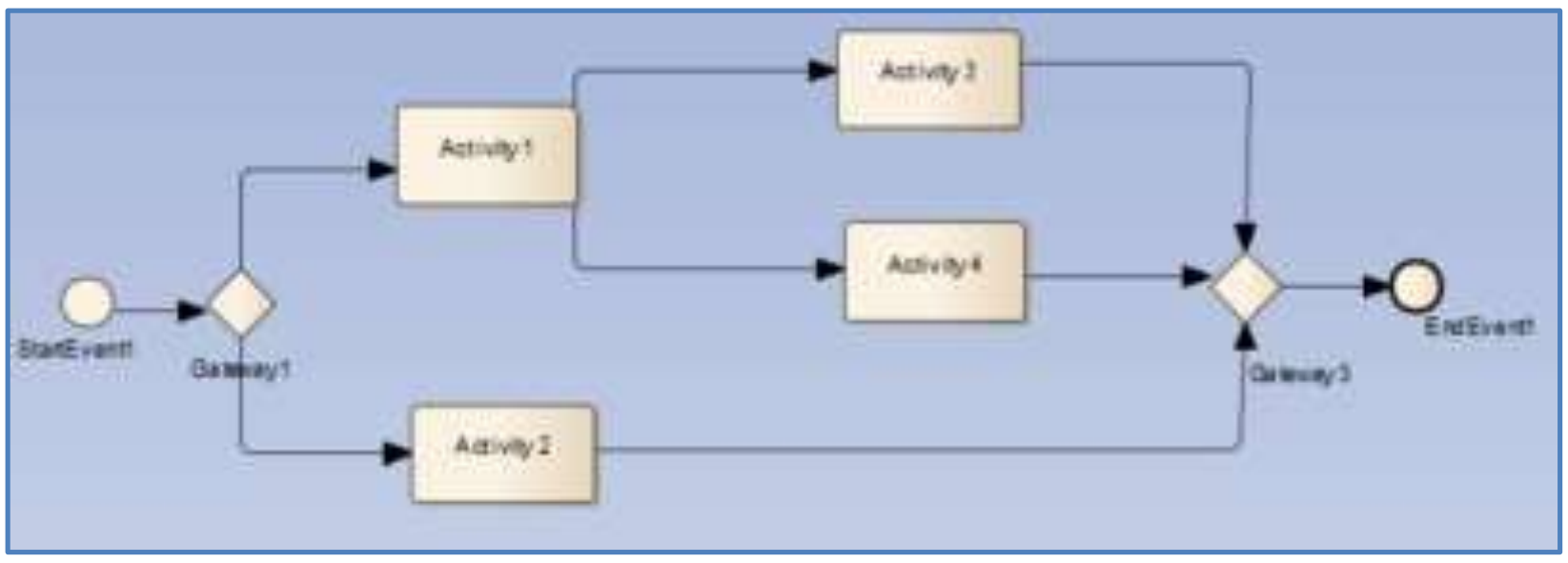

**Fig. 75.** – A join *Exclusive Gateway* must merge only exclusive *Sequence Flows* (II)

**Picture's interpretation**: Wrong: **(1)** A splitting *Parallel Gateway* (Gateway1) precedes a merging *Exclusive Gateway* (Gateway3) that cannot handle non-exclusive sequence flows (top); **(2)** A Parallel splitting *Event Based Gateway* (Gateway1) precedes a merging *Exclusive Gateway* (Gateway3) that cannot deal with non-exclusive sequence flows (2nd from top); **(3)** A splitting Inclusive *Gateway* (Gateway2) preceding a merging *Exclusive Gateway* (Gateway3), can generate non-exclusive tokens that Gateway3 is unable to deal with (3rd from top); **(4)** Parallel sequence flows are generated by the splitting task Activity1. However Gateway3, a merging *Exclusive Gateway*, can only deal with exclusive tokens (bottom).

The well-formedness rule regarding Exclusive Gateways can be enforced by attaching the following invariants to the *Gateway* element of the BPMN metamodel.

```
context Gateway
  inv mergingExclGatewayIsPrecededBySplitWithExclGateway:
    (self.isJoin() and self.oclIsTypeOf(ExclusiveGateway))
      implies
    precedentSplitElementIsExclusive()
```

**Listing 70 – A join *Exclusive Gateway* must merge only exclusive *Sequence Flows* (I)**

The previous invariant ensures that the *Sequence Flows*, coming from the previous *Flow Nodes* of an *Exclusive Gateway*, only carry on an exclusive token.

The following invariant is more general since it covers loops situations, as well as it ensures that on the path of an exclusive gateway, there are no other kinds of gateways that generate non-exclusive tokens.

```
context Gateway
  inv joinExclusiveGatewayHasOnlyExclusiveTokens:
    joinExclusiveGatewayIsTargetOfExclusiveTokens()
```

**Listing 71 – A join *Exclusive Gateway* must merge only exclusive *Sequence Flows* (II)**

### 5.1.60. A *Data-Based Exclusive Gateway* must have exclusive outgoing *Sequence Flows*

The outgoing *Sequence Flows* in *Data-Based Exclusive Gateway* must have conditions. Alternatively, if the *Sequence Flows* do not have conditions, the first element in each outgoing sequence flow must be a catching Intermediate Event. This kind of events can be replaced by a *Receive Activity*. However, *Intermediate Events* and *Receive Tasks* cannot be mixed.

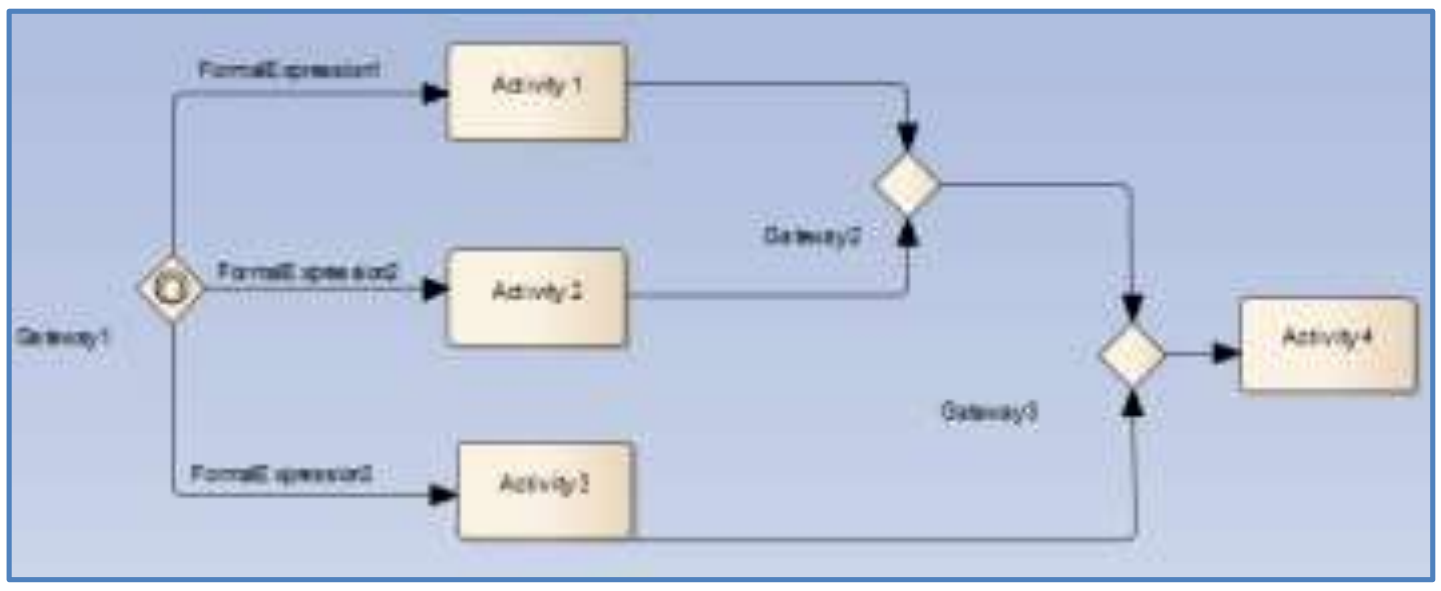

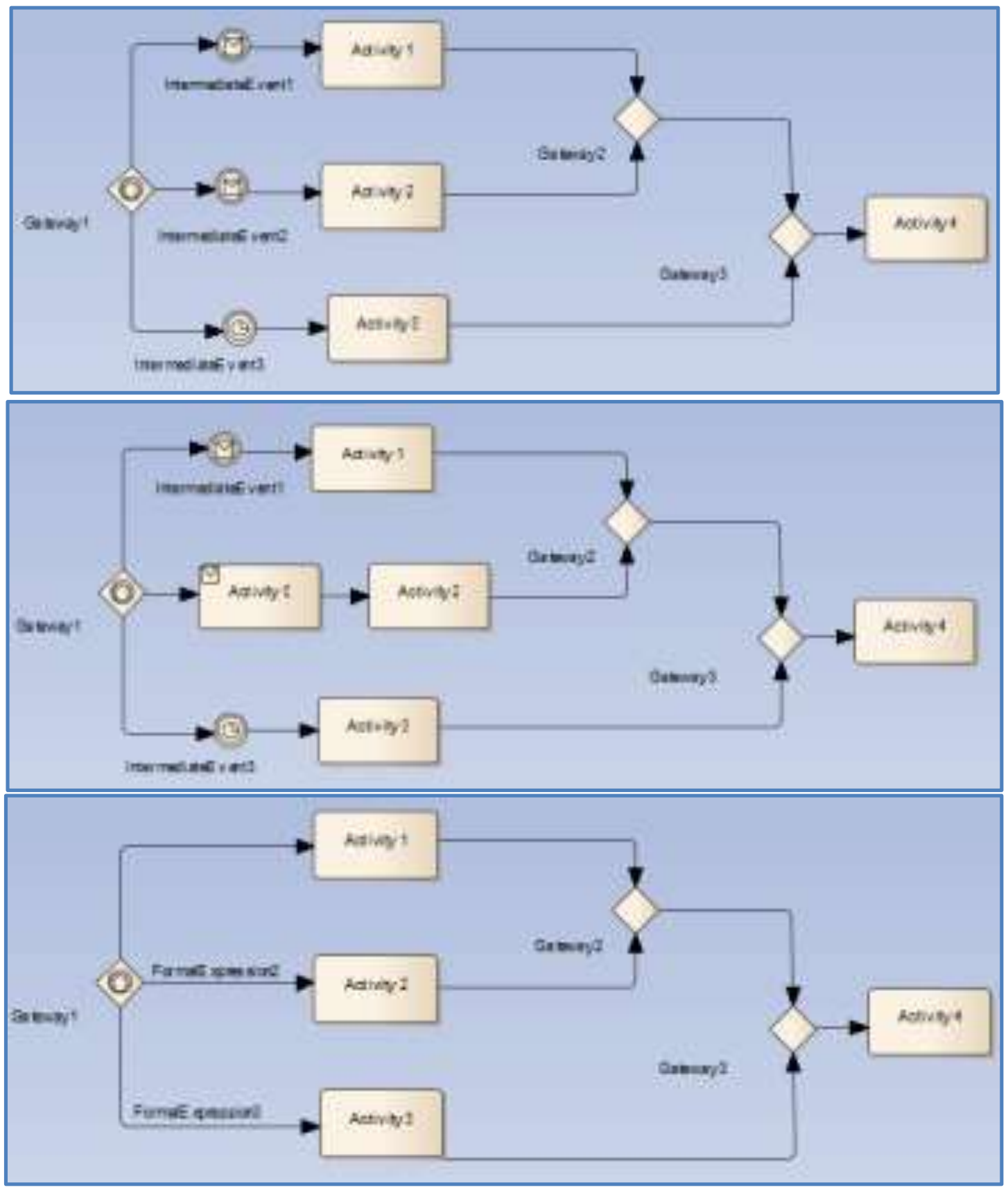

**Fig. 76.** – A *Data-Based Exclusive Gateway* must have exclusive outgoing *Sequence Flows*

**Picture's interpretation**: Correct: **(1)** A *Data-Based Exclusive Gateway* with outgoing Sequence Flows that have conditions (top); **(2)** A *Data-Based Exclusive Gateway* where the first element in each outgoing sequence flow is a catching I*ntermediate Event* (2nd from top); Wrong: **(1)** A *Data-Based Exclusive Gateway* where catching *Intermediate Event* are mixed with *Receive Activity* (3rd from top); **(2)** A *Data-Based Exclusive Gateway* with outgoing *Sequence Flows* without conditions (bottom).

The well-formedness rule regarding Gateways can be enforced by attaching the following invariants to the *Gateway* element of the BPMN metamodel.

```
context Gateway
  inv multipleExitsDataBasedExclGatewayMustHaveCond:
    dataBasedExclusiveGatewayHasConditions()
```

**Listing 72 – A *Data-Based Exclusive Gateway* must have exclusive outgoing *Sequence Flows***

#### 5.1.61. A *Gateway* must have either multiple incoming *Sequence Flow* or multiple outgoing Seque*nce Flow*

A Gateway must either merge or split the flows.

The well-formedness rule regarding *Gateways* can be enforced by attaching the following invariants to the *Gateway* element of the BPMN metamodel.

```
context Gateway
  inv mergeOrFork:
    self.isJoin() or isSplit()
```

**Listing 73 – A *Gateway* must have either multiple incoming *Sequence Flow* or multiple outgoing Seque*nce Flow***

#### 5.1.62. A *Gateway* with a *gatewayDirection* property of converging must have multiple incoming *Sequence Flow*

A Gateway with a *gatewayDirection* property of converging must have multiple incoming *Sequence Flow*. However, it must not have multiple outgoing *Sequence Flow*.

The well-formedness rule regarding *Gateways* can be enforced by attaching the following invariants to the *Gateway* element of the BPMN metamodel.

```
context Gateway
  inv convergingGateway:
    self.gatewayDirection =
    GatewayDirection::Converging
  implies
    (self.numberInputSequenceFlows() > 1 and
    self.numberOutputSequenceFlows() <= 1)
```

**Listing 74 – A *Gateway* with a *gatewayDirection* property of converging must have multiple incoming *Sequence Flow***

### 5.1.63. A *Gateway* with a *gatewayDirection* property of diverging must have multiple outgoing *Sequence Flow*

A *Gateway* with a *gatewayDirection* property of diverging must have multiple outgoing *Sequence Flow*. However, it must not have multiple incoming *Sequence Flow*.

The well-formedness rule regarding *Gateways* can be enforced by attaching the following invariant to the *Gateway* element of the BPMN metamodel.

```
context Gateway
  inv divergingGateway:
    self.gatewayDirection =
      GatewayDirection::Diverging
  implies
    (self.numberOutputSequenceFlows()  > 1 and
         self.numberInputSequenceFlows() <= 1)
```

**Listing 75 – A *Gateway* with a *gatewayDirection* property of diverging must have multiple outgoing *Sequence Flow***

### 5.1.64. An *Event Gateway* must have two or more outgoing *Sequence Flow*

The well-formedness rule regarding *Gateways* can be enforced by attaching the following invariant to the *EventBasedGateway* element of the BPMN metamodel.

```
context EventBasedGateway
  inv twoOrMoreSequenceFlows:
    self.numberOutputSequenceFlows()  >= 2
```

**Listing 76 – An *Event Gateway* must have two or more outgoing *Sequence Flow***

### 5.1.65. A *Conditional Sequence Flow* must not be used if the source *Gateway* is of type *Event-Based*

The well-formedness rule regarding Gateways can be enforced by attaching the following invariant to the *SequenceFlow* element of the BPMN metamodel.

```
context SequenceFlow
  inv sourceMustNotBeEventBasedGateway:
   self.sourceRef.oclIsKindOf(EventBasedGateway)
    implies
    conditionExpression.isUndefined()
```

**Listing 77 – A *Conditional Sequence Flow* must not be used if the source *Gateway* is of type *Event-Based***

### 5.1.66. A *Condition Expression* must be defined if the source of the *Sequence Flow* is an *Exclusive* or *Inclusive Gateway*

The well-formedness rule regarding *Gateways* can be enforced by attaching the following invariant to the *SequenceFlow* element of the BPMN metamodel.

```
context SequenceFlow
  inv sourceMustBeExclusiveInclusiveGateway:
     (self.sourceRef.oclIsKindOf(ExclusiveGateway)
      or  self.sourceRef.oclIsKindOf(InclusiveGateway))
    implies
    conditionExpression.isDefined()
```

**Listing 78 – A *Condition Expression* must be defined if the source of the *Sequence Flow* is an *Exclusive* or *Inclusive Gateway***

### 5.1.67. The target of the *Event Based Gateway* must be *Receive Task* or specific Inter*mediate Catch Event*

The target of the Gateway's outgoing *Sequence Flow* must be one of the following elements: A *Receive Task*, an *Intermediate Catch Event* of type *Message*, *Timer*, *Signal*, *Conditional* or *Multiple*. The exceptions are *Choreography Activities*.

The well-formedness rule regarding *Gateways* can be enforced by attaching the following invariant to the *EventBasedGateway* element of the BPMN metamodel.

```
context EventBasedGateway
  inv targetMustBeSpecific:
    self.outgoing_a.targetRef
  ->forAll(oclIsTypeOf(ReceiveTask) or
      (oclIsTypeOf(IntermediateCatchEvent)
          and (oclAsType(CatchEvent).isMessageEvent() or
              oclAsType(CatchEvent).isTimerEvent() or
              oclAsType(CatchEvent).isSignalEvent() or
              oclAsType(CatchEvent).isConditionalEvent() or
              oclAsType(CatchEvent).isMultipleEvent() )))
```

**Listing 79 – The target of the *Event Based Gateway* must be *Receive Task* or specific Inter*mediate Catch Event***

### 5.1.68. *Receive Tasks* used in an *Event Based Gateway* configuration must not have any attached *Boundary Event*

The well-formedness rule regarding Gateways can be enforced by attaching the following invariant to the *EventBasedGateway* element of the BPMN metamodel.

```
context EventBasedGateway
  inv receiveTasksWithoutBoundaryEvent:
    self.outgoing_a.targetRef
      ->select(oclIsTypeOf(ReceiveTask))
      ->collect(oclAsType(Task).boundaryEventRefs)->isEmpty()
```

**Listing 80 – *Receive Tasks* used in an *Event Based Gateway* configuration must not have any attached *Boundary Event***

### 5.1.69. *Message Intermediate Catch Events* are used as alternative to *Receive Tasks*

If *Message Intermediate Catch Events* are used as Target for the Gateway's outgoing *Sequence Flow*, then *Receive Tasks* must not be used and vice versa.

The well-formedness rule regarding Gateways can be enforced by attaching the following invariant to the *EventBasedGateway* element of the BPMN metamodel.

```
context EventBasedGateway
  inv receiveTasksWithoutMessageEventVV:
    (self.outgoing_a.targetRef
      ->select(oclIsTypeOf(IntermediateCatchEvent))
      ->collect(oclAsType(CatchEvent).isMessageEvent())
      ->notEmpty()
    implies
    self.outgoing_a.targetRef
      ->select(oclIsTypeOf(ReceiveTask))->isEmpty())
    and
    (self.outgoing_a.targetRef
      ->select(oclIsTypeOf(ReceiveTask))->notEmpty()
    implies
    self.outgoing_a.targetRef
      ->select(oclIsTypeOf(IntermediateCatchEvent))
      ->collect(oclAsType(CatchEvent).isMessageEvent())
      ->isEmpty())
```

**Listing 81 – *Message Intermediate Catch Events* are used as alternative to *Receive Tasks***

### 5.1.70. Target elements in an *Event Gateway* must not have other incoming *Sequence Flow*

Target elements in an *Event Gateway* configuration must not have any additional incoming *Sequence Flow* (in addition to the one from the *Event Gateway*).

The well-formedness rule regarding *Gateways* can be enforced by attaching the following invariant to the *EventBasedGateway* element of the BPMN metamodel.

```
context EventBasedGateway
  inv targetElementsMustHaveOnlyOneIncomingSeqFlow:
        self.outgoing_a.targetRef
                ->forAll(numberInputSequenceFlows() = 1)
```

**Listing 82 – Target elements in an *Event Gateway* must not have other incoming *Sequence Flow***

### 5.1.71. A *Parallel* or *Complex Gateway* must not have outgoing *Conditional Sequence Flow*

The well-formedness rule regarding *Gateways* can be enforced by attaching the following invariant to the *SequenceFlow* element of the BPMN metamodel.

```
context SequenceFlow
  inv sourceMustNotBeParalellComplexGateway:
              self.sourceRef.oclIsKindOf(ParallelGateway) or
              self.sourceRef.oclIsKindOf(ComplexGateway)
              implies
              conditionExpression.isUndefined()
```

**Listing 83 – A *Parallel* or *Complex Gateway* must not have outgoing *Conditional Sequence Flow***

**Activity**

An *Activity* represents work performed during some period of time, using resources of the organization, in the context of a process. The *Activity* requires some input, and produces some output.

### 5.1.72. An Activity with multiple *Conditional Sequence Flows* must have at least two outgoing *Sequence Flows*

If an *Activity* has various outgoing *Sequence Flows*, the sequence flows leaving the activity should have conditions. Such *Activity* must have at least two outgoing *Sequence Flow* and at least one of the *Sequence Flow* must be guaranteed to occur (otherwise the *Process* might become stuck). The completion of an *Activity* activates the Sequence flows with true Conditions [70].

However, normal and conditional *Sequence Flows* can leave an activity together. In this case, each *Sequence Flow* without condition diamond generates a token — the others only if their conditions are true. In this case, however, none of the outgoing *Sequence Flows* should be marked as *default flow*, because it would never be selected, since there are other *Sequence Flows* that always generate a token [70].

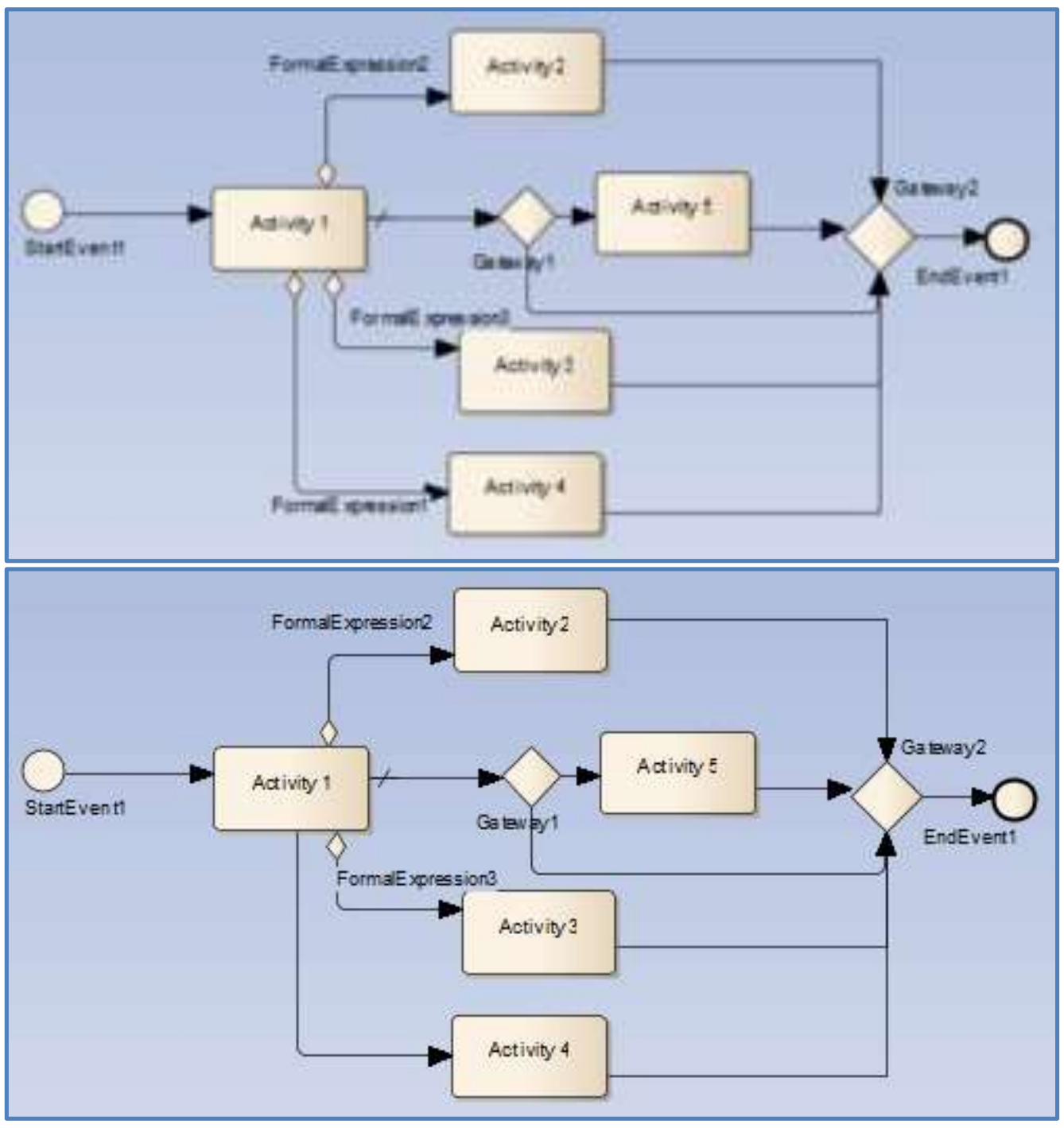

**Fig. 77.** – An Activity with multiple *Conditional Sequence Flows* must have at least two outgoing *Sequence Flows*

**Picture's interpretation**: Correct: An *Activity* with outgoing *Sequence Flows* that have conditions (top); Wrong: An *Activity* with one of the outgoing *Sequence Flows* marked as default flow, which will never be selected (bottom).

The well-formedness rule regarding Conditional Sequence Flows can be enforced by attaching the following invariant to the *Activity* element of the BPMN metamodel.

```
context Activity
  inv multipleExitsActivityMustHaveConditions:
    outgoingSequenceFlowsHaveConditions()
```

**Listing 84 – An Activity with multiple *Conditional Sequence Flows* must have at least two outgoing *Sequence Flows***

### 5.1.73. A *Compensation Activity* must not have any incoming or outgoing *Sequence Flow*

Only one *Compensation Activity* can be associated with the *Compensation Intermediate Event* and that *Compensation Activity* must not have any incoming or outgoing *Sequence Flow*.

The well-formedness rule regarding Compensation Activity can be enforced by attaching the following invariant to the *Activity* element of the BPMN metamodel.

```
context Activity
  inv compensationActivityNoIncomingAndOutgoing:
```

```
    self.isForCompensation() implies
    (hasNoIncomingAndOutgoingSequenceFlow() and
       self.incoming.sourceRef.outgoing.targetRef->size() = 1)
```

**Listing 85 – A *Compensation Activity* must not have any incoming or outgoing *Sequence Flow***

### 5.1.74. A *Compensation Activity* must reside within the *Process* inherent to the compensated *Activity*

The *Activity* must be within the *Process*, either at the top level *Process* or within a *SubProcess* of the *Compensation Activity*.

The well-formedness rule regarding *Compensation Activity* can be enforced by attaching the following invariant to the *Activity* element of the BPMN metamodel.

```
context Activity
  inv compensationActivityResideWithinSameProcess:
    self.isForCompensation() implies
    self.container.myProcess() =
    self.incoming.sourceRef.oclAsType(BoundaryEvent).
      attachedToRef.container.myProcess()
      ->asOrderedSet()->first()
```

**Listing 86 – A *Compensation Activity* must reside within the *Process* of the compensated *Activity***

### 5.1.75. A *Receive Task* must not have an outgoing *Message Flow*

The well-formedness rule regarding *Receive Task* can be enforced by attaching the following invariant to the *ReceiveTask* element of the BPMN metamodel.

```
context ReceiveTask
  inv mustNotHaveOutgoingMessageFlow:
    not self.hasOutputMessageFlows()
```

**Listing 87 – A *Receive Task* must not have an outgoing *Message Flow***

### 5.1.76. A *Send Task* must not have an incoming *Message Flow*

The well-formedness rule regarding *Send Task* can be enforced by attaching the following invariant to the *SendTask* element of the BPMN metamodel.

```
context SendTask
  inv mustNotHaveIncomingMessageFlow:
    not self.hasInputMessageFlows()
```

**Listing 88 – A *Send Task* must not have an incoming *Message Flow***

### 5.1.77. A *Script* or *Manual Task* must not have an incoming or an outgoing *Message Flow*

The well-formedness rule regarding *Script* and *Manual Task* can be enforced by attaching the following invariants to the *ScriptTask* and *ManualTask* elements of the BPMN metamodel.

```
context ScriptTask
  inv scriptTaskMustNotHaveIncomingOutgoingMessageFlow:
    not self.hasInputMessageFlows() and
    not self.hasOutputMessageFlows()

context ManualTask
  inv manualTaskMustNotHaveIncomingOutgoingMessageFlow:
    not self.hasInputMessageFlows() and
    not self.hasOutputMessageFlows()
```

**Listing 89 – A *Script* or *Manual Task* must not have an incoming or an outgoing *Message Flow***

**Sequence Flow**

A *Sequence Flow* connects and orders flow nodes (*Activities*, *Events*, and *Gateways*) in a container (*Process* or *SubProcess*). Variations of *Sequence Flow* include *Conditional Sequence Flow* and *Default Sequence Flow*.

### 5.1.78. A *Conditional Sequence Flow* cannot be used if there is only one *sequence flow* goes out of the element

The well-formedness rule regarding *Sequence Flow* can be enforced by attaching the following invariant to the *SequenceFlow* element of the BPMN metamodel.

```
context SequenceFlow
  inv conditionalCannotBeUsedIfOnlyOne:
   (self.sourceRef.outgoing_a->size() = 1)
    implies
    conditionExpression.isUndefined()
```

**Listing 90 – A *Conditional Sequence Flow* cannot be used if there is only one *sequence flow* goes out of the element**

### 5.1.79. *Sequence Flows* cannot cross the container boundaries

The well-formedness rule regarding *Sequence Flow* can be enforced by attaching the following invariant to the *SequenceFlow* element of the BPMN metamodel.

```
context SequenceFlow
  inv cannotCrossContainerBoundaries:
    if self.sourceRef.oclIsKindOf(BoundaryEvent) or
      self.targetRef.oclIsKindOf(BoundaryEvent) then
        self.sourceRef.oclAsType(BoundaryEvent).
           attachedToRef.container =
        self.targetRef.container or
      self.targetRef.oclAsType(BoundaryEvent).
           attachedToRef.container =
       self.sourceRef.container
    else
      self.sourceRef.container =
           self.targetRef.container
  endif
```

**Listing 91 – *Sequence Flows* cannot cross the container boundaries**

### 5.1.80. The source and target of a *Sequence Flow* must not be the same

The well-formedness rule regarding *Sequence Flow* can be enforced by attaching the following invariant to the *SequenceFlow* element of the BPMN metamodel.

```
context SequenceFlow
  inv sourceCannotBeTarget:
self.sourceRef <> self.targetRef
```

**Listing 92 – The source and target of a *Sequence Flow* must not be the same**

**Message Flow**

A Message Flow defines the interactions/communications between two separate participants (shown as Pools) of the collaboration diagram.

### 5.1.81. A *Message Flow* can only have as source a *Message End Event, Intermediate Throw Event, Send Task, User Task, Service Task, Subprocess* or a "black box" pool

The well-formedness rule regarding Message Flow can be enforced by attaching the following invariant to the MessageFlow element of the BPMN metamodel.

```
context MessageFlow
  inv allowedSource:
  (self.sourceRef.oclIsTypeOf(EndEvent)
  implies
  self.sourceRef.oclAsType(ThrowEvent).isMessageEvent()) or
  (self.sourceRef.oclIsTypeOf(IntermediateThrowEvent)
  implies
  self.sourceRef.oclAsType(ThrowEvent).isMessageEvent()) or
  self.sourceRef.oclIsTypeOf(SendTask) or
  self.sourceRef.oclIsTypeOf(UserTask) or
  self.sourceRef.oclIsTypeOf(ServiceTask) or
  self.sourceRef.oclIsTypeOf(SubProcess) or
  self.sourcePool()->asOrderedSet()->first().isBlackBox()
```

**Listing 93 – A *Message Flow* can only have as source a *Message End Event, Intermediate Throw Event, Send Task, User Task, Service Task, Subprocess* or a "black box" pool**

### 5.1.82. A *Message Flow* can only go to a *Message Start, Intermediate Catch Event, Boundary Event, Receive Task, User Task*, or *Service Task, Subprocess*, or "black box" *pool*

The well-formedness rule regarding *Message Flow* can be enforced by attaching the following invariant to the *MessageFlow* element of the BPMN metamodel.

```
context MessageFlow
  inv allowedTarget:
  (self.targetRef.oclIsTypeOf(StartEvent)
  implies
  self.targetRef.oclAsType(CatchEvent).isMessageEvent()) or
  (self.targetRef.oclIsTypeOf(IntermediateCatchEvent)
  implies
  self.targetRef.oclAsType(CatchEvent).isMessageEvent()) or
  (self.targetRef.oclIsTypeOf(BoundaryEvent)
  implies
  self.targetRef.oclAsType(CatchEvent).isMessageEvent()) or
  self.targetRef.oclIsTypeOf(SendTask) or
  self.targetRef.oclIsTypeOf(UserTask) or
  self.targetRef.oclIsTypeOf(ServiceTask) or
  self.targetRef.oclIsTypeOf(SubProcess) or
  self.targetPool()->asOrderedSet()->first().isBlackBox()
```

**Listing 94 – A Message Flow can only go to a Message Start, Intermediate Catch Event, Boundary Event, Receive Task, User Task, or Service Task, Subprocess, or "black box" pool**

### 5.1.83. A *Message Flow* must not be connected to the border of a "white Box" Pool

The well-formedness rule regarding *Message Flow* can be enforced by attaching the following invariant to the *MessageFlow* element of the BPMN metamodel.

```
context MessageFlow
  inv doNotConnectToWhiteBox:
  not (self.targetPool()
      ->asOrderedSet()->first().isWhiteBox() or
      self.sourcePool()
      ->asOrderedSet()->first().isWhiteBox())
```

**Listing 95 – A *Message Flow* must not be connected to the border of a "white Box" Pool**

### 5.1.84. A *Message* must be attached to a *Message Flow* or must be connected to an *Association* connected to a *Message Flow*, a *Send Task* a *Receive Task*, or a *Message Event Definition*

The well-formedness rule regarding *Message* can be enforced by attaching the following invariant to the *Message* element of the BPMN metamodel.

```
context Message
  inv allowedConnectors:
    self.messageFlow->size() > 0 or
    self.sendTask->size() > 0 or
    self.receiveTask->size() > 0 or
    self.messageEventDefinition->size() > 0
```

**Listing 96 – A *Message* must be attached to a *Message Flow* or must be connected to an *Association* connected to a *Message Flow*, a *Send Task* a *Receive Task*, or a *Message Event Definition***

### Artifacts

An *Artifact* provides a way of capturing additional information about a specific element or group of elements. There are three kinds of *Artifacts* in BPMN: Association, Group, and Text Annotation.

#### 5.1.85. An *Association* should not connect two *Text Annotations*

The well-formedness rule regarding *Text Annotation* can be enforced by attaching the following invariant to the *Association* element of the BPMN metamodel.

```
context Association
  inv connectTwoTextAnnotation:
    not (self.targetRef.oclIsTypeOf(TextAnnotation) and
  self.sourceRef.oclIsTypeOf(TextAnnotation))
```

**Listing 97 – An *Association* should not connect two *Text Annotations***

## 5.2. Data flow well-formedness rules

*Data flow* is a mechanism to move data between *Activities*. Thus, it represents the movement of data passing information to or from an *Activity*. An incoming data flow means that the source *Data Object* or *Data Store* must be available in order to an *Activity* be able to start. Graphically, directed *Data Input* and *Output Associations* connectors show the data flow between *Data Object*/*Data Stores* and *Activities*.

### 5.2.1. A reusable *SubProcess* has only self-contained data

Any data used by a *reusable SubProcess* is completely separate from the data of the invoking *parent* Process. The reuse capability of the *SubProcess* relies on the fact that its data is completely self-contained.

The well-formedness rule regarding *Data Element* in *SubProcess* can be enforced by attaching the following invariant to the *SubProcess* element of the BPMN metamodel.

```
context SubProcess
  inv selfContainedData:
      isDataSelfContained()
```

**Listing 98 – A reusable *SubProcess* has only self-contained data**

### 5.2.2. A *Data Object* must have at least one connected *Data Association*

The well-formedness rule regarding data objects can be enforced by attaching the following invariant to the *DataObject* element of the BPMN metamodel.

```
context DataObject
  inv connectedDataAssociation:
  (self.dataAssociation_a->notEmpty() or
  self.dataAssociation->notEmpty())
```

**Listing 99 – A *Data Object* must have at least one connected *Data Association***

### 5.2.3. A *Data Store* must have at least one connected *Data Association*

The well-formedness rule regarding data stores can be enforced by attaching the following invariant to the *DataStore* element of the BPMN metamodel.

```
context DataStore
  inv dataStoreconnectedDataAssociation:
  (self.dataAssociation_a->notEmpty() or
  self.dataAssociation->notEmpty())or self.dataStoreReference->notEmpty
```

**Listing 100 – A *Data Store* must have at least one connected *Data Association***

### 5.2.4. A *Data Object Reference* can only access a *Data Object* inside the same container or on its parent

The well-formedness rule regarding data reference store can be enforced by attaching the following invariant to the *DataObjectReference* element of the BPMN metamodel.

```
context DataObjectReference
  inv dataObjectAtSameContainerOrParent:
  isDataObjectAtSameContainerOrParent()
```

**Listing 101 – A *Data Object Reference* can only access a *Data Object* inside the same container or on its parent**

## 5.3. Best-practices rules

This section includes a set of rules aiming to suggest modeling good practices for BPMN. These rules are not mandatory, so the rules' adoption is dependent upon process modeling and methodological procedures set out by each organization.

In order to differentiate OCL constraints implementing best-practices rules from the rules enforced by the BPMN standard, the former are prefixed with "bp_".

**Process**

### 5.3.1. Use 7± 2 Flow Nodes per diagram

According to the so-called Miller's law [72], "the number of objects an average human can hold in working memory is 7 ± 2". There are modeling techniques that advocate this rule of thumb regarding the number of objects in a diagram. For instance Data Flow Diagram, use the "7 ± 2" principle in order to not clutter diagrams. IDEF also limits the number of *Activity* elements to 5 or 6 per page.

For BPMN process models, an invariant is suggested to notify the modeler whether the number of flow nodes (activity, event, or gateway) inside a container (*Process* or *SubProcess*) has surpassed the "7 ± 2" threshold.

The best-practice rule regarding flow nodes can be enforced by attaching the following invariant to the *FlowElementsContainer* element of the BPMN metamodel.

```
context FlowElementsContainer
  inv bp_limitOfActivitiesPerDiagram:
  withinLimitOfActivitiesPerDiagram()
```

**Listing 102 – Use 7± 2 Flow Nodes per diagram**

## Event

### 5.3.2. Use exclusively *Send/Receive Task* or *Throw/Catch Message Intermediate Events*

The modeler could choose to use only *Send* and *Receive Tasks*, or to use only the *Throw* and *Catch Message Intermediate Events* (not both). The rule intends to avoid mixing both elements' usage in the same model [71].

The best-practice rule regarding *Send/Receive Task* can be enforced by attaching the following invariant to the *FlowElementsContainer* element of the BPMN metamodel.

```
context FlowElementsContainer
  inv bp_useSendReceiveTaskOrThrowCatchMesInterEvents:
  isSendReceiveTaskUsedExclusively()
```

**Listing 103 – Use exclusively *Send/Receive Task* or *Throw/Catch Message Intermediate Events***

### 5.3.3. Use explicitly *Start Events* and *End Events*

BPMN process modeling best practices recommendations advise the explicit use of *Start* and *End Events* [71].

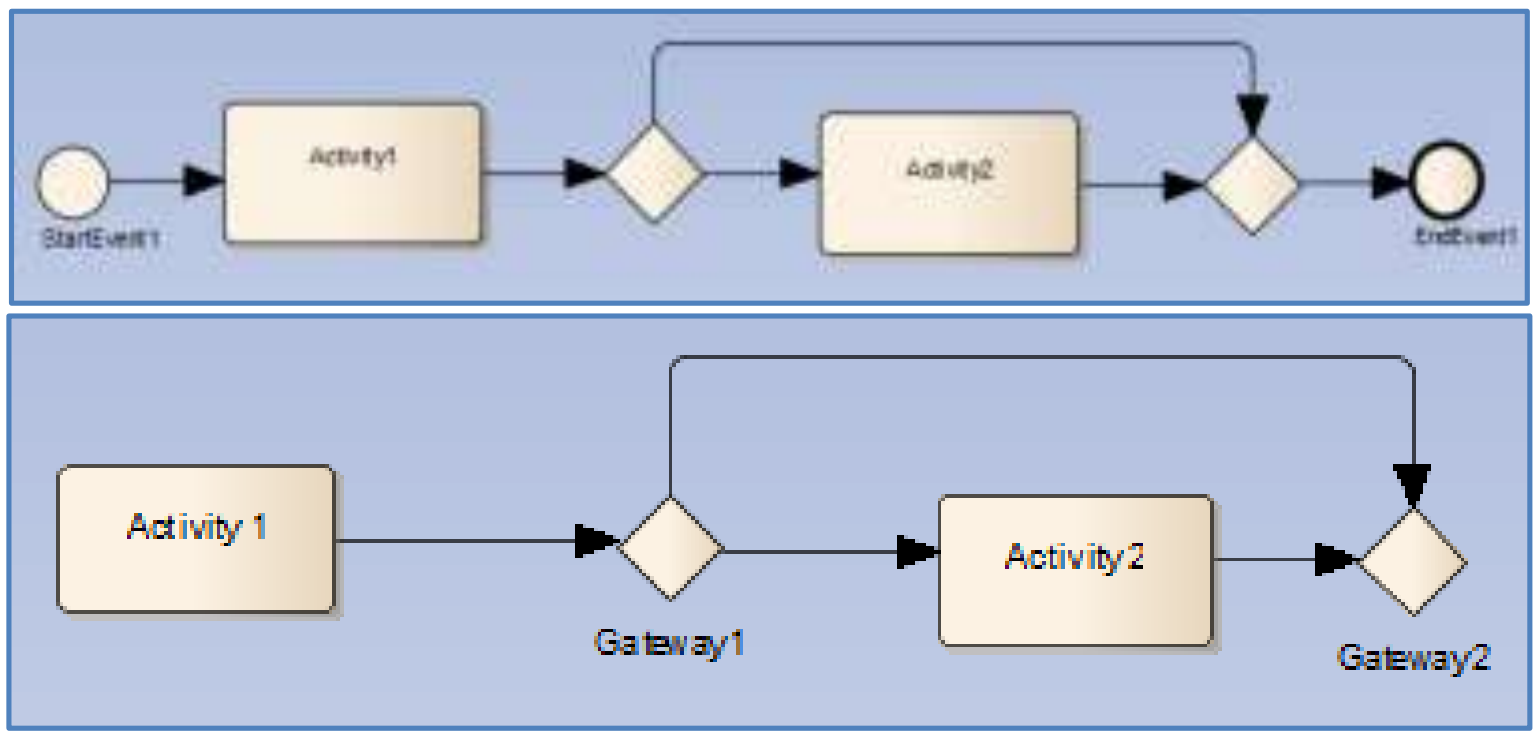

**Fig. 78.** – Use explicitly *Start Events* and *End Events*

**Picture's interpretation**: Recommendation: explicit of *Start* and *End Events*; Avoid: use of *Flow Nodes* as implicit *Start* and *End Events*.

The best-practice rule regarding *Start* and *End Event* can be enforced by attaching the following invariant to the *FlowElementsContainer* element of the BPMN metamodel.

```
context FlowElementsContainer
  inv bp_useExplicitStartAndEndEvents:
    existsExplicitStartAndEndEvents()
```

**Listing 104 – Use explicitly *Start Events* and *End Events***

### 5.3.4. Use only one *Start Event*

It is recommended that multiple *Start Events* should not be used.

The best-practice rule regarding Start Event can be enforced by attaching the following invariant to the *FlowElementsContainer* element of the BPMN metamodel.

```
context FlowElementsContainer
  inv bp_useOnlyOneStartEvent:
    existsOnlyOneStartEvent()
```

**Listing 105 –*Use only one Start Event***

### 5.3.5. Use a *Default Condition* always a *Conditional Sequence Flow* is used

One way to ensure that a Process does not become stuck, after an *Activity*, is to use a *Default Sequence Flow* whenever *Conditional Sequence Flow* is used [71].

The best-practice rule regarding conditional sequence flows can be enforced by attaching the following invariant to the *FlowNode* element of the BPMN metamodel.

```
context FlowNode
  inv bp_useDefaultCondition:
    existsDefaultCondition()
```

**Listing 106 – Use a *Default Condition* always a *Conditional Sequence Flow* is used**

### 5.3.6. Use a *Timer Intermediate Event* with an *Event Gateway*

One way to ensure a *Process* does not get stuck at an *Event Based Exclusive Gateway* is to use always a *Timer Intermediate Event* as one of the options exiting Gateway [71].

The best-practice rule regarding *Event Gateway* can be enforced by attaching the following invariant to the *Event-BasedGateway* element of the BPMN metamodel.

```
context EventBasedGateway
  inv bp_useTimerIntermediateEvent:
    existsTimerIntermediateEvent()
```

**Listing 107 – Use a *Timer Intermediate Event* with an *Event Gateway***

### 5.3.7. An *Event* should have at most one outgoing *Sequence Flow*

A *Gateway* or *Activity* is required if a split is needed after an event (not *EndEvent*), in order that only one outgoing sequence flow comes out the event [70].

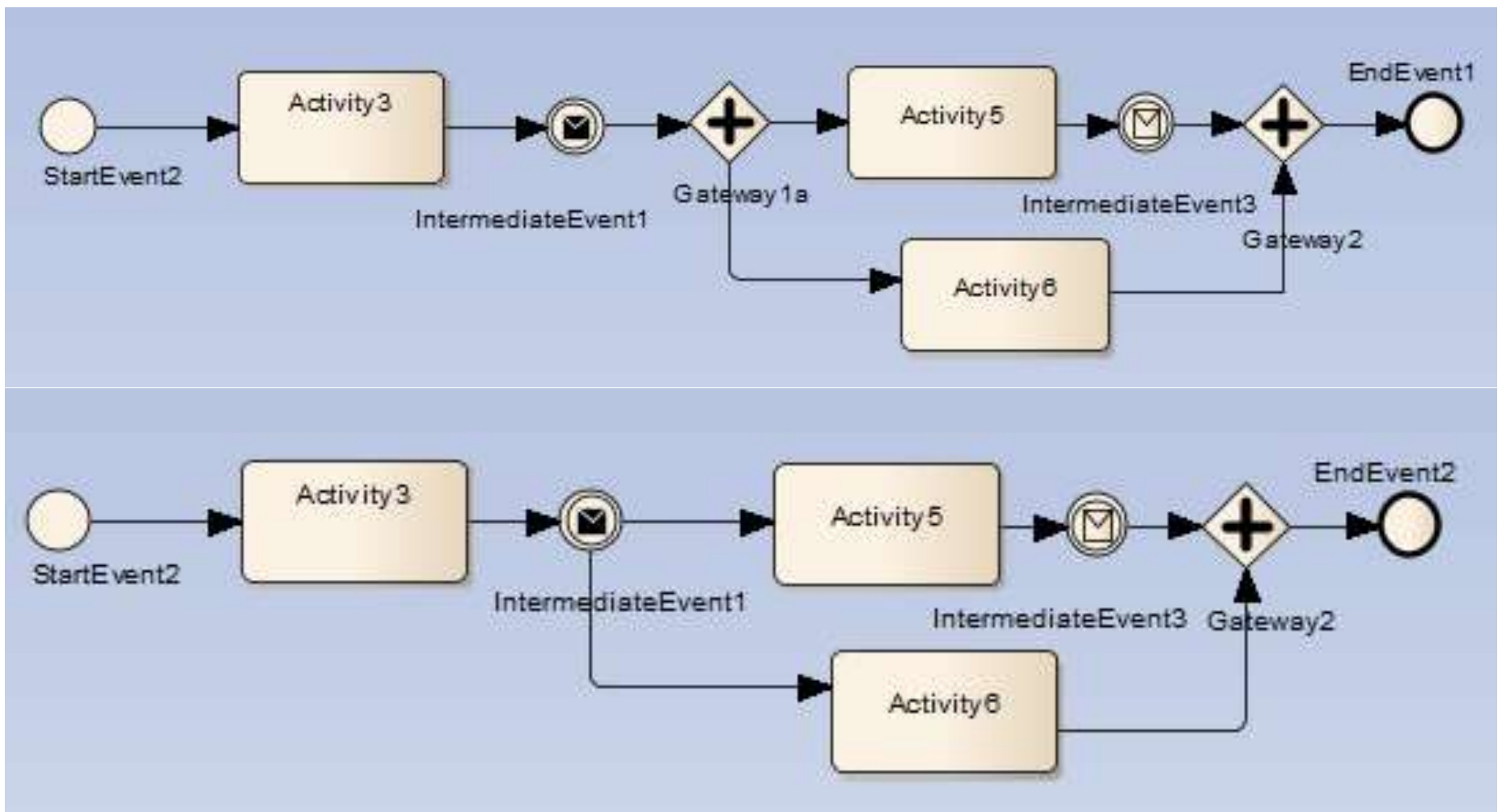

**Fig. 79.** – An *Event* should have most one outgoing *Sequence Flow*

**Picture's interpretation**: Recommendation: Since "IntermediateEvent1" is a split event, a Gateway or Activity should follow it.

The best-practice rule regarding events can be enforced by attaching the following invariant to the *Event* element of the BPMN metamodel.

```
context Event
  inv bp_splitingEventShouldNotOccurAndOneTargetIsNeeded:
    isNotSplitAndHasOneTarget()
```

**Listing 108 – An *Event* should have most one outgoing *Sequence Flow***

### 5.3.8. A *Start Event* should have a name

If a top-level process contains more than one start event, all of them should be labeled to identify the alternative start conditions. A *Timer Start Event* should be labeled to indicate the process schedule. A *Signal Start Event* should be labeled to indicate the signal name. A *Conditional Start Event* should be labeled to indicate the condition. A *None Start Event* at the top-level process should also be labeled [7].

The best-practice rule regarding start event can be enforced by attaching the following invariant to the *Start Event* element of the BPMN metamodel.

```
context StartEvent
```

```
    inv bp_shouldHaveName:
      hasName()
```

**Listing 109 – A *Start Event* should have a name**

### 5.3.9. A *Message Start Event* should have an incoming *Message Flow*

The best-practice rule regarding message start event can be enforced by attaching the following invariant to the *Start Event* element of the BPMN metamodel [7].

```
  context StartEvent
    inv bp_incomingMessageFlow:
      hasMessageFlow()
```

**Listing 110 – A *Message Start Event* should have an incoming *Message Flow***

### 5.3.10. A *Catching Intermediate Message Event* should have an incoming *Message flow*

The best-practice rule regarding message intermediate catch event can be enforced by attaching the following invariant to the *IntermediateCatchEvent* element of the BPMN metamodel [7].

```
  context IntermediateCatchEvent
    inv bp_incomingMessageFlow:
      hasMessageFlow()
```

**Listing 111 – A *Catching Intermediate Message Event* should have an incoming *Message flow***

### 5.3.11. A *Throwing Intermediate Message Event* should have an outgoing *Message Flow*

The best-practice rule regarding message intermediate catch event can be enforced by attaching the following invariant to the *IntermediateCatchEvent* element of the BPMN metamodel.

```
  context IntermediateCatchEvent
    inv bp_outgoingMessageFlow:
      hasMessageFlow()
```

**Listing 112 – A *Throwing Intermediate Message Event* should have an outgoing *Message Flow***

### 5.3.12. An *Intermediate Event* must have a name

A throwing intermediate event should be labeled. A catching intermediate event should also be labeled [7].

The best-practice rule regarding intermediate event can be enforced by attaching the following invariants to the *IntermediateThrowEvent* and *IntermediateCatchEvent* elements of the BPMN metamodel.

```
  context IntermediateThrowEvent
    inv bp_shouldHaveName:
      hasName()

  context IntermediateCatchEvent
    inv bp_shouldHaveName:
      hasName()
```

**Listing 113 – An *Intermediate Event* must have a name**

### 5.3.13. An *End Event* should be labeled with the name of the end state

Two *End Events* in a process level should not have the same name. If they mean the same end state, they should be combined; otherwise they should have different names. If there is more than one *End Event* in a process level, all should be labeled with the name of the end state [7].

The best-practice rule regarding start event can be enforced by attaching the following invariant to the Start Event element of the BPMN metamodel.

```
  context StartEvent
    inv bp_shouldHaveName:
      hasName()
```

**Listing 114 – An *End Event* should be labeled with the name of the end state**

#### 5.3.14. If a *SubProcess* is followed by a yes/no gateway, at least one *End Event* of the *SubProcess* should be labeled to match the gateway label

The best-practice rule regarding sub-process [7] can be enforced by attaching the following invariant to the *Sub-Process* element of the BPMN metamodel.

```
context SubProcess
  inv bp_matchGatewayName:
    matchGatewayName()
```

**Listing 115 – If a *SubProcess* is followed by a yes/no gateway, at least one *End Event* of the *SubProcess* should be labeled to match the gateway label**

**Gateway**

### 5.3.15. Use a *Default Condition* in an *Exclusive Gateway*

One way to ensure that a Process does not get stuck in an *Exclusive Gateway* is to use a default condition for one of the outgoing *Sequence Flow*. This creates a *Default Sequence Flow*. The default is chosen if all the other *Sequence Flow* conditions turn out to be false [71].

The best-practice rule regarding exclusive gateway can be enforced by attaching the following invariant to the *ExclusiveGateway* element of the BPMN metamodel.

```
context ExclusiveGateway
  inv bp_useDefaultConditionInExclusiveGateway:
    existsDefaultConditionInExclusiveGateway()
```

**Listing 116 – Use a *Default Condition* in an *Exclusive Gateway***

### 5.3.16. Use a *Default Condition* at an *Inclusive Gateway*

One way to ensure that a Process does not get stuck at an *Inclusive Gateway* is to use a default condition for one of the outgoing *Sequence Flow*. This *Default Sequence Flow* will always evaluate to true if all the other *Sequence Flow* conditions turn out to be false [71].

The best-practice rule regarding inclusive gateway can be enforced by attaching the following invariant to the *InclusiveGateway* element of the BPMN metamodel.

```
context InclusiveGateway
  inv bp_useDefaultConditionInInclusiveGateway:
    existsDefaultConditionInInclusiveGateway()
```

**Listing 117 – Use a *Default Condition* at an *Inclusive Gateway***

### 5.3.17. Use a *Default Condition* in an *Complex Gateway*

One way to ensure that the Process does not get stuck in a *Complex Gateway* is to use a default condition for one of the outgoing *Sequence Flow*. This *Default Sequence Flow* will always evaluate to true if all the other *Sequence Flow* conditions turn out to be false [71].

The best-practice rule regarding inclusive gateway can be enforced by attaching the following invariant to the *ComplexGateway* element of the BPMN metamodel.

```
context ComplexGateway
  inv bp_useDefaultConditionInComplexGateway:
    existsDefaultConditionInComplexGateway()
```

**Listing 118 – Use a *Default Condition* in an *Complex Gateway***

### 5.3.18. Match merging and splitting *Sequence Flow* in *Parallel Gateways*

The modeler must ensure that merging *Parallel Gateways* have the correct number of incoming *Sequence Flow*—especially when used in conjunction with other Gateways. As a guide, modelers should match merging and splitting *Parallel Gateways* (if the desired behavior is to merge them again) [71].

The best-practice rule regarding parallel gateway can be enforced by attaching the following invariant to the *ParallelGateway* element of the BPMN metamodel.

```
context ParallelGateway
  inv bp_matchMergingAndSplittingParallelGateway:
    isMergingAndSplittingMatchingParallelGateway()
```

**Listing 119 – Match merging and splitting *Sequence Flow* in *Parallel Gateways***

### 5.3.19. Match merging and splitting *Inclusive Gateways*

A way to avoid unexpected behavior of a process is to create models where a merging *Inclusive Gateway* follows a splitting *Inclusive Gateway* and that the number of *Sequence Flow*, between them, matches [71].

The best-practice rule regarding inclusive gateway can be enforced by attaching the following invariant to the *InclusiveGateway* element of the BPMN metamodel.

```
context InclusiveGateway
  inv bp_matchMergingAndSplittingInclusiveGateway:
    isMergingAndSplittingMatchingInclusiveGateway()
```

**Listing 120 – Match merging and splitting *Inclusive Gateways***

### 5.3.20. Use a *Gateway* as mediator when merging exclusive paths

Merging exclusive paths require a gateway as mediator, if an event follows. A gateway is also required if a split directly follows another gateway. A general rule is that *Gateways* are only required where *Sequence Flow* requires control [70].

The best-practice rule regarding gateway can be enforced by attaching the following invariant to the *FlowNode* element of the BPMN metamodel.

```
context FlowNode
  inv bp_exclusivePathsMergingIntermediateByGateway:
    isMediadorAGateway()
```

**Listing 121 – Use a *Gateway* as mediator when merging exclusive paths**

### 5.3.21. Simultaneous merging and splitting *Gateway* should be avoid

A *Gateway* with several inputs and several outputs at the same time may cause misunderstandings and should be avoided [70].

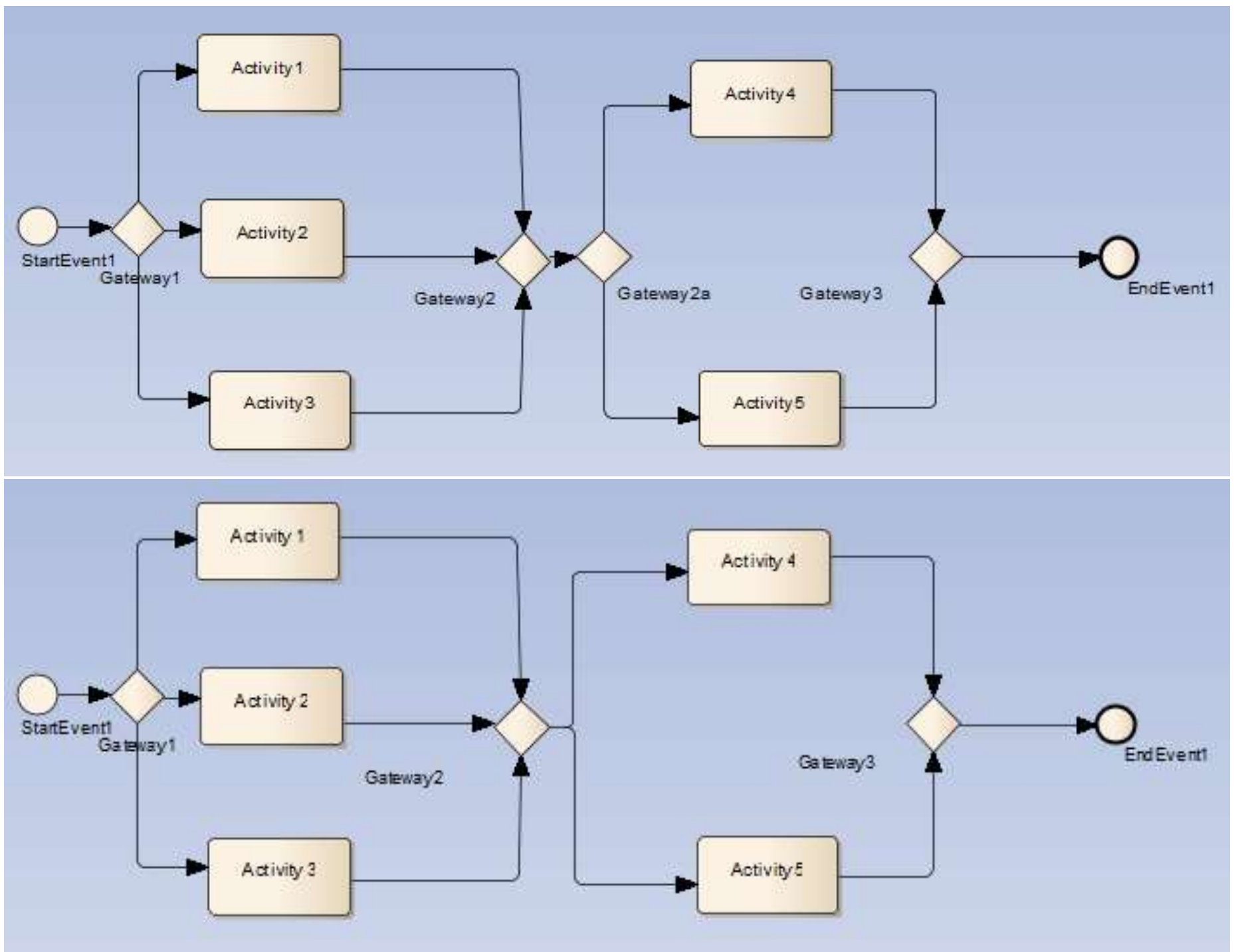

**Fig. 80.** – Simultaneous merging and splitting *Gateway* should be avoid

**Picture's interpretation**: Recommendation: Gateway2" and "Gateway2a" have distinct merging and splitting role; Avoid: "Gateway2" has simultaneous merging and splitting role.

The best-practice rule regarding gateway can be enforced by attaching the following invariant to the *FlowElementsContainer* element of the BPMN metamodel.

```
context FlowElementsContainer
 inv bp_gatewayWithSeveralInputsAndSeveralOutputs:
   noGatewayWithSeveralInputsAndSeveralOutputs()
```

**Listing 122 – Simultaneous merging and splitting *Gateway* should be avoid**

### 5.3.22. An *Exclusive Gateway* should have at most one unnamed outgoing *Sequence Flow*

The best-practice rule regarding *Gateways* can be enforced by attaching the following invariant to the *FlowElementsContainer* element of the BPMN metamodel [7].

```
context FlowElementsContainer
  inv bp_unnamedSequenceFlow:
    noMoreThanOneUnnamedSequenceFlow()
```

**Listing 123 – An *Exclusive Gateway* should have at most one unnamed outgoing *Sequence Flow***

### 5.3.23. An *Inclusive Gateway* should have all outgoing *Sequence Flow* named

The best-practice rule regarding gateway can be enforced by attaching the following invariant to the *InclusiveGateway* element of the BPMN metamodel [7].

```
context InclusiveGateway
  inv bp_noUnnamedSequenceFlow:
    notExistsUnnamedSequenceFlow()
```

**Listing 124 – An *Inclusive Gateway* should have all outgoing *Sequence Flow* named**

### 5.3.24. If a *SubProcess* is followed by a yes/no *Gateway*, at least one *End Event* of the *SubProcess* should be named to match the *Gateway* name

The best-practice rule regarding gateway can be enforced by attaching the following invariant to the *FlowElementsContainer* element of the BPMN metamodel [7].

```
context FlowElementsContainer
  inv bp_matchEndEventToGatewayName:
    matchEndEventAndGatewayName()
```

**Listing 125 – If a *SubProcess* is followed by a yes/no *Gateway*, at least one *End Event* of the *SubProcess* should be named to match the *Gateway* name**

**Activity**

### 5.3.25. *Activities* should be named

The best-practice rule regarding activities can be enforced by attaching the following invariant to the *FlowElementsContainer* element of the BPMN metamodel [7].

```
context FlowElementsContainer
  inv bp_activitiesMustBeName:
    isActivityNamed()
```

**Listing 126 – *Activities* should be named**

### 5.3.26. Two *Activities* inside the same *Process* should not have the same name

It is recommended that names be unique. A *Global Activity* should be used to reuse an activity in a process [7].
The best-practice rule regarding process can be enforced by attaching the following invariant to the Process element of the BPMN metamodel.

```
context Process
  inv bp_activitiesNameMustBeUnique:
    isActivityNameUnique()
```

**Listing 127 – Two *Activities* inside the same *Process* should not have the same name**

### 5.3.27. A *Send Task* should have an outgoing *Message Flow*

The best-practice rule regarding send task can be enforced by attaching the following invariant to the *SendTask* element of the BPMN metamodel [7].

```
context SendTask
  inv bp_hasOutgoingMessageFlow:
    self.hasOutputMessageFlows()
```

**Listing 128 – A *Send Task* should have an outgoing *Message Flow***

### 5.3.28. A *Receive Task* should have an incoming *Message Flow*

The best-practice rule regarding receive task can be enforced by attaching the following invariant to the *ReceiveTask* element of the BPMN metamodel [7].

```
context ReceiveTask
  inv bp_hasIncomingMessageFlow:
    self.hasInputMessageFlows()
```

**Listing 129 – A *Receive Task* should have an incoming *Message Flow***

### 5.3.29. If a *SubProcess* is followed by a *Gateway* labeled as a question, it must have more than one *End Event*

The best-practice rule regarding SubProcess can be enforced by attaching the following invariant to the *FlowElementsContainer* element of the BPMN metamodel [7].

```
context FlowElementsContainer
  inv bp_subProcessHasMoreThanOneEndEvent:
    hasMoreThanOneEndEvent()
```

**Listing 130 – If a *SubProcess* is followed by a *Gateway* labeled as a question, it must have more than one *End Event***

### Message Flow

#### 5.3.30. A *Message Flow* should be named with the name of the *Message*

The best-practice rule regarding message flow can be enforced by attaching the following invariant to the *MessageFlow* element of the BPMN metamodel [7].

```
context MessageFlow
  inv bp_hasMessageName:
    hasMessageName()
```

**Listing 131 – A *Message Flow* should be named with the name of the *Message***

### Artifact

#### 5.3.31. A *Text Annotation* should be connect by an *Association*

The well-formedness rule regarding text annotation can be enforced by attaching the following invariant to the *TextAnnotation* element of the BPMN metamodel.

```
context TextAnnotation
  inv bp_connectToAssociation:
  (self.incoming->notEmpty() or
  self.outgoing->notEmpty())
```

**Listing 132 – A *Text Annotation* should be connect by an *Association***

#### 5.3.32. An *Association* connected to a *Text Annotation* should have assigned none to the associationDirection property

The well-formedness rule regarding Text Annotation can be enforced by attaching the following invariant to the *Association* element of the BPMN metamodel.

```
context Association
  inv bp_connectedToTextAnnotationWithoutDirection:
  (self.targetRef.oclIsTypeOf(TextAnnotation) or
  self.sourceRef.oclIsTypeOf(TextAnnotation))
  implies
  self.associationDirection = AssociationDirection::None
```

**Listing 133 – An *Association* connected to a *Text Annotation* should have assigned none to the associationDirection property**

# 6. Rules verification

In this section we summarize the process by which, the rules proposed in previous section, were verified for correctness.

After having been accomplished the conversion of BPMN metamodel to the USE abstract syntax referred in section 4, the metamodel could be loaded by the USE tool environment. As mentioned before, this tool allows checking whether a set of objects and their links match the corresponding structural constraints model, namely regarding cardinality and type conformance, as well as OCL invariants concerning well-formedness rules.

Subsequently, we enriched the BPMN metamodel with the addition of well-formedness rules as OCL invariants (derived in section 5), translating the informally conveyed rules of the BPMN specification, and best practices collected from practitioners.

BPMN models snippets depicted in section 5, built with a CASE tool (Enterprise Architect) were exported in the USE concrete syntax (see [54, 73] for details). We were then able to instantiate the BPMN metamodel with BPMN models instances. By this time, syntactical errors were caught by the USE tool. Examples of such type of errors include among others, typeless instances and connections among elements not allowed in the metamodel, such as a *DataInputAssociation* linking two instances of *Task*, an instance of *MessageFlow* linking an instance of *Gateway* to an instance of *Task*.

In Fig. 81 we depicted the overall process model with the activities taking place to enhance the BPMN with the well-formedness rules. The JUSE-JUnit lane refers the role of a Java facade for the USE tool used for rules testing and debugging. After each rule was codified, added to the BPMN metamodel and syntactically validated (lanes *Researcher* and USE), a BPMN model snippet (test case) was build (lanes Enterprise Architect and EA2USE) for checking the correctness of the rule.

We elicited an overall of 121 invariants and implemented 610 operations, resulting in a total of 743 OCL expressions classified as follows:

- *Flow Control Well-formedness Rules*: rules related with the interaction among modeling elements;
- *Data Flow Well-formedness Rules*: rules related with sharing of data by activities;
- *Best-Practices Recommendations*: optional rules related with advised usage of BPMN elements in diagrams.

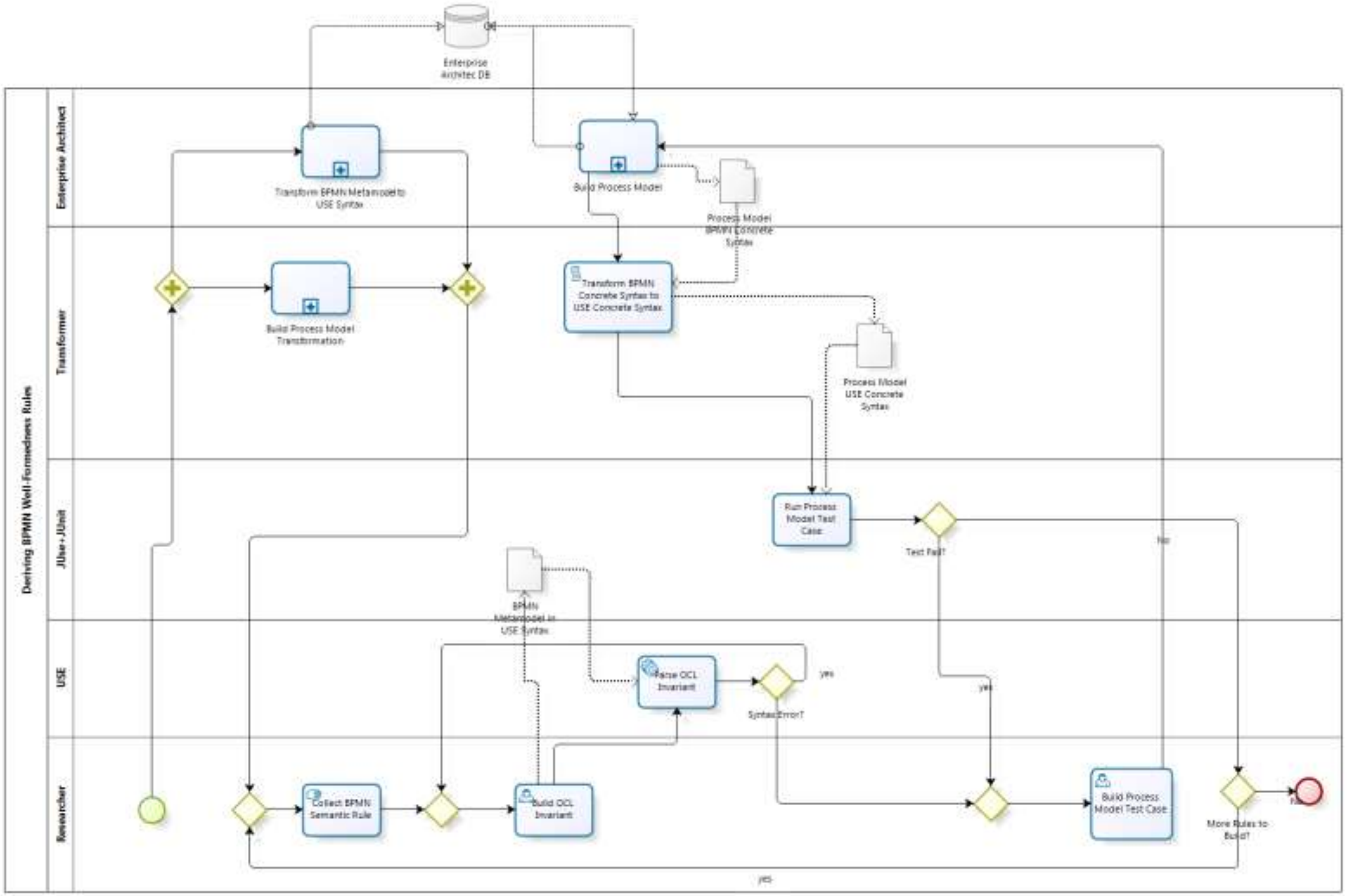

**Fig. 81.** – Building and verifying BPMN well-formedness rules using models snippets

# 7. Results' assessment

This technical report is concerned with rules formalization and the correct usage of BPMN diagrammatic rules by process modelers. To substantiate the relevance and empirical validate the presented work we conducted the two following surveys:

- On current BPMN tools – to ascertain the effectiveness of current BPMN tools to ensure the conformance of generated models with BPMN standard;
- On the correctness of produced BPMN models by users – to evaluate the users' knowledge of the BPMN standard and the ability to produced correct models.

## 7.1. Assessment of BPM tools

The set of tools from which the sample for assessment was chosen, came mainly from the OMG site[5]. From an initial list of 74 tools, a set of 11 units (15%) was chosen. In the selection process, we considered only process modeling tools. So we discarded graphical design tools, as well as those tools with main focus in processes' execution at runtime (BPMS). The selected convenience sample of BPMN tools, was a mix of proprietary and open source tools, according the criteria of popularity and maturity.

For testing the effectiveness of the sample of BPMN tools verifying the conformance of produced models with the BPMN standard, a faulty business process model (Fig. 82) was drew up in each of the tools. In Fig. 82 we can identified each of the violations, inserted on the business process model, by a label numbered near the BPMN element(s) where the violation occurs. The rules to be verified, from a total of 121 previously elicited (section 5), was a sample of 10 of the most used by every BPMN modeler in regular BPMN process modeling. Those 10 rules (8% of the total of rules) we chose a sample of 7 (8%), among the 89 well-formedness rules (described in sections 5.1 and 5.2) (see rules 1-7 in Table 1). The remaining were a sample of 3 (9%) of best-practices rules (see rules 8-10 in Table 1) advocated by BPMN practitioners and experienced users from a total of 32 (section 5.3). The data collect regarding caught violations by each of eleven BPMN tools are summarized in Table 2.

**Table 1** – Sample of Well-formedness Rules verified

| # | Source | Rule | Section |
|---|---|---|---|
| 1 | BPMN Standard | A top-level Process can only be instantiated by a restricted set of Start Event types. | 5.1.4 |
| 2 | | Outgoing Sequence Flow not allowed in an End Event. | 5.1.14 |
| 3 | | Outgoing Message Flow not allowed in a Catch Event. | 5.1.17 |
| 4 | | A Catch Event with incoming Message Flow must have Message or Multiple type. | 5.1.15 |
| 5 | | Explicit Start Event or End Event do not allow Activity or Gateway without incoming/outgoing Sequence Flow. | 5.1.27 |
| 6 | | A conditional Sequence Flow cannot be used if there is only one Sequence Flow out of the element. | 5.1.78 |
| 7 | | A Boundary Event must have exactly one outgoing Sequence Flow, unless it has the Compensation type. | 5.1.55 |
| 8 | Best Practices | Use a Timer intermediate event with an Event Gateway. | 5.3.6 |
| 9 | | Use a Default Condition at an Exclusive Gateway. | 5.3.16 |
| 10 | | Two activities in the same Process should not have the same name. | 5.3.26 |

[5] http://www.bpmn.org/ [accessed in Feb. 10, 2015]

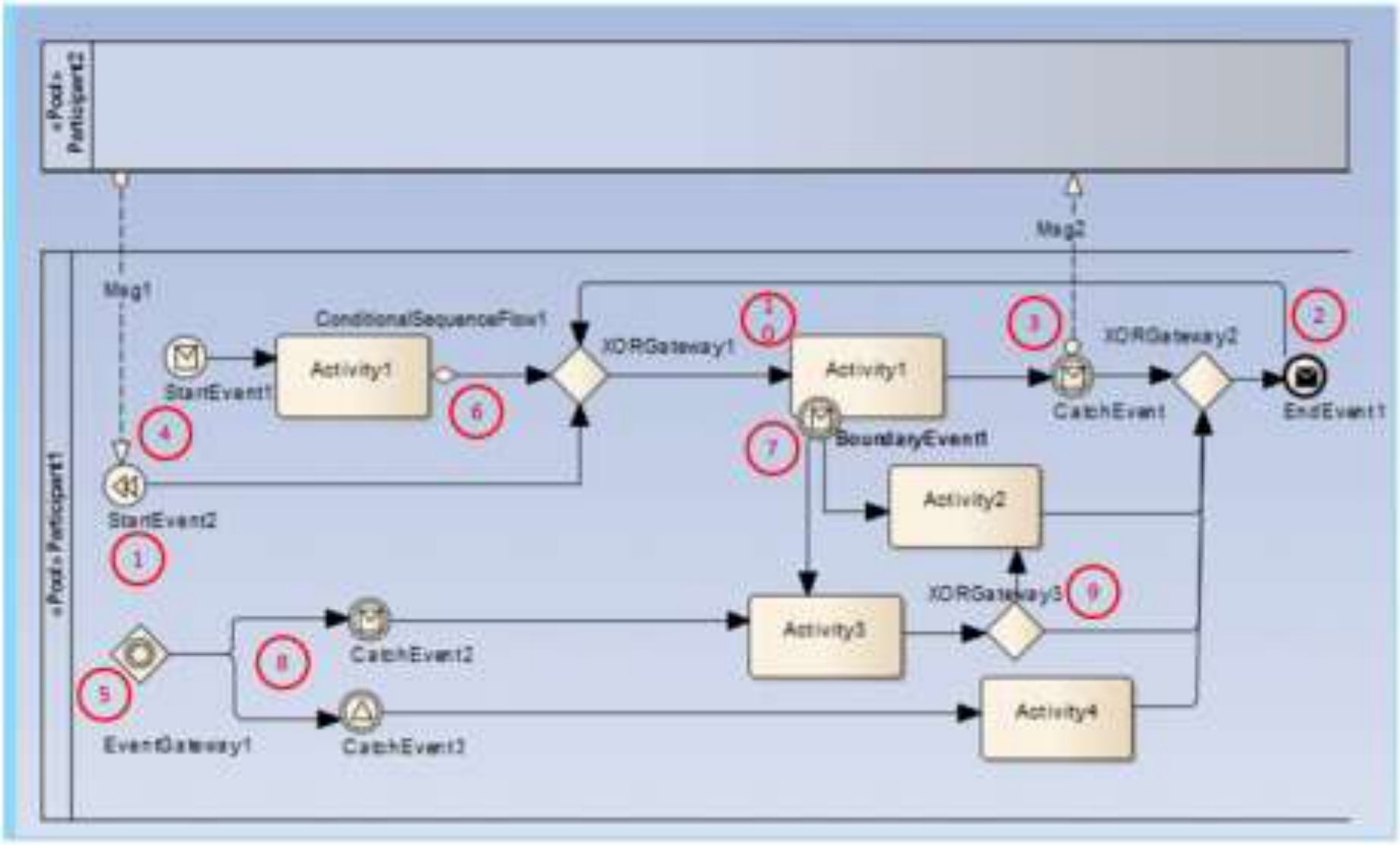

**Fig. 82.** – Model-snippet for assessment of the effectiveness of BPMN tools verification

**Table 2 –** Rules' violations detected on a model-snippet, by a sample of BPMN modeling tools

| Rule Id. / BPMN Tool | 1 | 2 | 3 | 4 | 5 | 6 | 7 | 8 | 9 | 10 | Std. # | Std. % | BP # | BP % | Total % |
|---|---|---|---|---|---|---|---|---|---|---|---|---|---|---|---|
| Adonis CE | x | x | x | | x | | x | | | x | 5 | 71% | 1 | 33% | 60% |
| Aris | | | | x | | | | | | | 1 | 14% | | 0% | 10% |
| Bizagi | x | x | | | | | | | | | 2 | 29% | | 0% | 20% |
| eClarus | | x | | | | | | | | | 1 | 14% | | 0% | 10% |
| EA | | | | | | | | | | | | 0% | | 0% | 0% |
| iGrafx | x | | x | x | x | | | | | | 4 | 57% | | 0% | 40% |
| MagicDraw | | x | | | | | | | | | 1 | 14% | | 0% | 10% |
| Modelio | x | x | x | | x | | | | x | | 4 | 57% | 1 | 33% | 50% |
| Signavio | | x | x | x | x | | x | | | | 5 | 71% | | 0% | 50% |
| TIBCO | x | x | x | x | | | x | | x | x | 5 | 71% | 2 | 66% | 70% |
| Visio+Modeler | x | x | x | x | | x | | | | | 5 | 71% | | 0% | 50% |

Given the simple model-snippet checked (Fig. 82), and the analysis of Table 2, one can conclude that none of the tools from the sample, fully detected the rules violations. The best score was attained by a unique tool that detected 70% (column Total %) of the violations inserted in the model-snippet. Regarding the rules prescribed by the BPMN standard the best score was given by tools that have detected 5 violations (71%) of the standard's rule violations. Concerning the violation of best-practices rules, the best score was given by a tool that detected 2 violations (66%) of the total of best-practices rules in the model-snippet.

The achieved results seems to corroborate the idea that there are limitations in current BPMN tools in terms of models' verification. This has a negative impact on the quality of produced BPMN process models as will be also verified in next section.

## 7.2. Assessment of BPM process models' correctness

The aim of the second empirical study [74] was to evaluate the correctness of produced BPMN process models by users. The *convenience sample* used was based upon available BPMN models stored in public repositories managed by two BPMN tool providers:

BizAgi[6] - a Business Process Management (BPM) solution provider, positioned in the 2010 Gartner's BPMS Magic Quadrant [75], which made available online 19 customizable templates of business process models;

Trisotech[7] - a provider of consulting services and BPM solutions, which runs an online resource repository, the Business Process Incubator, with almost 50 BPMN business process models collected from several sources.

[6] http://www.bizagi.com/

[7] http://www.businessprocessincubator.com/

We wanted to determine the conformance of the produced BPMN models with the BPMN specification, and by doing so evaluate the users' knowledge of the BPMN standard and their ability to produce correct models.

Table 3 summarizes the results attained by checking the publicly available BPMN models against the rules described in section 5. As can be seen, only 53.6% of the models were in conformance with specification rules that are part of the BPMN standard. Adding more strict requirements, by imposing the conformance of BPMN models with best-practices modeling rules, reduces to only 3.6% the percentage of models that comply with all kind of rules. The achieved results underline the effectiveness and importance of users having some sort of automatic support for checking BPMN models in order to attain correct results.

**Table 3 –** Number of Rule violations, by type, in BPMN Models

| # Errors | Standard Violation | Best-Practices Violation |
|---|---|---|
| 0 | 53,6% | 3,6% |
| 1 | 26,8% | 7,1% |
| 2 | 12,5% | 10,7% |
| 3 | 3,6% | 19,6% |
| 4 | 1,8% | 16,1% |
| 5 | 1,8% | 25,0% |
| 6 | | 7,1% |
| 7 | | 5,4% |
| 8 | | 3,6% |
| 9 | | 1,8% |

## 8. Conclusions and future work

An analysis was made to the recent evolution of the BPMN standard, as well as to the approaches for BPMN models validation.

OCL is a common practice in OMG specifications to express well-formedness rules. However, it is absent on the BPMN 2 specification. To overcome this limitation we have translated the informal (natural language) well-formedness rules presented in the current standard, as well as modeling best-practices, into a formalized format, based on OCL invariants upon the BPMN metamodel. We also depicted the rules usage through examples of correct and ill-formed process models.

The catalog of BPMN, proposed in this report, intends to contribute for reducing the complexity of BPMN, by systematizing set of rules for BPMN constructs' correct usage. These patterns can also be enforced as modeling best-practices easing the BPMN adoption by organizations. On the other hand, embedding the rules in BPMN tools, would reduce or even remove the discretionary use of constructs by process modelers.

In a survey on BPMN tools conformance to the standard, none of them complied with all formalized rules, an indicator that the BPMN tool market has still to technically evolve. Another survey on BPMN models available in public repositories allowed us to conclude that users have difficulty in produce correct models given the limitation of available tools.

At the moment we are working in a cloud-based validator of BPMN models, which includes the formalized rules in this document, to be made freely available.